\documentclass[a4paper,12pt]{article}

\usepackage{longtable}

\renewcommand{\theequation}{\arabic{equation}}
\usepackage{color}

\usepackage[T1]{fontenc}
\usepackage{amsmath}
\usepackage{graphicx}
\usepackage{titlesec}
\titleformat{\paragraph}
{\normalfont\normalsize\bfseries}{\theparagraph}{1em}{}
\titlespacing*{\paragraph}
{0pt}{3.25ex plus 1ex minus .2ex}{1.5ex plus .2ex}

\newcommand{\fett}[1]{\mbox{\boldmath$#1$}}
\usepackage{overcite}
\newcommand{\onlinecite}[1]{\hspace{-1 ex} \nocite{#1}\citenum{#1}}

\title{\vspace{-2cm}\bf Semiempirical Molecular Orbital Models based on 
the Neglect of Diatomic Differential Overlap Approximation}
\author{Tamara Husch, Alain C. Vaucher, and Markus Reiher\thanks{corresponding author: 
markus.reiher@phys.chem.ethz.ch.}
\vspace{10 mm}\\
ETH Z\"urich, Laboratorium f\"ur Physikalische Chemie,\\ Vladimir-Prelog-Weg 2, 8093 Z\"urich, Switzerland.
}

\begin{document}

\maketitle

\begin{center}
 \textbf{Abstract}
\end{center}

Semiempirical molecular orbital (SEMO) models
based on the neglect of diatomic differential overlap (NDDO) approximation efficiently solve the
self-consistent field equations by rather drastic approximations.
The computational efficiency comes at the cost of 
an error in the electron-electron
repulsion integrals. The error may be compensated by the 
introduction of parametric expressions to evaluate 
the electron-electron repulsion integrals, the one-electron integrals, 
and the core-core repulsion.
We review the resulting formalisms of popular NDDO-SEMO models
(such as the MNDO(/d), AM1, PM$x$, and OM$x$ models)
in a concise and self-contained manner.
We discuss the approaches to implicitly and 
explicitly describe electron correlation effects within NDDO-SEMO models 
and we dissect strengths and weaknesses of the different approaches 
in a detailed analysis. For this purpose, we consider the results of recent benchmark studies.
Furthermore, we apply bootstrapping to perform a sensitivity analysis 
for a selection of parameters in the MNDO model. 
We also identify systematic limitations of NDDO-SEMO models by drawing on an analogy 
to Kohn--Sham density functional theory. 

\newpage

\section{Introduction}
\label{sec:introduction}

The driving force for the development of semiempirical molecular orbital (SEMO) models 
has always been the desire to accelerate quantum chemical calculations.
At the outset of the development of SEMO models in the middle of the last century, 
\cite{Huckel1931,
Parr1952,Pople1953,Pariser1953a,
Pariser1953,Parr1960,Ohno1964,Pople1965,
Dewar1969,Pople1970}
the goal was to carry out electronic structure calculations 
for small molecules, which was not routinely possible with \textit{ab initio} 
electronic structure methods at that time.
Since then, theoretical chemistry has seen a remarkable development in terms of 
 computational resources, but also in terms of \textit{ab initio} methodology. \cite{Dykstra2005}
One must not forget that most electronic structure methods
which we apply routinely today, such as Kohn--Sham density functional theory (KS-DFT) \cite{Kohn1965}
and coupled cluster theory \cite{Cizek1966}, were developed concurrently with today's SEMO models. 
As a consequence of algorithmic and methodological developments, \cite{Dykstra2005}
accurate \textit{ab initio} electronic structure methods have long replaced SEMO models 
in their original areas of application (electronic structure calculations for small molecules).
Nevertheless, SEMO models did not become extinct.
Instead, they opened up different areas of application which can broadly be divided into 
three categories (see also Ref.~\onlinecite{Akimov2015} for a recent review):
(i) simulations of very large systems such as proteins \cite{Senn2006,Alexandrova2008a,
Senn2009a,Alexandrova2009,Stewart2009,Acevedo2010, Doron2011,Polyak2012} 
and those with thousands of small molecules, \cite{B.Gerber2014, Weber2015}
(ii) calculations for a large number of isolated and unrelated medium-sized molecules, e.g., 
 in virtual high-throughput screening schemes for materials discovery \cite{Husch2015a, Husch2015} and 
docking-and-scoring of potential drug candidates,
\cite{Lepsik2013,S.Brahmkshatriya2013, Yilmazer2015, Vorlova2015a,DuyguYilmazer2016,Sulimov2017}
and
(iii) entirely new applications such as real-time quantum chemistry where ultra-fast 
 SEMO models allow the perception of visual and haptic feedback in real time when manipulating 
medium-sized molecular structures. \cite{Marti2009,Haag2011,Haag2013,Haag2014,
Haag2014a,Vaucher2016a,Muhlbach2016,Heuer2018} 

In this work, we review, dissect, and analyze SEMO models 
which apply the neglect of diatomic differential overlap (NDDO) 
approximation \cite{Pople1965} (NDDO-SEMO models).
These models are currently among the most popular SEMO models \cite{Thiel2014} 
and all members of this class of 
SEMO models share the same conceptual framework.
Another class of semiempirical models, which is under continuous development, are 
tight-binding versions of KS-DFT. \cite{Elstner2000,Gaus2011,Seifert2012,Elstner2014,
Grimme2017,Bursch2017,Krishnapriyan2017} 
We will not discuss these density-functional 
tight-binding models as the focus of this work is on NDDO-SEMO models.

\begin{figure}[ht] 
 \centering
 \includegraphics[width=.9\textwidth]{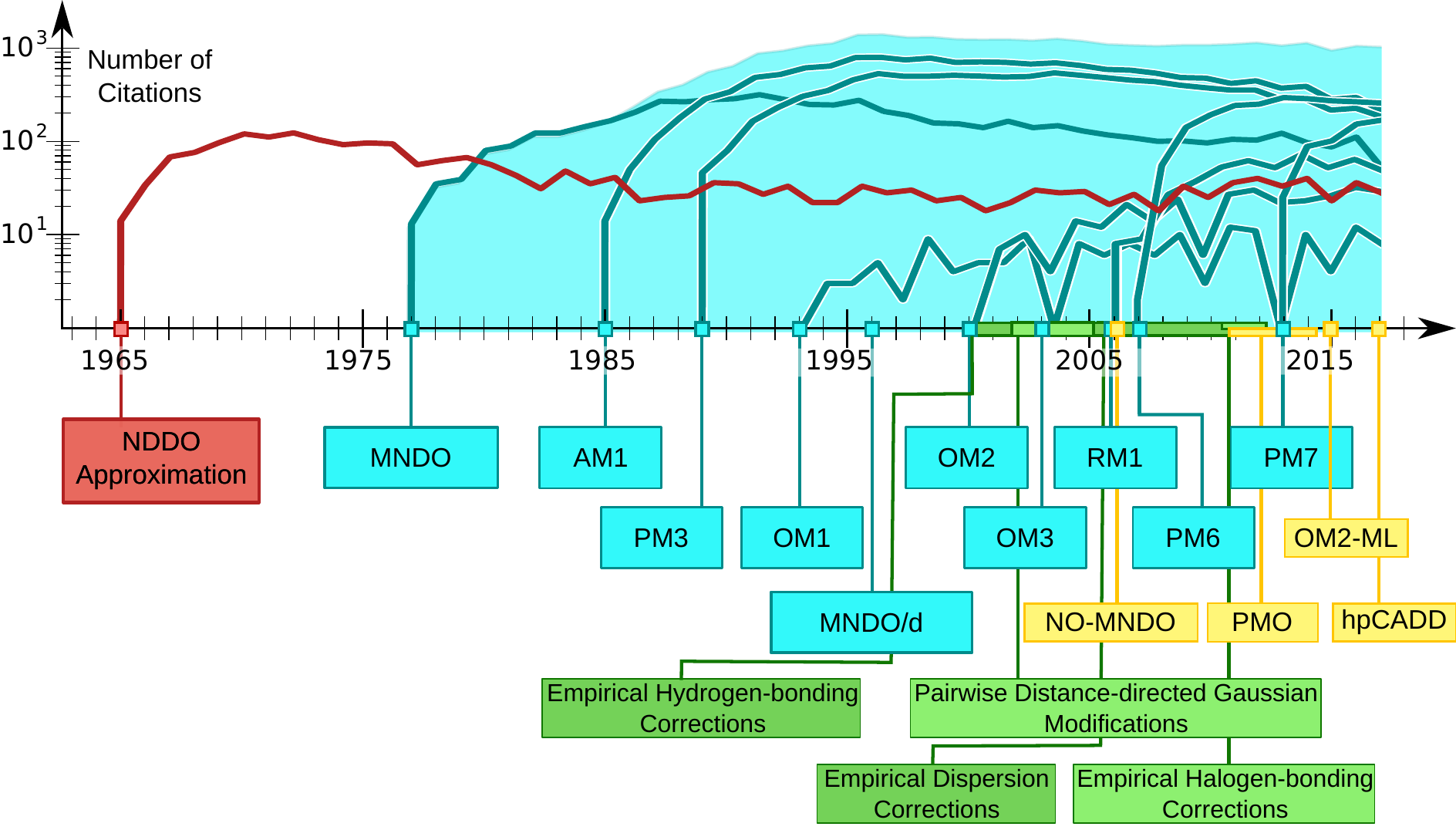}
 \caption{Chronology of key steps in the development of NDDO-SEMO models
starting with the introduction of the NDDO approximation in 1965 \cite{Pople1965} 
(highlighted in red)
and reaching until today. Popular
NDDO-SEMO models 
(MNDO\cite{Dewar1977}, AM1, \cite{Dewar1985} PM3, \cite{Stewart1989} MNDO/d, \cite{Thiel1991,Thiel1996c} 
RM1, \cite{Rocha2006} PM6, \cite{Stewart2007} PM7, \cite{Stewart2012} OM1, \cite{Kolb1993a} 
OM2, \cite{Weber2000a} and OM3 \cite{Dral2016b})
are highlighted in blue, 
novel suggestions 
(NO-MNDO, \cite{Sattelmeyer2006} PMO, 
\cite{Fiedler2011,Zhang2011,Zhang2012,Isegawa2013,Fiedler2014}
OM2-ML, \cite{Dral2015a}
and hpCADD \cite{Thomas2017})
in yellow, and semiclassical correction terms in green. 
The number of citations of the original publications 
\cite{Dewar1977,Dewar1985,Stewart1989,Thiel1991,Thiel1996c,
 Rocha2006,Stewart2007,Stewart2012,Kolb1993a,Weber2000a,Dral2016b} 
was assessed with Google Scholar. \cite{WOS} 
}
\label{fig:timeline}
\end{figure}

The central NDDO approximation drastically reduces 
the computational effort associated with the calculation of electron-electron 
repulsion integrals (ERIs), and hence, leads to a significant speed-up. \cite{Pople1965}
However, it took over ten years to successfully incorporate the NDDO approximation in a 
useful SEMO model (see Figure~\ref{fig:timeline}), 
the  \textit{Modified Neglect of Diatomic Overlap} (MNDO) model proposed in 1977. \cite{Dewar1976,Dewar1977} 

Since that time, small adjustments were made to the original MNDO model which 
gave rise to different closely related models such as 
AM1, \cite{Dewar1985} PM3, \cite{Stewart1989} MNDO/d, \cite{Thiel1991,Thiel1996c} 
RM1, \cite{Rocha2006} PM6, \cite{Stewart2007} and PM7. \cite{Stewart2012}
In the nineties of the last century, the development of another family of NDDO-SEMO models began, the 
\textit{Orthogonalization-Corrected Models} OM1, \cite{Kolb1993a} OM2, \cite{Weber2000a} and OM3.
\cite{Dral2016b}

Although the development efforts were consistently accompanied by 
articles, reviews, and books,    
\cite{Dewar1983,Dewar1992a,Thiel1988,Clark1993,Thiel1996a,Thiel1998,Clark2000,
Bredow2004,Lipkowitz2009,Clark2011a,Thiel2014,Jug2015,Lewars2016,Bredow2017}
there is no single resource which contains the detailed formalisms of the NDDO-SEMO models  
in a form which enables their facile implementation.
Currently, for example, the implementation of a modern NDDO-SEMO model such as PM7 requires 
the consultation of at least ten references 
\cite{Condon1959,Pelikan1974, Dewar1976,Dewar1977,Kumar1987, Glaeske1987,
Stewart1989,Thiel1991,Stewart2007,Stewart2012}.
Some of these references contain errors or misprints (which are clarified later on in this work) 
and some of them 
may be hard to obtain because they are books and dissertations written 
over forty years ago, some of them in German. 
Furthermore, the varying notation, adoption of jargon, and introduction of acronyms
 may hamper an in-depth understanding.
An in-depth understanding is, however, mandatory to be able to implement NDDO-SEMO models
following the original references.
As a concise and self-contained presentation of the formalism of these models 
is appropriate to understand their central ideas, we intend to provide such an overview in this work.

In recent years, a variety of semiclassical correction terms was designed to correct for 
specific flaws of NDDO-SEMO models, e.g., for hydrogen-bonding interactions, 
\cite{Rezac2009,Korth2010,Korth2010b, Rezac2012, Brahmkshatriya2013,Kromann2014a,Vorlova2015b}
dispersion interactions, \cite{Tuttle2008,Rezac2009,Rezac2012,Brahmkshatriya2013,Kromann2014a} 
halogen-bonding interactions, \cite{Rezac2011,Brahmkshatriya2013}
and other pairwise interactions. \cite{Repasky2002a,Tubert-Brohman2004}
The form of these terms is documented in a 
self-contained manner in the respective publications. The introduction of these terms does not explicitly
affect the electronic structure part of the models. 
Hence, we refer to a recent review by Christensen \textit{et al.} \cite{Christensen2016a} on this topic and we instead focus on the question of how NDDO-SEMO models 
attempt to 
approximate the electronic structure problem in this work.

This review is organized as follows:
We first briefly introduce the notation and the quantum chemical foundations necessary to discuss 
 NDDO-SEMO models which makes this review self-contained (Section~\ref{sec:basics}).
NDDO-SEMO models share, apart from the NDDO approximation, 
the application of a minimal basis set and the restriction of the number of explicitly considered 
electrons. We outline the formalism essential to all NDDO-SEMO models (Section~\ref{sec:formalism})
before moving on to specific NDDO-SEMO models (Sections~\ref{sec:mndo}--\ref{sec:othermethods}).
After discussing how NDDO-SEMO models are assembled, we discuss how 
(static and dynamical) electron correlation effects are captured.
Generally, there are two strategies to tackle this problem:
The calibration of parameters incorporated in NDDO-SEMO models 
against accurate reference data and the explicit description  
of electron correlation effects (Sections~\ref{subsec:param} and \ref{subsec:explicitcorr}).
We summarize the current state of knowledge with respect to both 
of these aspects and draw some more general 
conclusions for the prospects of NDDO-SEMO models (Section~\ref{sec:prospect}).

\section{Setting the Stage}
\label{sec:basics}

Electronic structure methods aim at the solution of the 
electronic Schr{\"odinger} equation, 
\begin{equation}
 \label{eq:schroedinger}
 \mathcal{H}_{\text{el}} \Psi_{\text{el}}^{\{\tilde{\fett{R}}_I\}}(\{\fett{r}_i\})=
E_{\text{el}}^{\{\tilde{\fett{R}}_I\}} 
 \Psi_{\text{el}}^{\{\tilde{\fett{R}}_I\}}(\{\fett{r}_i\}),
\end{equation}
which asserts that we can calculate the electronic energy 
$E_{\text{el}}^{\{\tilde{\fett{R}}_I\}}$ from the electronic wave function 
$\Psi_{\text{el}}^{\{\tilde{\fett{R}}_I\}}(\{\fett{r}_i\})$
by applying the electronic Hamiltonian operator $ \mathcal{H}_{\text{el}}$.
In the Born--Oppenheimer approximation, 
$E_{\text{el}}^{\{\tilde{\fett{R}}_I\}}$ and $\Psi_{\text{el}}^{\{\tilde{\fett{R}}_I\}}(\{\fett{r}_i\})$
depend parametrically on the fixed (indicated by the tilde and by giving them as superscripts) 
coordinates of the $N$ atomic nuclei 
($\{\tilde{\fett{R}}_I\}$) of a system. The electronic wave function  
$\Psi_{\text{el}}^{\{\tilde{\fett{R}}_I\}}(\{\fett{r}_i\})$ depends on the coordinates of 
 $n$ electrons ($\{\fett{r}_i\}$).
The Hamiltonian operator $\mathcal{H}_{\text{el}}$ contains operators for the $n$ 
kinetic energy contributions of the electrons
and for the electrostatic pair interaction energies of electrons and 
 the $N$ atomic nuclei (in Hartree atomic units (a.u.)),
\begin{equation}
 \label{eq:H}
\begin{split}
 \mathcal{H}_{\text{el}} =& \sum_{i=1}^n \left( -\frac{1}{2}\fett{\nabla}^2_i - 
 \sum_{I=1}^N \frac{Z_I}{|\fett{r}_i-\tilde{\fett{R}}_I|} \right) 
               + \sum_{i=1}^{n} \sum_{j>i}^n \frac{1}{|\fett{r}_i-\fett{r}_j|}  + \sum_{I=1}^N\sum_{J>I}^N 
                  \frac{Z_I Z_J}{|\tilde{\fett{R}}_I-\tilde{\fett{R}}_J|}.
\end{split}
\end{equation}
The gradient defined for the coordinates of electron $i$ is denoted as $\fett{\nabla}_i$ 
and  $Z_I$ denotes the nuclear charge of the $I$-th nucleus (note that capital letters 
denote quantities defined for atomic nuclei).
It is convenient to collect the first two terms in a one-electron operator $h_i$, 
the third term in a two-electron operator $g_{(i,j)}$, and the nuclear interaction 
energy in $V$,
\begin{equation}
 \label{eq:Hshort}
\mathcal{H}_{\text{el}} = \sum_{i=1}^n h_i + \sum_{i=1}^n\sum_{j>i}^n g_{(i,j)} + V.
\end{equation}

Since the early days of quantum mechanics, various approximations
 were developed to solve Eq.\ (\ref{eq:schroedinger}).
In the following, we focus on Hartree--Fock (HF) theory 
to lay the foundation for the discussion of NDDO-SEMO models 
(see, e.g., Ref.~\onlinecite{Szabo2012} for a detailed presentation of HF theory in 
a one-electron basis set).
The exact wave function $\Psi_{\text{el}}^{\{\tilde{\fett{R}}_I\}}(\{\fett{r}_i\})$ is approximated by 
the HF wave function $\Phi_{\text{el}}^{\text{HF},\{\tilde{\fett{R}}_I\}}(\{\fett{r}_i\})$,
which is constructed as the antisymmetrized product of one-particle functions 
$\psi_i^{\{\tilde{\fett{R}}_I\}}(\fett{r}_i)$ 
(i.e., molecular spin orbitals), 
\begin{equation}
 \Psi_{\text{el}}^{\{\tilde{\fett{R}}_I\}}(\{\fett{r}_i\}) \approx 
\Phi_{\text{el}}^{\text{HF},\{\tilde{\fett{R}}_I\}}(\{\fett{r}_i\})
= \mathcal{A}\ \prod_{i=1}^n
\psi_i^{\{\tilde{\fett{R}}_I\}}(\fett{r}).
\end{equation}
Antisymmetrization of the product of one particle-states by means of 
the antisymmetrization operator $\mathcal{A}$ implements the Pauli principle. \cite{Helgaker2014}
We approximate the spatial orbitals that enter the spin orbitals as  
linear combinations of $M$ atom-centered basis functions $\chi_\mu^I=\chi_\mu^{\tilde{\fett{R}_I}}(\fett{r})$ 
($\mu$-th basis function of type $\chi$ centered on atom $I$)
weighted with the expansion coefficients ${}^\chi\fett{C}^{\{\tilde{\fett{R}}_I\}}
=\{{}^\chi C^{\{\tilde{\fett{R}}_I\}}_{\mu i}\}$,
\begin{equation}
 \label{eq:bsexpansion}
 \psi_i^{\{\tilde{\fett{R}}_I\}}(\fett{r}) = \sum_{\mu=1}^M {}^\chi C^{\{\tilde{\fett{R}}_I\}}_{\mu i} 
\chi^{\fett{R}_I}_\mu(\fett{r}).
\end{equation}
For the sake of brevity, we drop the superscripts ${\{\tilde{\fett{R}}_I\}}$ and $\tilde{\fett{R}}_I$ in 
the following.
We require the $\chi$-basis to be \textit{locally orthogonal}, 
i.e., the overlap ${}^{\chi}\!{S}_{\mu\nu}=\left<\chi_{\mu}^I\middle|\chi_{\nu}^J\right>$ of 
different basis functions centered on the same atom must be 
zero,
\begin{equation}
\label{eq:locorth}
 \;{}^{\chi}\!{S}_{\mu\nu} = 
\begin{cases}
 \left<\chi_{\mu}^I\middle|\chi_{\nu}^J\right> & I\ne J,\ \forall \mu,\nu \\
 \delta_{\mu\nu} & I=J,\ \forall \mu,\nu \\
\end{cases},
\end{equation}
 in order to be able to apply the NDDO approximation. 
This is no general requirement for HF theory, but the introduction of another basis would 
 complicate the notation.
We will discuss this requirement in detail in Section~\ref{subsec:nddo}.

The   following   equations  are  given  
for  the spin-restricted formulation for the   sake   of   simplicity.
The central step underlying a canonical HF calculation in basis-set representation is then the iterative solution 
of the nonlinear Roothaan--Hall equation, 
\begin{equation}
 \label{eq:roothanhall}
 \;{}^{\chi}\!{\fett{F}} \;{}^{\chi}{\fett{C}} = 
\;{}^{\chi}\!{\fett{S}} \;{}^{\chi}{\fett{C}} \fett{\epsilon},
\end{equation}
for which we first need to calculate the Fock matrix ${}^{\chi}\!{\fett{F}}=
{}^{\chi}\!{\fett{F}}({}^{\chi}{\fett{C}})$ and the overlap matrix 
 ${}^{\chi}\!{\fett{S}}$. We obtain the 
matrix of basis set expansion coefficients ${}^{\chi}\!{\fett{C}}$ and the diagonal 
matrix of orbital energies $\fett{\epsilon}$ as the solution of this generalized eigenvalue equation.
The left superscript `$\chi$' continues to indicate that the calculations are carried out in the 
$\chi$-basis.
It is necessary to explicitly specify the basis because we will operate with different bases throughout 
this work. The matrix of orbital energies $\fett{\epsilon}$
is invariant under unitary matrix transformations by which one basis is transformed into 
another one. 
Consequently, $\fett{\epsilon}$ does not carry a superscript.

We can transform Eq.~(\ref{eq:roothanhall}) to read
\begin{equation}
 \begin{split}
 \;{}^{\chi}\!{\fett{S}}^{-\frac{1}{2}} \;{}^{\chi}\!{\fett{F}} 
\;{}^{\chi}\!{\fett{S}}^{-\frac{1}{2}} \;{}^{\chi}\!{\fett{S}}^{\frac{1}{2}} 
\;{}^{\chi}\!{\fett{C}} &= 
  \;{}^{\chi}\!{\fett{S}}^{\frac{1}{2}} \;{}^{\chi}\!{\fett{C}} \fett{\epsilon}, \\
 \end{split}
\end{equation}
which may be re-written in a simpler way as 
\begin{equation}
 \label{eq:roothanhallS}
 \begin{split}
 \;{}^{\phi}\!{\fett{F}} \;{}^{\phi}\!{\fett{C}} &= \;{}^{\phi}\!{\fett{C}} \fett{\epsilon}, \\
 \end{split}
\end{equation}
where 
\begin{equation}
 {}^{\phi}\!{\fett{F}}=\;{}^{\chi}\!{\fett{S}}^{-\frac{1}{2}} 
\;{}^{\chi}\!{\fett{F}} \;{}^{\chi}\!{\fett{S}}^{-\frac{1}{2}}
\end{equation}
 and 
\begin{equation}
 {}^{\phi}\!{\fett{C}}=\;{}^{\chi}\!{\fett{S}}^{\frac{1}{2}} 
\;{}^{\chi}\!{\fett{C}}.
\end{equation}
This constitutes a transformation of the Fock and coefficient matrices 
to the L\"owdin orthogonalized \cite{Lowdin1970} 
$\phi$-basis (indicated by a left superscript `$\phi$').
 The L\"owdin orthogonalized basis functions $\fett{\phi}=\{\phi_\mu\}$
and the locally orthogonal basis functions $\fett{\chi}=\{\chi^I_\mu\}$  are related through 
\begin{equation}
 \label{eq:oao}
  {\phi_\nu} = \sum_{\mu=1}^M (\;{}^{\chi}\!{S}^{-\frac{1}{2}})_{\mu\nu}\ 
\chi_\mu^I.
\end{equation}
Obviously, the basis functions ${\phi_\mu}$ are not centered 
on a single atom
and therefore do not carry a superscript `$I$'.
The solution of the Roothaan--Hall equations in either basis (Eqs.~(\ref{eq:roothanhall}) and 
(\ref{eq:roothanhallS})) requires a
calculation of one-electron integrals and of ERIs.
In this work, we employ Dirac's bra-ket notation
 for the one-electron integrals, 
\begin{equation}
\begin{split}
 & \left<\chi_\mu^I|  h|\chi_\nu^J\right> =  \int 
\chi_\mu^{*,I}(\fett{r}_1)
\left[ 
-\frac{1}{2}\fett{\nabla}^2_1 - 
 \sum_{I=1}^N \frac{Z_I}{|\fett{r}_1-\tilde{\fett{R}}_I|}
\right]
\chi_\nu^{J}(\fett{r}_1) d^3r_1
\end{split}
\end{equation}
and the ERIs,   
\begin{equation}
\begin{split}
  & \left<\chi_\mu^I\chi_\nu^J  |\chi_\lambda^K\chi_\sigma^L\right> =
 \int \int 
\chi_\mu^{*,I}(\fett{r}_1) 
\chi_\nu^{J}(\fett{r}_1)
\frac{1}{|\fett{r}_1-\fett{r}_2|}
\chi_\lambda^{*,K}(\fett{r}_2)
\chi_\sigma^{L}(\fett{r}_2) d^3r_1 d^3r_2,
\end{split} 
\end{equation}
in the $\chi$-basis. 
A Fock matrix element in the $\chi$-basis is then evaluated as,
\begin{equation}
\label{eq:fockmatrix_chi}
\begin{split}
 \;{}^{\chi}\!{F}_{\mu\nu} =& \left<\chi_\mu^I|h|\chi_\nu^J\right>
+ \sum_{\lambda=1}^M \sum_{\sigma=1}^M \;{}^{\chi}\!{P}_{\lambda\sigma}
\left[ \left<\chi_\mu^I\chi_\nu^J|\chi_\lambda^K\chi_\sigma^L\right>                   -\frac{1}{2} \left<\chi_\mu^I\chi_\sigma^L|\chi_\lambda^K\chi_\nu^J\right> \right], \\
\end{split}
\end{equation}
where the ERIs are contracted with elements of the density matrix  
${}^{\chi}\!{\fett{P}}$.
The elements of the density matrix in closed-shell systems are given by 
\begin{equation}
\begin{split}
 \;{}^{\chi}\!{P}_{\mu\nu} &= 2 \sum_{i=1}^{n/2}\;{}^{\chi}{C}_{\mu i} \;{}^{\chi}{C}_{\nu i}, \\
\end{split}
\end{equation}
(assuming real expansion coefficients).
The Fock matrix elements in the $\phi$-basis are assembled analogously, 
\begin{equation}
\label{eq:fockmatrix_phi}
\begin{split}
 \;{}^{\phi}\!{F}_{\mu\nu} =&  \left<\phi_\mu|h|\phi_\nu\right> 
+ \sum_{\lambda=1}^M \sum_{\sigma=1}^M \;{}^{\phi}\!{P}_{\lambda\sigma}
\left[ \left<\phi_\mu\phi_\nu|\phi_\lambda\phi_\sigma\right> 
                  -\frac{1}{2} \left<\phi_\mu\phi_\sigma|\phi_\lambda\phi_\nu\right> \right]. \\
\end{split}
\end{equation}
Eqs.~(\ref{eq:roothanhall}) and (\ref{eq:roothanhallS}) must be solved iteratively because 
the Fock matrix elements depend on the elements of the density matrix which is why 
the Roothaan--Hall equations are also known as the self-consistent field (SCF) equations.

For the following discussion, it is convenient to divide the Fock matrices into 
one-electron matrices $\fett{H}$ and two-electron matrices $\fett{G}$.
The two-electron matrices $\fett{G}$ can be further divided into the Coulomb matrices $\fett{J}$ and 
the exchange matrices $\fett{K}$, so that 
\begin{equation}
 \begin{split}
  {}^\chi\!{\fett{F}}&={}^\chi\!{\fett{H}}+{}^\chi\!{\fett{G}} ={}^\chi\!{\fett{H}}+{}^\chi\!{\fett{J}}
+{}^\chi\!{\fett{K}}, \\
 \end{split}
\end{equation}
and
\begin{equation}
 \begin{split}
  {}^\phi\!{\fett{F}}&={}^\phi\!{\fett{H}}+{}^\phi\!{\fett{G}} ={}^\phi\!{\fett{H}}+{}^\phi\!{\fett{J}}
+{}^\phi\!{\fett{K}}. \\
 \end{split}
\end{equation}

After reaching self consistency, the total electronic HF energy 
$E_{\text{el}}^{\{\tilde{\fett{R}}_I\},\text{HF}}= E_{\text{el}}^{\text{HF}}$
 is  calculated  from 
the resulting density matrices, Fock matrices, and the nucleus-nucleus repulsion  
energy,
\begin{equation}
\begin{split}
 E_{\text{el}}^{\text{HF}} &= 
\frac{1}{2} \sum_{\mu=1}^M \sum_{\nu=1}^M 
\;{}^{\chi}\!{P}_{\nu\mu} \left( {}^\chi\!{H}_{\mu\nu}+{}^\chi\!{F}_{\mu\nu} \right)  + V, \\
\end{split}
\end{equation}
and 
\begin{equation}
\begin{split}
  E_{\text{el}}^{\text{HF}} &= \frac{1}{2} \sum_{\mu=1}^M \sum_{\nu=1}^M 
\;{}^{\phi}\!{P}_{\nu\mu} \left( {}^\phi\!{H}_{\mu\nu}+{}^\phi\!{F}_{\mu\nu} \right)  + V. \\
\end{split}
\end{equation}

\section{General Considerations for the Formalism of NDDO-SEMO Models}
\label{sec:formalism}

All NDDO-SEMO models describe a way to efficiently 
approximate the Fock matrix.
Formally, the assembly of the Fock matrix in the course of the iterative solution 
of the SCF equations 
requires the calculation,
repeated processing, and (if possible) storage of $M^4$ ERIs.
Consequently, a lot of effort was put into the development of strategies to reduce the computational 
cost associated with this step. \cite{Helgaker2014}

\subsection{Neglect of Diatomic Differential Overlap}
\label{subsec:nddo}

\begin{figure}[ht] 
 \centering
  \includegraphics[width=\textwidth]{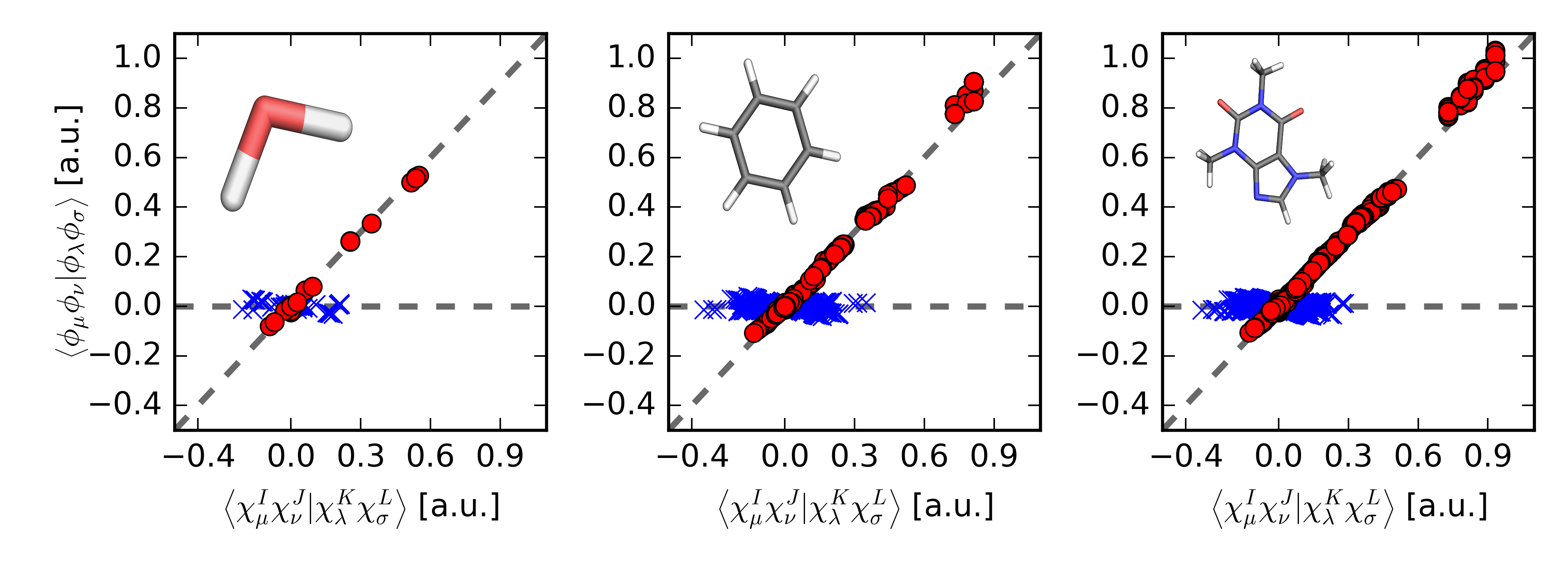}
 \caption{Comparison of the analytical values of 
$\left<\phi_\mu \phi_\nu | \phi_\lambda \phi_\sigma \right>$ 
and $\left<\chi_\mu^I \chi_\nu^J | \chi_\lambda^K \chi_\sigma^L\right>$ in a.u.\
for water (left), benzene (middle), and caffeine (right) in the OM2-3G basis set 
\cite{Stevens1984, Kolb1993a, Dral2016b}.
The gray dashed lines indicate $\left<\phi_\mu \phi_\nu | \phi_\lambda \phi_\sigma \right>=0.0$
and $\left<\phi_\mu \phi_\nu | \phi_\lambda \phi_\sigma \right>=
\left<\chi_\mu^I \chi_\nu^J | \chi_\lambda^K \chi_\sigma^L\right>$.
Red circles encode ERIs where $I=J$ and $K=L$ in the $\chi$-basis, and 
blue crosses encode ERIs where $I\ne J$ or $K\ne L$ in the $\chi$-basis. 
}
\label{fig:aovsoaonddo1}
\end{figure}

One of these strategies is the NDDO approximation \cite{Pople1965}
which drastically reduces the number of ERIs that must be calculated explicitly 
 to assemble ${}^{\phi}{\fett{G}}$.
The NDDO approximation,
\begin{equation}
\label{eq:nddo}
\left<\phi_\mu \phi_\nu | \phi_\lambda \phi_\sigma\right>
\approx \delta_{IJ} \delta_{KL} \left<\chi^I_\mu \chi_\nu^J | \chi_\lambda^K \chi_\sigma^L\right>,
\end{equation}
specifies how ERIs in the $\phi$-basis may be approximated based on the values of the 
respective ERIs in the $\chi$-basis. 
It is not immediately obvious why Eq.~(\ref{eq:nddo}) should hold true, especially 
 in view of Eq.~(\ref{eq:oao}), but numerical data suggest that there is some 
merit to the NDDO approximation. \cite{Fischer-Hjalmars1966,Cook1967,
Roby1969,Roby1971, Roby1972,Brown1971,Brown1973,Weinhold1988,Neymeyr1995a,Neymeyr1995b,Neymeyr1995c,
Neymeyr1995d,Neymeyr1995e,
Koch2014,Husch2018}
We additionally illustrate the NDDO approximation in Figure~\ref{fig:aovsoaonddo1} 
for three examples. 
Eq.~(\ref{eq:nddo}) asserts that $\left<\phi_\mu \phi_\nu | \phi_\lambda \phi_\sigma\right>$ is 
negligibly small if $I\ne J$ or $K\ne L$, i.e., if  $\chi_\mu^I$ and $\chi_\nu^J$ are centered on 
different atoms or if $\chi_\lambda^K$ and $\chi_\sigma^L$ are centered on different atoms.
As a consequence, the formal scaling of the ERI evaluation step is reduced from $\mathcal{O}(M^4)$ 
to $\mathcal{O}(M^2)$.
We see that this statement holds true 
for the three examples in Figure~\ref{fig:aovsoaonddo1}
because all blue crosses are located 
close to the horizontal dashed lines.
Furthermore, Eq.~(\ref{eq:nddo}) states that 
$\left<\phi_\mu \phi_\nu | \phi_\lambda \phi_\sigma\right>$ is  
approximately equal to $\left<\chi_\mu^I \chi_\nu^J | \chi_\lambda^K \chi_\sigma^L\right>$ 
when $I=J$ and $K=L$, 
e.g., the value of $\left<\phi_1\phi_1 | \phi_1 \phi_1\right>$ is 
approximately equal to the value of  
$\left<\chi^I_1\chi^I_1 | \chi^I_1 \chi^I_1\right>$. 
We see that this part of the NDDO approximation also holds true for the 
three examples that we considered for calculations in Figure~\ref{fig:aovsoaonddo1}
because the red circles are located close to the diagonal dashed lines.

Evidently, the NDDO approximation 
 \textit{emulates} a basis transformation for the ERIs, and hence, also for the two-electron matrix, i.e.,
\begin{equation}
\label{eq:gnddo}
 {}^{\chi}{\fett{G}}^{\text{NDDO}} \approx \;{}^{\phi}{\fett{G}}
\end{equation}
and 
\begin{equation}
 {}^{\chi}{\fett{G}}^{\text{NDDO}} \not\approx \;{}^{\chi}{\fett{G}}.
\end{equation}
This means that the matrix elements ${}^{\phi}{G}_{\mu\nu}$ can be 
approximately determined based on ERIs in the $\chi$-basis by 
\begin{equation}
\begin{split}
\;{}^{\phi}\!{G}_{\mu\nu} \approx& \;{}^{\chi}\!{G}^{\text{NDDO}}_{\mu\nu} \\ = &  
\sum_{\lambda=1}^M \sum_{\sigma=1}^M \;{}^{\chi}\!{P}_{\lambda\sigma}^{\text{NDDO}}
 \left( \delta_{IJ} \delta_{KL} \left<\chi_\mu^I \chi_\nu^J | \chi_\lambda^K \chi_\sigma^L\right>  -
 \delta_{IK} \delta_{JL}\; \frac{1}{2} \left<\chi_\mu^I \chi_\lambda^K  | \chi_\nu^J \chi_\sigma^L\right>\right). \\
\end{split}
\end{equation}

The NDDO approximation may also be formulated in the $\phi$-basis, 
\begin{equation}
\label{eq:nddophi}
\left<\phi_\mu \phi_\nu | \phi_\lambda \phi_\sigma\right>
\approx \delta_{IJ} \delta_{KL} \left<\phi_\mu \phi_\nu | \phi_\lambda \phi_\sigma\right>,
\end{equation}
which is, however, not very illuminating in the context of NDDO-SEMO models
as it is not obvious what the meaning of $I$, $J$, $K$, and $L$ in 
Eq.~(\ref{eq:nddophi})
actually is for the basis functions $\phi$ that 
are not centered on single atoms (see Eq.~(\ref{eq:locorth})).
Additionally, it is important to note that all NDDO-SEMO models 
calculate the ERIs in the $\chi$-basis and do not carry out 
an explicit basis transformation to the $\phi$-basis. 
In order to understand how NDDO-SEMO models work, we need to understand how this implicit basis 
transformation occurs, i.e., Eq.~(\ref{eq:nddo}). 

We emphasize that the NDDO statement is not that the 
ERIs in the $\chi$-basis are close to zero when $I\ne J$ or $K\ne L$. 
\textit{Only} their corresponding ERIs in the $\phi$-basis are approximately zero.
Figure~\ref{fig:aovsoaonddo1} also illustrates this statement as 
several ERIs for which $I\ne J$ or $K\ne L$ are as large as 0.3 a.u.\ in the $\chi$-basis. 
It is therefore misleading to formulate the NDDO approximation in the $\chi$-basis, 
\begin{equation}
\label{eq:nddofalse}
\left<\chi_\mu^I \chi_\nu^J | \chi_\lambda^K \chi_\sigma^L\right> 
\not\approx \delta_{IJ} \delta_{KL} \left<\chi_\mu^I \chi_\nu^J | \chi_\lambda^K \chi_\sigma^L\right>.
\end{equation}

The NDDO approximation leads to 
uncontrollable errors for the ERIs in the $\phi$-basis 
\cite{Fischer-Hjalmars1966,Cook1967,
Roby1969,Roby1971, Roby1972,Brown1971,Brown1973,Weinhold1988,Neymeyr1995a,Neymeyr1995b,Neymeyr1995c,
Neymeyr1995d,Neymeyr1995e,
Koch2014,Husch2018}
which propagate to all quantities based on these erroneous ERIs (most importantly to electronic energies).
\cite{Roby1969,Sustmann1969, Koster1972,Birner1974,Chandler1980,Duke1981,Neymeyr1995a,Neymeyr1995b,Neymeyr1995c,
Neymeyr1995d,Neymeyr1995e,Tu2003,Husch2018}
Most likely due to the uncontrollable errors and the lack of systematic improvability, 
the NDDO approximation has never found any use in \textit{ab initio} theories. 

In NDDO-SEMO models, the NDDO approximation is coupled to multiple other approximations,
not least to correct for the errors introduced by the NDDO approximation itself. \cite{Pople1965,
Dewar1977,Thiel1988,Kolb1993a,Weber2000a}
These additional approximations generally concern 
the calculation of the nonzero ERIs in the $\chi$-basis, the elements of the one-electron matrix 
in the $\phi$-basis, and $V$.
 Hence, we must specify for each NDDO-SEMO model which parametrized
expressions were applied to evaluate these three quantities. 
Throughout this work, we mostly adhered to the original parameter abbreviations. We present 
a comparison of the parameter abbreviations which we chose to the ones in the original literature 
in Tables~\ref{tab:parameter_names} and \ref{tab:parameter_names2} in 
Section~\ref{subsec:basicspecifications}. 

\subsection{Restriction to an Effective Valence Shell}
\label{subsec:effectiveH}

NDDO-SEMO models further reduce the computational effort by 
restricting the number of explicitly considered electrons.
When restricting the number of explicitly considered electrons, it is  necessary to 
specify for each atom $I$ 
which of its $n_I$ electrons are considered valence (`$v$') electrons ($n_{v,I}$)
and which ones are considered core (`$c$') electrons ($n_{c,I}=n_I-n_{v,I}$).
Accordingly, each atomic core then exhibits a core charge $Q_{I}$,
\begin{equation}
 Q_{I}=Z_{I}-n_{c,I}.
\end{equation}
Note that a rigorous distinction between core charge $Q_I$ and nuclear charge 
$Z_I$ is crucial for NDDO-SEMO models. Both quantities are required in 
parametric expressions in the formalism of some NDDO-SEMO models 
(see, e.g., Eq.~(\ref{eq:pm6ecc}) in Section~\ref{subsubsec:pmx}).
No rigorous method exists to justify a specific choice for $n_v=\sum_I^N n_{v,I}$ and 
$n_c=\sum_I^N n_{c,I}$. 
Within the NDDO-SEMO models, $n_{v,I}$ is restricted drastically 
so that $n_{v,I}=1$, $n_{v,I}=4$, $n_{v,I}=5$, and $n_{v,I}=6$ 
for hydrogen, carbon, nitrogen, and oxygen atoms, respectively.
\cite{Dewar1977,Stewart2007,Stewart2012,Dral2016b}.
In fact, $n_{v,I} \le 12$ for all elements, and in principle,
no more than two $s$ electrons, six $p$ electrons, and ten $d$ electrons 
could be considered per atom within the current formalism of the popular 
NDDO-SEMO models. \cite{Dewar1977,Stewart2007,Stewart2012,Dral2016b}
We specify $n_{v,I}$ for all elements up to $Z_I=83$ in Table~\ref{tab:nv} 
in Section~\ref{subsec:basicspecifications}.
We note that usually it is considered better to avoid too large $n_{c,I}$ in \textit{ab initio}
calculations (see, e.g., Ref.~\onlinecite{Cao2011a} in which it was shown that a choice of $n_{v,I}=4$
for lead and for titanium (as in PM6 and PM7) 
is not adequate to yield accurate electronic energies). 

In \textit{ab initio} theories, one may approximate the effects of the core 
electrons by an effective core potential (ECP), \cite{Cao2011a,
Dolg2012a} so that 
for $n_v$ valence electrons,
the full Hamiltonian (Eq.~(\ref{eq:H})) is replaced by an approximate 
 valence-only Hamiltonian $\mathcal{H}_{\text{el},v}$,
\begin{equation}
 \label{eq:vsH}
\begin{split}
 \mathcal{H}_{\text{el},v} =& \sum_{i=1}^{n_v} \left( -\frac{1}{2}\fett{\nabla}^2_i + 
\sum_I^N \left(-\frac{Q_I}{|\fett{r}_i-\tilde{\fett{R}}_I|} + 
\text{ECP}_I\right) \right)  \\ & 
               + \sum_{i=1}^{n_v} \sum_{j>i}^{n_v} \frac{1}{|\fett{r}_i-\fett{r}_j|} + 
\sum_{I=1}^N\sum_{J>I}^N \frac{Q_I Q_J}{|\tilde{\fett{R}}_I-\tilde{\fett{R}}_J|}, \\
 \end{split}
\end{equation}
or in short notation,
\begin{equation}
 \label{eq:vsH_short}
\begin{split}
 \mathcal{H}_{\text{el},v} &= \sum_{i=1}^{n_v} h_{v,i} +  \sum_{i=1}^{n_v}\sum_{j>i}^{n_v} g_{(i,j)} + 
  V_{v}. \\
 \end{split}
\end{equation}
The core-core repulsion energy is denoted as $V_v$ and 
the effective one-electron operator is denoted as $h_{v,i}$. The 
effective one-electron operator $h_{v,i}$ incorporates an 
effective core potential describing the interaction with the core electrons 
of atom $I$, $\text{ECP}_I$. 
We will not specify the different functional forms for the ECPs which are applied in 
\textit{ab initio} theories here (see, e.g., Refs.~\onlinecite{Cao2011a,Dolg2012a} for recent reviews)
because they differ strongly from the ones which are applied in NDDO-SEMO models today.\cite{Zerner1972,Ridley1973,
Freed1983a,Freed1995}
We will discuss the form of $\text{ECP}_I$ for each individual NDDO-SEMO model below. 

\subsection{Restriction of the Basis Set Expansion} 
\label{subsec:minbas}

Another way of reducing the computational effort is the restriction of 
the number of basis functions $M$ in Eq.~(\ref{eq:bsexpansion}).
Generally, the number of basis function which are activated for an atom $I$, $M_I$, is 
less than or equal to nine. For each atom, at most, one $s$-type, three $p$-type, and five $d$-type 
basis functions may be considered, i.e., the number of basis functions activated for 
a molecule is  
\begin{equation}
 M=\sum_{I=1}^N M_I \le 9 N.
\end{equation}
All NDDO-SEMO models apply only one $s$-type basis function for hydrogen.
For carbon, nitrogen, and oxygen, the NDDO-SEMO models activate one $s$-type basis function and 
three $p$-type basis function.
The basis functions which are considered for an atom of a certain element type are given in 
Table~\ref{tab:nv} 
in Section~\ref{subsec:basicspecifications}.

The application of such a minimal valence-shell basis set has the 
 practical advantage that the basis functions are inherently locally orthogonal 
(Eq.~(\ref{eq:locorth})), i.e., the NDDO approximation is straightforwardly applicable. 
\cite{Chandler1980,Duke1981,Neymeyr1995a,Neymeyr1995b,Neymeyr1995c,
Neymeyr1995d,Neymeyr1995e,Husch2018}
A minimal basis set is, however, generally unsuitable for the description of atoms in
molecules. Conceptually, molecules are characterized by interacting atoms
which polarize each other through additional external fields 
exerted by electrons and nuclei of the other atoms 
that distort the
spherical symmetry of an atom. 
The description of the polarization of the electron density requires basis functions 
with a higher angular momentum. 
Consequently, it is common knowledge that calculations with a 
minimal basis sets do not yield reliable relative energies, force constants, 
electric dipole moments, static dipole polarizabilities, and other properties. 
\cite{Kolos1979,Francl1982,Davidson1986,
Giese2005,Li2014} 
More specifically, it would, for example, not be possible to obtain satisfactory results 
for polarizabilities or for the description of non-covalent interactions 
at the full-configuration-interaction (FCI) limit when applying a minimal basis set. \cite{Helgaker2014}
A minimal basis set may, however, be sufficient to predict 
reasonable molecular equilibrium structures \cite{Pople1977,Kolos1979,Francl1982,Davidson1986}
which led to a re-consideration of minimal-basis-set HF  
as a quick preliminary structure optimization method in recent years \cite{Kulik2012, Sure2013}.
Hence, the application of minimal basis sets may be adequate if the area of application of the 
NDDO-SEMO model is restricted accordingly.

\section{The Modified Neglect of Diatomic Overlap (MNDO) Model}
\label{sec:mndo}

The MNDO model is the first successful SEMO model which is based on the NDDO approximation. \cite{Dewar1977,
Dewar1977b,Dewar1976}
The MNDO model activates one $s$- and three $p$-type basis functions ($\chi$-basis) for 
carbon, nitrogen, oxygen, and fluorine. \cite{Dewar1977}
Only one $s$-type basis function is retained for hydrogen. \cite{Dewar1977}
The MNDO model was later on extended to a larger part of the 
periodic table (see, e.g, Ref.~\onlinecite{Stewart2004}
and also Section~\ref{subsubsec:mndod}).
The Slater-type basis functions incorporate a parameter, the 
exponent $\zeta^{Z_I}$, which depends on the element type (here indicated by the 
superscript atomic number $Z_I$) of the atom $I$ on which the  
basis function $\chi$ is centered.
The same $\zeta^{Z_I}$ is applied for $s$- and $p$-type basis functions, i.e., 
$\zeta^{Z_I}_s=\zeta^{Z_I}_p$
. \cite{Dewar1977}
We explicitly indicate for each parameter 
its dependencies (e.g., $Z_I$ of the atom $I$ on which a $\chi$-basis function is centered or the 
orbital-type of the basis function $\chi$, i.e., $s$, $p$, or more generally, the angular momentum of the $\mu$-th $\chi$-basis function $l(\mu)$).

\subsection{Parametrization of One-Center ERIs}
\label{subsec:mndo1ceri}

At most, one $s$-type basis function and three $p$-type basis functions are centered 
on one atom.
Altogether, six unique nonzero one-center ERIs 
$\left<\chi_\mu^I \chi_\nu^I | \chi_\lambda^I \chi_\sigma^I\right>$ may
arise.
Five of the six ERIs are substituted for element- and orbital-type-dependent parameters,
\begin{equation}
\label{eq:gss}
   \gamma_{ss}^{Z_I} = \left<s^Is^I\middle|s^Is^I\right>,
\end{equation}
\begin{equation}
\label{eq:gpp}
   \gamma_{pp}^{Z_I} = \left<p^Ip^I\middle|p^Ip^I\right>,
\end{equation}
\begin{equation}
\label{eq:gsp}
   \gamma_{sp}^{Z_I} = \left<s^Is^I\middle|p^Ip^I\right>,
\end{equation}
\begin{equation}
\label{eq:gp2}
   \gamma_{pp'}^{Z_I} = \left<p^Ip^I\middle|p'^Ip'^I\right>,
\end{equation}
and
\begin{equation}
\label{eq:gsp2}
   \tilde{\gamma}_{sp}^{Z_I} = \left<s^Ip^I\middle|s^Ip^I\right>, 
\end{equation}
where we only indicate the orbital type ($s$ or $p$) of the basis functions when specifying the 
ERIs.
The magnetic quantum numbers do not need to be explicitly specified
because otherwise rotational invariance could not be guaranteed. \cite{Nanda1977} 
Hence, it is only necessary to indicate 
for the $p$-type basis functions whether the magnetic quantum number is the same (no prime, e.g., 
$\left<p_x^Ip_x^I\middle|p_x^Ip_x^I\right> =$ $\left<p^Ip^I\middle|p^Ip^I\right>$) or different (prime, e.g.,
$\left<p_x^Ip_x^I\middle|p_y^Ip_y^I\right> =$ \\ $\left<p^Ip^I\middle|p'^Ip'^I\right>$). 
The sixth one-center ERI is calculated from $\gamma_{pp}^{Z_I}$ and $\gamma_{pp'}^{Z_I}$,
\begin{equation}
\label{eq:hpp}
 \left<p^Ip'^I\middle|p^Ip'^I\right> = \frac{1}{2} 
\left(\gamma_{pp}^{Z_I} - \gamma_{pp'}^{Z_I}\right).
\end{equation}
If $\gamma_{pp}^{Z_I} < \gamma_{pp'}^{Z_I}$, $\left<p^Ip'^I\middle|p^Ip'^I\right>$ will 
be negative which leads to issues in the determination of the two-center ERIs
(see also Section~\ref{subsec:param1ceri}). 
In the original MNDO model, $\left<p^Ip'^I\middle|p^Ip'^I\right>$ is always positive so 
that no issues arise, 
but $\gamma_{pp}^{Z_I}$ may be smaller than 
$\gamma_{pp'}^{Z_I}$ for extensions of the MNDO model. \cite{Stewart2004,Stewart2007,Stewart2012}
All other one-center ERIs in the minimal valence-shell basis (e.g., 
$\left<s^Ip^I|s^Is^I\right>$,
 $\left<s^Ip^I|p^Ip^I\right>$, $\left<s^Ip^I|p^Ip'^I\right>$, or 
$\left<p^Ip^I|p^Ip'^I\right>$) are exactly zero.

\subsection{Approximating the Two-Center ERIs}
\label{subsec:mndo2ceri}
\begin{figure}[ht]
 \centering
 \includegraphics[width=.9\textwidth]{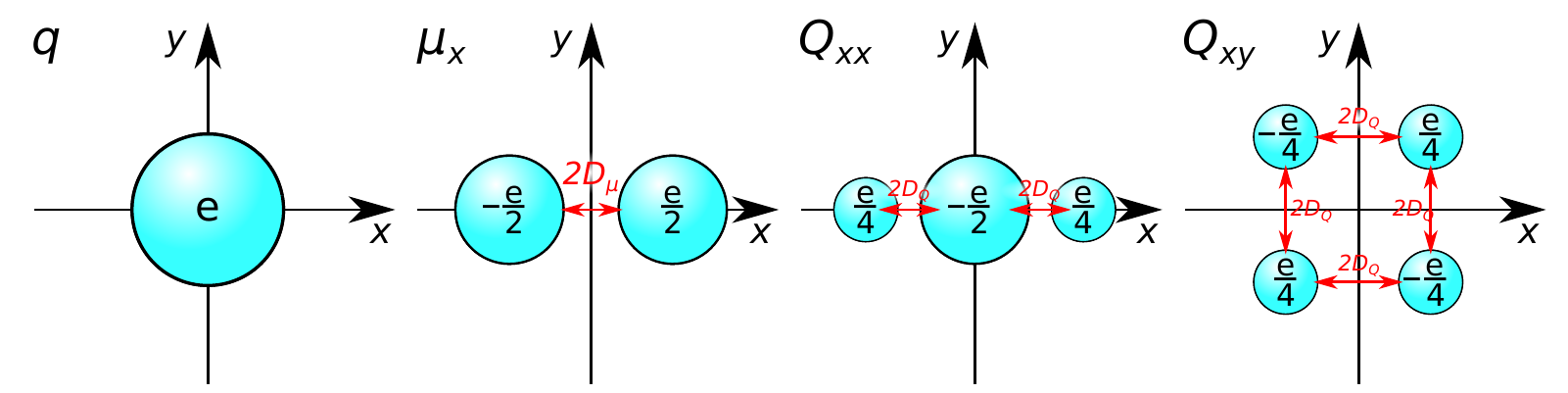}
 \caption{Illustration of the configuration of point charges (blue spheres) for the monopole $q$, 
the dipole $\mu_x$, the linear quadrupole $Q_{xx}$, and the square quadrupole $Q_{xy}$.
The charge for each point charge is given in units of the 
elementary charge. The point charges in $\mu_x$
are $2D_\mu$ apart, the ones in $Q_{xx}$ $2D_Q$, and the ones in $Q_{xy}$ $2D_Q$.}
\label{fig:pointChargeConfigurations}
\end{figure}

The two-center ERIs
$\left<\chi_\mu^I \chi_\nu^I | \chi_\lambda^J \chi_\sigma^J\right>,\ I\ne J$, can be interpreted as the 
electrostatic interaction between a charge distribution $\chi_\mu^I \chi_\nu^I$ centered on atom $I$ and 
a charge distribution $\chi_\lambda^J \chi_\sigma^J$ centered on atom $J$.
Each possible charge distribution $\chi_\mu^I \chi_\nu^I$ in the $s$, $p$ minimal 
valence-shell basis is approximately represented as a truncated classical multipole expansion 
\cite{Dewar1976, Dewar1977, Thiel1991, Thiel1996c}  
(see Ref.~\onlinecite{Dewar1976} or Section~\ref{subsec:sp2ceris}  
for explicit formulae).
Dewar and Thiel decided to specify four individual arrangements of discrete point charges for this purpose; 
a monopole $q$, a dipole $\mu_{x,y,z}$, a linear quadrupole $Q_{xx,yy,zz}$, and a
square quadrupole $Q_{xy,xz,yz}$ (see Figure~\ref{fig:pointChargeConfigurations}).
The distances $D_\mu$ and $D_Q$ between the point charges in each configuration were chosen such that 
the multipole moment of each point charge configuration approximated
 the one of the corresponding charge distribution which is ensured by 
calculating $D_\mu$ and $D_Q$ based on $\zeta^{Z_I}$
(see Ref.~\onlinecite{Dewar1976} or Section~\ref{subsec:sp2ceris}). 
As soon as the positions of the point charges in space have been specified (by defining $D_\mu$ and $D_Q$
and the local arrangements around the atomic nuclei), 
one can readily calculate the electrostatic potential energy as a sum 
over all possible electrostatic interactions of the point charges.
The approximation of a two-center ERI as the electrostatic interaction of 
discrete point charges will, however, break 
down when the distance between the atomic nuclei,
\begin{equation}
 \tilde{R}_{IJ} = |\tilde{\fett{R}}_I-\tilde{\fett{R}}_J|,
\end{equation}
 becomes small.
\begin{figure}[ht]
 \centering
 \includegraphics[width=.45\textwidth]{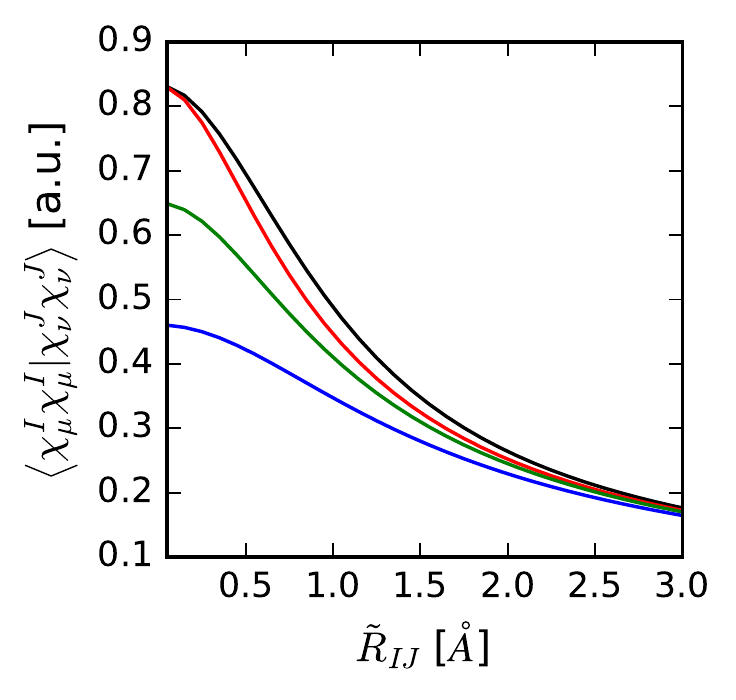}
 \caption{Dependence of $\left<\chi_\mu^I \chi_\mu^I | \chi_\nu^J \chi_\nu^J\right>$ (black line) and 
$\left[\chi_\mu^I \chi_\mu^I | \chi_\nu^J \chi_\nu^J\right]$ on 
$\tilde{R}_{IJ}=|\tilde{\fett{R}}_{\text{H}_1}-\tilde{\fett{R}}_{\text{H}_2}|$ in 
an H$_2$ molecule described by an MNDO-3G basis. 
We applied a Gaussian-type 
basis set (denoted MNDO-3G) to calculate the 
analytical ERI values instead of the Slater-type basis set inherent to the MNDO model. 
We generated the MNDO-3G basis based on $\zeta^{Z_I}$ \cite{Dewar1977}.
We chose different values for $\gamma_{ss}^{1}$ in Eq.~(\ref{eq:klopman}): 
$\gamma_{ss}^1$=0.83 a.u.\ (analytical one-center ERI, red line), 
$\gamma_{ss}^1$=0.65 a.u.\ (green line), and $\gamma_{ss}^1$=0.45 a.u.\ (MNDO value \cite{Dewar1972},
blue line).
}
\label{fig:semiemiricalTwoElectronIntegrals}
\end{figure}
We illustrate this at the example of the two-center ERI 
$\left<s^Is^I|s^Js^J\right>$ which is approximated by 
the electrostatic interaction of a point charge located on $I$ and 
one located on $J$.
If $\tilde{R}_{IJ}=0$, a singularity would arise for the electrostatic interaction between 
the two point charges.
This would not be the case if we considered the electrostatic interaction 
of a charge distribution centered on $I$ and one centered on $J$.
As a consequence, the expression to calculate the electrostatic interaction between point charges is 
modified in an empirical manner 
in such a way that it yields the \textit{one-center} ERIs 
(Eqs.~(\ref{eq:gss})--(\ref{eq:hpp})) in the limit $\tilde{R}_{IJ}=0$,
(`Klopman formula', Eq.~(14) in Ref.~\onlinecite{Klopman1964a}),
\begin{equation}
\label{eq:klopman}
\begin{split}
\left<s^Is^I\middle|s^Js^J\right>  \approx & 
 \left[s^Is^I\middle|s^Js^J\right] = 
  \left[ \tilde{R}_{IJ}^2 + 
\left(\frac{1}{2 \gamma_{ss}^{Z_I}} 
+  \frac{1}{2 \gamma_{ss}^{Z_J}} \right)^2 \right]^{-\frac{1}{2}}.
\end{split}
\end{equation}
In the following, we denote an ERI calculated in the Klopman approximation 
by square brackets to easily distinguish them from analytically calculated ERIs 
(denoted in angle brackets).
From Eq.~(\ref{eq:klopman}) and Figure~\ref{fig:semiemiricalTwoElectronIntegrals},
 we see that for large $\tilde{R}_{IJ}$
(and constant $ \gamma_{ss}^{Z_I}$ and $\gamma_{ss}^{Z_J}$),
the term approaches the regular Coulomb interaction of two elementary point charges,
$\left[s^Is^I\middle|s^Js^J\right] \approx 1 / \tilde{R}_{IJ} $.
In the limit $\tilde{R}_{IJ}=0$ (where $\gamma_{ss}^{Z_I}
= \gamma_{ss}^{Z_J}$), 
the expression reduces to 
$\left[s^Is^I\middle|s^Is^I\right] = \gamma_{ss}^{Z_I}$.
The value of $\gamma_{ss}^{Z_I}$ determines how closely 
the approximate ERI follows the analytical one (see also Figure~\ref{fig:semiemiricalTwoElectronIntegrals}).
Usually $\gamma_{ss}^{Z_I}$ is chosen such that the semiempirical two-center ERIs are smaller than
the analytical values. E.g., Pariser and Parr \cite{Pariser1953}, 
Dewar and Klopman \cite{Dewar1967}, and 
Voigt \cite{Voigt1973} argued that in this way dynamic electron correlation effects 
can be emulated. We will analyze this claim in detail in Section~\ref{subsec:param}.

\begin{figure}[ht]
 \centering
 \includegraphics[width=.5\textwidth]{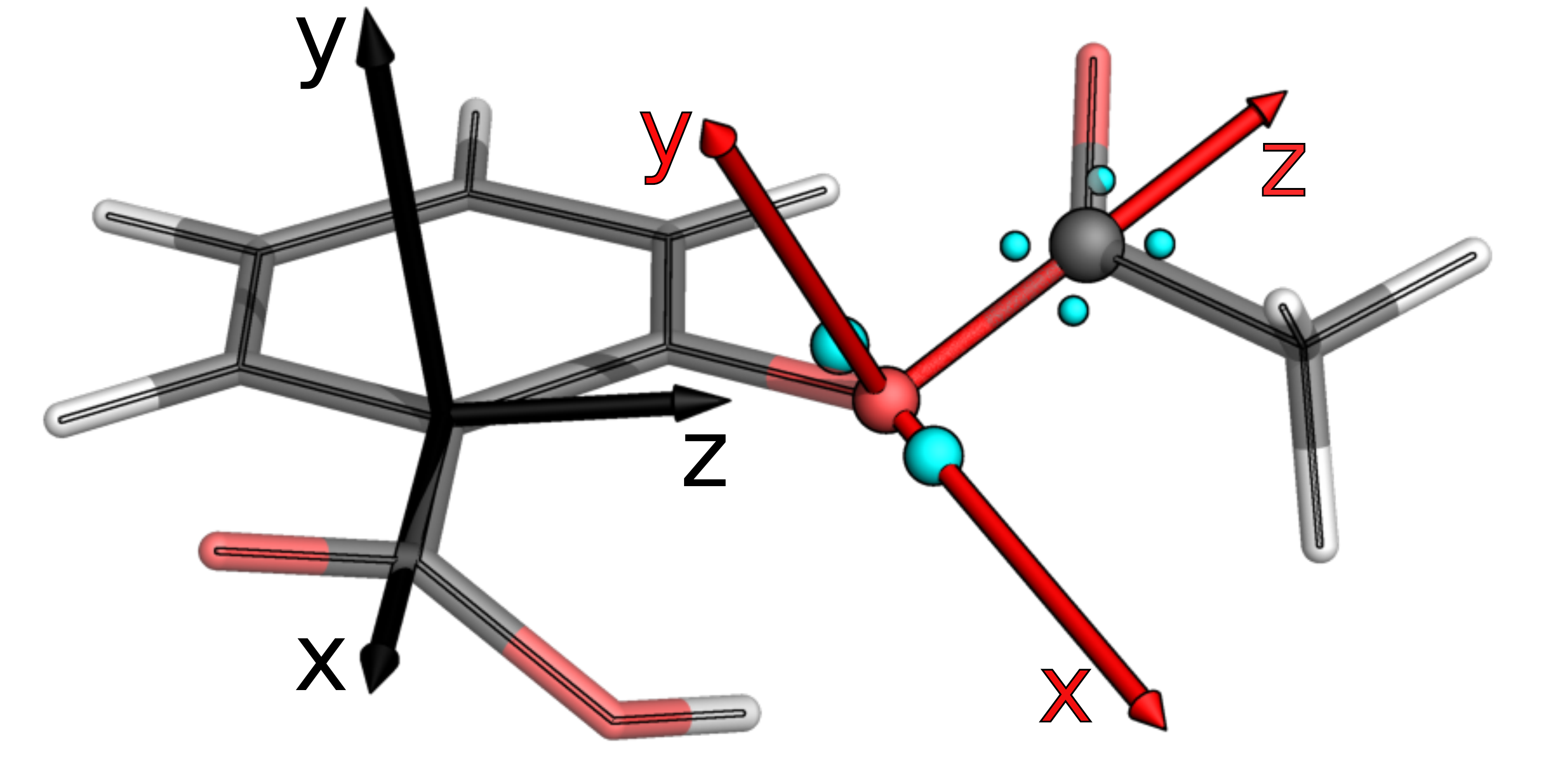}
 \caption{Illustration of the difference of the local (red axes) and global (black axes) 
coordinate systems when calculating $\left<s^Ip_x^I\middle|p^J_xp^J_z\right>$  
in aspirin. The positions of the point charges with which the charge distributions 
$s^Ip^I_x$ and $p^J_xp^J_z$ are approximated are indicated by blue spheres.}
\label{fig:illustrationCoordinateSystems}
\end{figure}

For the other two-center ERIs, similar formulae as the one in Eq.~(\ref{eq:klopman}) 
can be derived which yield the respective 
one-center ERI in the limit $\tilde{R}_{IJ}=0$ (see Ref.~\onlinecite{Dewar1976} or 
Section~\ref{subsec:sp2ceris}).
For the calculation of two-center ERIs, which involve at least one $p$-type basis function, 
a local coordinate system is adopted. \cite{Dewar1976} 
This local coordinate system is defined based on $\tilde{\fett{R}}_I$ and $\tilde{\fett{R}}_J$
(see Figure~\ref{fig:illustrationCoordinateSystems} and Section~\ref{sec:localcoordinate}).
The results obtained in this local coordinate system have to be transformed  
to yield the ERIs in the global coordinate system. 
The necessary transformations can be formulated in terms of 
rotation matrices \cite{Glaeske1987} which are outlined in Section~\ref{sec:localcoordinate}. 

Issues with the presented approach were detected years after the introduction 
of the MNDO model. 
It was remarked \cite{Denton1988,Thiel1991} that rotational invariance was 
not satisfied for the ERIs $\left[p_x^Ip_y^I\middle|p_x^Jp_y^J\right]$
due to the chosen point charge configurations.
It was then suggested \cite{Thiel1991} to impose rotational invariance by setting 
\begin{equation}
 \left[p_x^Ip_y^I\middle|p_x^Jp_y^J\right]= 0.5 \left(
\left[p_x^Ip_x^I\middle|p_x^Jp_x^J\right]-\left[p_x^Ip_x^I\middle|p_y^Jp_y^J\right] \right).
\end{equation}
Moreover, the Klopman approximation 
causes distinct errors in the ERIs \cite{Dewar1993, Holder1994, Holder1994a} which culminates in an infinite error 
in periodic electronic structure calculations. \cite{Stewart2008}
To be able to apply the MNDO in periodic electronic structure calculations, an additional 
scaling factor has to be introduced to yield the exact limit for large $\tilde{R}_{IJ}$. 
\cite{Stewart2008,Margraf2015}

\subsection{Assembling the Symmetrically Orthogonalized One-Electron Matrix}
\label{subsec:mndoH}

\begin{figure}[ht] 
 \centering
 \includegraphics[width=.9\textwidth]{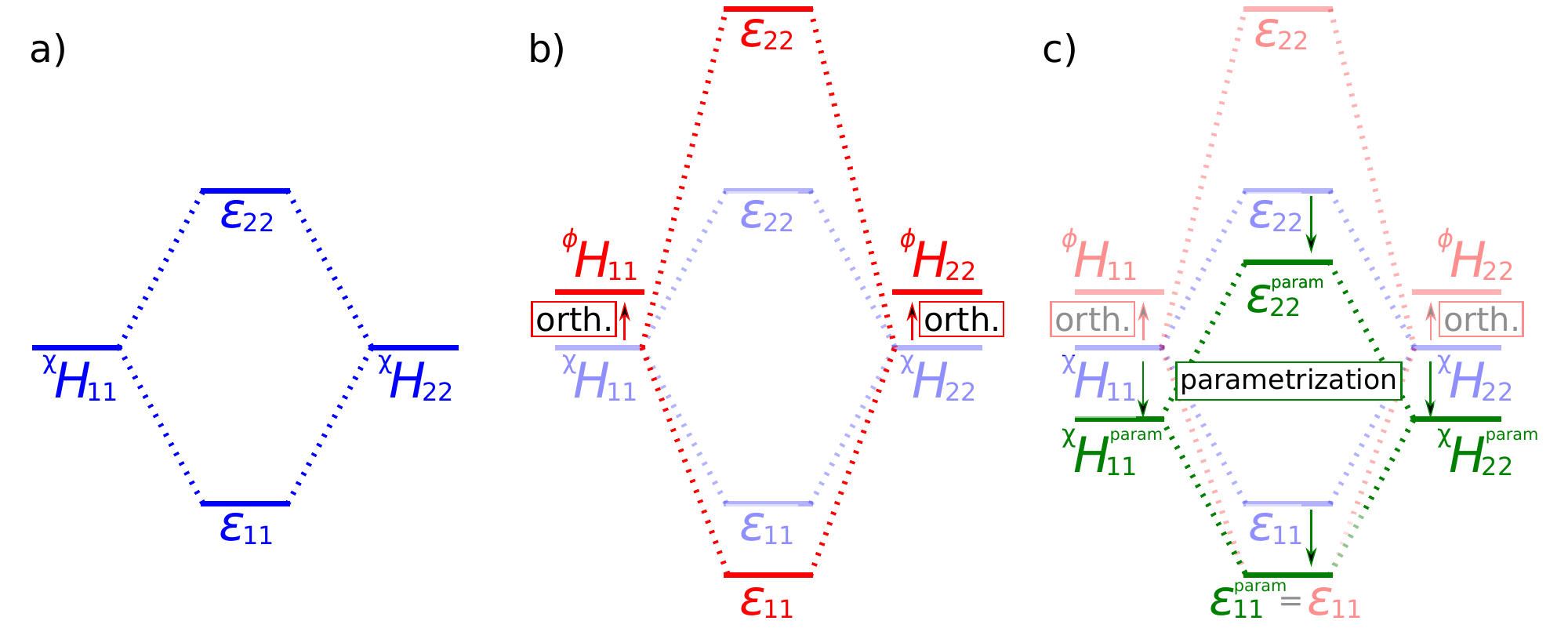} 
 \caption{Illustration of the effect of parametrization when orthogonalization effects are not 
considered at the example of dihydrogen ($\fett{\chi}=\left\{\chi_1^{H_1}, \chi_2^{H_2}\right\}$).
 a) Values of $\epsilon_{11}$ and $\epsilon_{22}$ when solving the eigenvalue 
equation ${}^\chi\!\fett{H}\;{}^\chi\!\fett{C}= {}
^\chi\!\fett{C}\fett{\epsilon}$ without considering ${}^\chi\!\fett{S}$. b) 
Value of $\epsilon_{11}$ and $\epsilon_{22}$
obtained when solving the eigenvalue equation 
$^\phi\!\fett{H}\;{}^\phi\!\fett{C}= \;{}
^\phi\!\fett{C}\fett{\epsilon}$. The eigenvalues are split asymmetrically 
with respect to the $^\chi\!H_{11}= {}
^\chi\!H_{22}$ reference.
c) Effect of parametrization to reproduce the lowest energy eigenvalue when solving the equation 
$^\chi\!\fett{H}^{\text{param}}\;{}^\chi\!\fett{C}= {}
^\chi\!\fett{C}\fett{\epsilon}$.
}
\label{fig:breakdownZDO}
\end{figure}

The MNDO model does not provide an explicit way to account for the 
 change from the $\chi$-basis 
to the $\phi$-basis for the one-electron matrix $\fett{H}$. \cite{Dewar1977}
It is assumed that the elements of ${}^{\phi}\!{\fett{H}}$ 
are approximately equal to ${}^{\chi}\!{\fett{H}}$ when an appropriate parametrization 
is chosen, \cite{Thiel1988}
\begin{equation}
 \;{}^{\phi}\!{\fett{H}} \approx \;{}^{\chi}\!{\fett{H}}^{\text{param}}.
\end{equation}
An example  
discussed in Refs.~\onlinecite{Kolb1993a,Weber2000a, WeberPhD, ScholtenDiss} explains 
why this is not generally possible 
(see Figure~\ref{fig:breakdownZDO}).
We may consider the dihydrogen molecule ($H_1$---$H_2$) in a minimal basis set consisting 
of two $1s$ orbitals $\fett{\chi}=\{\chi_1^{H_1}, \chi_2^{H_2}\}$.
For the moment, we neglect 
electron-electron interactions, i.e., we solve the 
 eigenvalue equations, 
\begin{equation}
\label{eq:horth1}
 {}^{\phi}\!{\fett{H}}\;{}^{\phi}\!{\fett{C}}=
\;{}^{\phi}\!{\fett{C}}\fett{\epsilon},
\end{equation}
in the $\phi$-basis, or analogously in the $\chi$-basis, 
\begin{equation}
\label{eq:horth2}
 {}^{\chi}\!{\fett{H}}\;{}^{\chi}\!{\fett{C}}= \;{}^{\chi}\!{\fett{S}}
\;{}^{\chi}\!{\fett{C}}\fett{\epsilon}.
\end{equation}
We now want to know whether it is possible to obtain the same $\fett{\epsilon}=\fett{\epsilon}^\text{param}$ when solving 
an eigenvalue equation of the type
\begin{equation}
\label{eq:horth3}
  {}^{\chi}\!{\fett{H}}^{\text{param}}
\;{}^{\chi}\!{\fett{C}}= \;{}^{\chi}\!{\fett{C}}\fett{\epsilon}^\text{param},
\end{equation}
where we neglect ${}^\chi\!\fett{S}$.
When solving Eqs.~(\ref{eq:horth1}) and (\ref{eq:horth2}), we see that 
$|{}^{\phi}\!{H_{11}}-\epsilon_{11}|$ and
$|{}^{\phi}\!{H_{11}}-\epsilon_{22}|$ are equally large while 
$|{}^{\chi}\!{H_{11}}-\epsilon_{11}|$ and
$|{}^{\chi}\!{H_{11}}-\epsilon_{22}|$ differ.
Independently of the parametrization, it is not possible to 
obtain different $|{}^{\chi}\!{H^\text{param}_{11}}-\epsilon^\text{param}_{11}|$ and 
$|{}^{\chi}\!{H^\text{param}_{11}}-\epsilon^\text{param}_{22}|$
when solving Eq.~(\ref{eq:horth3}) because 
${}^{\chi}\!{\fett{S}}$ is neglected in Eq.~(\ref{eq:horth3}) (see Figure~\ref{fig:breakdownZDO}).

Apart from this example, it is evident that orthogonalization effects, 
\begin{equation}
\label{eq:explicit_orth}
 \;{}^{\phi}\!{H}_{\mu\nu} = 
\sum_{\lambda=1}^M 
\sum_{\sigma=1}^M
\;\left({}^{\chi}\!{S}^{-\frac{1}{2}}\right)_{\mu\lambda}
\;{}^{\chi}\!{H}_{\lambda\sigma}
\;\left({}^{\chi}\!{S}^{-\frac{1}{2}}\right)_{\sigma\nu},
\end{equation}
cannot be captured by introducing element-dependent parameters
which was, for instance, pointed out in 
Refs.~\onlinecite{Brown1970,Chandler1980,Spanget-Larsen1980,deBruijn1984}. 
The matrix element ${}^{\phi}\!{H}_{\mu\nu}$ depends on contributions from 
all matrix elements of $^\chi\fett{H}$.
Consequently, the parametrization would need to depend on the chemical environment of each atom in 
some manner. The MNDO parameters, however, are only element-dependent and do not 
depend on the chemical environment.
Despite this inherent limitation, MNDO has been a very successful model and we will continue to discuss 
 how the contributions to ${}^{\phi}\!{\fett{H}}^{\text{MNDO}}$ are evaluated in the following Sections
(Sections~\ref{subsubsec:mndoh1c} and \ref{subsubsec:mndoh2c}). 
We assume that the inclusion of empirical parameters accounts for orthogonalization effects 
in some average manner, and hence, retain the superscript $\phi$ for ${}^{\phi}\!{\fett{H}}^{\text{MNDO}}$.

The parametric expressions applied for the evaluation of the 
matrix elements ${}^{\phi}\!{H}_{\mu\nu}^{\text{MNDO}}$ differ 
depending on the number of atoms on which the corresponding basis functions
$\chi_\mu^I$ and $\chi_\nu^J$ are centered: (i) $\chi_\mu^I$ and $\chi_\nu^I$ 
are centered on a single atom $I$ (one-center one-electron 
matrix elements) and (ii) $\chi_\mu^I$ and $\chi_\nu^J$ 
are centered on different atomic nuclei $I\ne J$ (two-center one-electron 
matrix elements).

\subsubsection{One-Center One-Electron Matrix Elements}
\label{subsubsec:mndoh1c}

In the case that $\chi_\mu^I$ and $\chi_\nu^I$ are centered on the same atom, 
the analytical matrix elements in the $\chi$-basis are 
 given by 
\begin{equation}
\label{eq:href_mndo}
 \begin{split}
  {}^{\chi}\!{H}_{\mu\nu} =& \left<\chi_\mu^I\middle|-\frac{1}{2}\fett{\nabla}^2\middle|\chi_\nu^I\right>
- \left<\chi_\mu^I\middle|\frac{Q_I}{|\fett{r}_i-\tilde{\fett{R}}_I|}\middle|\chi_\nu^I\right>  - 
\sum_{\substack{J=1\\ J\ne I}}^N \left<\chi_\mu^I\middle|\frac{Q_J}{|\fett{r}_i-\tilde{\fett{R}}_J|}\middle|\chi_\nu^I\right>. \\
 \end{split}
\end{equation}
These first two terms in the right hand side of Eq.~(\ref{eq:href_mndo}) only refer to the atom $I$ 
 (`one-center' one-electron contributions to ${}^{\chi}\!{H}_{\mu\nu}$).
The remaining 
`two-center' one-electron contributions to ${}^{\chi}\!{H}_{\mu\nu}$ 
describe the electrostatic attraction between the 
charge distribution $\chi_\mu^I \chi_\nu^I$ and the atomic cores $J\ne I$.

In view of Eq.~(\ref{eq:href_mndo}), it is apparent why 
Dewar and Thiel suggested \cite{Dewar1977} 
to calculate ${}^{\phi}\!{H}_{\mu\nu}^{\text{MNDO}}$ by means of  
\begin{equation}
\label{eq:hsameatom_mndo}
 \begin{split}
  {}^{\phi}\!{H}_{\mu\nu}^{\text{MNDO}} &= 
U_{l(\mu)l(\nu)}^{Z_I} - 
\sum_{\substack{J=1\\ J\ne I}}^N Q^J \left[ \chi_\mu^I\chi_\nu^I\middle|s^Js^J\right]. \\
 \end{split}
\end{equation}
The element- and orbital-type-dependent 
parameter $U_{l(\mu)l(\nu)}^{Z_I}$ comprises all one-center 
one-electron contributions in Eq.~(\ref{eq:href_mndo}).
The one-center one-electron terms are exactly zero in a locally orthogonal basis when 
$\mu\ne\nu$,
\begin{equation}
U_{l(\mu)l(\nu)}^{Z_I} = 
 \begin{cases}
 \text{const.} & \mu=\nu \\
 0 & \mu\ne\nu \\
\end{cases}.
\end{equation} 
The parameter $U_{l(\mu)l(\nu)}^{Z_I}$ may not depend on the magnetic quantum number 
to ensure rotational invariance, \cite{Pople1965,Nanda1977} and hence, 
at most two parameters, $U_{ss}^{Z_I}$ and $U_{pp}^{Z_I}$, arise  
per element.
Within a given $\chi$-basis, $U_{ss}^{Z_I}$
and $U_{pp}^{Z_I}$ can be calculated exactly and are transferable between molecules. 
The MNDO model, however, attempts to approximate the matrix element in the 
$\phi$-basis, 
\begin{equation}
  \begin{split}
  {}^{\phi}\!{H}_{\mu\nu} =& \left<\phi_\mu\middle|-\frac{1}{2}\fett{\nabla}^2\middle|\phi_\nu\right>
- \left<\phi_\mu\middle|\frac{Q_I}{|\fett{r}_i-\tilde{\fett{R}}_I|}\middle|\phi_\nu\right> -  
\sum_{\substack{J=1\\ J\ne I}}^N \left<\phi_\mu\middle|\frac{Q_J}{|\fett{r}_i-\tilde{\fett{R}}_J|}\middle|\phi_\nu\right>. \\
 \end{split}
\end{equation}
In the $\phi$-basis, the first two terms obviously depend on the chemical environment of the atom $I$
which is neglected by introducing constant $U_{ss}^{Z_I}$
and $U_{pp}^{Z_I}$. 
It is assumed that the calibration of 
$U_{ss}^{Z_I}$ and $U_{pp}^{Z_I}$
will implicitly lead to a modeling of average orthogonalization effects and will 
also absorb the effects from the core electrons. \cite{Weber2000a} 

A two-center contribution to Eq.~(\ref{eq:href_mndo}) is approximated 
by the negative electrostatic interaction energy of   
$\chi_\mu^I\chi_\nu^I$ with a model charge distributions $s^Js^J,\ J\ne I$ 
 scaled with $Q_J$
in Eq.~(\ref{eq:hsameatom_mndo}).
Pople and Segal proposed \cite{Pople1966} to apply this so-called Goeppert-Mayer--Sklar approximation 
(named for its relation to an equation proposed in Ref.~\onlinecite{Goeppert-Mayer1938}) 
after observing that the application of the analytical expression in Eq.~(\ref{eq:href_mndo}) 
led to far too short bond lengths for several diatomic molecules. \cite{Pople1966}
A decade after their proposal, Coffey analyzed the Goeppert-Mayer--Sklar approximation in more detail and concluded 
that a fortunate error cancellation occurs, so that \cite{Coeffey1974}
\begin{equation}
\label{eq:gms}
\begin{split}
 \left<\phi_\mu\middle|\frac{Q_J}{|\fett{r}_i-\tilde{\fett{R}}_J|}\middle|\phi_\nu\right> =& 
 \left<\chi_\mu^I\middle|\frac{Q_J}{|\fett{r}_i-\tilde{\fett{R}}_J|}\middle|\chi_\nu^I\right>  \\ & 
 -Q^J \left< \chi_\mu^I\chi_\nu^I\middle|s^Js^J\right>+
 Q^J \left< \chi_\mu^I\chi_\nu^I\middle|s^Js^J\right> \\ &+ 
 \left( \left<\phi_\mu\middle|\frac{Q_J}{|\fett{r}_i-\tilde{\fett{R}}_J|}\middle|\phi_\nu\right>-
   \left<\chi_\mu^I\middle|\frac{Q_J}{|\fett{r}_i \tilde{\fett{R}}_J|}\middle|\chi_\nu^I\right>  
 \right) \\ \approx &~   Q^J \left< \chi_\mu^I\chi_\nu^I\middle|s^Js^J\right>.
\end{split}
\end{equation}
The first two terms in Eq.~(\ref{eq:gms}) are the 
so-called `penetration integrals',  \cite{Goeppert-Mayer1938}
\begin{equation}
 \label{eq:gmsapprox1}
 \left<\chi_\mu^I\chi_\nu^I, J\right> 
\approx \left<\chi_\mu^I\middle|\frac{Q^J}{|\fett{r}_{i}-\tilde{\fett{R}}_{J}|}\middle|\chi_\nu^I\right> - Q^J
 \left< \chi_\mu^I\chi_\nu^I\middle|s^Js^J\right>,
\end{equation}
approximately cancel the orthogonalization effects, 
$ \left<\phi_\mu\middle|\frac{Q_J}{|\fett{r}_i-\tilde{\fett{R}}_J|}\middle|\phi_\nu\right> - $ \\ $
  \left<\chi_\mu^I\middle|\frac{Q_J}{|\fett{r}_i-\tilde{\fett{R}}_J|}\middle|\chi_\nu^I\right>  
$
 which are required to transform 
${}^{\chi}\!{H}_{\mu\nu}$ from the 
$\chi$- into the $\phi$-basis (see Figure~1 in Ref.~\onlinecite{Coeffey1974}). 
Hence, the Goeppert-Mayer--Sklar approximation entails an implicit basis set transformation 
from the $\chi$- to the $\phi$-basis.
The orthogonalization effects, which Coffey considered \cite{Coeffey1974}, also included orthogonalization of 
the core orbitals to the valence orbitals. Hence, the application of Eq.~(\ref{eq:gms})
may also be interpreted as the emulation of the application of an approximate effective core potential.
Unfortunately, Coffey's analysis was restricted to the C$_2$ molecule 
and included several additional approximations (such as an averaging of one- and two-electron 
integrals). \cite{Coeffey1974}
It is not evident whether (and appears improbable\cite{deBruijn1984,Kollmar1995} that) 
Coffey's analysis can easily be 
generalized to arbitrary polyatomic molecules.
The success of the MNDO model indicates, however, that --- at least in the context of all other 
invoked approximations --- the Goeppert-Mayer--Sklar approximation is a satisfactory one.

\subsubsection{Two-Center One-Electron Matrix Elements}
\label{subsubsec:mndoh2c}
 \begin{figure}[ht]
  \centering 
  \includegraphics[width=.75\textwidth]{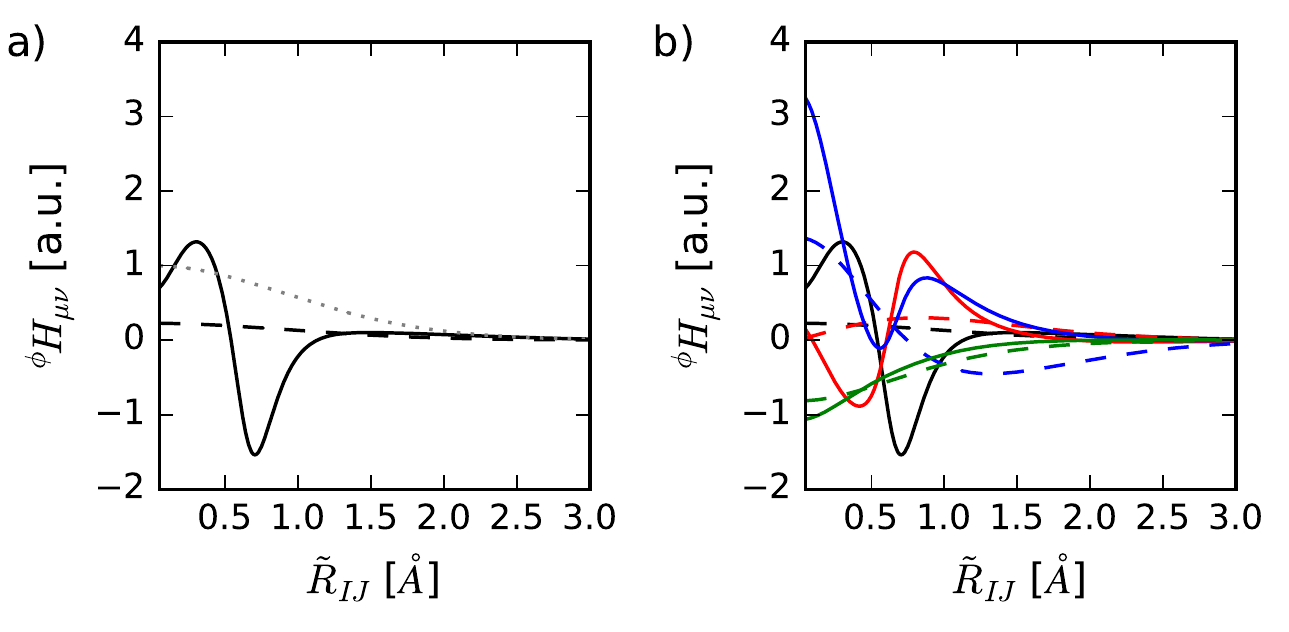}
  \caption{
Dependence of ${}^{\phi}\!{H_{\mu\nu}}$ (solid lines)
and ${}^{\phi}\!{H^{\text{MNDO}}_{\mu\nu}}$ (dashed lines)
on the distance $\tilde{R}_{IJ}$ in a C$_2$ molecule described in an MNDO-3G 
basis 
($\fett{\chi}=\left\{\chi_1^{C_1}, \chi_2^{C_1},
\chi_3^{C_1},\chi_4^{C_1},\chi_5^{C_2}, \chi_6^{C_2},
\chi_7^{C_2},\chi_8^{C_2}\right\}$ where the first and fifth basis functions 
are $2s$, the second and sixth are $2p_x$, the third and seventh are $2p_y$,
and the fourth and eighth are $2p_z$ basis functions).
a) ${}^{\phi}\!{H_{15}}^{\text{(MNDO)}}$ 
with the best-fit parameter $\beta_s^{6}=0.1$ a.u.\ (black line) and 
${}^{\chi}\!{S_{15}}$ (gray dotted line).
b) ${}^{\phi}\!{H_{15}}^{\text{(MNDO)}}$ with $\beta_s^{6}=0.1$ a.u.\ (black lines),
${}^{\phi}\!{H_{18}}^{\text{(MNDO)}}$ with $\beta_s^{6}=\beta_p^{6}=-0.3$ a.u.\ 
(red lines),
${}^{\phi}\!{H_{48}}^{\text{(MNDO)}}$ with $\beta_s^{6}=1.3$ a.u.\ (blue lines), and
${}^{\phi}\!{H_{26}}^{\text{(MNDO)}}$ with $\beta_s^{6}=-0.2$ a.u.\ (green lines).
}
 \label{fig:resonanceIntsM}
 \end{figure}
In the case that $\chi_\mu^I$ and $\chi_\nu^J$ are centered on different atomic nuclei ($I\ne J$),
the matrix elements ${}^{\phi}\!{H}_{\mu\nu}^{\text{MNDO}}$ are taken to 
be proportional to ${}^{\chi}\!{S}_{\mu\nu}$, 
\begin{equation}
 \label{eq:resS}
 \;{}^{\phi}\!{H}^{\text{MNDO}}_{\mu\nu} = 
\frac{\beta_{l(\mu)}^{Z_I} + 
 \beta_{l(\nu)}^{Z_J}}{2} \;{}^{\chi}\!{S}_{\mu\nu}.
\end{equation}
The mean of two element- and orbital-type-dependent parameters 
$\beta_{l(\mu)}^{Z_I}$ and $\beta_{l(\nu)}^{Z_J}$
yields the proportionality factor.
In analogy to the parameters $U_{ss}^{Z_I}$ and $U_{pp}^{Z_I}$, 
at most two parameters arise per element, $\beta_{s}^{Z_I}$ and 
$\beta_{p}^{Z_I}$.
Taking ${}^{\phi}\!{H}^{\text{MNDO}}_{\mu\nu}$ to be proportional to 
${}^{\chi}\!{S}_{\mu\nu}$ has a long history \cite{Pople1965, Dewar1977}
and the initial idea is ascribed to Mulliken. \cite{Mulliken1949}
Generally, Eq.~(\ref{eq:resS}) was, however, found to be a poor approximation to the analytical value of 
${}^{\phi}\!{H}_{\mu\nu}$,  
irrespective of the chosen values for $\beta_{l(\mu)}^{Z_I}$ 
\cite{Jug1971,Coffey1973,DeBruijn1978a,deBruijn1984}. 
This can be attributed to the fact that ${}^{\phi}\!{H}_{\mu\nu}$  is not necessarily proportional to 
${}^{\chi}\!{S}_{\mu\nu}$  \cite{Jug1971,Coffey1973,DeBruijn1978a,deBruijn1984}
(for an example, see Figure~\ref{fig:resonanceIntsM}).
Hence, not even the nodal structure of ${}^{\phi}\!{H}_{\mu\nu}={}^{\phi}\!{H}_{\mu\nu}(\tilde{R}_{IJ})$ is 
captured correctly. 
This finding appears puzzling in view of the success of the MNDO model,
apparently Eq.~(\ref{eq:resS}) suffices to obtain satisfactory results, e.g., for heats of 
formation in this context.

\subsection{Empirical Modification of Core-Core Repulsion Energy}
\label{subsec:mndorepulsion}

\begin{figure}[ht]
 \centering 
 \includegraphics[width=.75\textwidth]{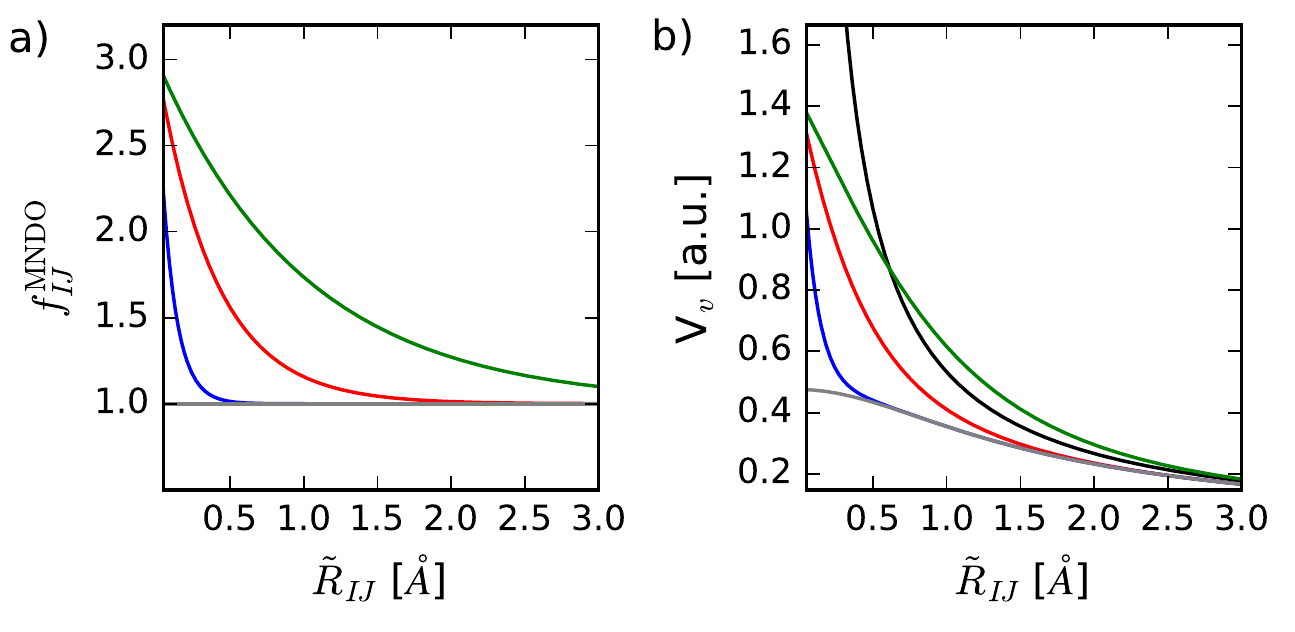}
 \caption{Dependence of a) $f_{IJ}^{\text{MNDO}}$ and b) 
$V^{\text{MNDO}}_{v}$ on $\tilde{R}_{IJ}$ in an H$_2$ molecules.
We calculated the MNDO scaling factors $f_{IJ}^{\text{MNDO}}$ and 
$V_{v}^{\text{MNDO}}$ with 
 $\alpha^{1}=1.0$ \AA$^{-1}$ (green lines), 
$\alpha^{1}=2.5$ \AA$^{-1}$ (red lines),
$\alpha^{1}=10.0$ \AA$^{-1}$ (blue lines), and 
$\alpha^{1}=\infty$ (i.e., $f_{IJ}^{\text{MNDO}}=1$, gray lines).
We compare $V_{v}^{\text{MNDO}}$ to $V_{v}$ (black line).
}
\label{fig:eccmndo}
\end{figure}

The core-core repulsion energy in the MNDO model, $V_{v}^{\text{MNDO}}$, is 
also determined from a parametric expression.
The substitution of the analytical expression,
\begin{equation}
\label{eq:nuclearEnergy}
 V_{v} = \sum_{I=1}^N \sum_{J>I}^N \frac{Q_I Q_J}{\tilde{R}_{IJ}},
\end{equation}
 for a parametric one cannot be physically motivated. 
Empirically, it was determined that a parametric expression 
needs to be introduced to decrease the average core-core repulsion energy to 
define a useful NDDO-SEMO model. \cite{Pople1965, Dewar1977}
The parametric expression to evaluate $V_{v}^{\text{MNDO}}$, 
\begin{equation}
 \label{eq:mndoNuclearEnergy}
 \begin{split}
 V_{v}^{\text{MNDO}} =& \sum_{I=1}^N \sum_{J>I}^N Q_I Q_J \left[s^Is^I\middle|s^Js^J\right]  \cdot 
 f_{IJ}^{\text{MNDO}} \\
  =& \sum_{I=1}^N \sum_{J>I}^N Q_I Q_J
\left[ \tilde{R}_{IJ}^2 + 
\left(\frac{1}{2 \gamma_{ss}^{Z_I}} 
+  \frac{1}{2 \gamma_{ss}^{Z_J}} \right)^2 \right]^{-\frac{1}{2}}  \cdot   
 f_{IJ}^{\text{MNDO}}, \\
 \end{split}
\end{equation}
features two key modifications with respect to Eq.~(\ref{eq:nuclearEnergy}).
Firstly, the pairwise point-charge interaction is substituted 
by a scaled interaction of the charge distributions $s^Is^I$ and $s^Js^J$ which 
is evaluated in the Klopman approximation (cf.\ Eq.~(\ref{eq:klopman})).
Secondly, each core-core interaction energy is scaled by $f_{IJ}^{\text{MNDO}}$,
\begin{equation}
 \label{eq:mndoscal}
 \begin{split}
f_{IJ}^{\text{MNDO}} =& 
 1+  
\exp{\left( - \alpha^{Z_I} \tilde{R}_{IJ}\right)} + 
\exp{\left( - \alpha^{Z_J} \tilde{R}_{IJ}\right)}, \\
 \end{split}
\end{equation}
where $\alpha^{Z_I}$ is an element-dependent parameter. 
The introduction of these modifications provides a large flexibility for the MNDO model,
but  this flexibility comes at a high price.
Most strikingly, $V_{v}^{\text{MNDO}}$ is finite for $\tilde{R}_{IJ}=0$.
More specifically, $f_{IJ}^{\text{MNDO}}=3.0$ and $V_{v}^{\text{MNDO}}=3 Q_I^2\gamma_{ss}^{Z_I}$
in the limit $\tilde{R}_{IJ}=0$ for a homonuclear diatomic system (see also Figure~\ref{fig:eccmndo}).
Obviously, this limit is entirely artificial.
For intermediate values of $\tilde{R}_{IJ}$, the 
parameter $\alpha^{Z_I}$ determines how fast $f_{IJ}^{\text{MNDO}}$ 
declines from three ($\tilde{R}_{IJ}=0$) to one ($\tilde{R}_{IJ}\rightarrow\infty$)
if $\alpha^{Z_I}$ is not negative.
This appears to be a constraint invoked during the calibration of $\alpha^{Z_I}$.
Hence, in the limit of large $\tilde{R}_{IJ}$, 
$V_{v}^{\text{MNDO}}$ will tend toward $V_{v}$, as it should. 
Depending on the choice of $\alpha^{Z_I}$, 
$V_{v}^{\text{MNDO}}$ may be larger or smaller 
than $V_{v}$ for a given $\tilde{R}_{IJ}$.

The scaling factor is not calculated according to Eq.~(\ref{eq:mndoscal})
when the element pair is N--H ($Z_I = 7$ and $Z_J = 1$) or O--H
($Z_I = 8$ and $Z_J = 1$). 
Dewar and Thiel found \cite{Dewar1977} that they could achieve a better agreement with experimental 
data when they instead applied the scaling factor $f_{IJ}^{',\text{MNDO}}$ for these 
element pairs,
\begin{equation}
 \label{eq:mndoscal2}
 \begin{split}
f_{IJ}^{',\text{MNDO}} =& 
 1+  \tilde{R}_{IJ}
\exp{\left( -  \alpha^{Z_I=7,8} \tilde{R}_{IJ}\right)}  + 
\exp{\left( - \alpha^{Z_J=1} \tilde{R}_{IJ}\right)}. \\
 \end{split}
\end{equation}
Because there is no theoretical foundation for the introduction of Eqs.~(\ref{eq:mndoNuclearEnergy}) and 
(\ref{eq:mndoscal}), 
it remains unclear why the application of Eq.~(\ref{eq:mndoscal2}) yields a better 
agreement with experimental data.
We note that $f_{IJ}^{',\text{MNDO}}=2$ in the limit $\tilde{R}_{IJ}=0$.
The modification does, hence, not rectify the theoretically unsatisfactory situation 
of finite core-core repulsion energies in the limit $\tilde{R}_{IJ}=0$.
The scaling factor $f_{IJ}^{',\text{MNDO}}$ also tends to one for  
large $\tilde{R}_{IJ}$.
For given $\alpha^{Z_I}$, $f_{IJ}^{',\text{MNDO}}<f_{IJ}^{\text{MNDO}}$ for all $\tilde{R}_{IJ}$.

\subsection{Direct Descendants of the MNDO Model}
\label{subsec:mndofamily}

\subsubsection{Extension to $d$ Orbitals: The MNDO/d Model}
\label{subsubsec:mndod}

The acronym `MNDO/d' denotes the extension of the MNDO model from an $s,p$ basis to an $s,p,d$ basis. 
\cite{Thiel1991, Thiel1996c} The consideration of $d$-type basis functions requires, on the one hand, 
the specification of additional orbital-type-dependent parameters per element, and on the other hand,
an adjustment of the parametric expressions themselves.

The number of unique nonzero one-center ERIs increases from six  
(see Eqs.~(\ref{eq:gss})--(\ref{eq:hpp})) for an $s, p$ basis 
to 58 for an $s, p, d$ basis. 
These 58 one-center ERIs are determined analytically \cite{Pelikan1974}
from a set of auxiliary orbital exponents ${\zeta'_{s}}^{Z_I}$, 
${\zeta'_{p}}^{Z_I}$, and ${\zeta'_{d}}^{Z_I}$. \cite{Thiel1991, Thiel1996c}
The auxiliary orbital exponents  are derived from the 
fitted parameters $\gamma_{ss}^{Z_I}$, $\gamma_{pp}^{Z_I}$, and the newly 
introduced parameter $\gamma_{dd}^{Z_I}$.
Note that these auxiliary orbital exponents are different from the set of 
Slater exponents $\zeta_{s}^{Z_I}$ , $\zeta_{p}^{Z_I}$ , and 
$\zeta_{d}^{Z_I}$ which are, e.g., applied to calculate the overlap integrals. 
The formulae for calculating the one-center ERIs from the auxiliary Slater exponents ${\zeta'_{s}}^{Z_I}$, 
${\zeta'_{p}}^{Z_I}$, and ${\zeta'_{d}}^{Z_I}$ are 
given in Refs.~\onlinecite{Pelikan1974, Kumar1987}.
It appears that several of the formulae presented in Ref.~\onlinecite{Pelikan1974} 
(Eqs.~(17), (51), (53), (54), (56), and (57))
contain typographical mistakes which we correct in Section~\ref{sec:mndod}
(Eqs.~(\ref{eq:17}), (\ref{eq:51}), (\ref{eq:53}), (\ref{eq:54}), 
(\ref{eq:56}), and (\ref{eq:57})). 
These errors affect the one-center ERIs of 
the types
$\left<p_z^Id_{z^2}^I|p_x^Id_{xz}^I\right>$,
 $\left<p_z^Id_{z^2}^I|p_y^Id_{yz}^I\right>$,
$\left<s^Id_{z^2}^I|p_x^Ip_x^I\right>$,
$\left<s^Id_{z^2}^I|p_y^Ip_y^I\right>$,
$\left<p_y^Ip_y^I|s^Id_{x^2-y^2}^I\right>$,
$\left<p_z^Is^I|p_z^Id_{z^2}^I\right>$,
$\left<p_z^Ip_z^I|s^Id_{z^2}^I\right>$,
$\left<p_z^Ip_x^I|s^Id_{xz}^I\right>$, \\
$\left<p_z^Ip_y^I|s^Id_{yz}^I\right>$, 
$\left<p_x^Ip_x^I|s^Id_{x^2-y^2}^I\right>$, and
$\left<p_x^Ip_z^I|s^Id_{xy}^I\right>$.

Thiel and Voityuk also extended the formalism to approximate two-center ERIs in 
a point-charge model to charge distributions including $d$-type orbitals. \cite{Thiel1991,Thiel1996c} 
For this purpose, they introduced a new quadrupole point charge configuration $\tilde{Q}_{xy,xz,yz}$
 (see also Figure~\ref{fig:multipoles} in Section~\ref{sec:mndod}).
One can then straightforwardly apply the concepts of the multipole expansion introduced 
in Section~\ref{subsec:mndo2ceri} and derive the necessary formulae for all 
possible combinations of arising multipoles (see Refs.~\onlinecite{Thiel1991,Thiel1996c,Horn2005,Horn2007} 
and Section~\ref{sec:mndod}). 

The two-center ERIs are also applied to calculate the core-core repulsion energy 
(Eq.~(\ref{eq:mndoNuclearEnergy})) and contributions to the one-electron matrix 
(Eq.~(\ref{eq:hsameatom_mndo})).
In these equations, the atomic core $I$ was described by a charge distribution 
$s^Is^I$ which involves the parameter $\gamma_{ss}^{Z_I}$.
The MNDO/d formalism makes these expressions independent from the parameter
$\gamma_{ss}^{Z_I}$. Therefore, the atomic core $I$ is described by a 
spherical charge distribution $\varrho_\text{core}^{I}$. The electrostatic interaction energy with this charge distribution is computed within the Klopman--Ohno 
approximation, so that Eq.~(\ref{eq:mndoNuclearEnergy}) is substituted for 
\begin{equation}
 \label{eq:mndoNuclearEnergy2}
 \begin{split}
 V_{v}^{\text{MNDO/d}} = & \sum_{I=1}^N \sum_{J>I}^N Q_I Q_J [\varrho_\text{core}^I|\varrho_\text{core}^J] \cdot 
 f_{IJ}^{\text{MNDO}} \\
 = &
\sum_{I=1}^N \sum_{J>I}^N Q_I Q_J
\left[ \tilde{R}_{IJ}^2 + 
\left(\vartheta^{Z_I}_{\text{core}} 
+  \vartheta^{Z_J}_{\text{core}} \right)^2 \right]^{-\frac{1}{2}} \cdot 
 f_{IJ}^{\text{MNDO}}, \\
 \end{split}
\end{equation}
where $\vartheta^{Z_J}_{\text{core}}= 1/(2\gamma_{ss}^{Z_I})$ 
for elements which do not activate $d$-type orbitals
and $\vartheta^{Z_J}_{\text{core}}\ne 1/(2\gamma_{ss}^{Z_I})$ for elements which activate $d$-type orbitals.
Similar adjustments are necessary for the formula to calculate the one-electron matrix (Eq.~(\ref{eq:hsameatom_mndo})) which is now evaluated as 
\begin{equation}
\label{eq:hsameatom_mndo2}
 \begin{split}
  {}^{\phi}\!{H}_{\mu\nu}^{\text{MNDO}} &= 
U_{l(\mu)l(\nu)}^{Z_I} - 
\sum_{\substack{J=1\\ J\ne I}}^N Q^J \left[ \chi_\mu^I\chi_\nu^I\middle|\varrho_\text{core}^J\right]. \\
 \end{split}
\end{equation} 
Additionally, we have to specify the parameters 
 $U_{dd}^{Z_I}$, $\beta_{d}^{Z_I}$, and $\zeta_{d}^{Z_I}$
to assemble the one-electron matrix.

\subsubsection{The Austin Models (AM$x$)}
\label{subsubsec:am1}

The \textit{Austin model 1} (AM1) \cite{Dewar1985}
 differs from the MNDO model in the way in which the scaling factor for the pairwise
core-core repulsion energies is determined,
\begin{equation}
 \label{eq:am1NuclearEnergy}
 \begin{split}
 V_{v}^{\text{AM1}} =& \sum_{I=1}^N \sum_{J>I}^N Q_I Q_J
\left[s^Is^I\middle|s^Js^J\right] \cdot 
 f_{IJ}^{\text{AM1}}. \\
 \end{split}
\end{equation}
The scaling factor $f_{IJ}^{\text{AM1}}$ is defined as the sum 
of the  scaling factor $f_{IJ}^{\text{MNDO}}$ (Eq.~(\ref{eq:mndoscal})) 
and an element-specific number of additional Gaussian functions
$A^{Z_I}$, \cite{Dewar1985,Burstein1984}
\begin{equation}
 \label{eq:am1scal}
 \begin{split}
f_{IJ}^{\text{AM1}} =& f_{IJ}^{\text{MNDO}} +
  \sum_{a=1}^{A^{Z_I}} K_a^{Z_I} 
               \exp{\left(-  L_a^{Z_I} 
\left( \tilde{R}_{IJ} - 
 M_a^{Z_I} \right)^2 \right)}   \\ & +  
\sum_{a=1}^{A^{Z_{\!J}}} K_a^{Z_J} 
               \exp{\left(-  L_a^{Z_J} 
\left( \tilde{R}_{IJ} - 
 M_a^{Z_J} \right)^2 \right)}. \\
 \end{split}
\end{equation}
The shape of the $a$-th Gaussian function
is characterized by the element-dependent parameters 
$K_a^{Z_I}$, $ L_a^{Z_I}$, 
and $ M_a^{Z_I}$. 
The sign of $ K_a^{Z_I}$ determines whether the $a$-th Gaussian increases (positive sign) 
or decreases (negative sign) $f_{IJ}^{\text{AM1}}$ at a given $\tilde{R}_{IJ}$. 
$ L_a^{Z_I}$ determines the width of the Gaussian; it 
must be positive because $f^{\text{AM1}}$ would otherwise tend to infinity for large $\tilde{R}_{IJ}$. 
$ M_a^{Z_I}$ specifies where the $a$-th Gaussian is centered and, hence,
for which $\tilde{R}_{IJ}$ additional repulsive or attractive interactions are added.  
\begin{figure}[ht]
 \centering 
 \includegraphics[width=.75\textwidth]{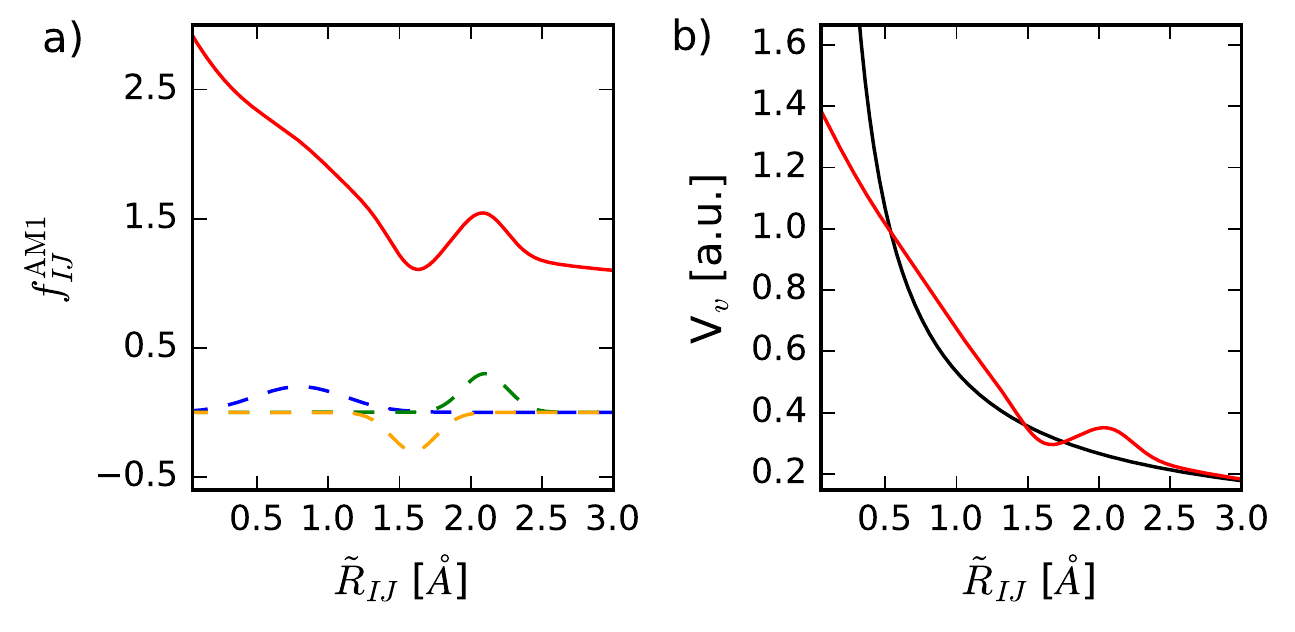}
 \caption{
Dependence of a) $f_{IJ}^{\text{AM1}}$ (red line) and b) 
$V^{\text{AM1}}_{v}$ (red line) on $\tilde{R}_{IJ}$ in H$_2$.
Diagram a) illustrates the contributions of the 
$3$  Gaussian functions (dashed lines) with 
$K_0^1=0.10$, $L_0^1=5.0$ \AA$^{-2}$, $M_0^1=0.80$ \AA\ (blue dashed line),
$K_1^1=-0.15$, $L_1^1=20.0$ \AA$^{-2}$, $M_1^1=1.6$ \AA\ (orange dashed line), and
$K_2^1=0.15$, $L_2^1=20.0$ \AA$^{-2}$, $M_2^1=2.1$ \AA\ (green dashed line).
b) Comparison of $V_{v}^{\text{AM1}}$ (red line) with $V_{v}$ (black line).
}
 \label{fig:scalingAM1}
\end{figure}
If more than one Gaussian is added to the scaling factor ($A^{Z_I}>1$), 
$f_{IJ}^{\text{AM1}}$ becomes a quite 
involved function. It can decrease the core-core repulsion energy at certain distances
(e.g., at $\tilde{R}_{IJ}=1.6$ \AA\ in Figure~\ref{fig:scalingAM1})
and increase it at 
other distances
(e.g., at $\tilde{R}_{IJ}=2.1$ \AA\ in Figure~\ref{fig:scalingAM1}).
Hence, $f_{IJ}^{\text{AM1}}$ offers a tremendous flexibility and allows for tightly focused fine-tuning to 
achieve a better agreement with reference data.
Simultaneously, the addition of Gaussian functions introduces a high degree of arbitrariness, 
which has already been noted by Dewar and co-workers when they introduced 
this modification. \cite{Dewar1985}

A popular reparameterization of AM1 \cite{Dewar1985} was presented by 
Rocha \textit{et al.} under the 
name \textit{Recife model 1} (RM1). \cite{Rocha2006}
Its formalism is identical to that of AM1. \cite{Rocha2006}

The AM1 model was also generalized to include $d$ orbitals (AM1/d 
\cite{Voityuk2000}) in the same way in which MNDO was generalized to MNDO/d.
Additionally, the scaling factor $f_{IJ}^{\text{AM1}}$ is usually slightly modified for heavier elements, 
\cite{Voityuk2000}
\begin{equation}
\label{eq:fam1mod}
\begin{split}
f_{IJ}^{',\text{AM1}} =&  1 + {  2 x^{Z_I,Z_J}} 
\exp{\left(- {  {\alpha'}^{Z_I,Z_J}} \tilde{R}_{IJ}\right)}   + \sum_{a=1}^{A^{Z_I}} K_a^{Z_I} 
               \exp{\left(-  L_a^{Z_I} 
\left( \tilde{R}_{IJ} - 
 M_a^{Z_I} \right)^2 \right)}   \\ & +  
\sum_{a=1}^{A^{Z_J}} K_a^{Z_J} 
               \exp{\left(-  L_a^{Z_J} 
\left( \tilde{R}_{IJ} - 
 M_a^{Z_J} \right)^2 \right)}, \\
\end{split}
\end{equation}
so that it contains element-pair-dependent parameters 
$x^{Z_I,Z_J}$ (denoted as $\delta$ in 
Ref.~\onlinecite{Voityuk2000}) and ${\alpha'}^{Z_I,Z_J}$. \cite{Voityuk2000}

The acronym AM1* denotes a popular reparameterization of the AM1/d model
\cite{Winget2003, Winget2005, Kayi2007,
Kayi2009b,Kayi2009c,Kayi2010c,Kayi2010d,Kayi2010e,Kayi2011a} which is implemented in the 
\textsc{Empire} suite of programs. \cite{Margraf2015}

\subsubsection{The Parametrized Models (PM$x$)}
\label{subsubsec:pmx}

Stewart introduced three popular NDDO-SEMO models, the  
\textit{parametrized models} (PM$x$, $x=3, 6, 7$) \cite{Stewart1989,
Stewart2007,Stewart2012}. 
The parametrized models regard
\textit{all} element-dependent parameters as independent \cite{Stewart1989} and 
the element-dependent parameters are calibrated 
individually, hence the name. 
The MNDO/d model specifies a set of auxiliary orbital 
exponents ${\zeta'_{s}}^{Z_I}$, ${\zeta'_{p}}^{Z_I}$, and 
${\zeta'_{d}}^{Z_I}$ which are deduced from the parameters
$\gamma_{ss}^{Z_I}$, $\gamma_{pp}^{Z_I}$, and $\gamma_{dd}^{Z_I}$, respectively. 
In the PM$x$ models, the conceptual relation of the auxiliary orbital exponents 
to $\gamma_{ss}^{Z_I}$, $\gamma_{pp}^{Z_I}$, and $\gamma_{dd}^{Z_I}$ is ignored
for main-group elements. \cite{Stewart2007}
For several transition metals, the one-center ERIs 
$\left<s^Is^I|d_{z^2}^Id_{z^2}^I\right>=
\left<s^Is^I|d_{x^2-y^2}^Id_{x^2-y^2}^I\right>=
\left<s^Is^I|d_{xy}^Id_{xy}^I\right>=
\left<s^Is^I|d_{xz}^Id_{xz}^I\right>=
\left<s^Is^I|d_{yz}^Id_{yz}^I\right>$
and $\left<s^Id_{z^2}^I|s^Id_{z^2}^I\right>= 
\left<s^Id_{x^2-y^2}^I|s^Id_{x^2-y^2}^I\right>=
\left<s^Id_{xy}^I|s^Id_{xy}^I\right>=
\left<s^Id_{xz}^I|s^Id_{xz}^I\right>=
\left<s^Id_{yz}^I|s^Id_{yz}^I\right>$
are also considered parameters independent of ${\zeta'_{s}}^{Z_I}$ and 
${\zeta'_{d}}^{Z_I}$. \cite{Stewart2007}

The formalism of the PM3 model ($s,p$ basis) is identical to that of the AM1 model.
It differs from the AM1 model only in the values of the parameters, and in the way in which 
they are determined \cite{Stewart1989}. 

\begin{figure}[ht]
 \centering 
 \includegraphics[width=.45\textwidth]{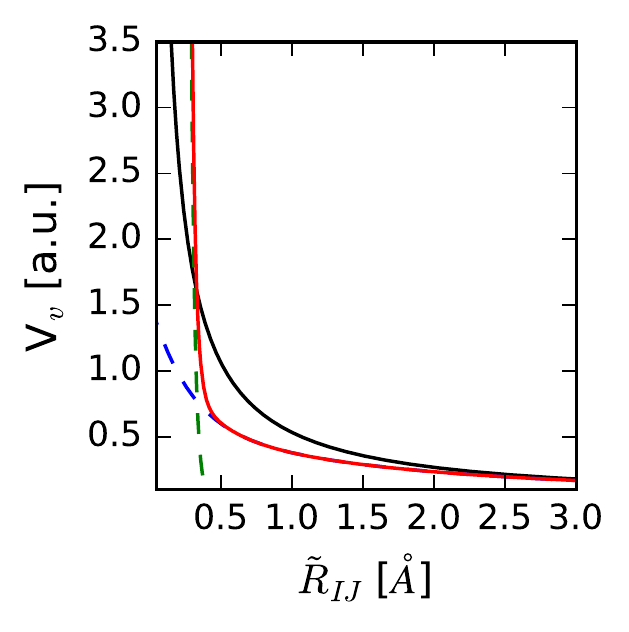}
 \caption{
$V^{\text{PM6}}_{v}$ (red line) and 
$V_{v}$ (black line) in a.u.\ for H$_2$ 
for various distances $\tilde{R}_{IJ}$ in \AA.
$V^{\text{PM6}}_{v}$ is decomposed in the contribution 
of the first term in Eq.~(61) (blue dashed line) and that of the 
second term in Eq.~(61) (green dashed line).}
 \label{fig:eccpm62}
\end{figure}

The PM6 model ($s, p, d$ basis), by contrast, features several modifications with respect to PM3 and, hence, 
AM1 \cite{Stewart2007}.
Most prominently, the parametric expression which is applied to calculate  
$V_{v}$, was modified even further. 
In general, the PM6 core-core repulsion energy, $V_{v}^{\text{PM6}}$, is given by  
\begin{equation}
\label{eq:pm6ecc}
\begin{split}
  V_{v}^{\text{PM6}} =& \sum_{I=1}^N \sum_{J>I}^N \left\{ 
Q_I Q_J
\left[\varrho_\text{core}^I\middle|\varrho_\text{core}^J\right]
 \cdot f_{IJ}^{\text{PM6}}  + 10^{-8} \text{eV}
\left(\frac{Z_I^{1/3} Z_J^{1/3}}{\tilde{R}_{IJ}} \right)^{12} \right\}.  \\
\end{split}
\end{equation}
Eq.~(\ref{eq:pm6ecc}) resembles 
Eq.~(\ref{eq:mndoNuclearEnergy2}) (MNDO/d core-core repulsion energy), 
but applies a different scaling factor, $f_{IJ}^{\text{PM6}}$,
and adds an additional term to each pairwise interaction. 
This additional term re-introduces a singularity for $\tilde{R}_{IJ}=0$ (see also Figure~\ref{fig:eccpm62}),
One could therefore conclude that the expression is physically more consistent.
The term was designed to resemble the repulsive part of the Lennard-Jones potential. \cite{Stewart2007}
The prefactor of $10^{-8}$ electron volt (eV) is an empirical choice which is not further commented on 
in Ref.~\onlinecite{Stewart2007}. It appears to be chosen such that 
$V_{v}^{\text{PM6}}$ is only affected by the Lennard-Jones-like term for very small $\tilde{R}_{IJ}$.

The scaling factor $f_{IJ}^{\text{PM6}}$ was constructed in analogy to the 
scaling factor which was proposed for heavier elements for the AM1/d model \cite{Voityuk2000} (Eq.~(\ref{eq:fam1mod})),
\begin{equation}
\label{eq:pm6scal}
\begin{split}
f_{IJ}^{\text{PM6}} =&  1 + 2 {  x^{Z_I,Z_J}} 
\exp{\left(- {  {\alpha'}^{Z_I,Z_J}} \left(\tilde{R}_{IJ}   + 
0.0003 \text{\AA}^{-5} \tilde{R}_{IJ}^6\right)\right)}  \\
& + K^{Z_I} 
               \exp{\left(-  L^{Z_I} 
\left( \tilde{R}_{IJ} - 
 M^{Z_I} \right)^2 \right)}    +  
K^{Z_J} 
               \exp{\left(-  L^{Z_J} 
\left( \tilde{R}_{IJ} - 
 M^{Z_J} \right)^2 \right)} \\
\end{split}
\end{equation}
It also contains element-pair-dependent parameters 
$x^{Z_I,Z_J}$ and ${\alpha'}^{Z_I,Z_J}$.
In comparison to Eq.~(6) in Ref.~\onlinecite{Stewart2007}, 
we replaced $x^{Z_I,Z_J}$ for $2x^{Z_I,Z_J}$; 
this is necessary to achieve an agreement with the implementation in \textsc{Mopac} \cite{mopac} with 
the parameter values reported for $x^{Z_I,Z_J}$ in Ref.~\onlinecite{Stewart2007}.
Stewart restricted the number of additional Gaussian functions to one per element, so that 
we do not have to specify an index anymore for the parameters characterizing 
the Gaussian functions (i.e., $K^{Z_I}$, $ L^{Z_I}$, 
and $ M^{Z_I}$ instead of $K_a^{Z_I}$, $ L_a^{Z_I}$, 
and $ M_a^{Z_I}$, respectively).
In comparison to $f_{IJ}^{',\text{AM1}}$, a 
 term $0.0003 \text{\AA}^{-5} \tilde{R}_{IJ}^6$ was added to the exponential scaling 
function. 
Apparently, this modification enabled a better agreement 
with reference data for rare-gas compounds. \cite{Stewart2007}
The PM6 model defines additional special expressions which are only applied 
for certain compound classes or for the evaluation of the scaling 
factors for certain atom pairs (i.e., C--H, N--H, O--H, C--C, and 
Si--O). We discuss these minor modifications in Section~\ref{subsec:pm6mod}. 

Disturbingly, the PM6 model contains special corrections to the heats of formations at 298 K,  $\Delta H_f^{\text{298K}}$,  for several compound classes.
In the PM6 model, the predicted $\Delta H_f^{\text{298K}}$ in kcal~mol$^{-1}$ is empirically modified 
depending on a measure for the non-planarity of the amine nitrogen atom, $\phi$, \cite{Stewart2007}
\begin{equation}
 \Delta H_f^{\text{298K}} = \Delta H_f^{\text{298K,PM6}} - 
0.5\ \text{kcal}~\text{mol}^{-1} \exp{(-10 \phi)}.
\end{equation}
The measure for the non-planarity of the amine nitrogen atom is determined as 
 2$\pi$ minus the sum of the three bond angles involving the 
amine nitrogen atom. 
For a perfectly planar amine, $\Delta H_f^{\text{298K}}$ is reduced by 0.5 kcal~mol$^{-1}$. 
With an increasing pyramidalization of the amine, $\Delta H_f^{\text{298K}}$ is reduced 
by a smaller amount.
Additionally, the PM6 model (as implemented in \textsc{Mopac}) includes an undocumented 
modification to $\Delta H_f^\text{298K}$ when 
the computed bond order for a carbon--carbon bond exceeds 2.5, as for example, in acetylenic bonds.
A contribution of 12.0 kcal~mol$^{-1}$ is added to $\Delta H_f^\text{298K}$ for every detected 
acetylenic bond, e.g., \textsc{Mopac} outputs $\Delta H_f^\text{298K}=$ 57.4 kcal~mol$^{-1}$ 
for acetylene. If one applies the formulae specified in Ref.~\onlinecite{Stewart2007} instead, one would obtain 
$\Delta H_f^\text{298K}=$ 45.4 kcal~mol$^{-1}$ for this molecule.

The PM7 model was introduced as the successor of the PM6 model in 2013 \cite{Stewart2012}.
The largest changes were again made to the core-core repulsion energy.
It became evident that it is essential that the two-center ERIs decrease 
to the exact value at large distances
when applying a SEMO model in periodic calculations. \cite{Stewart2008}
Hence, Eq.~(\ref{eq:klopman}) was modified  so that 
the ERI $\left[s^Is^I\middle|s^Js^J\right]$
is approximated as, 
\begin{equation}
\begin{split}
\left[s^Is^I\middle|s^Js^J\right]^{\text{PM7}}
  = &
\frac{1}{\tilde{R}_{IJ}}
\exp{\left(-0.22 \text{\AA}^{-2} (\tilde{R}_{IJ}-7.0 \text{\AA})^2\right)}  \\ & + \left(1-
\exp{\left(-0.22 \text{\AA}^{-2} (\tilde{R}_{IJ}-7.0 \text{\AA})^2\right)}\right) \\ &
 \times \left[ \tilde{R}_{IJ}^2 + 
\left(\frac{1}{2 {  \gamma_{ss}^{Z_I}}} 
+  \frac{1}{2 {  \gamma_{ss}^{Z_J}}} \right)^2 \right]^{-\frac{1}{2}}.
\end{split}
\end{equation} 
The value 7.0 \AA\ was apparently chosen as some random distance which is far 
larger than usual bond lengths. \cite{Stewart2012}
This equation is also consulted to evaluate $V_{v}^{\text{PM7}}$,
 \begin{equation}
 \label{eq:pm7ecc}
 \begin{split}
   V_{v}^{\text{PM7}} =& \sum_{I=1}^N \sum_{J>I}^N 
\left\{  
Q_I Q_J
\left[\varrho_\text{core}^I\middle|\varrho_\text{core}^J\right]^{\text{PM7}} \cdot f_{IJ}^{\text{PM6}}
       + 10^{-8} \text{eV} \left(\frac{Q_I^{1/3} Q_J^{1/3}}{\tilde{R}_{IJ}} \right)^{12} 
\right\}.  \\
 \end{split}
 \end{equation}
In Eq.~(\ref{eq:pm7ecc}), we did not include the additional 
 empirical corrections for hydrogen bonding 
and dispersion interactions which are inherent to the PM7 model and 
described in Refs.~\onlinecite{Korth2010b,Stewart2012}.
Note that the description of dispersion interactions in PM7 creates a conceptual 
problem as pointed out by Grimme \textit{et al.}: \cite{Grimme2016}
Ref.~\onlinecite{Stewart2012} states that the dispersion energy is damped down and truncated at longer distances.
This is obviously not sensible for dispersion interactions which are long-range interactions
and was also shown to cause significant errors for larger systems. \cite{Brandenburg2014,Sure2015}

\section{The Orthogonalization-Corrected Models (OM$x$)}
\label{sec:omx}

The OM$x$ ($x=1,2,3$) models activate one $s$-type basis functions for hydrogen and 
one $s$- and three $p$-type basis functions 
for carbon, nitrogen, oxygen, and fluorine. \cite{Kolb1993a,KolbPhD,
Weber2000a,WeberPhD,ScholtenDiss,Dral2016b}
Each of these basis function consists of three primitive Gaussian functions, 
\cite{Stevens1984,Kolb1993a,Dral2016b} and hence,  we  
denote the basis sets for the OM1, OM2, and OM3 models with OM1-3G, OM2-3G, and OM3-3G,
respectively.
The OM$x$-3G basis sets are based on the ECP-3G basis set. \cite{Stevens1984}
The exponents of the primitive Gaussian functions of the ECP-3G basis  
are scaled with $\left(\zeta^{Z_I}\right)^2$  to yield the OM$x$-3G basis sets.
The factor $\zeta^{Z_I}$ is a parameter of the respective OM$x$ model. \cite{Dral2016b}
The OM$x$ models currently do only provide parameters for  
hydrogen, carbon, nitrogen, oxygen, and fluorine. \cite{Dral2016b}

\subsection{Approximation of Electron-Electron Repulsion Integrals}
\label{subsec:omxeri}

In analogy to MNDO-type models, the five one-center ERIs 
$\left<\chi_\mu^I \chi_\nu^I\middle| \chi_\lambda^I \chi_\sigma^I\right>$ 
arising in the minimal $s,p$
basis are substituted for the parameters $ \gamma_{ss}^{Z_I}$, $\gamma_{pp}^{Z_I}$, $\gamma_{sp}^{Z_I}$,
$\gamma_{pp'}^{Z_I}$, and $\tilde{\gamma}_{sp}^{Z_I}$ 
(see also Eqs.~(\ref{eq:gss})--(\ref{eq:gsp2})).

Within the OM$x$ models, the value of the two-center 
ERIs $\left<\chi_\mu^I \chi_\nu^I\middle| \chi_\lambda^J \chi_\sigma^J\right>, 
I\ne J$ is determined analytically. 
The analytical values of the two-center ERIs 
are then scaled with the so-called Klopman--Ohno factor $f^{\text{KO}}_{IJ}$ 
when assembling ${}^\phi \fett{G} \approx {}^\chi \fett{G}^{\text{NDDO}}$,
\begin{equation}
 \left<\chi_\mu^I \chi_\nu^I\middle| \chi_\lambda^J \chi_\sigma^J\right>^{\text{OM}x} = f^{\text{KO}}_{IJ} \cdot 
\left<\chi_\mu^I \chi_\nu^I\middle| \chi_\lambda^J \chi_\sigma^J\right>
\end{equation}
The Klopman--Ohno factor $f^{\text{KO}}_{IJ}$ is given as the quotient of the 
MNDO-type ERI $\left[s^I s^I |s^J s^J\right]$ (Eq.~(\ref{eq:klopman})) and 
the analytical ERI $\left<s^I s^I |s^J s^J\right>$, \cite{Kolb1993a}
\begin{equation}
 \label{eq:fko}
 \begin{split}
f^{\text{KO}}_{IJ} &=
\frac{\left[s^I s^I |s^J s^J\right]}{\left<s^I s^I |s^J s^J\right>} = 
\frac{
\left[ \tilde{R}_{IJ}^2 + 
\left(\frac{1}{2 {  \gamma_{ss}^{Z_I}}} 
+  \frac{1}{2 {  \gamma_{ss}^{Z_J}}} \right)^2 \right]^{-\frac{1}{2}}
}{\left<s^I s^I |s^J s^J\right>}.\\
 \end{split}
\end{equation}
When $\chi_\mu^I$, $\chi_\nu^I$, $\chi_\lambda^J$, and $\chi_\sigma^J$
are $s$-type basis functions, the product of $f^{\text{KO}}_{IJ}$ 
and $\left<s^I s^I |s^J s^J\right>$ reduces to the MNDO-type model, 
\begin{equation}
  \label{eq:omxInts_ssss}
 \begin{split}
 \left<s^I s^I |s^J s^J\right>^{\text{OM}x} =& \frac{\left[s^I s^I |s^J s^J\right]}{\left<s^I s^I |s^J s^J\right>}
\left<s^I s^I |s^J s^J\right> \\ =& \left[s^I s^I |s^J s^J\right]. \\
 \end{split}
\end{equation}
In the case of one basis function, $\chi_\mu^I$, $\chi_\nu^I$, $\chi_\lambda^J$, or $\chi_\sigma^J$,
not being an $s$-type basis function, the two-center ERI is still scaled with 
$f^{\text{KO}}_{IJ}$.

\begin{figure}[ht]
 \centering
 \includegraphics[width=.75\textwidth]{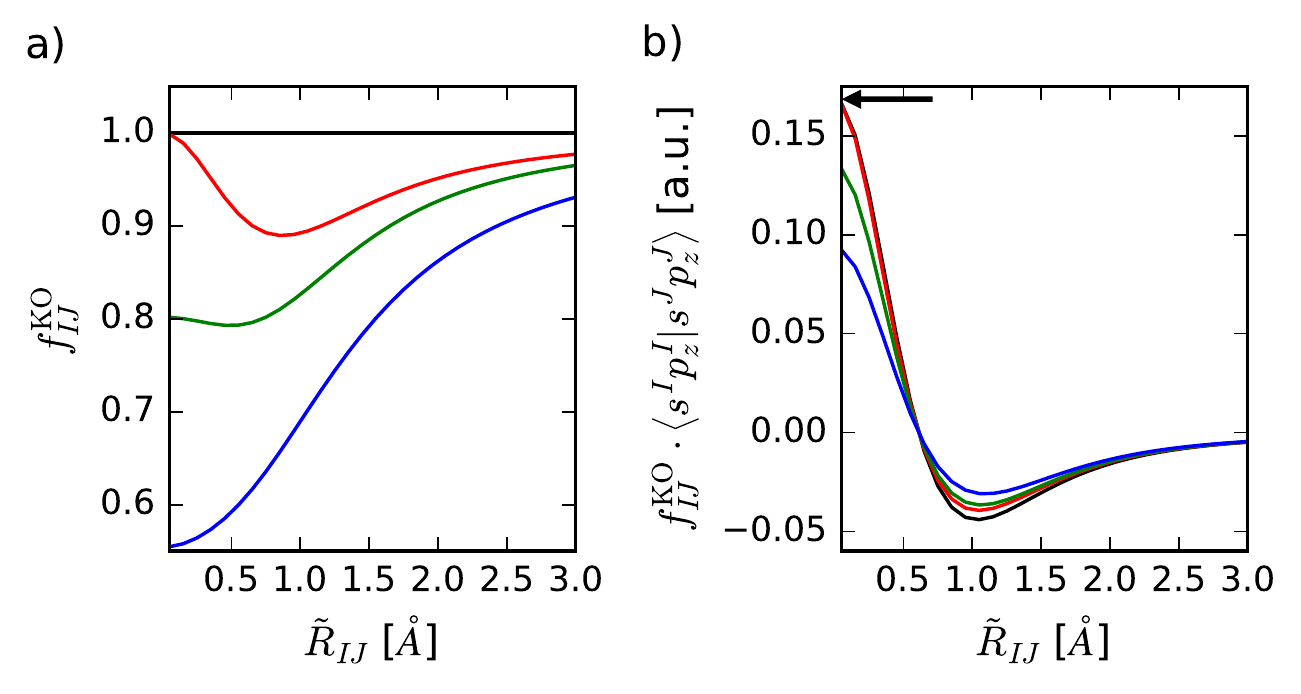}
 \caption{Dependence of a) the Klopman--Ohno factor 
$f^{\text{KO}}_{IJ}$ and b) the value of 
$f^{\text{KO}}$ multiplied by $\left<s^Ip_z^I | s^J p_z^J\right>$ on 
$\tilde{R}_{IJ}$ in 
a C$_2$ molecule described by an OM2-3G basis.
We show the values for $f^{\text{KO}}_{IJ}=1$ (black lines),
Eq.~(67) with $\gamma_{ss}^6 = \left<s^Is^I | s^I s^I\right> = $ 0.81 a.u.\ (analytical one-center limit, 
red lines), Eq.~(67) with $\gamma_{ss}^6 = $ 0.65 a.u.\ (green lines), and
Eq.~(67) with $\gamma_{ss}^6 = $ 0.45 a.u.\ (OM$x$ value \cite{Kolb1993a,Weber2000a,Dral2016b},
blue lines).
The analytical one-center limit $\left<s^Ip_z^I | s^I p_z^I\right>=0.17$ a.u.\ is highlighted 
by a black arrow and the OM$x$ one-center limit $\gamma_{sp}^{Z_I} =0.42 $ a.u.\ is not shown here.
}
\label{fig:fko}
\end{figure}

Let us now examine how this Klopman--Ohno scaling affects the value which enters 
the two-electron matrices.
Generally, $f^{\text{KO}}_{IJ}$ tends to one in 
the limit $\tilde{R}_{IJ}\rightarrow\infty$, i.e., the unscaled analytical value for 
$\left<\chi_\mu^I \chi_\nu^I\middle| \chi_\lambda^J \chi_\sigma^J\right>$ is applied.
It would be theoretically satisfactory if we had a smooth transition 
from the one-center ERIs 
$\left<\chi_\mu^I \chi_\nu^I\middle| \chi_\lambda^I \chi_\sigma^I\right>$ 
= $ \gamma_{ss}^{Z_I}$, $\gamma_{pp}^{Z_I}$, $\gamma_{sp}^{Z_I}$,
$\gamma_{pp'}^{Z_I}$, or $\tilde{\gamma}_{sp}^{Z_I}$
to 
the two-center ERIs 
$f^{\text{KO}}_{IJ} \cdot \left<\chi_\mu^I \chi_\nu^I\middle| \chi_\lambda^J \chi_\sigma^J\right>$.
This situation would occur if the analytical one-center limits were chosen for 
$ \gamma_{ss}^{Z_I}$, $\gamma_{pp}^{Z_I}$, $\gamma_{sp}^{Z_I}$,
$\gamma_{pp'}^{Z_I}$, and $\tilde{\gamma}_{sp}^{Z_I}$.
If $\gamma_{ss}^{Z_I}=\left<s^I s^I |s^I s^I\right>$, 
$f^{\text{KO}}$ would be one in the limit
$\tilde{R}_{IJ}=0$,
\begin{equation}
  \begin{split}
f^{\text{KO}}_{II} &=
\frac{\gamma_{ss}^{Z_I}}{\left<s^I s^I |s^I s^I\right>} = 
\frac{\left<s^I s^I |s^I s^I\right>}{\left<s^I s^I |s^I s^I\right>} = 1
.\\
 \end{split}
\end{equation}
Hence, all two-center ERIs are scaled with a factor of one in the limit 
$\tilde{R}_{IJ}=0$ and the analytical one-center limit is recovered.
In this case, the scaled two-center ERIs differ negligibly from the analytical two-center ERIs
 (see Figure~\ref{fig:fko}).

Usually, $\gamma_{ss}^{Z_I}$ is, however, chosen to be significantly  
smaller than the analytical one-center ERI limit ($\left<s^I s^I |s^I s^I\right> - \gamma_{ss}^{Z_I} = 0.36$
a.u. in our example in Figure~\ref{fig:fko}). 
Consequently, a theoretically unsatisfactory situation arises which we illustrate at the 
example of the $\left<s^Ip_z^I | s^J p_z^J\right>$ ERI in C$_2$.
The parametrized one-center ERI limits of interest are 
$\gamma_{ss}^{6} = 0.45$ a.u.\ and $\gamma_{sp}^{6} = 0.42$ a.u. \cite{Oleari1966}
We then observe a discontinuity from the point $\tilde{R}_{IJ}=0$ where 
$\left<s^Ip_z^I | s^J p_z^J\right>^{\text{OM}x} = \gamma_{sp}^{6} = 0.42$ a.u.\ 
to the point $\tilde{R}_{IJ}\ll1$ where $\left<s^Ip_z^I | s^J p_z^J\right>^{\text{OM}x} = 
f^{\text{KO}}_{IJ} \left<s^Ip_z^I | s^J p_z^J\right> = 0.09$
a.u.
While these discontinuities are unsatisfactory, they 
do not appear to lead to practical issues in the calculations.

\subsection{Approximation of the Symmetrically Orthogonalized One-Electron Matrix}
\label{subsec:omxH}

An exact orthogonalization of ${}^{\chi}\!{\fett{H}}$ 
(e.g., by the transformation ${}^{\chi}\!{\fett{S}}^{-\frac{1}{2}}
\;{}^{\chi}\!{\fett{H}}\;{}^{\chi}\!{\fett{S}}^{-\frac{1}{2}}$) 
 was initially attempted, but did not provide a useful model to claculate electroni energies \cite{Kolb1993a,
KolbPhD,Weber2000a} as confirmed by other studies. \cite{Spanget-Larsen1980, Duke1981, Gleghorn1982, 
Kollmar1995,Zhidomirov1996,Spanget-Larsen1997, Kollmar1997} 
Kolb and Thiel therefore decided to develop approximate orthogonalization 
corrections to be added to ${}^{\chi}\!{\fett{H}}$. \cite{Kolb1993a,
KolbPhD}
These \cite{Kolb1993a,KolbPhD,Weber2000a,WeberPhD}  and other 
\cite{Coffey1973,Nanda1980,Filatov1987,Jug1987} 
approximate orthogonalization corrections, 
are based on an expansion of ${}^{\chi}\!{\fett{S}}^{-\frac{1}{2}}$  
into a power series,
\begin{equation}
 \label{eq:powerseries}
\begin{split}
{}^{\chi}\!{\fett{S}}^{-\frac{1}{2}} = &
\left(\fett{1}+ \;{}^{\chi}\!{\fett{S'}} \right)^{-\frac{1}{2}}
 = 1- \frac{1}{2} {}^{\chi}\!{\fett{S'}} + \frac{3}{8} {}^{\chi}\!{\fett{S'}}^2
-\frac{5}{16} {}^{\chi}\!{\fett{S'}}^3 + \mathcal{O}(\;{}^{\chi}\!{\fett{S}}'^4),
\end{split}
\end{equation}
where $\;{}^{\chi}\!{\fett{S'}}$ is defined as
\begin{equation}
 \label{eq:sdash}
\;{}^{\chi}\!{\fett{S'}}= {}^{\chi}\!{\fett{S}}-\fett{1}.
\end{equation}
The transformation of $\;{}^{\chi}\!{\fett{H}}$ to $\;{}^{\phi}\!{\fett{H}}$, 
\begin{equation}
\label{eq:reihenentwicklung}
 \begin{split}
   \;{}^{\phi}\!{\fett{H}} =& 
(\;{}^{\chi}\!{\fett{S}})^{-\frac{1}{2}} \;{}^{\chi}\!{\fett{H}} 
(\;{}^{\chi}\!{\fett{S}})^{-\frac{1}{2}} = 
(\fett{1}+\;{}^{\chi}\!{\fett{S}}')^{-\frac{1}{2}} \;{}^{\chi}\!{\fett{H}} 
(\fett{1}+\;{}^{\chi}\!{\fett{S}}')^{-\frac{1}{2}} \\
 \end{split}
\end{equation}
is then approximated as 
\begin{equation}
\label{eq:reihenentwicklung2}
 \begin{split}
    \;{}^{\phi}\!{\fett{H}} \approx& {}\;{}^{\chi}\!{\fett{H}} - \frac{1}{2} ({}^{\chi}\!{\fett{S}'} 
\;{}^{\chi}\!{\fett{H}} + \;{}^{\chi}\!{\fett{H}} \;{}^{\chi}\!{\fett{S}}') + 
\frac{3}{8} ({}^{\chi}\!{\fett{S}}'^2\;{}^{\chi}\!{\fett{H}}  + 
\;{}^{\chi}\!{\fett{H}}\;{}^{\chi}\!{\fett{S}}'^2) \\ & + \frac{1}{4} \;{}^{\chi}\!{\fett{S}}' 
\;{}^{\chi}\!{\fett{H}} \;{}^{\chi}\!{\fett{S}}'+ \mathcal{O}(\;{}^{\chi}\!{\fett{S}}'^3).
\end{split}
\end{equation}
Accordingly, a matrix element ${}^{\phi}\!{H}_{\mu\nu}$ is approximated as 
\begin{equation}
 \label{eq:reihenentwicklung_matrixelements}
\begin{split}
  {}^{\phi}\!{H}_{\mu\nu} \approx& \;{}^{\chi}\!{H}_{\mu\nu} - \frac{1}{2} \sum_{\lambda=1}^M
\left( {}^{\chi}\!{S}'_{\mu\lambda}\;{}^{\chi}\!{H}_{\lambda\nu} + 
{}^{\chi}\!{H}_{\mu\lambda}\;
{}^{\chi}\!{S}'_{\lambda\nu}  \right) \\
& + \frac{1}{8} \sum_{\lambda=1}^M\sum_{\sigma=1}^M 
\left( 3 \;{}^{\chi}\!{S}'_{\mu\lambda}\;{}^{\chi}\!{S}'_{\lambda\sigma}\;{}^{\chi}\!{H}_{\sigma\nu} 
 +
3 \;{}^{\chi}\!{H}_{\mu\lambda}\;{}^{\chi}\!{S}'_{\lambda\sigma}\;{}^{\chi}\!{S}'_{\sigma\nu} +
2 \;{}^{\chi}\!{S}'_{\mu\lambda}\;{}^{\chi}\!{H}_{\lambda\sigma} \;{}^{\chi}\!{S}'_{\sigma\nu} 
\right).\\
\end{split}
\end{equation}
Gray and Stone showed \cite{Gray1970} that this power series expansion is 
nonconvergent in the general case. 
More specifically, the power series expansion fails to converge when the largest eigenvalue 
of ${}^{\chi}\!{\fett{S}'}$ exceeds $1.0$ \cite{Gray1970}
which is often the case (e.g., it is $1.3$ for methane 
and $2.1$ for benzene when applying an ECP-3G basis set).
Chandler and Grader \cite{Chandler1980} and Neymeyr \cite{Neymeyr1995a,Neymeyr1995b,Neymeyr1995c,
Neymeyr1995d,Neymeyr1995e} 
subsequently introduced alternative convergent 
power series expansions. These were, however, not applied to derive approximate 
orthogonalization corrections for the OM$x$ models, nor for any other NDDO-SEMO model.
The nonconvergence of Eq.~(\ref{eq:reihenentwicklung2}) does not appear to be a problem in practice, 
which may be attributed to the fact that it was only taken as a guideline to develop 
parametric expressions. Eq.~(\ref{eq:reihenentwicklung2}) is therefore not  directly
applied to carry out the basis transformation in the 
OM$x$ models (recall that the exact transformation of 
$^\chi\fett{H}$ from the $\chi$- to the $\phi$-basis does not yield a useful NDDO-SEMO model).

Analogously to the MNDO model, 
different parametric expressions are applied for the evaluation of the 
matrix elements ${}^{\phi}\!{H}_{\mu\nu}^{\text{OM}x}$ 
depending on the number of atoms on which the corresponding  basis functions $\chi$
are centered (one-center one-electron and two-center one-electron matrix elements).

\subsubsection{One-Center One-Electron Matrix Elements}
\label{subsubsec:omxH1c}
We first discuss how ${}^{\phi}\!{H}_{\mu\nu}^{\text{OM}x}$
is determined when the corresponding  basis functions $\chi$
are centered on the \textit{same} atom, i.e, $\chi_\mu^I$ and $\chi_\nu^I$ are both centered 
on the atom $I$.
In this case, the matrix elements ${}^{\phi}\!{H}_{\mu\nu}^{\text{OM}x}$ are given by 
\begin{equation}
 \label{eq:hsameatom_omx}
 \begin{split}
  {}^{\phi}\!{H}_{\mu\nu}^{\text{OM}x} =& 
  {}^{\chi}\!{H}_{\mu\nu}^{\text{OM}x} 
-
\frac{1}{2} {F_1^{Z_I}} 
\sum_{\lambda=1}^M 
\left( \;{}^{\chi}\!{S}'_{\mu\lambda} \theta(\chi_\nu^I,\chi_\lambda^J)
+ \theta(\chi_\mu^I,\chi_\lambda^J) \;{}^{\chi}\!{S}'_{\lambda\nu} \right) \\ & +
\frac{1}{8} {F_2^{Z_I}} \sum_{\lambda=1}^M \;{}^{\chi}\!{S}'
_{\mu\lambda} \;{}^{\chi}\!{S}'_{\lambda\nu}
\left( \eta(\chi_\mu^I,\chi_\lambda^J) + {\eta}(\chi_\nu^I,\chi_\lambda^J) - 
{\eta}(\chi_\lambda^J,\chi_\mu^I) - {\eta}(\chi_\lambda^J,\chi_\nu^I)
\right), \\
 \end{split}
\end{equation}
where $F_1^{Z_I}$ and $F_2^{Z_I}$ are element-dependent parameters. 
Additionally, Eq.~(\ref{eq:hsameatom_omx}) contains the functions 
 $\theta(\chi_\nu^I,\chi_\lambda^J)$ and $\eta(\chi_\mu^I,\chi_\lambda^J)$ which we will specify 
in the following paragraphs.
When comparing Eqs.~(\ref{eq:reihenentwicklung_matrixelements}) and (\ref{eq:hsameatom_omx}),
we notice several similarities and differences.
Firstly, both equations start with the corresponding matrix element in the $\chi$-basis.
The second contribution to ${}^{\phi}\!{H}_{\mu\nu}^{\text{OM}x}$
is similar to the second contribution to ${}^{\phi}\!{H}_{\mu\nu}$.
The analytical expression (Eq.~(\ref{eq:reihenentwicklung_matrixelements}))
does not contain the parameter $F_1^{Z_I}$ which implies that 
$F_1^{Z_I}$ should be close to one when the analytical expression 
is approximated (cf. Ref.~\onlinecite{Kolb1993a}).
The function $\theta(\chi_\nu^I,\chi_\lambda^J)$ appears to model 
an entry in ${}^{\chi}\!{\fett{H}}$.
More specifically, it models the matrix elements 
${}^{\chi}\!{H}_{\mu\lambda}$ when $\chi_\mu^I$ and $\chi_\lambda^J$ are centered on different 
atoms, i.e., $I\ne J$. 
This is apparent when studying ${}^{\chi}\!{S}'_{\mu\lambda}$
which is only different from zero when $\chi_\mu^I$ and $\chi_\lambda^J$ are centered on different atoms 
due to the condition of local orthogonality and the way in which ${}^{\chi}\!{\fett{S}'}$
is constructed (Eq.~(\ref{eq:sdash})).
In summary, the second contribution to Eq.~(\ref{eq:hsameatom_omx}) is identical to the 
analytical expression when $F_1^{Z_I}=1$ and $\theta(\chi_\mu^I,\chi_\lambda^J)=
{}^{\chi}\!{H}_{\mu\lambda}$.
The relation of the third contribution to Eqs.~(\ref{eq:hsameatom_omx}) and (\ref{eq:reihenentwicklung_matrixelements})
is harder to see. It includes the approximation that all four-center contributions are neglected, 
see Ref.~\onlinecite{Weber2000a} for a detailed derivation. Hence, it is impossible to 
establish a similar relationship between $F_2^{Z_I}$ and $\eta(\chi_\mu^I,\chi_\lambda^J)$ in 
the parametric expression
 and analogs in the analytical expression.
In the OM3 model, $F_2^{Z_I}=0$ whereas in the OM1 and OM2 models both 
orthogonalization corrections are considered.

We now examine how the contributions to ${}^{\phi}\!{H}_{\mu\nu}^{\text{OM}x}$
are evaluated.
Its first contribution, ${}^{\chi}\!{H}_{\mu\nu}^{\text{OM}x}$, on the right-hand side
of Eq.~(\ref{eq:hsameatom_omx}) 
is given by \cite{ScholtenDiss}
\begin{equation}
 \label{eq:hchisameatom_omx}
\begin{split}
{}^{\chi}\!{H}_{\mu\nu}^{\text{OM}x} =& 
U_{l(\mu)l(\nu)}^{Z_I} + 
\sum_{\substack{J=1\\ J\ne I}}^N \left<\chi_\mu^I\middle|\text{ECP}_J\middle|\chi_\nu^I\right> - 
\sum_{\substack{J=1\\ J\ne I}}^N \left\{
Q_J \left[\chi_\mu^I\chi_\nu^I|s^Js^J\right] \right. \\  &  \left. - f_{IJ}^{\text{KO}}
 Q_J \left<\chi_\mu^I\chi_\nu^I|s^Js^J\right> 
 + f^{\text{KO}}_{IJ}
\left<\chi_\mu^I\middle|\frac{Q_J}{|\fett{r}_{i}-\tilde{\fett{R}}_{J}|}\middle|\chi_\nu^I\right> \right\} 
, 
\end{split}
\end{equation} 
where the element- and orbital-type-dependent parameter $U_{l(\mu)l(\mu)}^{Z_I}$
replaces the calculation of $\left<\chi_\mu^I\middle|-\frac{1}{2}\nabla^2|\chi_\nu^I\right>-
\left<\chi_\mu^I\middle|\frac{Q_I}{|\fett{r}_{i}-\tilde{\fett{R}}_{I}|}|\chi_\nu^I\right>
+\left<\chi_\mu^I\middle|\text{ECP}_{I}\middle|\chi_\nu^I\right>$ 
(cf. the MNDO model). Kolb and Thiel explicitly stated that $U_{l(\mu)l(\mu)}^{Z_I}$ is also assumed 
to include all contributions from the core electrons. \cite{Kolb1993a}

The next term is the contribution from the ECP with all other atoms $J$. 
In the first OM$x$ model, OM1, the contributions to the 
one-electron matrix are evaluated analytically as 
presented in Refs.~\onlinecite{Stevens1984,Wadt1985}. 
The analytical results were then subjected to Klopman--Ohno scaling.\cite{Kolb1993a}
In the OM2 and OM3 models, the \textit{ab initio} effective core potential in the $\chi$-basis
was substituted for a semiempirical one, \cite{Weber2000a, Dral2016b}
\begin{equation}
\label{eq:omxecp}
\begin{split}
\left<\chi_\mu^I\middle|\text{ECP}_K\middle|\chi_\nu^J\right> \approx & 
 - \left<\chi_\mu^I|\omega^K\right> \theta(\chi_\nu^J,\omega^K) - 
\theta(\chi_\mu^I,\omega^K) \left<\omega^K|\chi_\nu^J\right>  \\  & 
-  \left<\chi_\mu^I|\omega^K\right> \left<\omega^K|\chi_\nu^J\right> 
W^{Z(K)}.\\ 
\end{split}
\end{equation}
In this expression, we introduced an auxiliary set of basis functions 
 $\fett{\omega}=\{\omega^I\}$ (no additional subscript index is necessary because there is at most 
one additional $s$-type basis function per atom). 
Each basis function $\omega$ is characterized by an orbital exponent $\zeta_{\omega}^{Z_I}$.
Note that an orbital $\omega^I$ is generally not locally orthogonal to $\chi_\mu^I$.
An additional element-dependent parameter $W^{Z(K)}$ enters 
Eq.~(\ref{eq:omxecp}).

The last contribution to Eq.~(\ref{eq:hchisameatom_omx}) describes the interaction 
of the charge distribution $\chi_\mu^I\chi_\nu^I$ with all other atomic nuclei $J\ne I$.
It is composed of three contributions: The first one is identical to the one 
applied to describe this interaction in the MNDO model (Eq.~(\ref{eq:gms})).
As stated in Section~\ref{subsubsec:mndoh1c}, this expression is the result of 
an error compensation between the so-called penetration integrals (Eq.~(\ref{eq:gmsapprox1}))
and the orthogonalization corrections. Because the OM$x$ models explicitly consider 
orthogonalization corrections, also the penetration integrals must be considered
which make up the last two terms (subjected to Klopman--Ohno scaling).
If $\chi_\mu^I$ and $\chi_\nu^I$ are $s$-type orbitals, 
Eq.~(\ref{eq:hchisameatom_omx}) will reduce to 
\begin{equation}
 \label{eq:hchisameatom_omx_wrong}
\begin{split}
{}^{\chi}\!{H}_{\mu\nu}^{\text{OM}x} =&
U_{ss}^{Z_I} -  
\sum_{\substack{J=1\\ J\ne I}}^N 
f^{\text{KO}}_{IJ}
\left<s^I\middle|\frac{Q_J}{|\fett{r}_{i}-\tilde{\fett{R}}_{J}|}\middle|s^I\right>   + 
\sum_{\substack{J=1\\ J\ne I}}^N  f^{\text{KO}}_{IJ}
\left<s^I\middle|\text{ECP}_J\middle|s^I\right>,
\end{split}
\end{equation} 
because the first two remaining terms in Eq.~(\ref{eq:hchisameatom_omx}) cancel out exactly,
\begin{equation}
\label{eq:hchisameatom_special}
\begin{gathered}
Q_J \left[s^Is^I|s^Js^J\right]- f_{IJ}^{\text{KO}}
 Q_J \left<s^Is^I|s^Js^J\right> = Q_J \left[s^Is^I|s^Js^J\right] \\ - Q_J \frac{\left[s^Is^I|s^Js^J\right]}
{\left<s^Is^I|s^Js^J\right>} 
  \left<s^Is^I|s^Js^J\right> 
=0. \\
\end{gathered}
\end{equation}
If either $\chi_\mu^I$ or $\chi_\nu^I$ are not $s$-type orbitals, 
Eq.~(\ref{eq:hchisameatom_special}) will not hold true and we will have a contribution 
$Q_J \left(
\left[\chi_\mu^I\chi_\nu^I|s^Js^J\right]- f_{IJ}^{\text{KO}} \left<\chi_\mu^I\chi_\nu^I|s^Js^J\right>
\right)$ to Eq.~(\ref{eq:hchisameatom_omx}). 
Generally, however, the differences in the results
of Eq.~(\ref{eq:hchisameatom_omx}) and Eq.~(\ref{eq:hchisameatom_omx_wrong})
are quite small. \cite{WeberPhD}

 \begin{figure}[ht]
  \centering 
  \includegraphics[width=.75\textwidth]{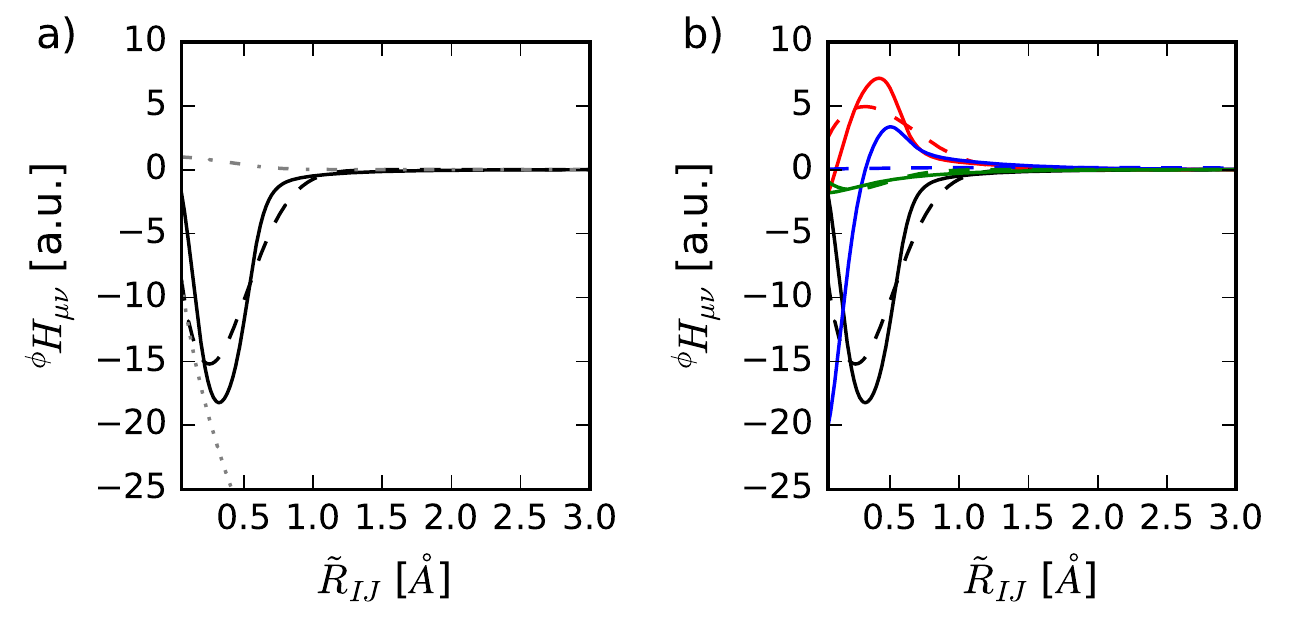}
  \caption{
Dependence of ${}^{\phi}\!{H_{\mu\nu}}$ (solid lines)
and $\theta(\chi_\mu^I,\chi_\nu^J)$ (dashed lines)
on the distance $\tilde{R}_{IJ}$ in a C$_2$ molecule described in an OM2-3G 
basis 
($\fett{\chi}=\left\{\chi_1^{C_1}, \chi_2^{C_1},
\chi_3^{C_1},\chi_4^{C_1},\chi_5^{C_2}, \chi_6^{C_2},
\chi_7^{C_2},\chi_8^{C_2}\right\}$ where the first and fifth basis functions 
are $2s$, the second and sixth are $2p_x$, the third and seventh are $2p_y$,
and the fourth and eighth are $2p_z$ basis functions).
a) $\theta(s^{C_1},s^{C_2})$ 
with the best-fit parameters $\beta_s^{6}=-38.9$ a.u.\ and 
 $a_s^{6}=1.0$ a.u. 
The contributions to $\theta(s^{C_1},s^{C_2})$ are divided into 
the scaled square-root contributions (dashed gray line) and 
the Gaussian contribution (dashed-dotted gray line). 
b) $\theta(s^{C_1},s^{C_2})$ with 
$\beta_s^{6}=-38.9$ a.u.\ and $a_s^{6}=1.0$ a.u.\ (black lines),
$\theta(s^{C_1},p_z^{C_2})$ with 
$\beta_s^{6}=
\beta_p^{6}=-11.2$ a.u.\ and $\alpha_s^{6}=\alpha_p^{6}=0.6$ a.u.\ (red lines),
$\theta(p_z^{C_1},p_z{C_2})$ with $\beta_p^{6}=-0.2$ a.u.\ 
and $\alpha_p^{6}=0.1$ a.u.\ (blue lines), and
$\theta(p_x^{C_1},p_x{C_2})$  with $\beta_p^{6}=-4.2$ a.u.\ 
and $\alpha_p^{6}=1.2$ a.u.\ (green lines). 
}
 \label{fig:resonanceIntsOMX}
 \end{figure}
The function $\theta(\chi_\mu^I,\chi_\lambda^J)$, \cite{Kolb1993a}
\begin{equation} 
 \label{eq:omxBeta}
\begin{split}
 \theta(\chi_\mu^I,\chi_\lambda^J) =& (-1)^{l(\lambda)+m(\lambda)}
\frac{{  b_{l(\mu)}^{Z_I}} + {  b_{l(\lambda)}^{
Z_J}}}{2} \sqrt{\tilde{R}_{IJ}} 
\exp{\left(-({ a_{l(\mu)}^{Z_I}} + 
{ a_{l(\lambda)}^{Z_J}}) \tilde{R}_{IJ}^2 \right)},
\end{split}
\end{equation}
 consists of the product of 
a phase vector, $(-1)^{l(\lambda)+m(\lambda)}$, 
 the scaled square-root of the interatomic distance $\tilde{R}_{IJ}$, 
and a Gaussian contribution depending on $\tilde{R}_{IJ}^2$ (see Figure~\ref{fig:resonanceIntsOMX}).
The function $\theta(\chi_\mu^I,\chi_\lambda^J)$ is evaluated in the same local coordinate 
system which is applied in MNDO-type methods (see Section~\ref{subsec:mndo2ceri}) and 
have to be transformed accordingly. 
The scaling factor for the square-root of the interatomic distance $\tilde{R}_{IJ}$ is 
determined from element- and orbital-type-dependent parameters 
${  b_{l(\mu)}^{Z_I}}$. 
The width of the Gaussian function is determined by element- and orbital-type-dependent parameters
${ a_{l(\mu)}^{Z_I}}$, it determines how fast the whole function 
approaches to zero.
The parameters ${ a_{l(\mu)}^{Z_I}}$ must be positive to obtain a sensible 
expression.
Within OM$x$ models, the parameters ${b_{l(\mu)}^{Z_I}}$ are 
unanimously negative. 
Hence, the phase vector determines the sign of $\theta(\chi_\mu^I,\chi_\lambda^J)$.
Consequently, the sign of $\theta(\chi_\mu^I,\chi_\lambda^J)$ does not depend on $\tilde{R}_{IJ}$. 
The function $\theta(\chi_\nu^I,\chi_\lambda^J)$ was designed to emulate 
${}^{\phi}\!{{H}}_{\mu\lambda}$. \cite{Kolb1993a} 
The matrix element ${}^{\phi}\!{H}_{\mu\lambda}$, 
which the function $\theta(\chi_\mu^I,\chi_\lambda^J)$ is supposed to 
model, may, however, have a different sign for different $\tilde{R}_{IJ}$.
In fact, Kolb and Thiel included an example where 
${}^{\phi}\!{H}_{\mu\lambda}$ changes its sign for different 
$\tilde{R}_{IJ}$
in Figure~2 of Ref.~\onlinecite{Kolb1993a} (see also Figure~\ref{fig:resonanceIntsOMX} in the 
present work).
Furthermore, it is interesting that no ECP is explicitly considered for the two-center 
one-electron matrix elements which is apparently assumed to be absorbed into Eq.~(\ref{eq:omxBeta}).

Finally, we need to evaluate the function $\eta(\chi_\mu^I,\chi_\lambda^J)$
to assemble Eq.~(\ref{eq:hsameatom_omx}).
It is evaluated similarly to the corresponding local one-electron matrix element 
in the MNDO model.
The function $\eta(\chi_\mu^I,\chi_\lambda^J)$ is evaluated as \cite{WeberPhD}
\begin{equation}
 \label{eq:eta_omx}
 \eta(\chi_\mu^I,\chi_\lambda^K) = U_{\mu\mu}^{Z_I} -
Q_K \left[\chi_\mu^I\chi_\mu^I|s^Ks^K\right] 
\end{equation}
(cf. Eq.~(\ref{eq:hsameatom_mndo})).
 Note that the order in which the basis functions are written 
in the function matters, i.e., $\eta(\chi_\mu^I,\chi_\lambda^K)\ne\eta(\chi_\lambda^K,\chi_\mu^I)$.
The function $\eta(\chi_\mu^I,\chi_\lambda^J)$ is not rotationally invariant, so that 
the functions have to be averaged when $\chi_\mu^I$ is a $p$-type basis function, \cite{WeberPhD}
\begin{equation}
 \label{eq:eta_omx2}
 \eta(p^I,\chi_\lambda^K) = \frac{1}{3} \left(
\eta(p_x^I,\chi_\lambda^K) + 
\eta(p_y^I,\chi_\lambda^K) + 
\eta(p_z^I,\chi_\lambda^K)
\right).
\end{equation}

\subsubsection{Two-Center One-Electron Matrix Elements}
\label{subsubsec:omxH2c}

While Kolb and Thiel explicitly pointed out that $\theta(\chi_\mu^I,\chi_\lambda^J)$ is assumed 
to contain orthogonalization corrections, \cite{Kolb1993a}
it turned out that it cannot accomplish this fully. \cite{Weber2000a}
As a remedy, Weber and Thiel \cite{Weber2000a} developed an orthogonalization correction 
for matrix elements $\;{}^{\phi}\!{H}_{\mu\nu}^{\text{OM}x}$ 
for which $\chi_\mu^I$ and $\chi_\nu^J$ are centered on different atoms,
\begin{equation}
\label{eq:omxHorth}
\begin{split}
 \;{}^{\phi}\!{H}_{\mu\nu}^{\text{OM}x} =& \theta(\chi_\mu^I,\chi_\nu^J) 
  - 
\frac{1}{2} {\frac{G_1^{Z_I}+G_1^{Z_J}}{2}} 
\sum_{\lambda=1}^M (1-\delta_{IK})(1-\delta_{JK})
\left( \;{}^{\chi}\!{S}'_{\mu\lambda} \theta(\chi_\nu^I,\chi_\lambda^K)
\right. \\ & \left. 
+ \theta(\chi_\nu^J,\chi_\lambda^K) \;{}^{\chi}\!{S}'_{\lambda\nu} \right) + 
\frac{1}{8} {\frac{G_2^{Z_I}+G_2^{Z_J}}{2}} \sum_{\lambda=1}^M 
(1-\delta_{IK})(1-\delta_{JK})
\;{}^{\chi}\!{S}'
_{\mu\lambda} \;{}^{\chi}\!{S}'_{\lambda\nu} \\ & \times 
\left( \eta(\chi_\mu^I,\chi_\lambda^K) + {\eta}(\chi_\nu^J,\chi_\lambda^K)  - 
{\eta}(\chi_\lambda^K,\chi_\mu^I) - {\eta}(\chi_\lambda^K,\chi_\nu^J)
\right), \\
\end{split}
\end{equation}
which contains the element-dependent parameters 
$G_1^{Z_I}$ and $G_2^{Z_I}$.
This equation significantly differs from the analytic expression for the 
transformation of the matrix elements (Eq.~(\ref{eq:reihenentwicklung_matrixelements})).
Most importantly, the corrections do \textit{only} include terms with basis functions 
which are centered on a third atom ($K\ne I\ne J$; 
indicated by $(1-\delta_{IK})(1-\delta_{JK})$ in Eq.~(\ref{eq:omxHorth})).
Other than that similar considerations apply as in the one-center case. 
The similarity of Eqs.~(\ref{eq:hchisameatom_omx}) and (\ref{eq:omxHorth}) might imply that 
$ F_1^{Z_I} \approx 2 \cdot G_1^{Z_I}$ 
and $ F_2^{Z_I} \approx 2 \cdot G_2^{Z_I}$
which is, however, generally not the case.

In the earliest variant, OM1, $G_1^{Z_I}=0$ and $G_2^{Z_I}=0$, i.e.,
orthogonalization corrections are only considered when  
$\chi_\mu^I$ and $\chi_\nu^J$ are centered on a single atom.
The latest  version, OM3, sets $F_2^{Z_I}=0$ and $G_2^{Z_I}=0$.
The OM2 model considers all orthogonalization corrections.

\subsection{Empirical Scaling of the Core-Core Repulsion Energy}
\label{subsec:omxrepulsion}

The contribution of each pairwise repulsion of atomic cores $I, J\ne I$
 is scaled with the Klopman--Ohno factor $f^{\text{KO}}_{IJ}$,
\cite{Kolb1993a,Weber2000a,ScholtenDiss,Dral2016b}
\begin{equation}
\label{eq:eccomx}
 V_{v}^{\text{OM}x} = \sum_{I=1}^N \sum_{J>I}^N f_{IJ}^{\text{KO}} \frac{Q_I Q_J}{\tilde{R}_{IJ}}.
\end{equation}
It is argued \cite{Kolb1993a,Weber2000a,ScholtenDiss,Dral2016b} 
that the core-core repulsion energy needs to be reduced 
for small interatomic distances to ensure a balance within the model 
(ERIs and contributions to the core-electron attraction terms are 
also scaled with $f^{\text{KO}}_{IJ}$).
In contrast to the MNDO-type core-core repulsion (Eqs.~(\ref{eq:mndoNuclearEnergy}), 
(\ref{eq:am1NuclearEnergy}), 
and (\ref{eq:pm6ecc})), 
Eq.~(\ref{eq:eccomx}) has a singularity for $\tilde{R}_{IJ}=0$.
When $\tilde{R}_{IJ}\rightarrow\infty$, $f^{\text{KO}}_{IJ}\rightarrow 1$, i.e, 
the core-core repulsion energy approximates the point-charge model asymptotically.

\section{Other NDDO-SEMO models}
\label{sec:othermethods}

Several other models, which have not found widespread popularity, 
introduce new conceptual ideas beyond the MNDO-type and OM$x$ models.
In the following sections, several of these ideas are reviewed and discussed without 
providing the complete formalism for these models.
This list is by no means complete. We selected models which feature 
conceptually large differences to the introduced models, but are still built around 
the NDDO approximation.
For other NDDO-SEMO models see also 
Refs.~\onlinecite{Roby1969,Chandrasekhar1976,Chandrasekhar1979,
Dewar1993, Holder1994, Holder1994a, Zhidomirov1996,
Laikov2007,Chang2008a,  Laikov2011, Laikov2011a}. 

\subsection{The Nonorthogonalized Modified Neglect of Differential Overlap (NO-MNDO) Model}
\label{subsec:nomndo}

\begin{figure}[ht] 
 \centering 
 \includegraphics[width=\textwidth]{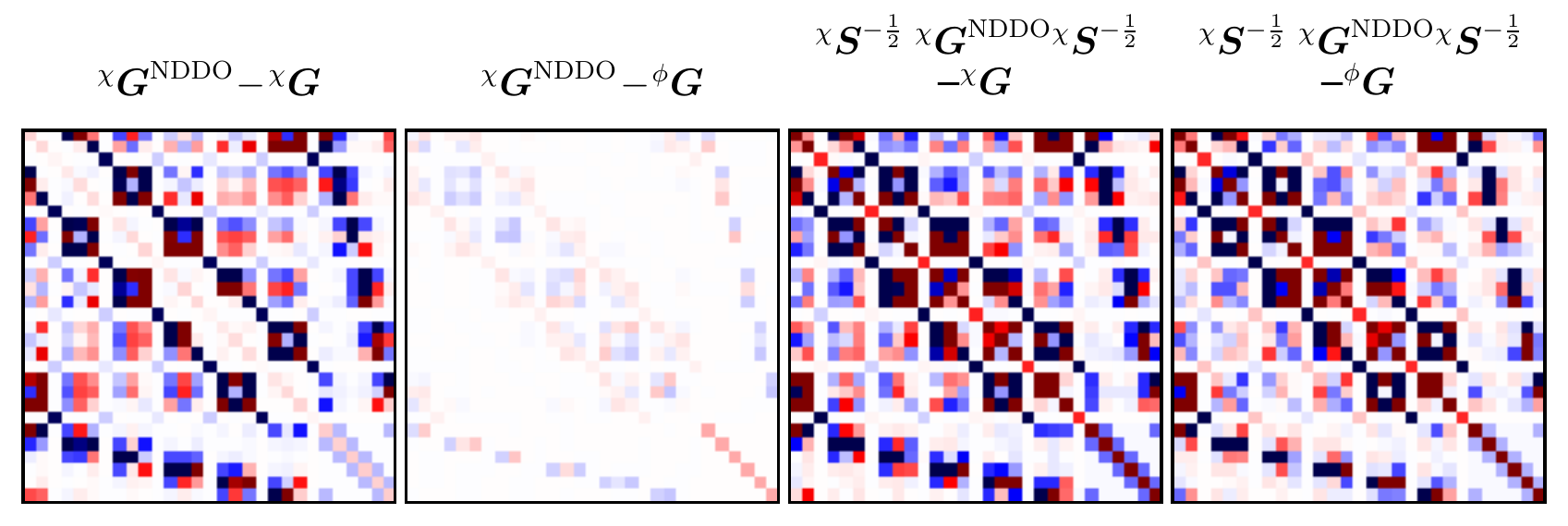}
 \caption{Graphical representation of  
${}^{\chi}{\fett{G}}^{\text{NDDO}}\!-\!{}^{\chi}{\fett{G}}$ (left), 
${}^{\chi}{\fett{G}}^{\text{NDDO}}\!-\!{}^{\phi}{\fett{G}}$ (middle left), 
${}^{\chi}\!{\fett{S}}^{-\frac{1}{2}} {}^{\chi}{\fett{G}}^{\text{NDDO}}
 \;{}^{\chi}\!{\fett{S}}^{-\frac{1}{2}}\!-{}^{\chi}{\fett{G}}$ (middle right), and  
${}^{\chi}\!{\fett{S}}^{-\frac{1}{2}} {}^{\chi}{\fett{G}}^{\text{NDDO}}
 \;{}^{\chi}\!{\fett{S}}^{-\frac{1}{2}}\!-{}^{\phi}{\fett{G}}$ (right)
for a benzene molecule described in an MNDO-3G 
basis.
The entries are colored according to their values from negative (blue) to zero (white) to positive (red).
}
 \label{fig:nomndo}
\end{figure}

An obvious weakness of the MNDO model is the lack of explicit orthogonalization 
corrections to $\fett{H}$ (Section~\ref{subsec:mndoH}). 
Sattelmeyer \textit{et al.} claimed that this problem can be addressed by  
introducing the overlap matrix into the SCF equations \cite{Sattelmeyer2006} and  
coined the name \textit{Nonorthogonalized Modified Neglect of Differential Overlap} (NO-MNDO) model
for this procedure. \cite{Sattelmeyer2006}
Let us neglect the presence of parameters for the moment and assume that all integrals 
are calculated analytically. Their suggestion then reads, in our notation,
\begin{equation}
\begin{split}
  \left(\;{}^{\chi}\!{\fett{H}} + \;{}^{\chi}\!{\fett{G}}^{\text{NDDO}} \right)
\;{}^{\chi}\!{\fett{C}} &= 
  \;{}^{\chi}\!{\fett{S}} \;{}^{\chi}\!{\fett{C}} \fett{\epsilon}, \\
\end{split}
\end{equation}
which can be reformulated to 
\begin{equation}
\begin{split}
  \left( {}^{\phi}\!{\fett{H}}
+ \;{}^{\chi}\!{\fett{S}}^{-\frac{1}{2}} \;{}^{\chi}{\fett{G}}^{\text{NDDO}}
 \;{}^{\chi}\!{\fett{S}}^{-\frac{1}{2}} \right)
\;{}^{\phi}\!{\fett{C}} &= 
  \;{}^{\phi}\!{\fett{C}} \fett{\epsilon}, \\
\end{split}
\end{equation}
when applying Eq.~(\ref{eq:gnddo}).
While it is obvious that the one-electron matrix is explicitly orthogonalized, a
problem arises.
The NDDO approximation emulates a basis transformation for 
$\fett{G}$ (cf. Eq.~(\ref{eq:gnddo})).
The application of a matrix transformation to ${}^{\chi}{\fett{G}}^{\text{NDDO}}$ is therefore  
not sensible.
When applying the matrix transformation again, 
we end up with a situation where 
\begin{equation}
 \;{}^{\chi}\!{\fett{S}}^{-\frac{1}{2}} {}^{\chi}{\fett{G}}^{\text{NDDO}}
 \;{}^{\chi}\!{\fett{S}}^{-\frac{1}{2}} \not\approx \;{}^{\phi}{\fett{G}}
\end{equation}
 and
\begin{equation}
 \;{}^{\chi}\!{\fett{S}}^{-\frac{1}{2}} {}^{\chi}{\fett{G}}^{\text{NDDO}}
 \;{}^{\chi}\!{\fett{S}}^{-\frac{1}{2}} \not\approx \;{}^{\chi}{\fett{G}}
\end{equation} 
(see Figure~\ref{fig:nomndo}).

According to Sattelmeyer \textit{et al.} \cite{Sattelmeyer2006} the NO-MNDO model appeared to 
significantly improve on the MNDO model .
We speculate that the good performance of the NO-MNDO model might be due to a combination of two reasons:
(i) The parameters in the NO-MNDO model might provide a sufficient flexibility to remedy the conceptual 
shortcomings and 
(ii) the explicit orthogonalization of $^\chi\fett{H}$ might outweigh 
the conceptual error to a certain extent. The improvements of NO-MNDO in areas 
which are typically 
associated with orthogonalization errors for MNDO  \cite{Sattelmeyer2006} (e.g., wrong 
barriers of rotations about single bonds) might be taken as an indicator for this statement.

In any case, it appears promising to attempt the construction of a similar (MNDO-type) model 
 which does not share the conceptual difficulties of NO-MNDO.
Without considering the parametrization, the corresponding SCF equations might read,  
\begin{equation}
\begin{split}
  \left( \;{}^{\chi}\!{\fett{S}}^{-\frac{1}{2}} \;{}^{\chi}\!{\fett{H}} 
\;{}^{\chi}\!{\fett{S}}^{-\frac{1}{2}}
+ \;{}^{\chi}\!{\fett{G}}^{\text{NDDO}} \right)  
\;{}^{\phi}\!{\fett{C}} =& 
  \;{}^{\phi}\!{\fett{C}} \fett{\epsilon}.  \\ 
\end{split}
\end{equation}
It might be possible that such an attempt could result in a model 
which is more accurate than NO-MNDO, and hence, significantly more accurate than MNDO.

\subsection{The Polarized Molecular Orbital (PMO$x$) Models}
\label{subsec:pmox}

The \textit{Polarized Molecular Orbital} (PMO$x$, $x=1,2$) models 
\cite{Fiedler2011, Zhang2011,Zhang2012,Isegawa2013,   Fiedler2014} 
were developed in an attempt to provide a more accurate 
description of noncovalent interactions and polarization 
effects than possible with the standard NDDO-SEMO models.
It is built upon the MNDO model, but features a key difference in its formalism:
 The PMO$x$ models activate one $s$- and three $p$-type basis functions for 
hydrogen (compared to only one $s$-type basis function for MNDO).
Truhlar and co-workers determined that the activation of diffuse $p$-type  
basis functions for hydrogen atoms is already sufficient to obtain a significant improvement 
in the description of polarization effects in \textit{ab initio} studies. 
\cite{Fiedler2011,Zhang2011,Zhang2012}
Similar results were also published before in a different context. \cite{Jug1976,Jug1993,Zuber2004}
Such a basis set, nevertheless, fulfills the condition of local orthogonality. 
The addition of $p$-type basis functions for hydrogen atoms was, furthermore, accompanied by 
 changes to the parametric expressions applied to evaluate the 
one-electron matrix elements and the core-core repulsion 
energy. \cite{Isegawa2013}

\subsection{The Machine Learning OM2 (ML-OM2) Model}
\label{subsec:mlom2}

Dral, von Lilienfeld, and Thiel suggested \cite{Dral2015a} to combine machine 
learning techniques with 
NDDO-SEMO models which resulted in the machine learning OM2 (ML-OM2) model.
The formalism of the ML-OM2 model is identical to that of the OM2 model.
It differs from the OM2 model only in the value of the parameter 
$\left(\zeta^6\right)^2$ with which the exponents of the primitive Gaussian functions of the ECP-3G basis functions 
for carbon are scaled. \cite{Dral2015a}
Dral \textit{et al.} applied \cite{Dral2015a} kernel ridge regression to predict 
$\left(\zeta^6\right)^2$ for individual molecules, i.e., $\left(\zeta^6\right)^2$ was 
not assumed to be a constant element-dependent parameter in ML-OM2. As a consequence, 
the resulting model offers a much greater flexibility. 
The mean absolute error in prediced atomization 
enthalpies could be reduced from 26.4 kJ~mol$^{-1}$ with OM2 to 7.1 
kJ~mol$^{-1}$ with ML-OM2 for a test set of organic molecules. \cite{Dral2015a}

\subsection{The High-Performance Computer-Aided Drug Design (hpCADD) Model}
\label{subsec:hpCADD}

Very recently, Thomas \textit{et al.} introduced the 
\textit{High-Performance Computer-Aided Drug Design} (hpCADD) model \cite{Thomas2017}
which differs from an MNDO-type model in the dependence of the parameters.
In the MNDO model, all parameters are element-dependent.
Thomas \textit{et al.} proposed to adopt 
the concept of `atom types' (well-known for force fields) 
into an MNDO-type model, i.e., they proposed to 
make the parameters in the MNDO model \textit{environment-dependent}.
\cite{Thomas2017}
E.g., hpCADD does not only comprise one parameter set for sulfur, but 
 separate sets of 
parameters for a sulfur atom which is part of a $\pi$-system (such as the one in 
thiophene) and for a sulfur atom which is part of a thiol group. \cite{Thomas2017} 
Hence, this conceptually follows the introduction of \textit{valence states}
of atoms in molecules, which is known to advance parametrized concepts such as 
electronegativity. \cite{Hinze1961}
We will come back to the advantages and disadvantages which are associated with such 
an approach in Section~\ref{subsubsec:domain}.

\section{Implicit Description of Electron Correlation Effects through Parametrization}
\label{subsec:param}

So far, we have discussed how NDDO-SEMO models approximate 
 the SCF equations in the $\phi$-basis (Eq.~(\ref{eq:roothanhallS})).
Historically, NDDO-SEMO models were developed to reproduce experimental data 
rather than, e.g., HF data. 
\cite{Dewar1969,Ridley1973,Dewar1977,Dewar1985} 
Consequently, NDDO-SEMO models have to be able to capture electron correlation 
effects in some manner.
The most popular way to describe 
electron correlation effects is implicit, i.e., through the calibration of the parameters
incorporated in the NDDO-SEMO model.

\subsection{Parallels to Kohn--Sham Density-Functional Theory}
\label{par:ksdft}

A comparison to correlation functional derivations of KS-DFT 
is likely to highlight insufficiencies in the description of electron correlation 
 in a parametrized  single-determinant approach. 
The comparison of the elements of the Fock matrix for KS-DFT in the $\chi$-basis, 
\begin{equation}
\label{eq:fockmatrix_dft}
\begin{split}
 \;{}^{\chi}\!{\fett{F}}^{\text{KS-DFT}} &= 
\;{}^{\chi}\!{\fett{H}}
+ \;{}^{\chi}\!{\fett{J}}
+ (1-\Lambda) \;{}^{\chi}\!{\fett{K}}
 + \Lambda \fett{V}^{x} + \fett{V}^c, \\
\end{split}
\end{equation}
highlights the connection between KS-DFT and HF (through $\Lambda$).
The parameter $\Lambda$ quantifies the amount of exact (HF) exchange ${}^{\chi}\!{\fett{K}}$.
For $0< \Lambda \le 1$, we have a contribution of 
a (approximate) DFT contribution to the exchange potential, 
$\fett{V}^{x}$, to $\;{}^{\chi}\!{\fett{F}}^{\text{KS-DFT}}$. \cite{Cohen2012}
Additionally, $\fett{V}^c$ is the DFT description of the correlation potential. HF theory does not consider a correlation potential so 
that when $\Lambda=0$ and no $\fett{V}^c$ is considered then ${}^{\chi}\!{\fett{F}}^{\text{KS-DFT}} = \;{}^{\chi}\!{\fett{F}}^{\text{HF}}$.

Yang and co-workers classified the most severe drawbacks in 
KS-DFT at the example of fractional electrons and of fractional spins
for the prototypical molecules H$_2$ and H$_2^+$. 
\cite{Perdew1982,Cohen2008, Cohen2012,Mori2014}
The energy for a system with a fractional number of electrons (or a fractional spin) 
is given by the straight line connecting the energies for the system with integer electron numbers 
(or integer spins). \cite{Perdew1982,Cohen2008, Cohen2012,Mori2014}
Approximate density functionals and HF are not able to correctly reproduce this behavior which 
may be interpreted as the source of many failures of approximate KS-DFT and HF models 
(such as delocalization and static correlation errors). \cite{Perdew1982,Cohen2008, Cohen2012,Mori2014}

\begin{figure}[ht]
 \centering 
 \includegraphics[width=.45\textwidth]{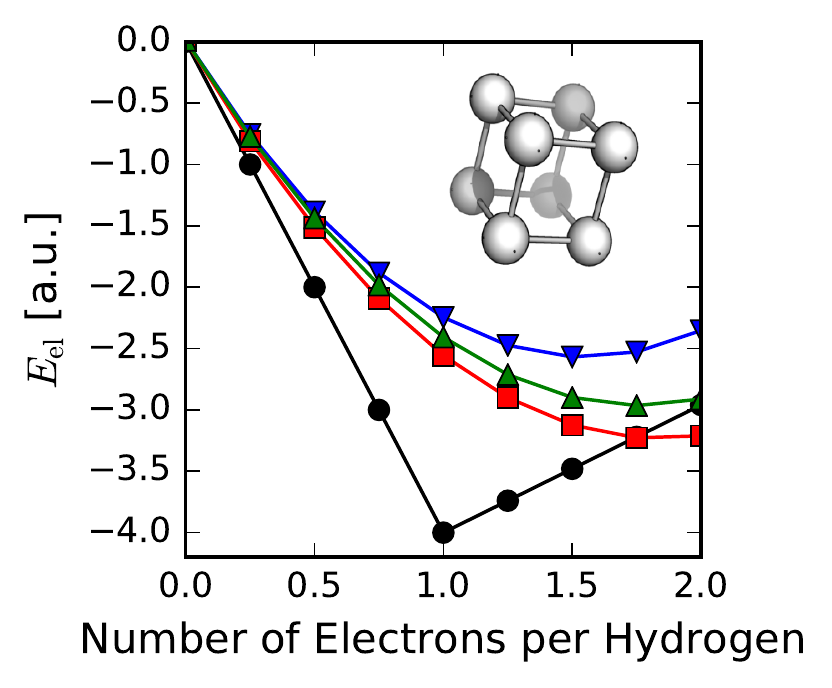}
 \caption{MNDO (red squares), AM1 (green triangles), and PM6 (blue triangles) electronic energies in a.u.\ 
for a closed-shell H$_8$ cube with an edge length of 1000 \AA\ and 0, 2, 4, 6, 8, 
10, 12, 14, and 16 electrons. The FCI reference energies (black circles) were taken from
Ref.~\onlinecite{Mori2014}. 
}
\label{fig:flatplane}
\end{figure}

We may also study NDDO-SEMO models in this respect for which 
we choose as an example an H$_8$ cube, whose structure is  
described in Ref.~\onlinecite{Mori2014}
(see Figure~\ref{fig:flatplane}).
Not surprisingly, the NDDO-SEMO models are not able to describe the discontinuities in the 
energy for integer electron numbers per hydrogen atom (see Figure~\ref{fig:flatplane}). 
The NDDO-SEMO models can also not be reparametrized to yield such a behavior.
We emphasize that the significance of this result is not the inability
of NDDO-SEMO models to accurately describe H$_8$
with different electron numbers.
Rather, it shows that NDDO-SEMO models fail to describe 
the quantum mechanical interaction of electrons 
in the same way as HF and approximate KS-DFT fail to do this.
We may therefore take this failure as an indication that NDDO-SEMO models share the same systematic errors as 
approximate KS-DFT and HF models. Hence, 
these systematic errors \textit{cannot} be alleviated through 
parametrization of the existing NDDO-SEMO models.
However, these fundamental errors may not severely affect the equilibrium 
structures of organic molecules, \cite{Cohen2012} 
but they will have a larger effect for non-equilibrium structures of organic molecules 
and for molecules with a more complicated electronic structure such as transition-metal complexes.  \cite{Cohen2012}

\subsection{NDDO-SEMO Models for Isolated Atoms}
\label{par:atomicdata}

Historically, NDDO-SEMO models are built upon considerations 
for isolated atoms. \cite{Pople1965,Oleari1966}
Studying isolated atoms has two distinct advantages: (i)
We do not have to consider orthogonalization effects 
($\fett{\chi}=\fett{\phi}$ for an isolated atom) and 
(ii) the NDDO approximation is no approximation in this special case. 
The one-electron matrix elements $^\phi H_{\mu\nu} = {}^\chi H_{\mu\nu}$ 
are equal to the corresponding one-center parameters $U_{l(\mu)l(\nu)}^{Z_I}$.
The one-center ERIs ($\gamma_{ss}^{Z_I}$, 
$\gamma_{pp}^{Z_I}$,
$\gamma_{sp}^{Z_I}$, 
$\gamma_{pp'}^{Z_I}$, and 
$\tilde{\gamma}_{sp}^{Z_I}$,
and if $d$-type orbitals are activated, the additional parameters specified 
in Section~\ref{subsubsec:mndod})
will enter the two-electron matrix.
For each element, the one-center parameters originally are calibrated \cite{Oleari1966,Sichel1967,DiSipio1971,Dewar1972}
with respect 
to reference electronic energies $E_\text{el}^\text{ref}$ for isolated atoms 
and monatomic ions (e.g., the one-center parameters for carbon (C) are calibrated with respect to $E_\text{el}^\text{ref}$
 for C$^{3+}$ ($n_v=1$), C$^{2+}$ ($n_v=2$), C$^{+}$ ($n_v=3$), 
C ($n_v=4$), and C$^{-}$ ($n_v=5$) in Ref.~\onlinecite{Oleari1966}).
$E_\text{el}^\text{ref}$ may be approximately determined 
from atomic ionization energies \cite{Oleari1966,Sichel1967,DiSipio1971,Dewar1972}, 
or, quite recently, from coupled cluster data. \cite{Margraf2016}
Refs.~\onlinecite{Oleari1966} and \onlinecite{Margraf2016} showed that it is not possible 
to achieve a good agreement between $E_\text{el}^\text{ref}$ and 
$E_\text{el}^\text{NDDO-SEMO}$ with a single element-dependent parameter set for a range of 
monatomic ions.
Margraf and co-workers, however, achieved \cite{Margraf2016} a good agreement with the first 
ionization potentials and electron affinities for the neutral atoms. 

\begin{figure}[ht]
 \centering 
 \includegraphics[width=.45\textwidth]{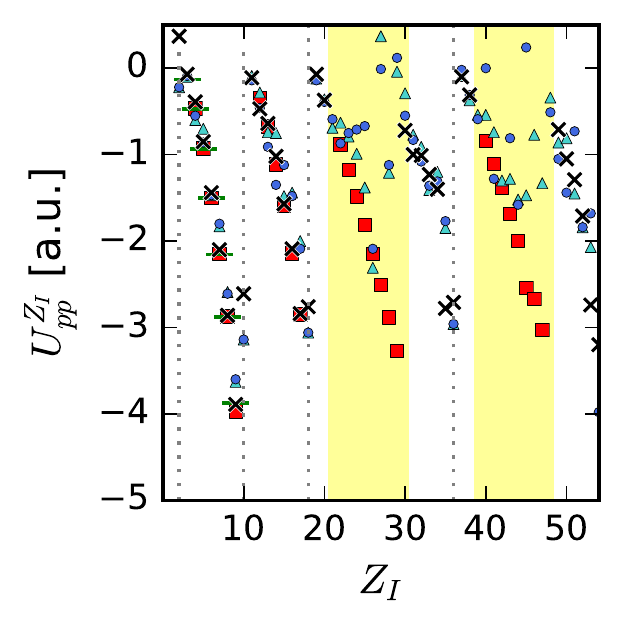}
 \caption{$U_{pp}^{Z_I}$ determined in a fit to \textit{atomic} data
in Refs.~\onlinecite{Oleari1966,DiSipio1971} (red squares) and 
Ref.~\onlinecite{Margraf2016} (green dashes) 
and $U_{pp}^{Z_I}$ determined in a fit to \textit{molecular} data
for MNDO \cite{Dewar1977,Stewart2004} (black crosses), 
 PM6  \cite{Stewart2007} (light blue triangles), and
 PM7  \cite{Stewart2012} (dark blue circles) for $Z_I = 2, 3, ..., 54$.
We highlight the transition metal blocks ($21\le Z_I\le 30$, 
and $39\le Z_I\le 48$) by a yellow background
and we indicate $Z_I$ of noble gases by vertical gray dashed lines.
}
\label{fig:parameter_upp}
\end{figure}

Oleari \textit{et al.} \cite{Oleari1966}, and subsequent studies, \cite{Sichel1967,DiSipio1971,Margraf2016}
found that the one-center parameters vary in a remarkably regular manner with respect to $Z_I$
(quadratic dependence of $U_{ss}^{Z_I}$, $U_{pp}^{Z_I}$, and 
$U_{dd}^{Z_I}$ on $Z_I$ \cite{Margraf2016} and linear dependence for the one-center 
ERI parameters on $Z_I$; \cite{Margraf2016} see also Figure~\ref{fig:parameter_upp}).
In NDDO-SEMO models, the one-center parameters are, however, not determined
with respect to data for atoms,  
but with respect to data for molecules.
Interestingly, the regularity of the one-center parameters with respect to $Z_I$ 
disappears for heavier elements ($Z_I > 23$) when taking molecular data as reference data
(compare, e.g., the red squares (atomic data as reference data) and blue 
circles (molecular data as reference data) in Figure~\ref{fig:parameter_upp}).
We may take this as a direct and method-inherent hint that the description of 
transition-metal complexes will be more challenging than the description of organic compounds 
with NDDO-SEMO models.

\subsection{General Parametrization Procedure}
\label{subsubsec:domain}

In general, parameters $\fett{p}=(p_1,p_2, ..., p_p)^T$ are calibrated against a reference data 
set $\mathcal{D}$ which comprises $D$ data triples, 
\begin{equation}
 \mathcal{D}=\{(y_d,\fett{x}_d,w_d)\},~~\mbox{with}~~ d=(1, 2, ..., D),
\end{equation}
consisting of 
(i) target observables $y_d$, 
(ii) input variables $\fett{x}_d$ (e.g., atomic coordinates, charge, and spin multiplicity of a molecule), 
and (iii) weights $w_d$.
Traditionally, the target observables are measured heats of formation 
$\Delta H_{f}^{\text{298K}}$ at 298 K, \cite{Bingham1975, Dewar1977,Jug2015}
structural variables (bond distances or bond angles), dipole 
moments, and first vertical ionization potentials for a variety of molecules, \cite{Dewar1977, Stewart2007, Stewart2012}
or calculated electronic energy differences. \cite{Hicks1986,Rossi1995a} 
The prediction of an observable by an NDDO-SEMO model, $f(\fett{x}_d, \fett{p})$,
is determined by $\fett{x}_d$ and $\fett{p}$. 
The parameter set $\fett{p}$ is then calibrated through minimization of an error function $\mathcal{E}$
which is evaluated from the sum of weighted 
square differences between 
$y_d$ and $f(\fett{x}_d, \fett{p})$, \cite{Dewar1977,ScholtenDiss,Stewart2007,
Stewart2012}
\begin{equation}
 \label{eq:errorFunction}
 \mathcal{E} = \mathcal{E}_{\mathcal{D}}(\fett{p}) = 
\sum_{d=1}^{D} w_d \left[ y_d-f(\fett{x}_d,\fett{p}) \right]^2,
\end{equation}
where $w_d$ are the weights.
The minimization of $\mathcal{E}$ with respect to $\fett{p}$ in a 
nonlinear ordinary least squares fit, 
\begin{equation}
 \label{eq:minErr}
 \frac{\partial \mathcal{E}_{\mathcal{D}}(\fett{p})}{\partial p_p} =0, \ \forall p_p\in \fett{p},
\end{equation}
yields an optimal parameter set $\fett{p}_{\mathcal{D}}$ with respect to the
reference data set $\mathcal{D}$.
Different optimization algorithms, e.g., 
the Levenberg--Marquardt algorithm, \cite{ScholtenDiss}
gradient-based methods, \cite{Stewart1989,Stewart2007,Stewart2012} 
genetic algorithms, 
\cite{Brothers2002,Rossi1995a,Hutter1998,Zhang2011,Fiedler2014}  and
line-search algorithms  \cite{Dewar1977}
can be straightforwardly applied for this task. 

\subsubsection{Applying Molecular Data Including Nuclear Effects as Reference Data}
\label{par:hof}

As we already noted, $\mathcal{D}$ traditionally incorporates measured $\Delta H_{f}^{\text{298K}}$
for a variety of molecules. \cite{Dewar1977,Dewar1985,Stewart2007,Stewart2012}
When applying an NDDO-SEMO model, $\Delta H_{f}^{\text{298K}}$ is usually predicted 
based on the electronic energies of the molecule and the constituent atoms, 
and the heats of formation of the atoms at 298 K, 
$\Delta H_{f,I}^{\text{298K}}$, \cite{Bingham1975, Dewar1977,Jug2015}
\begin{equation}
 \label{eq:hof}
 \Delta H_f^{\text{298K}} \approx E_{\text{el}}^{\{\tilde{\fett{R}}_I\}} + 
\sum_{I=1}^N \left( \Delta H_{f,I}^{\text{298K}} -  E^{\tilde{\fett{R}}_I}_{\text{el}}\right).
\end{equation}
The heat of formation of the atom at 298 K 
is taken from experimental data (for instance from Ref.~\onlinecite{Cox1989}).
We can examine which approximations are included in Eq.~(\ref{eq:hof}) by comparing it 
with the standard expression to calculate $\Delta H_{f}^{\text{298K}}$ 
from first principles, \cite{McQuarrie2000}
\begin{equation}
 \label{eq:hofc_298K}
\begin{split}
 \Delta H_f^{\text{298K}} =& E_{\text{el}}^{\{\tilde{\fett{R}}_I\}} + \text{ZPE}  
+ H_{\text{rest}}(T)    +
\sum_{I=1}^N \left( \Delta H_{f,I}^{\text{298K}}  
- E^{\tilde{\fett{R}}_I}_{\text{el}}\right).
\end{split}
\end{equation}
Compared to Eq.~(\ref{eq:hof}), Eq.~(\ref{eq:hofc_298K})  
incorporates the zero-point energy (ZPE) and the 
temperature-dependent translational, rotational, and vibrational contributions 
(if coupling of degrees of freedom is neglected), 
$H_{\text{rest}}(T)$.
Hence, the parameters of an NDDO-SEMO model must account for the neglect 
of ZPE and $H_{\text{rest}}(T)$ when calculating $\Delta H_f^{\text{298K}}$
according to Eq.~(\ref{eq:hof}). 
Consequently, an NDDO-SEMO electronic energy in a traditional parameterization 
\textit{cannot} be considered a \textit{pure} electronic energy.
This is a very unsatisfactory situation from a theoretical point of view (as, e.g., also noted 
in Refs.~\onlinecite{Hicks1986,Thiel1988, Repasky2002}).
We would like to emphasize that, in principle, the standard protocol (Eq.~(\ref{eq:hofc_298K})) 
and specialized approaches tailored toward SEMO models \cite{Repasky2002,Kromann2018}
could be readily applied instead of Eq.~(\ref{eq:hof}). 

Hicks and Thiel studied \cite{Hicks1986} the 
severity of this conceptual inconsistency by reparametrizing MNDO with 
respect to electronic atomization energies ($\Delta E^\text{at}_{\text{el}}$), 
\begin{equation}
 \Delta E^\text{at}_{\text{el}} = E_{\text{el}}^{\{\tilde{\fett{R}}_I\}} -  
\sum_{I=1}^N E^{\tilde{\fett{R}}_I}_{\text{el}}.
\end{equation}
Hicks and Thiel found \cite{Hicks1986} that the errors between  
reference and predicted $\Delta H_f^{\text{298K}}$ and the errors between  
reference and predicted $\Delta E^\text{at}_{\text{el}}$ are similarly large. 
They therefore  concluded \cite{Hicks1986} that the errors are dominated by the error in 
the MNDO electronic energies rather than by the error caused by applying Eq.~(\ref{eq:hof}).
Their study was, however, limited to 36 medium-sized hydrocarbon  
compounds.
Later, it was found that the application of Eq.~(\ref{eq:hof}) in the parametrization process 
is the reason for poor results for very small (e.g., diatomic) and large 
compounds in comparison to medium-sized compounds \cite{Repasky2002a,Winget2003b} (see, e.g., Figure~2 in Ref.~\onlinecite{Winget2003b}).
This is not surprising because $\mathcal{D}$ is dominated by medium-sized 
organic compounds. 
The opinion that the most severe errors stem from the NDDO-SEMO model itself 
and not from the application of Eq.~(\ref{eq:hof}), however, 
persisted in the literature. \cite{Winget2003b}

\subsubsection{Dependence of $\fett{p}_\mathcal{D}$ on $\mathcal{D}$}
\label{par:measureddata}

\begin{figure}[ht]
 \centering 
 \includegraphics[width=.75\textwidth]{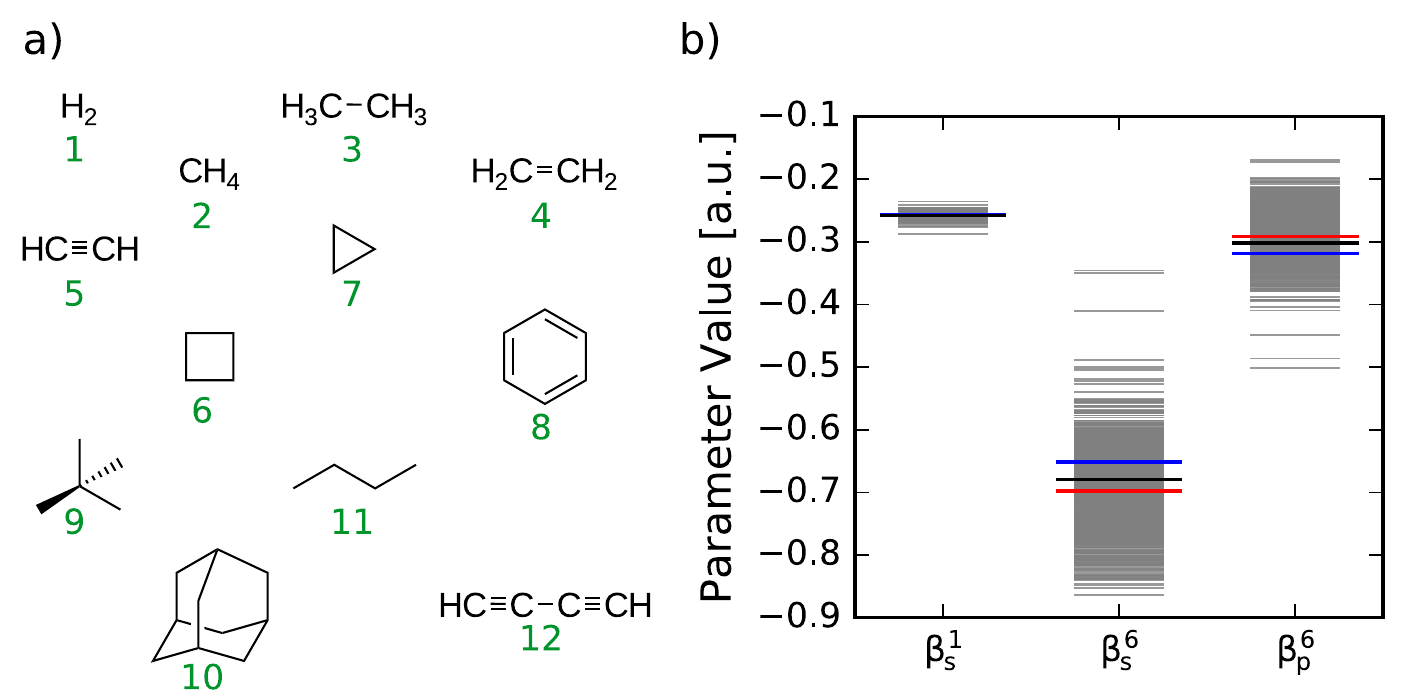}
 \caption{a) Model reference data set $\mathcal{D}^m$ consisting of dihydrogen (1), methane (2), 
 ethane (3), ethene (4), ethyne (5), cyclopropane (6), cyclobutane (7), 
 benzene (8), neopentane (9), n-butane (10), adamantane (11), and 1,3-butadyne (12). 
b) MNDO parameter values (red lines), the $\fett{p}^m_{\mathcal{D}^m}$ values 
(blue lines), the $B$ $\fett{p}^m_{\mathcal{D}^m_b}$ values (gray lines), 
and the $\bar{\fett{p}}^m_{\mathcal{D}^m}$ values (black lines) in a.u.}
\label{fig:bootstrapsampling}
\end{figure}

For the prediction of properties for molecules not included in $\mathcal{D}$, 
one needs to estimate the uncertainties of $\fett{p}$.
We recently demonstrated \cite{Proppe2017,Weymuth2018} 
how to apply nonparametric bootstrapping \cite{Efron1979,Hastie2016} in order to calibrate physicochemical 
property models with a limited amount of data and to determine the uncertainties 
of the incorporated parameters.
Here, we re-optimize $\fett{p}^m=\{\beta_s^1, \beta_s^6, \beta_p^6\}$ 
for the MNDO model
with respect to a model data set $\mathcal{D}^m$ containing 
twelve measured $\Delta H_{f}^{\text{298K}}$ of hydrocarbon compounds 
(see Figure~\ref{fig:bootstrapsampling}a).
Starting from the MNDO values for $\fett{p}^m$, 
 we determine the optimal parameter set $\fett{p}^m_{\mathcal{D}^m}$ 
by minimizing $\mathcal{E}_{\mathcal{D}^m}(\fett{p}^m)$ with the Nelder--Mead simplex algorithm.
The application of nonparametric bootstrap sampling 
now enables the quantification of the dependence $\fett{p}^m$ 
on the choice of $\mathcal{D}^m$.
We generate $B = 1000$ bootstrap samples $\{\mathcal{D}^M_b\}, \ b=(1,2,...,B)$ 
by drawing $D$ elements with replacement at random from $\mathcal{D}^m$.
For each bootstrap sample $\mathcal{D}^m_b$, we determine the optimal 
parameter set $\fett{p}^m_{\mathcal{D}^m_b}$ by minimizing $\mathcal{E}_{\mathcal{D}^m_b}(\fett{p^m})$.
We then determine the mean of the parameters $\bar{\fett{p}}^m_{\mathcal{D}^m}$ from all bootstrap samples, 
\begin{equation}
 \bar{\fett{p}}^m_{\mathcal{D}^m} = \frac{1}{B} \sum_{b=1}^B \fett{p}^m_{\mathcal{D}^m_b}.
\end{equation}
Overall, $\bar{\fett{p}}^m_{\mathcal{D}^m_b}$ coincides nicely with 
$\fett{p}^m_{\mathcal{D}^m}$ and with the MNDO values.
This means that we arrive at a very similar final parameter set,
but we have gained significantly more knowledge from the parametrization procedure than 
from a fit to  $\mathcal{D}^m$ alone.
Figure~\ref{fig:bootstrapsampling}b shows the distribution of 
$\fett{p}^m_{\mathcal{D}^m_b}$ which we obtained for the $B$ bootstrap samples $\{\mathcal{D}^m_b\}$.
The parameters $\beta_s^6$ and $\beta_p^6$ differ significantly when they are calibrated with 
respect to different $\mathcal{D}^m_b$ ($-0.87$ a.u.\ $<\beta_s^6<-0.33$ a.u.\ and $-0.51$ a.u.\ 
$<\beta_p^6<-0.17$ a.u.).
The parameter $\beta_s^1$, by contrast, hardly varies for different $\mathcal{D}^M_b$
($-0.29$ a.u.\ $<\beta_s^1<-0.24$ a.u.).
Simply put, this means that we were not able to identify a single value for 
$\beta_s^6$ and $\beta_p^6$ which minimizes all different $\mathcal{E}_{\mathcal{D}^m_b}$.
Rather, very different values for $\beta_s^6$ and $\beta_p^6$ are ideal to describe 
different $\mathcal{D}^M_b$.
From the bootstrap samples, we can then also sample the model prediction uncertainty for 
the target property, $\Delta H_{f}^{\text{298K}}$ 
which yields very large 95\% confidence intervals for all molecules in $\mathcal{D}^m$ 
($>20.0$ kJ~mol$^{-1}$, see Section~\ref{sec:mndoparam}).

The fact that there is no single transferable parameter set has been noted before.
\cite{ScholtenDiss,Thomas2017,Oreluk2018} 
Scholten remarked that different parameter values are well-suited 
to describe different properties 
for the same set of reference molecules. \cite{ScholtenDiss}
The parametrization of the HpCADD model demonstrated that the parameters for hydrogen 
atoms vary by 406\% when considering different environments. \cite{Thomas2017}
Very recently, Oreluk \textit{et al.} systematically assessed the variability of the 
PM7 parameters for a set of linear alkanes and 
 came to the conclusion that no single set of parameters is consistent with the entire 
data set. \cite{Oreluk2018} 
Oreluk \textit{et al.} propagated the uncertainties for the PM7 parameters to the 
prediction of heats of formations which then enables the attachment of an 
error bar to it. \cite{Oreluk2018}

\subsubsection{Insights from Benchmark Studies}
\label{par:benchmark}

Not surprisingly, NDDO-SEMO models are unable to describe systems with strong electron correlation.
Such systems are, however, present in $\mathcal{D}$ for some NDDO-SEMO models 
(see, e.g., the chromium dimer and CrO$_3$ 
which both exhibit a very strong multiconfigurational character \cite{Jiang2012} are contained 
 in the PM6 and PM7 reference data sets\cite{pm7referenceset}).
The inclusion of systems with strong electron correlation in $\mathcal{D}$ may
lead to a bias in $\fett{p}$ which would at least partially explain the 
generally poor accuracy for transition-metal complexes.
Despite significant efforts, it was not yet possible to create an NDDO-SEMO model
which achieves a similar accuracy with respect to the reference data for
transition-metal complexes as for organic compounds.
\cite{Nieke1985,Nieke1986,Filatov1992,Bosque2000,Stewart2007,
Stewart2012,Minenkov2018}

\begin{figure}[ht] 
 \centering 
 \includegraphics[width=\textwidth]{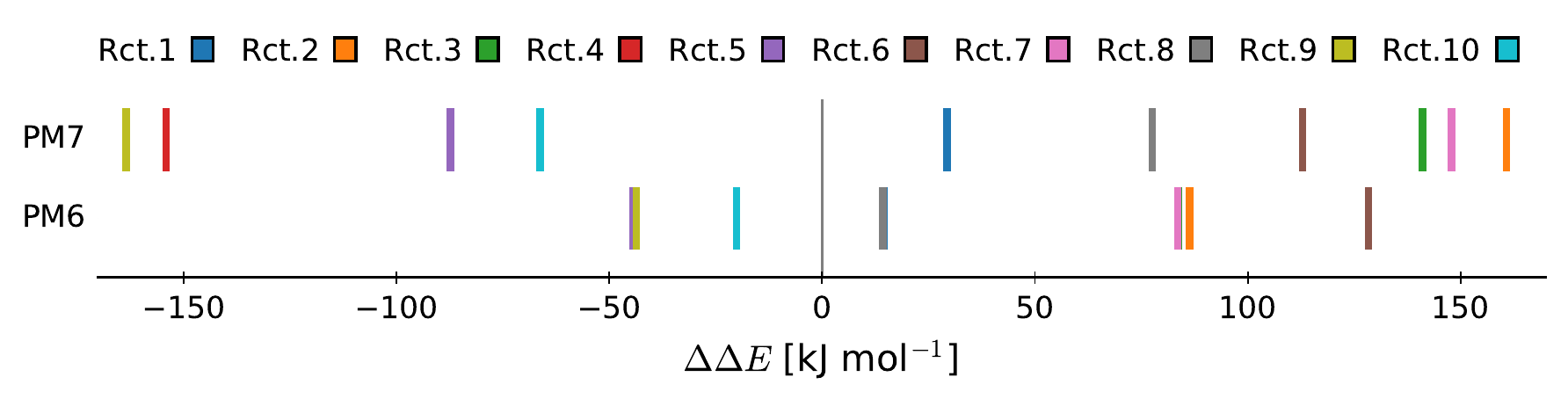}
 \caption{Deviation of electronic ligand dissociation energies $\Delta E$ in kJ~mol$^{-1}$ 
calculated with PM6 and PM7 from DLPNO-CCSD(T) energies\cite{Husch2018a} for the 
ten reactions in the WCCR10 set\cite{Weymuth2014}. 
We did not include $\Delta\Delta E = 786.8$ kJ~mol$^{-1}$ for reaction 4 for PM6
in this Figure.
}
\label{fig:wccr10}
\end{figure}
In this respect, we assess the performance of PM6 and PM7 for the WCCR10 set. 
The WCCR10 set\cite{Weymuth2014} contains ten ligand dissociation energies of 
large transition-metal complexes which 
 feature different transition metals (Au, Ag, Pt, Ru, Cu, Pd) and a
diverse selection of ligand environments.
The PM6 and PM7 ligand dissociation energies 
deviate significantly (on average $130.6$ kJ~mol$^{-1}$ 
and $114.1$ kJ~mol$^{-1}$, respectively) 
from reference DLPNO-CCSD(T) ligand dissociation energies \cite{Husch2018a} 
(see Figure~\ref{fig:wccr10}). 
While a deviation of PM6 and PM7 energies from DLPNO-CCSD(T) data is not particularly 
surprising, the severeness of the failure of PM6 and PM7 might be. 
The PM6 ligand dissociation energy for reaction 4, for instance, is strongly negative 
($\Delta E = - 579.8$ kJ~mol$^{-1}$).

\begin{figure}[ht]
 \centering 
 \includegraphics[width=.75\textwidth]{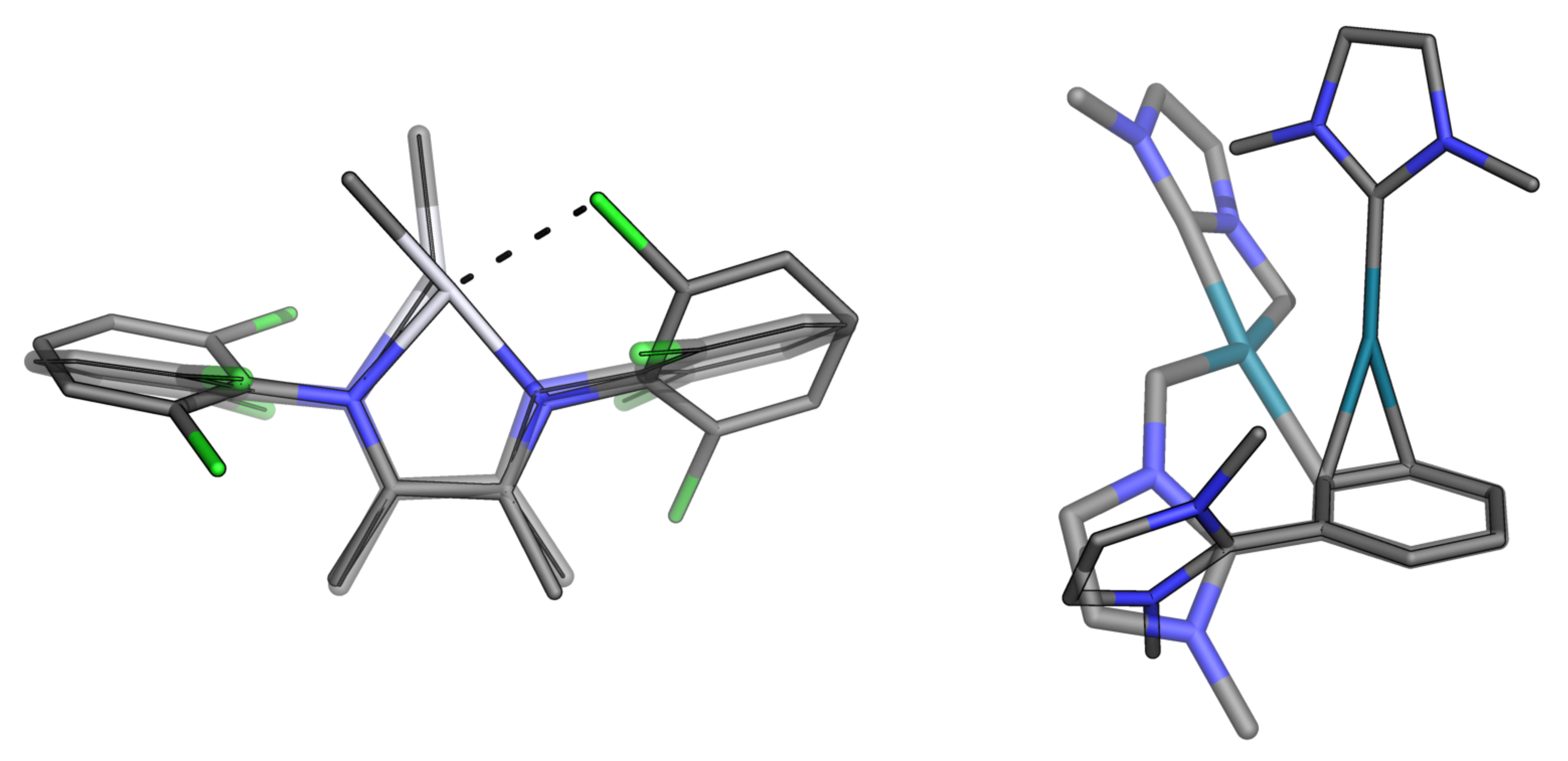}
 \caption{Overlay of the BP86/def2-QZVPP structure taken from Ref.~\onlinecite{Weymuth2014} (solid) 
and the PM7 structure (translucent) of the 
charged product of reaction 1 (left) and of the reactant of reaction 9 (right) 
Element color code: carbon, gray; nitrogen, blue; chlorine, green; 
palladium, teal; platinum, silver. 
Hydrogen atoms are omitted for clarity.
}
\label{fig:wccr10structures}
\end{figure}

Figure~\ref{fig:wccr10structures} shows that 
the PM7 structure of the charged product of reaction 1 
is strongly distorted compared 
to the BP86/def2-QZVPP reference structure taken from Ref.~\onlinecite{Weymuth2014}
even though the deviation of the PM7 ligand dissociation energy from the DLPNO-CCSD(T) energy 
is only $29.4$ kJ~mol$^{-1}$ for this reaction. 
In fact, the structures may be so severely distorted that 
a re-optimization with BP86/def2-QZVPP starting from 
the PM6 or PM7 optimized structures does not yield 
the original BP86/def2-QZVPP minimum-energy structures from which 
the PM6 and PM7 optimizations were started 
(e.g., reactant of reaction 9 in 
Figure~\ref{fig:wccr10structures}).
Great caution is therefore in order when applying NDDO-SEMO models 
to transition-metal complexes in general.

NDDO-SEMO models are mostly applied to study organic compounds which do not 
exhibit strong electron correlation.   
Recent benchmark studies show \cite{Korth2011,Dral2016a}
that OM$x$ models with dispersion corrections are slightly superior to MNDO-type models.
The performance of NDDO-SEMO models in extensive benchmark sets such as the GMTKN24 database \cite{Korth2011} 
is quite impressive considering their high 
computational efficiency (mean absolute deviation $< 33$ kJ~mol$^{-1}$ for OM3 at the 
GMTKN24 database\cite{Korth2011}). 
Nevertheless, it is insightful to take a closer look at the distribution of the individual errors:
It is not rare that a given NDDO-SEMO model either over- or underestimates 
relative electronic energies by over 80 kJ~mol$^{-1}$ (see, e.g., Figures~1--4 in Ref.~\onlinecite{Dral2016a}).
In special cases, the errors can be attributed to the insufficiency of the 
basis set (e.g., to explain the 
failure to describe nitro compounds \cite{Winget2003b})
or to the absence of orthogonalization corrections in MNDO-type models 
(which is, e.g., assumed to be responsible 
for wrong rotation barriers \cite{Winget2003b}).
However, it is basically impossible to rationalize why some error occurs in general  
due to the diversity of the approximations invoked in an NDDO-SEMO model.

\subsubsection{Focused Reparameterization}

A way to reduce the error is the restriction of the domain of applicability.
Rossi and Truhlar proposed to adjust the parameters to 
describe specific reactions yielding specific reaction parameters. \cite{Rossi1995a}
This approach has become more popular in recent years and  
parameters were adjusted to study specific compound classes 
and specific reactions (see, e.g., 
Refs.~\onlinecite{Lopez2003,Giese2005a, Nam2007,  Tejero2007, Wu2013, Liang2013, Zhou2014a,Dral2015a,
Saito2016,Fredin2016}
and other references citing Ref.~\onlinecite{Rossi1995a}).

A focused reparameterization is, however, plagued by problems.
Obviously, it cannot resolve systematic errors (cf. Section~\ref{par:ksdft}).
Additionally, it may be difficult to curtail the domain of applicability
adequately, i.e., to decide for a specific system whether it is 
similar enough to the ones for which it was parametrized.
To define such a structure-based metric has been a long-standing goal in machine learning 
applied to chemistry \cite{Bartok2013} 
and is related to the present problem.
Strictly speaking, parameters are only valid for one arrangement of the 
atomic nuclei because they implicitly encode orthogonalization effects which, obviously, 
depend on the atomic nuclei. This statement is 
valid for all NDDO-SEMO models (even for the OM$x$ models) 
because the parametric expressions are tuned to compensate for errors 
in $^\phi\fett{G}$.
Additionally, we might need different parameters for two 
atoms of the same element type in the same molecule (e.g., when they are encountered in  
different local  environments represented by different valence states).
In the worst case, this means that we need to have separate parameters for 
every valence state of an atom in a molecule.
It may be possible to define `atom types' as proposed in Ref.~\onlinecite{Thomas2017}.
However, as a matter of principle it remains challenging to divide the atoms in a molecule 
into different atom types in a meaningful way.
Moreover, when defining atom types, one obviously would inherit all of the problems associated with the definition 
of atom types from force-field development which does not appear particularly 
appealing for a method rooted in the first principles of quantum mechanics.
The practical consequence of these considerations is that the number of parameters 
which we have to determine increases dramatically, e.g., 
in the case of MNDO with its six adjustable parameters per element to $6N$ parameters 
per molecule in the worst case which would bring SEMO models very close to machine-learning 
approaches. 
For standard static benchmark approaches that apply a fixed amount of pre-defined 
reference data, it is hardly imaginable 
how one could achieve a similar increase in the number of reference data so 
that we can determine $\fett{p}$ in a well-defined manner (i.e., so that $D\gg p$).

\subsection{Improving Parametric Functions}
\label{subsubsec:improve}

Contemporary NDDO-SEMO models have limitations which cannot be addressed 
by reparameterization and overcoming these limitations requires the adoption of novel 
parametric functions. 
Unfortunately, the NDDO approximation causes large and uncontrollable 
errors in the ERIs in the $\phi$-basis and these errors propagate to all quantities calculated 
on the basis of the ERIs. \cite{Roby1969,Sustmann1969, Koster1972,Birner1974,Chandler1980,Duke1981,Neymeyr1995a,Neymeyr1995b,Neymeyr1995c,
Neymeyr1995d,Neymeyr1995e,Tu2003,Husch2018} 
Contemporary NDDO-SEMO models counteract the errors 
by introducing parametric expressions 
to evaluate the one-electron matrix and the core-core repulsion energy, i.e., 
they rely on error cancellation.

This raises the question why one does not directly correct 
${}^{\phi}\fett{G}$ or the ERIs in the $\phi$-basis.
A reason might be that the ERIs in the $\phi$-basis,
${}^{\phi}\fett{G}$, and also ${}^{\chi}\fett{G}$ encode information 
on the whole molecule (cf. Eqs.~(\ref{eq:oao}), (\ref{eq:fockmatrix_chi}), and 
(\ref{eq:fockmatrix_phi}), respectively). 
By contrast, the contributions to ${}^{\chi}\fett{H}$ and $V_v$ are straightforwardly
transferable from molecule to molecule. It is therefore comparatively easy 
to develop transferable parametric expressions to model ${}^{\chi}\fett{H}$ and $V_v$ on 
the examples of simple model systems (e.g., by considering diatomic systems).
Unfortunately, approximating ${}^{\chi}\fett{H}$ and $V_v$ well is not sufficient 
for the design of a reliable NDDO-SEMO model.  
Instead, the parameteric expressions applied to approximate ${}^{\chi}\fett{H}$ and $V_v$
 need to be flexible enough 
to compensate for the errors in ${}^{\phi}\fett{G}$. 
Hence, we may anticipate that the improvement of the parametric expressions 
is as complicated as the direct correction for the error in ${}^{\phi}\fett{G}$.

We recently introduced \cite{Husch2018} a strategy to directly correct for the error caused by the NDDO approximation in ${}^{\phi}\fett{G}$ 
which we call the correction inheritance for semiempirics (CISE) 
approach. We drew inspiration from the work carried out by Roby and Sinano\v{g}lu who suggested \cite{Roby1969} to scale $^\chi\fett{G}^{\text{NDDO}}$ 
with a scaling matrix $\fett{\Gamma}$ to obtain a better estimate for $^\phi\fett{G}$,
\cite{Roby1969}
\begin{equation}
^\phi\fett{G} \approx \fett{\Gamma} {}^\chi\fett{G}^{\text{NDDO}}.
\end{equation}
The goal of Roby and Sinano\v{g}lu in 1969 was to speed up single-point HF calculations for a diverse set of
small molecules, and hence, they attempted to define universal rules to assemble $\fett{\Gamma}$ 
which, not surprisingly, 
turned out to be impossible. \cite{Roby1969}
It is, however, possible to  exactly determine $\fett{\Gamma}(\{\tilde{\fett{R}}_I^n\})$ 
for a given structure $\{\tilde{\fett{R}}_I^n\}$
from a reference self-consistent field (i.e., HF, KS-DFT, or general multi-configurational 
SCF) calculation (yielding the exact ${}^\phi\fett{G}(\{\tilde{\fett{R}}_I^n\})$),
\begin{equation}
 \begin{split}
   \fett{\Gamma}(\{\tilde{\fett{R}}_I^n\}) &= {}^\phi\fett{G}(\{\tilde{\fett{R}}_I^n\}) \cdot 
\left( {}^\chi\fett{G}^{\text{NDDO}}(\{\tilde{\fett{R}}_I^n\})\right)^{-1}. \\
 \end{split}
\end{equation}
Obviously, we will then not achieve a speed-up with 
respect to the reference calculation.
We found \cite{Husch2018} that $\fett{\Gamma}(\{\tilde{\fett{R}}_I^n\})$ 
is transferable to a certain degree in a sequence of related structures, i.e., 
for two similar structures $\{\tilde{\fett{R}}_I^n\}$ and $\{\tilde{\fett{R}}_I^{(n+1)}\}$, 
\begin{equation}
\label{eq:correct}
 ^\phi \fett{G}(\{\tilde{\fett{R}}_I^{(n+1)}\}) \approx \fett{\Gamma}(\{\tilde{\fett{R}}_I^n\}) \cdot  
{}^\chi\fett{G}^{\text{NDDO}}(\{\tilde{\fett{R}}_I^{(n+1)}\}),
\end{equation}
for which we achieved a speed-up at a negligible loss of accuracy. 
We also showed \cite{Husch2018} that a correction to 
${}^\chi\fett{G}^{\text{NDDO}}(\{\tilde{\fett{R}}_I^{(n+1)}\})$ 
can be constructed in different ways, departing from a Roby--Sinano\v{g}lu-type approach.
We proposed \cite{Husch2018} to construct  additive corrections $\fett{\Gamma_J}$ and 
$\fett{\Gamma_K}$ to the matrices 
$^\chi\fett{J}^{\text{NDDO}}$ and 
to $^\chi\fett{K}^{\text{NDDO}}$, respectively,
\begin{equation}
\label{eq:correct2}
\begin{split}
 {}^\phi\fett{G}(\{\tilde{\fett{R}}_I^{(n+1)}\}) \approx& \fett{\Gamma_J}(\{\tilde{\fett{R}}_I^n\}) +  
{}^\chi\fett{J}^{\text{NDDO}}(\{\tilde{\fett{R}}_I^{(n+1)}\}) \\ &
+\fett{\Gamma_K}(\{\tilde{\fett{R}}_I^n\})  +  
{}^\chi\fett{K}^{\text{NDDO}}(\{\tilde{\fett{R}}_I^{(n+1)}\}).
\end{split}
\end{equation}

The CISE approach has a potential for application whenever we are interested in obtaining electronic energies for sequences 
of related structures, e.g., 
in the context of kinetic modeling, \cite{Kee1980,Glowacki2012,Proppe2017a, Proppe2018} in
real-time \cite{Haag2014,
Haag2013,Vaucher2016a}  and automated \cite{Maeda2013,Magoon2013,
Rappoport2014a,Zimmerman2015a,Habershon2016a,
Bergeler2015,Simm2017} reaction-mechanism explorations, 
or in reaction and first-principles \cite{Marx2009,Wang2014a,Martinez2017} molecular dynamics simulations. 
The CISE approach differs conceptually from the existing NDDO-SEMO models insofar as that no determination of parameters in a statistical calibration is required. 
Instead, we maintain complete error control on the resulting model
because we could straightforwardly determine $\fett{\Gamma}(\{\tilde{\fett{R}}_I^{(n+1)}\})$ 
for a given molecule 
with nuclear coordinates $\{\tilde{\fett{R}}_I^{(n+1)}\}$. 
By contrast, we cannot 
straightforwardly determine the best $\fett{p}(\{\tilde{\fett{R}}_I^{(n+1)}\})$ 
for the molecule $n+1$ for contemporary
NDDO-SEMO models. 

\section{Explicit Description of Electron Correlation Effects} 
\label{subsec:explicitcorr}

The last question which we address in this work is whether 
one could, in principle, obtain FCI quality 
results in a given one-electron basis when applying an NDDO-SEMO reference wave function.
It was suggested that all of the developed wave function methods can be 
(and many have been \cite{Schweig1980,Thiel1981,
Clark1993b,Clark1993,Liotard1999,Toniolo2002,Toniolo2003,Koslowski2003,Toniolo2004,Lei2010,
Dral2011a,Liu2018})
 straightforwardly applied after carrying out an NDDO-SEMO calculation 
which then essentially substitutes 
the HF calculation.
It is, however, important to recognize that the NDDO approximation affects the ERIs over
the molecular orbitals,
\begin{equation}
\label{eq:erimo}
\begin{split}
 \left<\psi_i\psi_j|\psi_k\psi_l\right> &= \sum_{\mu=1}^M\sum_{\nu=1}^M
\sum_{\lambda=1}^M\sum_{\sigma=1}^M {}^\phi C_{\mu i} {}^\phi C_{\nu j}
\left<\phi_\mu\phi_\nu|\phi_\lambda\phi_\sigma\right> 
{}^\phi C_{\lambda k} {}^\phi C_{\sigma l} \\  &\approx 
  \sum_{\mu=1}^M\sum_{\nu=1}^M \sum_{\lambda=1}^M\sum_{\sigma=1}^M {}^\phi C_{\mu i} {}^\phi C_{\nu j}\;
\delta_{IJ} \delta_{KL} 
\left<\chi_\mu^I\chi_\nu^J|\chi_\lambda^K\chi_\sigma^L\right>  
{}^\phi C_{\lambda k} {}^\phi C_{\sigma l}.  
\end{split}
\end{equation}
Previous results by Thiel and co-workers and Clark and co-workers 
showed that the correlation energy calculated 
with single-reference perturbation theories evaluated for an NDDO-SEMO reference is about one 
order of magnitude too small,  \cite{Thiel1981,Clark1993b,Clark1993}
which was also corroborated by our recent results. \cite{Husch2018} 
If we do not explicitly correct for the errors caused by the NDDO approximation, 
we will not able to adequately capture 
dynamic electron correlation effects and we must rely on the proper calibration of the parameters to 
achieve this.

It is no surprise that single-determinantal NDDO-SEMO models do not adequately capture static 
electron correlation effects and that static electron correlation effects 
have to be considered explicitly. \cite{Schweig1980,Thiel1981b}
Static electron correlation effects may be described through 
a multi-reference configuration interaction (MR-CI) procedure 
(including single and double excitations) 
using the graphical unitary group approach (GUGA).
\cite{Liotard1999,Toniolo2002,Koslowski2003,Toniolo2003,Toniolo2004}
Another approach is the application of an unrestricted natural orbital complete active space 
(UNO-CAS) or configuration interaction (UNO-CI) ansatz. \cite{Dral2011a}
Such methods are usually applied to describe excited states and the dynamics of excited states.
Note that all NDDO-SEMO models apply a valence-shell minimal basis sets which prevents 
a description of Rydberg states. \cite{Dewar1977}
Usually, MNDO-type models tend to underestimate excited-state energies due to 
the symmetric splitting of
 bonding and antibonding orbitals (see also Section~\ref{subsec:mndoH}), whereas OM$x$ models 
showed an overall good performance. \cite{Schweig1980,
Silva2010, Tuna2016}

An open question in the NDDO-SEMO/MR-CI approaches is whether 
contributions to the correlation energy may be doubly counted. 
We may draw the parallel to KS-DFT as it was combined with MR-CI
where a similar issue arises. \cite{Grimme1999,Silva2008}
 The main problem of KS-DFT/MR-CI is the double counting of
the correlation energy which can be alleviated through the introduction of empirical parameters. 
\cite{Grimme1999,Silva2008}
Similar measures have apparently not been taken when combining NDDO-SEMO models with 
MR-CI approaches.
Recent benchmarks show, however, that double counting and the error caused by the NDDO approximation 
in the ERIs over molecular orbitals appear not to be an issue in practice. \cite{Tuna2016}

\section{The Future of NDDO-SEMO Models}
\label{sec:prospect}

The success of NDDO-SEMO models is largely based on the effectiveness with which they allow one to 
 solve the SCF equations. 
Contemporary NDDO-SEMO models yield electronic energies 
about three orders of magnitude faster than HF or KS-DFT models. \cite{Thiel2014} 
The acceleration is largely due to the NDDO approximation
which drastically reduces the number of ERIs to be computed and processed 
in the course of a calculation.
The price to pay for the acceleration are significant errors in the ERIs 
in the L{\"o}wdin orthogonalized basis.
As a consequence, the NDDO approximation must be tied to many other 
approximations in the one-electron matrix and in the 
core-core repulsion energy to define a meaningful NDDO-SEMO model. 
In this work, we presented a comprehensive overview of the parametric expressions applied  
in the MNDO-type (MNDO, MNDO/d, AM1, PM3, PM6, PM7, and RM1) and OM$x$ models. 

We outlined the systematic limitations which NDDO-SEMO model face.
First, severe limitations are caused by the application of a small basis set.
The drastic restriction of the basis set size will, in general, prohibit the 
determination of accurate relative electronic energies, force constants, 
and polarizabilities. 
The increase of the basis set size is, however, challenging within the framework of 
contemporary NDDO-SEMO models for practical reasons. 
Second, systematic errors are caused by the adoption of a parametrized mean-field 
framework. 
The examination of the parallels to KS-DFT 
revealed that NDDO-SEMO models fail to describe the general behavior of 
electronic energy as a function of the electron number, in the same way as in HF and KS-DFT models.
We may therefore anticipate that NDDO-SEMO models will be plagued by the same difficulties in describing 
 electron correlation, irrespective of their specific parametrization.

In general, the parameters of NDDO-SEMO models are 
calibrated with respect to experimental reference data. 
When calibrating the parameters, one first encounters difficulties associated with the current practice of 
calculating heats of formation which leads to a contamination of 
the parameters with nuclear-motion contributions.
Consequently, the electronic energy calculated with an NDDO-SEMO model cannot be considered 
a pure electronic energy which, however, does not appear to have severe practical consequences. 
This conceptual inconsistency could simply be alleviated by adjusting the parametrization 
procedure.
Recent benchmark studies \cite{Korth2011,Dral2016a} 
showed that NDDO-SEMO models are notoriously 
unreliable. Large errors are observed  \cite{Korth2011,Dral2016a}  for molecules 
which do not show any apparent strong electron correlation.
This may be partially explained by the fact that it is highly unlikely that there 
is a single parameter set which is suited to describe all molecules. 
We believe that bootstrap sampling \cite{Efron1979,Simm2017a,Proppe2017,Weymuth2018,
Proppe2018}  offers an interesting insight 
into the parameterization of NDDO-SEMO models. 
When recalibrating a selection of parameters of the MNDO model in this work, we 
discovered that the parameters have to adopt significantly different values 
to describe different molecules well.

We briefly reviewed our recent proposal for system-focused NDDO-SEMO models 
that yield accurate results for structures related to a reference structure. 
Our CISE approach has the advantage that we are able to determine the 
parametrization of a corrective matrix directly for a given structure from a reference  
calculation. We, hence, do not have to apply a statistical procedure to calibrate 
parameters. This convenience obviously comes at the cost that 
the approach is restricted to the investigation of sequences of related structures 
which, however, are key areas of application for NDDO-SEMO models
(e.g., structure optimization, Born--Oppenheim molecular dynamics, and real-time reactivity 
exploration).

To conclude this overview, we would like to stress that the age of NDDO-SEMO models is far from 
being over. Although we pointed out several (conceptual and practical) difficulties, 
we want to highlight 
again that contemporary NDDO-SEMO models achieve, overall, a remarkably high accuracy 
with respect to experimental data.
To make NDDO-SEMO models useful for predictive work, we, however, have to know when, and 
why, they fail. 
This may, for instance, be achieved through statistical learning 
models. \cite{Ramakrishnan2015, Simm2018}

\section*{Acknowledgements}

This work was supported by the Schweizerischer Nationalfonds.
We are grateful to Professor Tim Clark for drawing our attention to the undocumented 6 kcal/mol
correction in \textsc{Mopac} and for providing information on AM1* that allowed us to implement this approach.
The authors thank Professors Walter Thiel, Alexander A. Voityuk, and Jens Spanget-Larsen for helpful discussions.

\section{Appendix}
\subsection{Computational Methodology}
\label{sec:methodology}

We implemented the MNDO(/d), AM1(/d), PM3, PM6, OM1, OM2, and OM3 models in our cross-platform 
quantum chemistry package \textsc{SCINE}. \cite{scine}
This new module of \textsc{SCINE}, \textsc{SCINEsemo}, will be made available on 
our Web page and can be applied as a stand-alone SEMO program or within 
the \textsc{SCINE} framework.  

We evaluated ERIs in the $\chi$-basis with \textsc{PySCF} (version 1.4) 
\cite{Sun2017,Sun2015}. 
The ERIs in the $\chi$-basis were transformed to the corresponding ERIs in the 
$\phi$-basis with the \textsc{ao2mo} integral transformation module of 
\textsc{PySCF}. 

Lastly, we evaluated heat of formations at 298 K with \textsc{Mopac} 2016 \cite{mopac}
in the course of the re-optimization of the parameters. 
We specified nonstandard parameters given in Table~\ref{tab:dm_par} through the keyword `\textsc{external}'.

\subsection{Basic Specifications}
\label{subsec:basicspecifications}
The MNDO-type models, MNDO\cite{Dewar1977}, MNDO/d \cite{Thiel1991,Thiel1996c}, 
AM1, \cite{Dewar1985} PM3, \cite{Stewart1989}
RM1, \cite{Rocha2006} PM6, \cite{Stewart2007} and PM7
are freely available in the \textsc{Mopac} program. \cite{mopac}
Throughout this work, \textsc{Mopac} served as our reference implementation 
for these NDDO-SEMO models because the parameters for the PM$x$ \cite{Stewart1989,Stewart2007,
Stewart2012} and RM1 \cite{Rocha2006} models and 
for many elements for the MNDO(/d) and AM1 models \cite{Stewart2004}
were determined with \textsc{Mopac}. 

In this work, we uncovered inconsistencies in the equations 
which we found implemented in many programs, also in \textsc{Mopac}. 
Note that it is not easily possible 
altering the implementation because the parameterization of the NDDO-SEMO models was carried out with 
a specific set of equations. 
Instead, one would have to determine a new set of parameters when implementing another set of  
equations.
Implementations of MNDO-type models are also available in other programs. 
If one wishes to check if an MNDO model is implemented in the same way as in 
\textsc{Mopac}, 
one can compare the parameter values and the values of the one- and two-center ERIs to 
the ones provided by \textsc{Mopac} when invoking the keyword \texttt{Hcore}.

The parameters for the OM$x$ models were determined with the (not freely available)
\textsc{MNDO2005} program. \cite{mndo2005}
We verified our implementation of the OM$x$ models by comparison to numerical data provided 
in Refs.~\onlinecite{Kolb1993a, Weber2000a, WeberPhD, ScholtenDiss, Dral2016b, Dral2016a}.

A calculation with an NDDO-SEMO model requires the specification of 
the number of explicitly considered electrons $n_v$, of the 
basis functions which are activated, and of a set of parameters 
for every element in the system of interest.
We specify these quantities in Table~\ref{tab:nv}
for the MNDO\cite{Dewar1977}, MNDO/d \cite{Thiel1991,Thiel1996c}, AM1, \cite{Dewar1985} 
AM1*, \cite{Dewar1985,Dewar1988b,Dewar1988c,Winget2003,Winget2005,Voityuk2000,
Kayi2007a,Kayi2009b,Kayi2009c,Kayi2010c,Kayi2010d,Kayi2010e,Kayi2011a}
PM3, \cite{Stewart1989}
RM1, \cite{Rocha2006} PM6, \cite{Stewart2007} PM7, \cite{Stewart2012} OM1, \cite{Kolb1993a} 
OM2, \cite{Weber2000a} and OM3 \cite{Dral2016b} models.

\LTcapwidth=\textwidth
\begin{center}
\begin{longtable}{ccccc}
  \caption{Nuclear charge $Z$, number of explicitly considered 
electrons $n_v$, and type of basis functions activated for each element
($1\le Z\le 57$ and $71\le Z\le 83$) in semiempirical models.
} \label{tab:nv} \\
\hline
\hline  
Element & Availability of Parameters & $Z$ & $n_v$ & Basis Functions \\ 
\hline 
\endfirsthead 
\hline
\hline 
Element & Availability of Parameters & $Z$ & $n_v$ & Basis Functions \\ 
\hline 
\endhead 
\hline \hline \endfoot
H  & MNDO(/d), AM1, AM1*, PM3,  &  &  &       \\
   & PM6, PM7, RM1, OM1, OM2, OM3 & 1 & 1 & $1s$      \\
He & MNDO(/d), AM1, PM3, PM6, PM7 & 2 & 2 & $1s$, $2p$  \\
Li & MNDO(/d), AM1, PM3, PM6, PM7 & 3 & 1 & $2s$, $2p$  \\
Be & MNDO(/d), AM1, PM3, PM6, PM7 & 4 & 2 & $2s$, $2p$  \\
B  & MNDO(/d), AM1*, PM3, PM6, PM7 & 5 & 3 & $2s$, $2p$  \\
C  & MNDO(/d), AM1, AM1*, PM3, &  &  &  \\
   & PM6, PM7, RM1, OM1, OM2, OM3 & 6 & 4 & $2s$, $2p$  \\
N  & MNDO(/d), AM1, AM1*, PM3, &  &  &   \\
   & PM6, PM7, RM1, OM1, OM2, OM3 & 7 & 5 & $2s$, $2p$  \\
O  & MNDO(/d), AM1, AM1*, PM3, &  &  &  \\
   & PM6, PM7, RM1, OM1, OM2, OM3 & 8 & 6 & $2s$, $2p$  \\
F  & MNDO(/d), AM1, AM1*, PM3, &  &  &   \\ 
   & PM6, PM7, RM1, OM1, OM2, OM3 & 9 & 7 & $2s$, $2p$  \\ 
Ne & MNDO(/d), AM1, PM3, PM6, PM7  & 10 & 6 & $2p$, $3s$  \\
Na & MNDO(/d), AM1, PM3, PM6, PM7  & 11 & 1 & $3s$, $3p$  \\
Mg & MNDO(/d), AM1, PM3, PM6, PM7  & 12 & 2 & $3s$, $3p$  \\
Al & MNDO, AM1, PM3   & 13 & 3 & $3s$, $3p$  \\
   & MNDO/d, AM1*, PM6, PM7  & 13 & 3 & $3s$, $3p$, $3d$  \\
Si & MNDO, AM1, PM3  & 14 & 4 & $3s$, $3p$   \\
   & MNDO/d, AM1*, PM6, PM7  & 14 & 4 & $3s$, $3p$, $3d$  \\
P  & MNDO, AM1, PM3, RM1  & 15 & 5 & $3s$, $3p$  \\
   & MNDO/d, AM1*, PM6, PM7  & 15 & 5 & $3s$, $3p$, $3d$  \\
S  & MNDO, AM1, PM3, RM1 & 16 & 6 & $3s$, $3p$  \\
   & MNDO/d, AM1*, PM6, PM7 & 16 & 6 & $3s$, $3p$, $3d$  \\
Cl & MNDO, AM1, PM3, RM1 & 17 & 7 & $3s$, $3p$  \\
   & MNDO/d, AM1*, PM6, PM7 & 17 & 7 & $3s$, $3p$, $3d$  \\
Ar & MNDO, AM1, PM3, PM6, PM7 & 18 & 6 & $3p$, $4s$  \\
K  & MNDO, AM1, PM3, PM6, PM7  & 19 & 1 & $4s$, $4p$  \\
Ca & MNDO, AM1, PM3, PM6, PM7  & 20 & 2 & $4s$, $4p$  \\
Sc & PM6, PM7 & 21 & 3 & $3d$, $4s$, $4p$ \\
Ti & AM1*, PM6, PM7 & 22 & 4 & $3d$, $4s$, $4p$ \\
V  & AM1*, PM6, PM7 & 23 & 5 & $3d$, $4s$, $4p$ \\ 
Cr & AM1*, PM6, PM7 & 24 & 6 & $3d$, $4s$, $4p$ \\
Mn & AM1*, PM6, PM7 & 25 & 7 & $3d$, $4s$, $4p$ \\
Fe & AM1*, PM6, PM7 & 26 & 8 & $3d$, $4s$, $4p$ \\
Co & AM1*, PM6, PM7 & 27 & 9 & $3d$, $4s$, $4p$ \\
Ni & AM1*, PM6, PM7 & 28 & 10 & $3d$, $4s$, $4p$ \\
Cu & AM1*, PM6, PM7 & 29 & 11 & $3d$, $4s$, $4p$ \\
Zn & MNDO, AM1, PM3, PM6, PM7  & 30 & 2 & $4s$, $4p$ \\
   & AM1*  & 30 & 12 & $3d$, $4s$, $4p$ \\
Ga & MNDO, AM1, PM3, PM6, PM7  & 31 & 3 & $4s$, $4p$ \\
Ge & MNDO, AM1, PM3, PM6, PM7  & 32 & 4 & $4s$, $4p$ \\
As & MNDO, AM1, PM3  & 33 & 5 & $4s$, $4p$ \\
   & PM6, PM7  & 33 & 5 & $4s$, $4p$, $4d$ \\
Se & MNDO, AM1, PM3, PM6, PM7  & 34 & 6 & $4s$, $4p$, $4d$ \\ 
Br & MNDO, AM1, PM3, RM1 & 35 & 7 & $4s$, $4p$ \\
   & MNDO/d, AM1*, PM6, PM7 & 35 & 7 & $4s$, $4p$, $4d$ \\
Kr & MNDO, AM1, PM3, PM6, PM7  & 36 & 6 & $4p$, $5s$ \\
Rb & MNDO, AM1, PM3, PM6, PM7  & 37 & 1 & $5s$, $5p$ \\
Sr & MNDO, AM1, PM3, PM6, PM7  & 38 & 2 & $5s$, $5p$ \\
Y  & PM6, PM7 & 39 & 3 & $4d$, $5s$, $5p$ \\
Zr & AM1*, PM6, PM7 & 40 & 4 & $4d$, $5s$, $5p$ \\
Nb & PM6, PM7 & 41 & 5 & $4d$, $5s$, $5p$ \\
Mo & AM1*, PM6, PM7, AM1 & 42 & 6 & $4d$, $5s$, $5p$ \\
Tc & PM6, PM7 & 43 & 7 & $4d$, $5s$, $5p$ \\
Ru & PM6, PM7 & 44 & 8 & $4d$, $5s$, $5p$ \\
Rh & PM6, PM7 & 45 & 9 & $4d$, $5s$, $5p$ \\
Pd & AM1*, PM6, PM7 & 46 & 10 & $4d$, $5s$, $5p$ \\
Ag & AM1*, PM6, PM7 & 47 & 11 & $4d$, $5s$, $5p$ \\
Cd & MNDO/d, PM3, PM6, PM7 & 48 & 2 & $5s$, $5p$ \\
In & MNDO, AM1, PM3, PM6, PM7  & 49 & 3 & $5s$, $5p$ \\
Sn & MNDO, AM1, PM3, PM6, PM7  & 50 & 4 & $5s$, $5p$ \\
Sb & MNDO, AM1, PM3  & 51 & 5 & $5s$, $5p$ \\
   & PM6, PM7  & 51 & 5 & $5s$, $5p$, $5d$ \\
Te & MNDO, AM1, PM3, PM6, PM7  & 52 & 6 & $5s$, $5p$, $5d$ \\
I  & MNDO, AM1, PM3, RM1 & 53 & 7 & $5s$, $5p$ \\
   & MNDO/d, AM1*, PM6, PM7 & 53 & 7 & $5s$, $5p$, $5d$ \\
Xe & MNDO, AM1, PM3, PM6, PM7  & 54 & 6 & $5p$, $6s$ \\
Cs & MNDO, AM1, PM3, PM6, PM7  & 55 & 1 & $6s$, $6p$ \\
Ba & MNDO, AM1, PM3, PM6, PM7  & 56 & 2 & $6s$, $6p$ \\
La & PM6, PM7 & 57 & 3 & $5d$, $6s$, $6p$\\
Lu & PM6, PM7 & 71 & 3 & $5d$, $6s$, $6p$\\
Hf & PM6, PM7 & 72 & 4 & $5d$, $6s$, $6p$\\
Ta & PM6, PM7 & 73 & 5 & $5d$, $6s$, $6p$\\
W  & PM6, PM7 & 74 & 6 & $5d$, $6s$, $6p$\\
Re & PM6, PM7 & 75 & 7 & $5d$, $6s$, $6p$\\
Os & PM6, PM7 & 76 & 8 & $5d$, $6s$, $6p$\\
Ir & PM6, PM7 & 77 & 9 & $5d$, $6s$, $6p$\\
Pt & PM6, PM7 & 78 & 10 & $5d$, $6s$, $6p$\\
Au & AM1*, PM6, PM7 & 79 & 11 & $5d$, $6s$, $6p$\\
Hg & MNDO(/d), AM1, PM3, PM6, PM7  & 80 & 2 & $6s$, $6p$\\
Tl & MNDO, AM1, PM3, PM6  & 81 & 3 & $6s$, $6p$\\
   & PM7  & 81 & 3 & $6s$, $6p$, $6d$\\
Pb & MNDO, AM1, PM3, PM6, PM7  & 82 & 4 & $6s$, $6p$\\
Bi & MNDO, AM1, PM3, PM6  & 83 & 5 & $6s$, $6p$ \\
   & PM7  & 83 & 5 & $6s$, $6p$, $6d$ \\
\end{longtable}
\end{center}

While we mostly adhered to the original parameter abbreviations, we chose to 
re-name several parameters to avoid confusion with other quantities.
The NDDO-SEMO models were developed independently of each other, and hence, they also sometimes 
apply different parameter names.
We indicate in Tables~\ref{tab:parameter_names} and \ref{tab:parameter_names2} how 
the parameter abbreviations introduced in the main text 
relate to the ones chosen in several popular publications.  
{\scriptsize 
\begin{center}
\begin{longtable}{ccccccc}
  \caption{Relation of the parameter abbreviations introduced in the 
main text for MNDO-type models to the ones chosen in several popular publications.
} \label{tab:parameter_names} \\
\hline \hline 
Main &  &  &  & & & PM6\cite{Stewart2007},\\
Text & \textsc{Mopac}\cite{mopac} & MNDO\cite{Dewar1977} & AM1\cite{Dewar1985} & AM1*\cite{Winget2003,Winget2005}& MNDO/d \cite{Thiel1991,Thiel1996c}
& PM7\cite{Stewart2012}\\
\hline 
\endfirsthead 
\hline \hline 
Main Text & \textsc{Mopac}\cite{mopac} & MNDO\cite{Dewar1977} & AM1\cite{Dewar1985} & AM1* & MNDO/d \cite{Thiel1991,Thiel1996c}
& PM6\cite{Stewart2007},PM7\cite{Stewart2012}\\
\hline 
\endhead \hline 
\hline \endfoot
$U_{ss}$ & USS & $U_{ss}$  & $U_{ss}$ & $U_{ss}$& $U_{ss}$& $U_{ss}$\\
$U_{pp}$ & UPP & $U_{pp}$  & $U_{pp}$ & $U_{pp}$& $U_{pp}$& $U_{pp}$\\
$U_{dd}$ & UDD & ---  & --- & $U_{dd}$ & $U_{dd}$& $U_{dd}$\\
$\zeta_{s}$ & ZS & $\zeta$ & $\zeta_s$ & $\zeta_s$ &$\zeta_{s}$& $\zeta_{s}$\\
$\zeta_{p}$ & ZP & $\zeta$ & $\zeta_p$ & $\zeta_p$ &$\zeta_{p}$& $\zeta_{p}$\\
$\zeta_{d}$ & ZD & ---& ---&$\zeta_{d}$&$\zeta_{d}$& $\zeta_{d}$\\
$\beta_s$ & BETAS & $\beta_s$& $\beta_s$ &$\beta_s$&$\beta_s$& $\beta_s$\\
$\beta_p$ & BETAP & $\beta_p$& $\beta_p$ &$\beta_p$&$\beta_p$& $\beta_p$\\
$\beta_d$ & BETAD & ---& --- & $\beta_d$&$\beta_d$& $\beta_d$\\
$\gamma_{ss}$ & GSS & $g_{ss}$& $g_{ss}$&$g_{ss}$&$g_{ss}$ & $g_{ss}$\\
$\gamma_{pp}$ & GPP & $g_{pp}$& $g_{pp}$&$g_{pp}$&$g_{pp}$& $g_{pp}$\\
$\gamma_{sp}$ & GSP & $g_{sp}$& $g_{sp}$&$g_{sp}$&---& $g_{sp}$\\
$\gamma_{pp'}$ & GP2 & $g_{p2}$& $g_{p2}$&$g_{p2}$&---& $g_{p2}$\\
$\tilde{\gamma}_{sp}$ & HSP & $h_{sp}$& $h_{sp}$&$h_{sp}$&---& $h_{sp}$\\
$\zeta'_{s}$ & ZSN & ---& --- &$z_{sn}$& $\tilde{\zeta}_s$& $z_{sn}$\\
$\zeta'_{p}$ & ZPN & ---& --- &$z_{pn}$&$\tilde{\zeta}_p$& $z_{pn}$\\
$\zeta'_{d}$ & ZDN & ---& --- &$z_{dn}$&$\tilde{\zeta}_d$& $z_{dn}$\\
$\vartheta$ & P09 & --- & --- &$\rho$(core)& $\varrho_{core}$& $\rho(\text{core})$\\
$K_a$ & FN1$a$ & --- & $K_a$&---&---& $a$\\
$L_a$ & FN2$a$ & --- & $L_a$&---&---& $b$\\
$M_a$ & FN3$a$ & --- & $M_a$&---&---& $c$\\
$\alpha$ & ALP & $\alpha$ &$\alpha$ &---&$\alpha$& ---\\
$\alpha'$ & ALPB & --- & ---&$\alpha_{ij}$&---& $\alpha$\\
$x$ & XFAC & --- & ---&$\delta_{ij}$&---& $\chi$\\
\end{longtable}
\end{center}}
\begin{center}
\begin{longtable}{cccc}
  \caption{Relation of the parameter abbreviations introduced in the 
main text for OM$x$ models to the ones chosen in several popular publications.
} \label{tab:parameter_names2} \\
\hline \hline 
Main Text & OM1\cite{Kolb1993a} & OM2\cite{Weber2000a} & OM3\cite{Dral2016b} \\
\hline 
\endfirsthead 
\hline \hline 
Main Text & OM1\cite{Kolb1993a} & OM2\cite{Weber2000a} & OM3\cite{Dral2016b} \\
\hline 
\endhead \hline 
\hline \endfoot
$U_{ss}$ & $U_{ss}$  & $U_{ss}$ & $U_{ss}$\\
$U_{pp}$ & $U_{pp}$  & $U_{pp}$ & $U_{pp}$\\
Scaling factor & $\zeta$ & $\zeta$ & $\zeta$ \\
Scaling factor for core orbitals & --- & $\zeta_\alpha$ & $\zeta_\alpha$ \\
$b_{l(\mu)}$ & $\beta_{l(\mu)}$ & $\beta_{l(\mu)}$ & $\beta_{l(\mu)}$ \\
$a_{l(\mu)}$ & $\alpha_{l(\mu)}$ & $\alpha_{l(\mu)}$ & $\alpha_{l(\mu)}$ \\
$W$ & --- & $F_{\alpha\alpha}$ & $F_{\alpha\alpha}$  \\
$F_1$ & $\gamma_1$ & $F_1$ & $F_1$ \\
$F_2$ & $\gamma_2$ & $F_2$ & --- \\
$G_1$ & --- & $G_1$ & $G_1$ \\
$G_2$ & --- & $G_2$ & --- \\
\end{longtable}
\end{center}
Furthermore, Thiel and co-workers denote $\theta(\chi_\mu^I,\chi_\lambda^J)$
 with $\beta_{\mu\lambda}$
in Refs.~\onlinecite{Kolb1993a, Weber2000a, ScholtenDiss, Dral2016b}
and $\eta({\chi_\mu^I,\chi_\lambda^J})$
is denoted  as \cite{Kolb1993a} $H_{\mu\mu}^{IJ}$ 
or as \cite{Weber2000a,ScholtenDiss,Dral2016b} $H_{\mu\mu,J}^{\text{loc}}$.

\subsection{Parametrization of One-Center ERIs}
\label{subsec:param1ceri}
The one-center ERI $\left<p^Ip'^I\middle|p^Ip'^I\right>$ 
 is calculated from $\gamma_{pp}^{Z_I}$ and $\gamma_{pp'}^{Z_I}$ with Eq.~(\ref{eq:hpp}).
A practical issue with Eq.~(\ref{eq:hpp}) arises when 
$\gamma_{pp}^{Z_I} < \gamma_{pp'}^{Z_I}$ so that $\left<p^Ip'^I\middle|p^Ip'^I\right>$
is negative. 
In this case, it will not be possible to determine a parameter necessary to calculate 
the distance $D_Q$ in the quadrupole moments according to Eq.~(\ref{eq:varthetaQpp}) 
because the applied numerical procedure will not converge. 
The search for the parameter value was terminated after 5 iterations 
in \textsc{Mopac} \cite{Stewart1990} even if it had not converged yet which is likely 
the reason why this failure has not been detected yet. 

This is not a practical issue in the original MNDO model \cite{Dewar1977}, 
in the original MNDO/d model, \cite{Thiel1991} in the original AM1 model \cite{Dewar1985}, 
the RM1 model, \cite{Rocha2006} or the OM$x$ models. \cite{Kolb1993a, Weber2000a, Dral2016b}

In the following, we list the elements and NDDO-SEMO models for which we encountered 
negative $\left<p^Ip'^I\middle|p^Ip'^I\right>$ in \textsc{Mopac}: 
\begin{itemize}
 \item MNDO --- Ga, Sr, Xe, 
Ba, and Tl \cite{Stewart2004}
 \item AM1 --- Li, Be, Mg, Ga, Sr, Sb, Xe, Ba, and Tl \cite{Stewart2004}
 \item PM3 --- Be, \cite{Stewart1989} Mg, \cite{Stewart1991} Rb, \cite{Stewart2004} 
Sr, \cite{Stewart2004} Xe, Cs, \cite{Stewart2004} Ba, \cite{Stewart2004} Hg, \cite{Stewart1991} 
and Tl \cite{Stewart1991} 
 \item PM6 --- He, Be, F, Ne, Na, Mg, Ar, K, Ca, Ga, Kr, Sr, In, Xe,  Hg, and Tl \cite{Stewart2007}
 \item PM7 --- He, B, Be, F, Ne, Na, Mg, Ar, K, Ca, Ga, Kr, Sr, In, Xe,  Hg, and Pb \cite{Stewart2012}
\end{itemize}

\subsection{Evaluation of Two-Center ERIs in MNDO-type Models with an $s,p$ basis set}
\label{subsec:sp2ceris}

The two-center ERIs
$\left<\chi_\mu^I \chi_\nu^I | \chi_\lambda^J \chi_\sigma^J\right>,\ I\ne J$ can be interpreted as the 
electrostatic interaction between a charge distribution $\chi_\mu^I \chi_\nu^I$ centered on atom $I$ and 
a charge distribution $\chi_\lambda^J \chi_\sigma^J$ centered on atom $J$.
The different possible charge distributions $\chi_\mu^I \chi_\nu^I$ in the $s$, $p$ minimal 
valence-shell basis are listed in Table~\ref{tab:multipoles}.
Each charge distribution is approximately represented as a truncated classical multipole expansion
of $T_{\mu\nu}$ multipoles $\Theta^{\mu\nu}_t,\ t=(1, 2, ..., T_{\mu\nu})$. \cite{Dewar1976}
The two-center ERI is then approximated as the electrostatic interaction energy 
$U(\Theta^{\mu\nu}_{t},\Theta^{\lambda\sigma}_{s})$
 of the 
$T_{\mu\nu}$ multipoles $\Theta^{\mu\nu}_{t}$ specified for $\chi_\mu^I \chi_\nu^I$ with 
the $T_{\lambda\sigma}$ multipoles $\Theta^{\lambda\sigma}_{s}$ specified for 
$\chi_\lambda^J \chi_\sigma^J$,
\begin{equation}
\label{eq:2ceri}
\left<\chi_\mu^I\chi_\nu^I|\chi_\lambda^J\chi_\sigma^J\right> \approx
\sum_{t=1}^{T_{\mu\nu}} \sum_{s=1}^{T_{\lambda\sigma}} 
U(\Theta^{\mu\nu}_{t},\Theta^{\lambda\sigma}_{s}).
\end{equation}
The multipoles $\Theta^{\mu\nu}_t$ may be a monopole $q^I$, a dipole $\mu^I_{x,y,z}$, 
a linear quadrupole $Q^I_{xx,yy,zz}$, 
and a
square quadrupole $Q^I_{xy,xz,yz}$  (see also Figure~3 in the main text).
Table~\ref{tab:multipoles} indicates which multipoles $\Theta^{\mu\nu}_t$ appear in the multipole 
expansion for the charge distribution $\chi_\mu^I \chi_\nu^I$. 
\begin{longtable}{ccc}
\caption{Number of multipoles $T_{\mu\nu}$ and 
types of multipoles applied to represent the 
 charge distribution $\chi_\mu^I\chi_\nu^J$ in an $s,p$ basis.
} \label{tab:multipoles} \\
\hline \hline 
Charge Distribution & $T_{\mu\nu}$ &  Multipoles\\
\hline 
\endfirsthead 
\hline \hline 
Charge Distribution & $T_{\mu\nu}$ & Multipoles\\
\hline 
\endhead \hline 
\hline \endfoot
$s^Is^I$ & 1 & $q^I$ \\
$p_x^Ip_x^I$ & 2& $q^I,Q^I_{xx}$ \\
$p_y^Ip_y^I$ & 2 & $q^I,Q^I_{yy}$ \\
$p_z^Ip_z^I$ & 2 & $q^I,Q^I_{zz}$ \\
$s^Ip_x^I$ & 1 & $\mu^I_x$ \\
$s^Ip_y^I$ & 1 & $\mu^I_y$ \\
$s^Ip_z^I$ & 1 & $\mu^I_z$ \\
$p_x^Ip_y^I$ & 1 & $Q^I_{xy}$ \\
$p_x^Ip_z^I$ & 1 & $Q^I_{xz}$ \\
$p_y^Ip_z^I$ & 1 & $Q^I_{yz}$ \\
\end{longtable}

The specification of the positions of the point charges which make up the dipoles and 
the quadrupoles necessitates the specification of 
$D_{\mu,sp}$ and $D_{Q,pp}$ (see also Figure~3 in the main text).
In the main text, we denoted $D_{\mu,sp}$ and $D_{Q,pp}$ as $D_{\mu}$ and $D_{Q}$, respectively, 
to keep the notation uncluttered. 
When an $s,p,d$ basis set is considered (see Section~\ref{sec:mndod}), we need 
to specify additional subscript identifiers indicating which kind of charge distribution is 
approximated.
The distances $D_{\mu,sp}$ and $D_{Q,pp}$ between the point charges are chosen such that 
the multipole moment of the point charge configuration approximates
 the one of the corresponding charge distribution. \cite{Dewar1976}
We first need to introduce the function $A(\chi_\mu^I, \chi_\nu^I,a)$, \cite{Thiel1991}
\begin{equation}
 \label{eq:defA}
\begin{split}
A(\chi_\mu^I, \chi_\nu^I,a) =& 
(2\zeta_\mu)^{n_\mu+\frac{1}{2}} 
(2\zeta_\nu)^{n_\nu+\frac{1}{2}} 
(\zeta_\mu + \zeta_\nu)^{-n_\mu-n_\nu-a-1}  
\left[(2n_\mu)!(2n_\nu)!\right]^{-\frac{1}{2}} 
(n_\mu+n_\nu+a)!,
\end{split}
\end{equation}
where $\zeta_\mu$ is the orbital exponent of $\chi_\mu^I$,  $n_\mu$ is 
the principal quantum number associated with  $\chi_\mu^I$, 
and $a$ characterizes the angular momentum 
of the multipole ($a\in\{0,1,2,...\}$).
The distances $D_{\mu,sp}$ and $D_{Q,pp}$ are then given by\cite{Dewar1976,Thiel1991}
\begin{equation}
 D_{\mu,sp} = 3^{-\frac{1}{2}} A(\chi_\mu^I, \chi_\nu^I,1)
\end{equation}
and
\begin{equation}
 D_{Q,pp} = 5^{-\frac{1}{2}} \sqrt{A(\chi_\mu^I, \chi_\nu^I,2)},
\end{equation}
respectively. 
The implementation of the formulae to calculate $D_{\mu,sp}$ and $D_{Q,pp}$
can be verified by comparison of $D_{\mu,sp}$ and $D_{Q,pp}$ to 
 \texttt{DD2} and \texttt{DD3}, respectively, which 
\textsc{Mopac} provides when specifying the 
 keyword \texttt{HCORE}.

When comparing our implementation to \textsc{Mopac},
we noticed that the MNDO-type models, in which $s$- and $p$-type basis functions 
with different principal 
quantum numbers ($n_s\ne n_p$)
are activated for an atom 
(i.e., for He, Ne, Ar, Kr, and Xe; see Table~\ref{tab:nv}),
do not actually apply the different principal quantum numbers to evaluate  $D_{\mu,sp}$ and $D_{Q,pp}$.
Instead, only the lower principal quantum number is applied.

With the help of $D_{\mu,sp}$ and $D_{Q,pp}$, we can specify the positions and 
charges of the $C$ individual point charges $q_c,\ c=(1,2,...,C)$ of the multipoles   
relative to the atom $I$ on which the charge distribution $\chi_\mu^I \chi_\nu^I$ is centered
(see Table~\ref{tab:multipoles_pos}).
\begin{center}
\begin{longtable}{ccccc}
  \caption{Specification of the position $\fett{r}_c = (r_{c,x},r_{c,y}, r_{c,z})$ 
of the $C$ point charges $q_c,\ c=(1,2,...,C)$ 
chosen to represent a specific multipole (Table~\ref{tab:multipoles}). 
The total number of point charges $C$ is given in brackets after the specification of the multipole.
The positions $\fett{r}_c$ are given in relation to the 
origin of the local coordinate system (Section~\ref{sec:localcoordinate}). 
The charges $q_c$ are given in atomic units, i.e., as multiples of the elementary 
charge.
} \label{tab:multipoles_pos} \\
\hline \hline 
Multipole & $r_{c,x}$ & $r_{c,y}$& $r_{c,z}$ & $q_c$ \\
\hline 
\endfirsthead
\hline \hline  
Multipole & $r_{c,x}$ & $r_{c,y}$& $r_{c,z}$ & $q_c$ \\
\hline 
\endhead 
\hline \hline \endfoot
$q$ ($C=1$)& $0.0$ & $0.0$ & $0.0$ & $+1.00$ \\
\hline 
$\mu_x$ ($C=2$)
        & $-D_{\mu,sp}$ & $0.0$ & $0.0$ & $-0.50$ \\
        & $+D_{\mu,sp}$ & $0.0$ & $0.0$ & $+0.50$ \\
\hline 
$\mu_y$ ($C=2$)
        & $0.0$ & $-D_{\mu,sp}$ & $0.0$ & $-0.50$ \\
        & $0.0$ & $+D_{\mu,sp}$ & $0.0$ & $+0.50$ \\
\hline 
$\mu_z$ ($C=2$)
        & $0.0$ & $0.0$ & $-D_{\mu,sp}$ & $-0.50$ \\
        & $0.0$ & $0.0$ & $+D_{\mu,sp}$ & $+0.50$ \\
\hline 
$Q_{xx}$ ($C=3$)
         & $-2D_{Q,pp}$ & $0.0$ & $0.0$ & $+0.25$ \\
         & $+2D_{Q,pp}$ & $0.0$ & $0.0$ & $+0.25$ \\
         & $0.0$ & $0.0$ & $0.0$& $-0.50$ \\
\hline 
$Q_{yy}$ ($C=3$)
         & $0.0$ & $-2D_{Q,pp}$ & $0.0$ & $+0.25$ \\
         & $0.0$ & $+2D_{Q,pp}$ & $0.0$ & $+0.25$ \\
         & $0.0$ & $0.0$ & $0.0$ & $-0.50$ \\
\hline 
$Q_{zz}$ ($C=3$)
         & $0.0$ & $0.0$ & $-2D_{Q,pp}$ & $+0.25$ \\
         & $0.0$ & $0.0$ & $+2D_{Q,pp}$ & $+0.25$ \\
         & $0.0$ & $0.0$ & $0.0$ & $-0.50$ \\
\hline 
$Q_{xy}$ ($C=4$)
         & $+D_{Q,pp}$ & $+D_{Q,pp}$ & $0.0$ & $+0.25$ \\
         & $-D_{Q,pp}$ & $-D_{Q,pp}$ & $0.0$ & $+0.25$ \\
         & $+D_{Q,pp}$ & $-D_{Q,pp}$ & $0.0$ & $-0.25$ \\
         & $-D_{Q,pp}$ & $+D_{Q,pp}$ & $0.0$ & $-0.25$ \\
\hline 
$Q_{xz}$ ($C=4$)
         & $+D_{Q,pp}$ & $0.0$ & $+D_{Q,pp}$ & $+0.25$ \\
         & $-D_{Q,pp}$ & $0.0$ & $-D_{Q,pp}$ & $+0.25$ \\
         & $+D_{Q,pp}$ & $0.0$ & $-D_{Q,pp}$ & $-0.25$ \\
         & $-D_{Q,pp}$ & $0.0$ & $+D_{Q,pp}$ & $-0.25$ \\
\hline 
$Q_{yz}$ ($C=4$)& 0.0 & $+D_{Q,pp}$ & $+D_{Q,pp}$ & $+0.25$ \\
         & 0.0 & $-D_{Q,pp}$ & $-D_{Q,pp}$ & $+0.25$ \\
         & 0.0 & $+D_{Q,pp}$ & $-D_{Q,pp}$ & $-0.25$ \\
         & 0.0 & $-D_{Q,pp}$ & $+D_{Q,pp}$ & $-0.25$ \\
\end{longtable}
\end{center}
After the specification of the position and value of the point charges, 
we can straightforwardly assess the 
electrostatic potential energy $ U(\Theta^{\mu\nu}_{t},\Theta^{\lambda\sigma}_{s})$
of the interaction of the $C_t$ and $C_s$ point charges making up $\Theta^{\mu\nu}_{t}$ 
and $\Theta^{\lambda\sigma}_{s}$, respectively (in atomic units),
\begin{equation}
\label{eq:epot}
  U(\Theta^{\mu\nu}_{t},\Theta^{\lambda\sigma}_{s}) = 
 \sum_{c=1}^{C_t} \sum_{\substack{d=1}}^{C_s} 
\frac{q^I_c q^J_d}{|\fett{r}^I_c-\fett{r}^J_d|}. 
\end{equation}
As stated in the main text, this interaction is not calculated 
analytically, but within the empirical Klopman approximation. \cite{Klopman1964a, Dewar1977, Dewar1976}
The Klopman approximation is introduced to be able to recover the respective one-center 
ERIs $\left<\chi_\mu^I\chi_\nu^I|\chi_\lambda^J\chi_\sigma^J\right>$ 
in the limit $\tilde{R}_{IJ}=0$.
In general, this means that the denominator in Eq.~(\ref{eq:epot}) 
is modified in such a way that we obtain the correct one-center limit,  
\begin{equation}
\label{eq:epot2}
U(\Theta^{\mu\nu}_{t},\Theta^{\lambda\sigma}_{s}) =  
\sum_{c=1}^{C_t} \sum_{\substack{d=1}}^{C_s}  \frac{q^I_c q^J_d}
{\sqrt{ |\fett{r}^I_c-\fett{r}^J_d|^2 + (\vartheta_c^I(\chi_\mu^I\chi_\nu^I) + 
\vartheta_d^J(\chi_\lambda^J\chi_\sigma^J))^2} }. 
\end{equation}
The term $\vartheta_c^I$ depends on the multipole to which the point charge $q_c$ 
belongs (see Table~\ref{tab:multipoles_pos}).
If $q_c$ is part of a monopole $q^I$ representing the charge distribution $s^Is^I$, 
$\vartheta^I_{q,ss}$ is applied in Eq.~(\ref{eq:epot2}).
The four terms $\vartheta^I_{q,ss}$, $\vartheta^I_{q,pp}$, $\vartheta^I_{\mu,sp}$,
and $\vartheta^I_{Q,pp}$ are calculated with reference to the one-center ERIs
 $\gamma_{ss}^{Z_I}$, 
$\tilde{\gamma}_{sp}^{Z_I}$, 
$\gamma_{pp}^{Z_I}$, 
and $\gamma_{pp'}^{Z_I}$, \cite{Dewar1976}
\begin{equation}
\label{eq:varthetaqss}
 \vartheta^I_{q,ss} = \frac{1}{2 \gamma_{ss}},
\end{equation}
\begin{equation}
\label{eq:varthetaqpp}
 \vartheta^I_{q,pp} = \vartheta^I_{q,ss},
\end{equation}
\begin{equation}
\label{eq:varthetampp}
 \left(\vartheta^I_{\mu,sp}\right)^{-1} - \left[\left(\vartheta^I_{\mu,sp}\right)^2 + 
(D_{\mu,sp})^2\right]^{-\frac{1}{2}} = \frac{4}{3} \tilde{\gamma}_{sp}, 
\end{equation}
and
\begin{equation}
\label{eq:varthetaQpp}
\begin{gathered}
 \left(\vartheta^I_{Q,pp}\right)^{-1} - 2 \left[
\left(\vartheta^I_{Q,pp}\right)^{2} +(D_{Q,pp})^2
\right]^{-\frac{1}{2}}
+ 
\left[
\left(\vartheta^I_{Q,pp}\right)^{2} +2 (D_{Q,pp})^2
\right]^{-\frac{1}{2}} 
= \frac{24}{25} \left<p^Ip'^I\middle|p^Ip'^I\right>, 
\end{gathered}
\end{equation}
The parameters cannot be calculated analytically, but have to be determined
in an iterative numerical procedure.
The values for $\vartheta^I_{q,ss}$, $\vartheta^I_{q,pp}$,  $\vartheta^I_{\mu,sp}$, 
and $\vartheta^I_{Q,pp}$ can be 
compared to those provided for \texttt{P01}, \texttt{P07}, \texttt{P02}, and 
\texttt{P03}, respectively, when specifying the keyword \texttt{Hcore} in \textsc{Mopac}. 

The implementation of the procedure to calculate the two-center ERIs 
can be compared to the implementation in \textsc{Mopac}
when specifying the keyword \texttt{Hcore} for the calculation of the electronic energy 
of a diatomic molecule which is aligned along the $z$-axis.
The values of the two-center ERIs are then listed under \texttt{TWO-ELECTRON MATRIX IN HCORE}
(note that these values really are the values for the ERIs and not the two-electron matrix entries).
The first one hundred entries are the one-center ERIs for the first atom 
(i.e., $\gamma_{ss}$, $\gamma_{sp}$, 
$\gamma_{pp}$, $\gamma_{pp'}$, $\tilde{\gamma}_{sp}$, $\tilde{\gamma}_{pp'}$, and zeros).
The next one hundred entries are the two-center ERIs which 
arise between the first and the second atom.
The order in which the two-center ERIs are given is described in Ref.~\onlinecite{mopac2} and is also given in the following:\\
$\left<ss|ss\right>$ $\left<ss|sp_x\right>$ $\left<ss|p_xp_x\right>$ $\left<ss|sp_y\right>$ $\left<ss|p_xp_y\right>$ $\left<ss|p_yp_y\right>$ $\left<ss|sp_z\right>$ $\left<ss|p_xp_z\right>$ $\left<ss|p_yp_z\right>$ $\left<ss|p_zp_z\right>$
$\left<sp_x|ss\right>$ $\left<sp_x|sp_x\right>$ $\left<sp_x|p_xp_x\right>$ $\left<sp_x|sp_y\right>$ $\left<sp_x|p_xp_y\right>$ $\left<sp_x|p_yp_y\right>$ $\left<sp_x|sp_z\right>$ $\left<sp_x|p_xp_z\right>$ $\left<sp_x|p_yp_z\right>$ $\left<sp_x|p_zp_z\right>$
$\left<p_xp_x|ss\right>$ $\left<p_xp_x|sp_x\right>$ $\left<p_xp_x|p_xp_x\right>$ $\left<p_xp_x|sp_y\right>$ $\left<p_xp_x|p_xp_y\right>$ $\left<p_xp_x|p_yp_y\right>$ $\left<p_xp_x|sp_z\right>$ $\left<p_xp_x|p_xp_z\right>$ $\left<p_xp_x|p_yp_z\right>$ $\left<p_xp_x|p_zp_z\right>$
$\left<sp_y|ss\right>$ $\left<sp_y|sp_x\right>$ $\left<sp_y|p_xp_x\right>$ $\left<sp_y|sp_y\right>$ $\left<sp_y|p_xp_y\right>$ $\left<sp_y|p_yp_y\right>$ $\left<sp_y|sp_z\right>$ $\left<sp_y|p_xp_z\right>$ $\left<sp_y|p_yp_z\right>$ $\left<sp_y|p_zp_z\right>$
$\left<p_xp_y|ss\right>$ $\left<p_xp_y|sp_x\right>$ $\left<p_xp_y|p_xp_x\right>$ $\left<p_xp_y|sp_y\right>$ $\left<p_xp_y|p_xp_y\right>$ $\left<p_xp_y|p_yp_y\right>$ $\left<p_xp_y|sp_z\right>$ $\left<p_xp_y|p_xp_z\right>$ $\left<p_xp_y|p_yp_z\right>$ $\left<p_xp_y|p_zp_z\right>$
$\left<p_yp_y|ss\right>$ $\left<p_yp_y|sp_x\right>$ $\left<p_yp_y|p_xp_x\right>$ $\left<p_yp_y|sp_y\right>$ $\left<p_yp_y|p_xp_y\right>$ $\left<p_yp_y|p_yp_y\right>$ $\left<p_yp_y|sp_z\right>$ $\left<p_yp_y|p_xp_z\right>$ $\left<p_yp_y|p_yp_z\right>$ $\left<p_yp_y|p_zp_z\right>$
$\left<sp_z|ss\right>$ $\left<sp_z|sp_x\right>$ $\left<sp_z|p_xp_x\right>$ $\left<sp_z|sp_y\right>$ $\left<sp_z|p_xp_y\right>$ $\left<sp_z|p_yp_y\right>$ $\left<sp_z|sp_z\right>$ $\left<sp_z|p_xp_z\right>$ $\left<sp_z|p_yp_z\right>$ $\left<sp_z|p_zp_z\right>$
$\left<p_xp_z|ss\right>$ $\left<p_xp_z|sp_x\right>$ $\left<p_xp_z|p_xp_x\right>$ $\left<p_xp_z|sp_y\right>$ $\left<p_xp_z|p_xp_y\right>$ $\left<p_xp_z|p_yp_y\right>$ $\left<p_xp_z|sp_z\right>$ $\left<p_xp_z|p_xp_z\right>$ $\left<p_xp_z|p_yp_z\right>$ $\left<p_xp_z|p_zp_z\right>$
$\left<p_yp_z|ss\right>$ $\left<p_yp_z|sp_x\right>$ $\left<p_yp_z|p_xp_x\right>$ $\left<p_yp_z|sp_y\right>$ $\left<p_yp_z|p_xp_y\right>$ $\left<p_yp_z|p_yp_y\right>$ $\left<p_yp_z|sp_z\right>$ $\left<p_yp_z|p_xp_z\right>$ $\left<p_yp_z|p_yp_z\right>$ $\left<p_yp_z|p_zp_z\right>$
$\left<p_zp_z|ss\right>$ $\left<p_zp_z|sp_x\right>$ $\left<p_zp_z|p_xp_x\right>$ $\left<p_zp_z|sp_y\right>$ $\left<p_zp_z|p_xp_y\right>$ $\left<p_zp_z|p_yp_y\right>$ $\left<p_zp_z|sp_z\right>$ $\left<p_zp_z|p_xp_z\right>$ $\left<p_zp_z|p_yp_z\right>$ $\left<p_zp_z|p_zp_z\right>$

\subsection{MNDO-type Models with an $s,p,d$ basis set}
\label{sec:mndod}

\subsubsection{Evaluation of One-Center ERIs}

The one-center ERIs $\left<\chi_\mu^I\chi_\nu^I|\chi_\lambda^I\chi_\sigma^I\right>$ 
will be calculated analytically if $s$-, $p$-, and $d$-type basis functions are activated 
for the atom $I$. 
For this purpose, the one-center ERIs are re-written as \cite{Condon1959}
\begin{equation}
\label{eq:erianalytical}
\begin{split}
 \left<\chi_\mu^I\chi_\nu^I|\chi_\lambda^I\chi_\sigma^I\right> =  & \sum_{k=0}^\infty 
C^k(l(\mu)m(\mu),l(\nu)m(\nu))
C^k(l(\sigma)m(\sigma),l(\lambda)m(\lambda)) \\ & \times 
R^k(\chi_\mu^I,\chi_\nu^I,\chi_\lambda^I,\chi_\sigma^I), \\ 
\end{split}
\end{equation}
where $C^k(l(\mu)m(\mu),l(\nu)m(\nu))$ denotes the 
so-called angular coefficients and 
$R^k(\chi_\mu^I,\chi_\nu^I,\chi_\lambda^I,\chi_\sigma^I)$
the radial integrals.
The radial integrals
$R^k(\chi_\mu^I,\chi_\nu^I,\chi_\lambda^I,\chi_\sigma^I)$ 
are calculated as follows (Eq.~(3) in Ref.~\onlinecite{Kumar1987}), 
\begin{equation}
\label{eq:slatercondon}
\begin{split}
 R^k&(\chi_\mu^I,\chi_\nu^I,\chi_\lambda^I,\chi_\sigma^I) = 
\frac{(2 \zeta'_{\lambda})^{n_\lambda+1/2} (2 \zeta'_{\sigma})^{n_\sigma+1/2} }
{\sqrt{(2n_\lambda)!(2n_\sigma)!)}}
\frac{(2 \zeta'_{\mu})^{n_\mu+1/2} (2 \zeta'_{\nu})^{n_\nu+1/2} }
{\sqrt{(2n_\mu)!(2n_\nu)!)}} \\
& \times 
\frac{( n_{\sigma} + n_{\lambda} + k) !}{ (\zeta'_{\sigma}+\zeta'_{\lambda})^{n_{\sigma}+n_{\lambda}+k+1}} 
\left\{\frac{(n_{\mu}+n_{\nu}-k-1)!}{(\zeta'_{\mu}+\zeta'_{\nu})^{n_{\mu}+n_{\nu}-k}} \right. \\
& \left. 
- \sum_{k'=1}^{n_{\sigma}+n_{\lambda}+k+1}  
\frac{ (\zeta'_{\sigma}+\zeta'_{\lambda})^{n_{\sigma}+n_{\lambda}+k-k'+1} }{ (n_{\sigma}+n_{\lambda}+k-k'+1)!}  
\frac{ [ n_{\mu}+n_{\nu}+n_{\lambda}+n_{\sigma}-k']! }{ [\zeta'_{\mu}+\zeta'_{\nu} + 
\zeta'_{\lambda}+\zeta'_{\sigma}]^
{n_{\mu}+n_{\nu}+n_{\lambda}+n_{\sigma}-k'+1} } \right. \\
& \left.
+ \sum_{k'=1}^{n_{\lambda}+n_{\sigma}-k} \left[
\frac{
(\zeta'_{\lambda}+\zeta'_{\sigma})^{n_{\lambda}+n_{\sigma}+k-k'+1} (n_{\lambda}+n_{\sigma}-k-1)! 
}
{
(n_{\lambda}+n_{\sigma}+k)!(n_{\lambda}+n_{\sigma}-k-k')!
} \right.
\right. \\ & \left. \left.
\times
\frac{
[n_{\mu}+n_{\nu}+n_{\lambda}+n_{\sigma}-k']!
}
{
[\zeta'_{\mu}+\zeta'_{\nu}+\zeta'_{\lambda}+\zeta'_{\sigma}]
^{n_{\mu}+n_{\nu}+n_{\lambda}+n_{\sigma}-k'+1}
}\right]
\right\},
\end{split}
\end{equation}
where $\zeta'_{\mu}$ is the auxiliary orbital exponent and $n_\mu$ 
the principal quantum number associated with the basis 
function $\chi_\mu^I$. 

Pelik\'{a}n and Nagy determined \cite{Pelikan1974} the values 
of the 58 unique nonzero one-center ERIs $\left<\chi_\mu^I\chi_\nu^I|\chi_\lambda^I\chi_\sigma^I\right>$  
in terms of $R^k(\chi_\mu^I,\chi_\nu^I,\chi_\lambda^I,\chi_\sigma^I)$ by 
explicitly evaluating Eq.~(\ref{eq:erianalytical}).
For this purpose, they used the angular coefficients\\ $C^k(l(\mu)m(\mu),l(\nu)m(\nu))$ which 
are presented in Table 1 on pp. 178--179 in Ref.~\onlinecite{Condon1959}.
In their work, the term $R^k(\chi_\mu^I,\chi_\nu^I,\chi_\lambda^I,\chi_\sigma^I)$
is also denoted as $R^k_{l(\mu)l(\nu)l(\lambda)l(\sigma)}$.
It is customary \cite{Condon1959, Pelikan1974} to introduce 
the quantities $F^k_{l(\mu)l(\nu)}$,
\begin{equation}
F^k_{l(\mu)l(\nu)} \equiv 
 R^k_{l(\mu)l(\mu)l(\nu)l(\nu)}  ,
\end{equation}
and $G^k_{l(\mu)l(\nu)}$,
\begin{equation}
G^k_{l(\mu)l(\nu)} \equiv
 R^k_{l(\mu)l(\nu)l(\mu)l(\nu)},
\end{equation}
to simplify the notation.
The formulae for the one-center ERIs are presented 
in Table 2 of Ref.~\onlinecite{Pelikan1974}, but some contain typographical mistakes
which we clarify here (corrected formulae are indicated by an asterisk appended to the equation number); 
for the 58 one-center ERIs, they read in our notation:
\begin{equation}
 \left<s^Is^I|s^Is^I\right> = F_{ss}^0
\end{equation}
\begin{equation}
 \left<s^Is^I|p^I_xp^I_x\right> = \left<s^Is^I|p^I_yp^I_y\right> = \left<s^Is^I|p^I_zp^I_z\right> =
F_{sp}^0
\end{equation}
\begin{equation}
 \left<s^Ip^I_x|s^Ip^I_x\right> = \left<s^Ip^I_y|s^Ip^I_y\right> = \left<s^Ip^I_z|s^Ip^I_z\right> = 
\frac{1}{3} G_{sp}^1
\end{equation}
\begin{equation}
 \left<p^I_xp^I_x|p^I_xp^I_x\right> = \left<p^I_yp^I_y|p^I_yp^I_y\right> = \left<p^I_zp^I_z|p^I_zp^I_z\right> =
F_{pp}^0 + \frac{4}{25} F_{pp}^2
\end{equation}
\begin{equation}
 \left<p^I_xp^I_x|p^I_yp^I_y\right> = \left<p^I_xp^I_x|p^I_zp^I_z\right> = \left<p^I_yp^I_y|p^I_zp^I_z\right> = 
F_{pp}^0 - \frac{2}{25} F_{pp}^2
\end{equation}
\begin{equation}
 \left<p^I_xp^I_y|p^I_xp^I_y\right> = \left<p^I_xp^I_z|p^I_xp^I_z\right> = \left<p^I_yp^I_z|p^I_yp^I_z\right> = 
\frac{3}{25} F_{pp}^2
\end{equation}
\begin{equation}
 \left<p^I_xd^I_{z^2}|p^I_xd^I_{z^2}\right> =\left<p^I_yd^I_{z^2}|p^I_yd^I_{z^2}\right> = 
\frac{1}{15} G_{pd}^1 + \frac{18}{245} G_{pd}^3
\end{equation}
\begin{equation}
 \left<p^I_xd^I_{z^2}|p^I_xd^I_{x^2-y^2}\right> =  \left<p^I_xd^I_{z^2}|p^I_yd^I_{xy}\right> =
 \left<p^I_xd^I_{xy}|d^I_{z^2}p^I_y\right> = 
-\frac{\sqrt{3}}{15} G_{pd}^1 - \frac{\sqrt{27}}{245} G_{pd}^3
\end{equation}
\begin{equation}
 \left<p^I_xd^I_{x^2-y^2}|p^I_yd^I_{xy}\right> = \frac{1}{5} G_{pd}^1 - \frac{21}{245} G_{pd}^3
\end{equation}
\begin{equation}
 \left<p^I_xd^I_{xy}|p^I_yd^I_{x^2-y^2}\right> = - \frac{1}{5} G_{pd}^1 + \frac{21}{245} G_{pd}^3
\end{equation}
\begin{equation}
 \left<p^I_xp^I_x|d^I_{z^2}d^I_{z^2}\right> =  \left<p^I_yp^I_y|d^I_{z^2}d^I_{z^2}\right> = 
F_{pd}^0 - \frac{2}{35} F_{pd}^2
\end{equation}
\begin{equation}
 \left<p^I_xp^I_x|d^I_{z^2}d^I_{x^2-y^2}\right> = \left<p^I_xp^I_y|d^I_{z^2}d^I_{xy}\right>=
-\frac{\sqrt{12}}{35} F_{pd}^2
\end{equation}
\begin{equation}
 \left<p^I_yd^I_{z^2}|p^I_yd^I_{x^2-y^2}\right> = \frac{\sqrt{3}}{15} G_{pd}^1 + \frac{\sqrt{27}}{245}
G_{pd}^3
\end{equation}
\begin{equation}
 \left<p^I_zd^I_{z^2}|p^I_zd^I_{z^2}\right> = \frac{4}{15} G_{pd}^1 + \frac{27}{245} G_{pd}^3
\end{equation}
\begin{equation}
 \begin{gathered}
  \left<p^I_zd^I_{x^2-y^2}|p^I_zd^I_{x^2-y^2}\right> = \left<p^I_zd^I_{xy}|p^I_zd^I_{xy}\right> = 
\left<p^I_xd^I_{yz}|p^I_xd^I_{yz}\right> = \left<p^I_yd^I_{xz}|p^I_yd^I_{xz}\right>  \\ = 
\left<p^I_zd^I_{x^2-y^2}|p^I_xd^I_{xz}\right> = \left<p^I_zd^I_{xy}|p^I_xd^I_{yz}\right> =
\left<p^I_zd^I_{xy}|p^I_yd^I_{xz}\right>  = \left<p^I_xd^I_{yz}|p^I_yd^I_{xz}\right> \\ = 
 \frac{3}{49} G_{pd}^3
 \end{gathered}
\end{equation}
\begin{equation}
\begin{gathered}
 \left<p^I_zd^I_{xz}|p^I_zd^I_{xz}\right> = \left<p^I_zd^I_{yz}|p^I_zd^I_{yz}\right> 
= \left<p^I_xd^I_{xy}|p^I_xd^I_{xy}\right> = 
\left<p^I_xd^I_{x^2-y^2}|p^I_xd^I_{x^2-y^2}\right> \\ = 
 \left<p^I_xd^I_{xz}|p^I_xd^I_{xz}\right> = \left<p^I_yd^I_{xy}|p^I_yd^I_{xy}\right> = 
\left<p^I_yd^I_{yz}|p^I_yd^I_{yz}\right> =  \left<p^I_yd^I_{x^2-y^2}|p^I_yd^I_{x^2-y^2}\right>\\ =
 \frac{1}{5} G_{pd}^1 + \frac{24}{245} G_{pd}^3
\end{gathered}
\end{equation}
\stepcounter{equation}
\begin{equation}
\label{eq:17}
 \left<p^I_zd^I_{z^2}|p^I_xd^I_{xz}\right> = \left<p^I_zd^I_{z^2}|p^I_yd^I_{yz}\right> = \frac{\sqrt{12}}{15}
G_{pd}^1 - \frac{\sqrt{243}}{245} G_{pd}^3
\tag{\theequation *}
\end{equation}
\begin{equation}
\begin{gathered}
 \left<p^I_zd^I_{xz}|p^I_xd^I_{x^2-y^2}\right> = \left<p^I_zd^I_{yz}|p^I_xd^I_{xy}\right> = 
\left<p^I_zd^I_{xz}|p^I_yd^I_{xy}\right>  =  
 \left<p^I_xd^I_{xz}|p^I_yp^I_{yz}\right> \\ = \frac{1}{5} G_{pd}^1 
- \frac{6}{245} G_{pd}^3
\end{gathered}
\end{equation}
\begin{equation}
 \left<p^I_zd^I_{x^2-y^2}|p^I_yd^I_{yz}\right> = -\frac{3}{49} G_{pd}^3
\end{equation}
\begin{equation}
 \left<p^I_zd^I_{yz}|p^I_yd^I_{z^2}\right> = \left<p^I_zd^I_{xz}|p^I_xd^I_{z^2}\right> = 
-\frac{\sqrt{3}}{15} G_{pd}^1 + \frac{\sqrt{432}}{245} G_{pd}^3
\end{equation}
\begin{equation}
 \left<p^I_zd^I_{yz}|p^I_yd^I_{x^2-y^2}\right> = -\frac{1}{5} G_{pd}^1 + \frac{6}{245} G_{pd}^3
\end{equation}
\begin{equation}
 \left<p^I_zp^I_z|d^I_{z^2}d^I_{z^2}\right> = F_{pd}^0 + \frac{4}{35} F_{pd}^2
\end{equation}
\begin{equation}
 \left<p^I_zp^I_x|d^I_{z^2}d^I_{xz}\right> = \left<p^I_zp^I_y|d^I_{z^2}d^I_{yz}\right> = \frac{\sqrt{3}}{35} F_{pd}^2 
\end{equation}
\begin{equation}
\begin{gathered}
  \left<p^I_zp^I_z|d^I_{x^2-y^2}d^I_{x^2-y^2}\right> = \left<p^I_zp^I_z|d^I_{xy}d^I_{xy}\right> = 
\left<p^I_xp^I_x|d^I_{yz}d^I_{yz}\right>  = 
  \left<p^I_yp^I_y|d^I_{xz}d^I_{xz}\right> \\ = F_{pd}^0 - \frac{4}{35} F_{pd}^2
\end{gathered}
\end{equation}
\begin{equation}
\begin{gathered}
\left<p^I_zp^I_x|d^I_{x^2-y^2}d^I_{xz}\right> = 
\left<p^I_zp^I_x|d^I_{xy}d^I_{yz}\right> = 
 \left<p^I_zp^I_y|d^I_{xy}d^I_{xz}\right>  = 
\left<p^I_xp^I_y|d^I_{xz}d^I_{yz}\right> = \frac{3}{35} F_{pd}^2
\end{gathered}
\end{equation}
\begin{equation}
 \left<p^I_zp^I_y|d^I_{x^2-y^2}d^I_{yz}\right> = -\frac{3}{35} F_{pd}^2
\end{equation}
\begin{equation}
 \begin{gathered}
  \left<p^I_zp^I_z|d^I_{xz}d^I_{xz}\right> = \left<p^I_zp^I_z|d^I_{yz}d^I_{yz}\right> = 
\left<p^I_xp^I_x|d^I_{xy}d^I_{xy}\right> = 
\left<p^I_xp^I_x|d^I_{x^2-y^2}d^I_{x^2-y^2}\right>  \\ = 
 \left<p^I_xp^I_x|d^I_{xz}d^I_{xz}\right> = \left<p^I_yp^I_y|d^I_{xy}d^I_{xy}\right> = 
\left<p^I_yp^I_y|d^I_{yz}d^I_{yz}\right> = 
 \left<p^I_yp^I_y|d^I_{x^2-y^2}d^I_{x^2-y^2}\right> \\= F_{pd}^0 + \frac{2}{35} F_{pd}^2
 \end{gathered}
\end{equation}
\begin{equation}
 \left<p^I_yp^I_y|d^I_{z^2}d^I_{x^2-y^2}\right> = \frac{\sqrt{12}}{35} F_{pd}^2
\end{equation}
\begin{equation}
 \begin{gathered}
  \left<s^Id^I_{z^2}|s^Id^I_{z^2}\right> = \left<s^Id^I_{x^2-y^2}|s^Id^I_{x^2-y^2}\right> = 
 \left<s^Id^I_{xy}|s^Id^I_{xy}\right> = \left<s^Id^I_{xz}|s^Id^I_{xz}\right> \\ =  
\left<s^Id^I_{yz}|s^Id^I_{yz}\right> = \frac{1}{5} G_{sd}^2
 \end{gathered}
\end{equation}
\begin{equation}
 \begin{gathered}
  \left<s^Is^I|d^I_{z^2}d^I_{z^2}\right> = \left<s^Is^I|d^I_{x^2-y^2}d^I_{x^2-y^2}\right> = 
\left<s^Is^I|d^I_{xy}d^I_{xy}\right> = \left<s^Is^I|d^I_{xz}d^I_{xz}\right> \\ = 
 \left<s^Is^I|d^I_{yz}d^I_{yz}\right> = F_{sd}^0
 \end{gathered}
\end{equation}
\begin{equation}
 \begin{gathered}
  \left<d^I_{z^2}d^I_{z^2}|d^I_{z^2}d^I_{z^2}\right> = 
\left<d^I_{x^2-y^2}d^I_{x^2-y^2}|d^I_{x^2-y^2}d^I_{x^2-y^2}\right> = 
\left<d^I_{xy}d^I_{xy}|d^I_{xy}d^I_{xy}\right>  \\= \left<d^I_{xz}d^I_{xz}|d^I_{xz}d^I_{xz}\right>   =
 \left<d^I_{yz}d^I_{yz}|d^I_{yz}d^I_{yz}\right> = F_{dd}^0 + \frac{4}{49} F_{dd}^2 + \frac{36}{441} F_{dd}^4
 \end{gathered}
\end{equation}
\begin{equation}
 \left<d^I_{z^2}d^I_{x^2-y^2}|d^I_{z^2}d^I_{x^2-y^2}\right> = 
\left<d^I_{z^2}d^I_{xy}|d^I_{z^2}d^I_{xy}\right> = 
\frac{4}{49} F_{dd}^2 + \frac{15}{441} F_{dd}^4
\end{equation}
\begin{equation}
\left<d^I_{z^2}d^I_{xz}|d^I_{z^2}d^I_{xz}\right> =
 \left<d^I_{z^2}d^I_{yz}|d^I_{z^2}d^I_{yz}\right> =
\frac{1}{49} F_{dd}^2 + \frac{30}{441} F_{dd}^4
\end{equation}
\begin{equation}
 \left<d^I_{z^2}d^I_{z^2}|d^I_{x^2-y^2}d^I_{x^2-y^2}\right> = 
 \left<d^I_{z^2}d^I_{z^2}|d^I_{xy}d^I_{xy}\right> = F_{dd}^0 - \frac{4}{49} F_{dd}^2 + \frac{6}{441} F_{dd}^4
\end{equation}
\begin{equation}
 \left<d^I_{z^2}d^I_{xz}|d^I_{x^2-y^2}d^I_{xz}\right> = \left<d^I_{z^2}d^I_{xz}|d^I_{xy}d^I_{yz}\right> = 
\left<d^I_{z^2}d^I_{yz}|d^I_{xy}d^I_{xz}\right> = \frac{\sqrt{3}}{49} F_{dd}^2 - \frac{\sqrt{75}}{441} F_{dd}^4
\end{equation}
\begin{equation}
 \left<d^I_{z^2}d^I_{yz}|d^I_{x^2-y^2}d^I_{yz}\right>=-\frac{\sqrt{3}}{49} F_{dd}^2 + \frac{\sqrt{75}}{441} F_{dd}^4
\end{equation}
\begin{equation}
 \left<d^I_{z^2}d^I_{z^2}|d^I_{xz}d^I_{xz}\right> = 
 \left<d^I_{z^2}d^I_{z^2}|d^I_{yz}d^I_{yz}\right> = F_{dd}^0 + \frac{2}{49} F_{dd}^2 - \frac{24}{441} F_{dd}^4 
\end{equation}
\begin{equation}
 \left<d^I_{z^2}d^I_{x^2-y^2}|d^I_{xz}d^I_{xz}\right> = \left<d^I_{z^2}d^I_{xy}|d^I_{xz}d^I_{yz}\right> =
-\frac{\sqrt{12}}{49} F_{dd}^2 + \frac{\sqrt{300}}{441} F_{dd}^4
\end{equation}
\begin{equation}
 \left<d^I_{z^2}d^I_{x^2-y^2}|d^I_{yz}d^I_{yz}\right> = \frac{\sqrt{12}}{49}F_{dd}^2 - \frac{\sqrt{300}}{441} F_{dd}^4
\end{equation}
\begin{equation}
 \left<d^I_{x^2-y^2}d^I_{xy}|d^I_{x^2-y^2}d^I_{xy}\right> = \frac{35}{441} F_{dd}^4
\end{equation}
\begin{equation}
 \begin{gathered}
  \left<d^I_{xy}d^I_{xz}|d^I_{xy}d^I_{xz}\right> = \left<d^I_{xy}d^I_{yz}|d^I_{xy}d^I_{yz}\right> = 
\left<d^I_{xz}d^I_{yz}|d^I_{xz}d^I_{yz}\right> = \left<d^I_{x^2-y^2}d^I_{xz}|d^I_{x^2-y^2}d^I_{xz}\right> \\ = 
 \left<d^I_{x^2-y^2}d^I_{yz}|d^I_{x^2-y^2}d^I_{yz}\right> = \frac{3}{49} F_{dd}^2 + \frac{20}{441} F_{dd}^4
 \end{gathered}
\end{equation}
\begin{equation}
 \left<d^I_{x^2-y^2}d^I_{x^2-y^2}|d^I_{xy}d^I_{xy}\right> = F_{dd}^0 + \frac{4}{49} F_{dd}^2 - \frac{34}{441}
F_{dd}^4
\end{equation}
\begin{equation}
 \left<d^I_{x^2-y^2}d^I_{xz}|d^I_{xy}d^I_{yz}\right> = \frac{3}{49} F_{dd}^2 - \frac{15}{441} F_{dd}^4
\end{equation}
\begin{equation}
 \left<d^I_{x^2-y^2}d^I_{yz}|d^I_{xy}d^I_{xz}\right> = - \frac{3}{49} F_{dd}^2 + \frac{15}{441} F_{dd}^4
\end{equation}
\begin{equation}
\begin{gathered}
\left<d^I_{xy}d^I_{xy}|d^I_{xz}d^I_{xz}\right> = \left<d^I_{xy}d^I_{xy}|d^I_{yz}d^I_{yz}\right> = 
\left<d^I_{xz}d^I_{xz}|d^I_{yz}d^I_{yz}\right> = \left<d^I_{x^2-y^2}d^I_{x^2-y^2}|d^I_{xz}d^I_{xz}\right> \\ = 
\left<d^I_{x^2-y^2}d^I_{x^2-y^2}|d^I_{yz}d^I_{yz}\right> = F_{dd}^0 - \frac{2}{49} F_{dd}^2 - 
\frac{4}{441} F_{dd}^4
\end{gathered}
\end{equation}
\begin{equation}
 \left<s^Id^I_{z^2}|d^I_{z^2}d^I_{z^2}\right> = \frac{2}{\sqrt{245}}R_{sddd}^2
\end{equation}
\begin{equation}
 \begin{gathered}
\left<s^Id^I_{x^2-y^2}|d^I_{z^2}d^I_{x^2-y^2}\right> = \left<s^Id^I_{xy}|d^I_{z^2}d^I_{xy}\right> = 
\left<s^Id^I_{z^2}|d^I_{x^2-y^2}d^I_{x^2-y^2}\right> = 
\left<s^Id^I_{z^2}|d^I_{xy}d^I_{xy}\right> \\ = - \frac{2}{\sqrt{245}} R_{sddd}^2
 \end{gathered}
\end{equation}
\begin{equation}
\begin{gathered}
\left<s^Id^I_{xz}|d^I_{z^2}d^I_{xz}\right> = \left<s^Id^I_{yz}|d^I_{z^2}d^I_{yz}\right> = 
\left<s^Id^I_{z^2}|d^I_{xz}d^I_{xz}\right>  = 
\left<s^Id^I_{z^2}|d^I_{yz}d^I_{yz}\right> = \frac{1}{\sqrt{245}} R_{sddd}^2
\end{gathered}
\end{equation}
\begin{equation}
 \begin{gathered}
  \left<s^Id^I_{xz}|d^I_{x^2-y^2}d^I_{xz}\right> = \left<s^Id^I_{xz}|d^I_{xy}d^I_{yz}\right> = 
\left<s^Id^I_{yz}|d^I_{xy}d^I_{xz}\right> = \left<s^Id^I_{xy}|d^I_{xz}d^I_{yz}\right> \\  = 
 \left<s^Id^I_{x^2-y^2}|d^I_{xz}d^I_{xz}\right> = \frac{\sqrt{3}}{\sqrt{245}}R_{sddd}^2
 \end{gathered}
\end{equation}
\begin{equation}
 \left<s^Id^I_{yz}|d^I_{x^2-y^2}d^I_{yz}\right> = \left<s^Id^I_{x^2-y^2}|d^I_{yz}d^I_{yz}\right> = 
- \frac{\sqrt{3}}{\sqrt{245}} R_{sddd}^2
\end{equation}
\stepcounter{equation}
\begin{equation}
\label{eq:51}
 \left<s^Id^I_{z^2}|p^I_xp^I_x\right> = \left<s^Id^I_{z^2}|p^I_yp^I_y\right> = - \frac{1}{\sqrt{125}} R_{sdpp}^2
\tag{\theequation *}
\end{equation}
\begin{equation}
 \left<s^Ip^I_y|p^I_yd^I_{x^2-y^2}\right> = -\frac{\sqrt{3}}{\sqrt{45}} R_{sppd}^1
\end{equation}
\stepcounter{equation}
\begin{equation}
\label{eq:53}
 \left<p^I_yp^I_y|s^Id^I_{x^2-y^2}\right> = -\frac{\sqrt{3}}{\sqrt{125}} R_{sdpp}^2
\tag{\theequation *}
\end{equation}
\stepcounter{equation}
\begin{equation}
\label{eq:54}
 \left<p^I_zs^I|p^I_zd^I_{z^2}\right> = \frac{2}{\sqrt{45}} R_{sppd}^1
\tag{\theequation *}
\end{equation}
\begin{equation}
\begin{gathered}
 \left<p^I_zs^I|p^I_xd^I_{xz}\right> = \left<p^I_zd^I_{xz}|p^I_xs^I\right> = \left<p^I_zs^I|p^I_yd^I_{yz}\right> = 
\left<p^I_zd^I_{yz}|p^I_ys^I\right>  = 
\left<p^I_xs^I|p^I_xd^I_{x^2-y^2}\right> \\ = \left<p^I_xs^I|p^I_yd^I_{xy}\right>  = \left<p^I_xd^I_{xy}|s^Ip^I_y\right> = 
\frac{\sqrt{3}}{\sqrt{45}} R_{sppd}^1
\end{gathered}
\end{equation}
\stepcounter{equation}
\begin{equation}
\label{eq:56}
 \left<p^I_zp^I_z|s^Id^I_{z^2}\right> = \frac{2}{\sqrt{125}} R_{sdpp}^2
\tag{\theequation *}
\end{equation}
\stepcounter{equation}
\begin{equation}
\label{eq:57}
\begin{gathered}
 \left<p^I_zp^I_x|s^Id^I_{xz}\right> = \left<p^I_zp^I_y|s^Id^I_{yz}\right> =
\left<p^I_xp^I_x|s^Id^I_{x^2-y^2}\right>  = 
 \left<p^I_xp^I_y|s^Id^I_{xy}\right> = \frac{\sqrt{3}}{\sqrt{125}} R_{sdpp}^2 \tag{\theequation *}
\end{gathered}
\end{equation}
\begin{equation}
 \left<p^I_xs^I|p^I_xd^I_{z^2}\right> =  \left<p^I_ys^I|p^I_yd^I_{z^2}\right> = -\frac{1}{\sqrt{45}} R_{sppd}^1
\end{equation}
The typographical mistakes in Ref.~\onlinecite{Pelikan1974} which we corrected here can be summarized 
as follows:
\begin{itemize}
 \item The factor 12 has to be substituted for $\sqrt{12}$ in Eq.~(17) of Ref.~\onlinecite{Pelikan1974}, 
see Eq.~(\ref{eq:17}).
 \item The radial integral $R_{sppd}^2$ has to be replaced by $R_{sdpp}^2$ in 
Eqs.~(51), (53), (56), and (57) of Ref.~\onlinecite{Pelikan1974},
see Eqs.~(\ref{eq:51}), (\ref{eq:53}), (\ref{eq:56}), and (\ref{eq:57}), respectively.
\item $R_{spdd}^1$ has to be replaced for $R_{sppd}^1$
in Eq.~(54) of Ref.~\onlinecite{Pelikan1974}, see Eq.~(\ref{eq:54}).
\end{itemize}
The implementation of the erroneous equations would 
affect all MNDO-type models which activate $d$-type basis functions 
 (i.e., MNDO/d, AM1, PM6, and PM7).
Specifically, the one-center ERIs 
$\left<p_z^Id_{z^2}^I|p_x^Id_{xz}^I\right>$,
 $\left<p_z^Id_{z^2}^I|p_y^Id_{yz}^I\right>$,
$\left<s^Id_{z^2}^I|p_x^Ip_x^I\right>$,
$\left<s^Id_{z^2}^I|p_y^Ip_y^I\right>$,
$\left<p_y^Ip_y^I|s^Id_{x^2-y^2}^I\right>$,
$\left<p_z^Is^I|p_z^Id_{z^2}^I\right>$,
$\left<p_z^Ip_z^I|s^Id_{z^2}^I\right>$, \\ 
$\left<p_z^Ip_x^I|s^Id_{xz}^I\right>$,
$\left<p_z^Ip_y^I|s^Id_{yz}^I\right>$,
$\left<p_x^Ip_x^I|s^Id_{x^2-y^2}^I\right>$, and
$\left<p_x^Ip_z^I|s^Id_{xy}^I\right>$ would then be erroneous. 

One can then find 2025 values for the one-center ERIs 
(their order is detailed in Ref.~\onlinecite{mopac2})  listed
under \texttt{TWO-ELECTRON MATRIX IN HCORE} in the output when carrying 
out a calculation with \textsc{Mopac} and invoking the keyword \texttt{Hcore}.

\subsubsection{Evaluation of Two-Center ERIs}
\begin{figure}[ht]
 \centering
 \includegraphics[width=\textwidth]{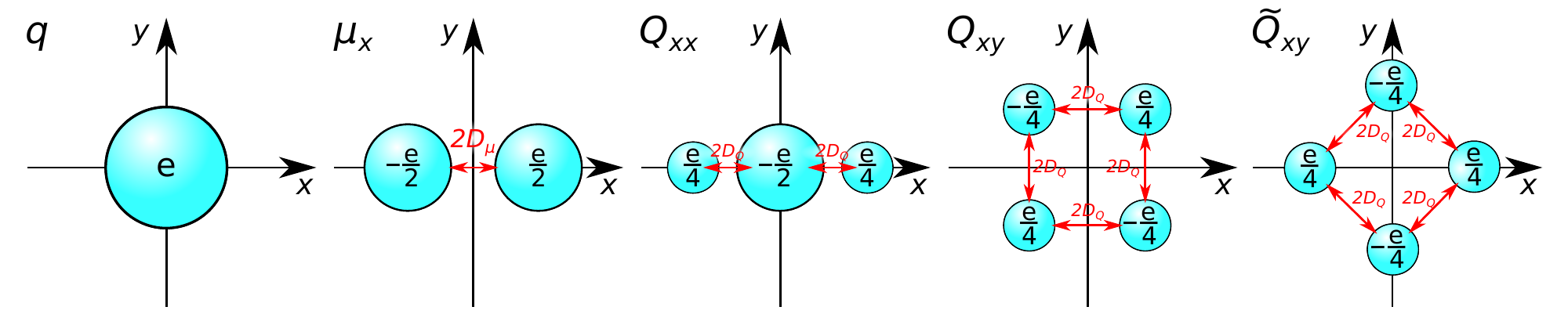}
 \caption{Illustration of the configuration of point charges (blue spheres) for the monopole $q$, 
the dipole $\mu_x$, the linear quadrupole $Q_{xx}$, the square quadrupole $Q_{xy}$, and 
the quadrupole $\tilde{Q}_{xy,xz,yz}$.
The charge for each point charge is given in units of the elementary charge. The point charges in $\mu_x$
are $2D_\mu$ apart and the ones in 
$Q_{xx}$, $Q_{xy}$, and $\tilde{Q}_{xy}$
are $2D_Q$ apart.
}
\label{fig:multipoles}
\end{figure}

Conceptually, the calculation of the two-center ERIs in the $s,p,d$ basis
is similar to the one in the $s,p$ basis. \cite{Thiel1991, Thiel1996c}
If at least \textit{one} of the basis functions contributing to a two-center ERI is 
a $d$-type basis function, the following equations are applied instead of the ones 
specified in Section~\ref{subsec:sp2ceris}. 

To fully describe the point charge interactions when a $d$-type orbital is involved 
in a two-center ERI, Thiel and Voityuk specified 
an additional point charge configuration $\tilde{Q}_{xy,yz,xz}$ 
(see Figure~\ref{fig:multipoles}).
Within an $s,p,d$-orbital basis, many more combinations of charge distributions are now possible 
\cite{Thiel1991,Thiel1996c}
than in the $s,p$ basis (see Table~\ref{tab:multipoles2}). These multipoles are applied 
when at least one $d$-type orbital is involved.
\begin{center}
\begin{longtable}{ccc}
  \caption{Specification of multipoles which are applied to represent the 
possible charge distributions $\chi_\mu^I\chi_\nu^J$ in an $s,p,d$ basis set
in a truncated multipole expansion.  
We indicate the center of the multipole by a superscript atom index.  
} \label{tab:multipoles2} \\
\hline \hline 
Charge Distribution & $T_{\mu\nu}$ & Multipoles\\
\hline 
\endfirsthead 
\hline \hline 
Charge Distribution & $T_{\mu\nu}$ & Multipoles\\
\hline 
\endhead \hline 
\hline \endfoot
$s^Is^I$ & 1& $q^I$ \\
$p_x^Ip_x^I$ &3& $q^I,-\frac{2}{3}\left(\tilde{Q}^I_{zx}-\frac{1}{2}\tilde{Q}^I_{xy}\right),
\tilde{Q}^I_{xy}$ \\
$p_y^Ip_y^I$ &3& $q^I,-\frac{2}{3}\left(\tilde{Q}^I_{zx}-\frac{1}{2}\tilde{Q}^I_{xy}\right),
-\tilde{Q}^I_{xy}$ \\
$p_z^Ip_z^I$ &2& $q^I,\frac{4}{3}\left(\tilde{Q}^I_{zx}-\frac{1}{2}\tilde{Q}^I_{xy}\right)$ \\
$s^Ip_x^I$ &1& $\mu^I_x$ \\
$s^Ip_y^I$ &1& $\mu^I_y$ \\
$s^Ip_z^I$ &1& $\mu^I_z$ \\
$p_x^Ip_y^I$ &1& $Q^I_{xy}$ \\
$p_x^Ip_z^I$ &1& $Q^I_{xz}$ \\
$p_y^Ip_z^I$ &1& $Q^I_{yz}$ \\
$s^Id_{z^2}^I$               &1&  $\sqrt{\frac{4}{3}}\left(\tilde{Q}^I_{zx}-\frac{1}{2}\tilde{Q}^I_{xy}\right)$ \\
$s^Id_{xz}^I$                &1&  $Q^I_{xz}$\\
$s^Id_{yz}^I$                &1&  $Q^I_{yz}$ \\
$s^Id_{x^2-y^2}^I$           &1&  $\tilde{Q}^I_{xy}$\\
$s^Id_{xy}^I$                &1&  $Q^I_{xy}$\\
$p_x^Id_{z^2}^I$             &1&  $-\sqrt{\frac{1}{3}}\mu^I_x$\\
$p_x^Id_{xz}^I$              &1&  $\mu^I_z$\\
$p_x^Id_{yz}^I$              &0&  --- \\
$p_x^Id_{x^2-y^2}^I$         &1&  $\mu^I_x$\\
$p_x^Id_{xy}^I$              &1&  $\mu^I_y$\\
$p_y^Id_{z^2}^I$             &1&  $-\sqrt{\frac{1}{3}}\mu^I_y$\\
$p_y^Id_{xz}^I$              &0&  ---\\
$p_y^Id_{yz}^I$              &1&  $\mu^I_z$\\
$p_y^Id_{x^2-y^2}^I$         &1&  $-\mu^I_y$\\
$p_y^Id_{xy}^I$              &1&  $\mu^I_x$\\
$p_z^Id_{z^2}^I$             &1&  $\sqrt{\frac{4}{3}}\mu^I_z$\\
$p_z^Id_{xz}^I$              &1&  $\mu^I_x$\\
$p_z^Id_{yz}^I$              &1&  $\mu^I_y$\\
$p_z^Id_{x^2-y^2}^I$         &0&  ---\\
$p_z^Id_{xy}^I$              &0&  ---\\
$d_{z^2}^Id_{z^2}^I$         &2&  $q^I,\frac{4}{3}\left(\tilde{Q}^I_{zx}-\frac{1}{2}\tilde{Q}^I_{xy}\right)$\\
$d_{z^2}^Id_{xz}^I$          &1&  $\sqrt{\frac{1}{3}}Q^I_{xz}$\\
$d_{z^2}^Id_{yz}^I$          &1&  $\sqrt{\frac{1}{3}}Q^I_{yz}$\\
$d_{z^2}^Id_{x^2-y^2}^I$     &1&  $-\sqrt{\frac{4}{3}}\tilde{Q}^I_{xy}$\\
$d_{z^2}^Id_{xy}^I$          &1&  $-\sqrt{\frac{4}{3}}Q^I_{xy}$ \\
$d_{xz}^Id_{xz}^I$           &3& $q^I,\frac{2}{3}\left(\tilde{Q}^I_{zx}-\frac{1}{2}\tilde{Q}^I_{xy}\right),
\tilde{Q}^I_{xy}$\\
$d_{xz}^Id_{yz}^I$           &1&  $Q^I_{xy}$\\
$d_{xz}^Id_{x^2-y^2}^I$      &1&  $Q^I_{xz}$\\
$d_{xz}^Id_{xy}^I$           &1&  $Q^I_{yz}$\\
$d_{yz}^Id_{yz}^I$           &3&  $q^I,\frac{2}{3}\left(\tilde{Q}^I_{zx}-\frac{1}{2}\tilde{Q}^I_{xy}\right),
-\tilde{Q}^I_{xy}$\\
$d_{yz}^Id_{x^2-y^2}^I$      &1&  $-Q^I_{yz}$\\
$d_{yz}^Id_{xy}^I$           &1&  $Q^I_{xz}$\\
$d_{x^2-y^2}^Id_{x^2-y^2}^I$ &2&  $q^I,-\frac{4}{3}\left(\tilde{Q}^I_{zx}-\frac{1}{2}\tilde{Q}^I_{xy}\right)$ \\
$d_{x^2-y^2}^Id_{xy}^I$      &0&  ---\\
$d_{xy}^Id_{xy}^I$           &2&  $q^I,-\frac{4}{3}\left(\tilde{Q}^I_{zx}-\frac{1}{2}\tilde{Q}^I_{xy}\right)$\\
\end{longtable}
\end{center}
For the charge distributions 
$p_x^Id_{yz}^I$, $p_y^Id_{xz}^I$, $p_z^Id_{x^2-y^2}^I$, $p_z^Id_{xy}^I$, and 
$d_{x^2-y^2}^Id_{xy}^I$
no multipoles (i.e., $T_{\mu\nu}=0$ in Table~\ref{tab:multipoles2}) are specified. 
This is due to the fact that octopole (or higher) moments are neglected. \cite{Thiel1991}
As a consequence, all two-center ERIs which involve at least one of these charge distributions are 
exactly zero for all $\tilde{R}_{IJ}$. 

Note how the multipoles specified for the charge distributions 
$p_x^Ip_x^I$, $p_y^Ip_y^I$, and $p_z^Ip_z^I$
are different than the 
ones specfied in Table~\ref{tab:multipoles} which means that the representation 
of $p_x^Ip_x^I$, $p_y^Ip_y^I$, and $p_z^Ip_z^I$ differs depending on whether the 
second charge distribution contains a $d$-type orbital or not. 

Additionally, we are now in a situation where a dipole moment $\mu_x^I$ is applied to describe different 
charge distributions, e.g., $s^Ip_x^I$ and $p_x^Id_{x^2-y^2}^I$.
As a consequence, we need to specify different distances for the multipoles which appear in the 
multipole expansions describing a charge distribution.
In addition to 
$D_{\mu,sd}$ and $D_{Q,pp}$, we must define the 
distances $D_{\mu,pd}$, $D_{Q,sd}$, and $Q_{\mu,dd}$.
These distances are again defined with respect to $A(\chi_\mu^I, \chi_\nu^I,a)$ (Eq.~(\ref{eq:defA})), 
\cite{Thiel1991}
\begin{equation}
 D_{\mu,pd} = 5^{-\frac{1}{2}} A(\chi_\mu^I, \chi_\nu^I,1),
\end{equation}
\begin{equation}
 D_{Q,sd} = 15^{-\frac{1}{4}} \sqrt{A(\chi_\mu^I, \chi_\nu^I,2)},
\end{equation}
and
\begin{equation}
 D_{Q,dd} = 7^{-\frac{1}{2}} \sqrt{A(\chi_\mu^I, \chi_\nu^I,2)}.
\end{equation}
The implementation of the formulae to calculate $D_{\mu,pd}$, $D_{Q,sd}$, and 
$D_{Q,dd}$
can be verified by comparison to the respective values
supplied by \textsc{Mopac} \cite{mopac} when specifying the 
 keyword \texttt{HCORE}. 
For each element, the value tabulated as \texttt{DD2} corresponds to 
$D_{\mu,sp}$ , \texttt{DD3} to $D_{Q,pp}$,
 \texttt{DD4} to $D_{Q,sd}$, 
\texttt{DD5} to $D_{\mu,pd}$, and 
\texttt{DD6} to $D_{Q,dd}$.

The positions of the point charges arising for a monopole $q$, a 
dipole $\mu_{x,y,z}$, a linear quadrupole $Q_{xx,yy,zz}$, and a square quadrupole $Q_{xy,yz,xz}$
are listed in Table~\ref{tab:multipoles_pos}.
The positions of the point charges for the quadrupole $\tilde{Q}_{xy,xz,yz}$
relative to atom $I$ on which the charge distribution $\chi_\mu^I \chi_\nu^I$  is centered are 
given in Table~\ref{tab:multipoles_pos2}.
\begin{center}
\begin{longtable}{ccccc}
  \caption{Specification of the position ($\fett{r}_C = (r_{C,x},r_{C,y}, r_{C,z})$) 
of the $C$ point charges chosen to represent a specific multipole (Table~\ref{tab:multipoles}). 
The number of point charges is given in brackets after the specification of the multipole 
moment.
The positions $\fett{r}_c$ are given in relation to the 
origin of the local coordinate system (Section~\ref{sec:localcoordinate}).
The charges $q_C$ are given in atomic units.
} \label{tab:multipoles_pos2} \\
\hline \hline 
Multipole & $r_{C,x}$ & $r_{C,y}$& $r_{C,z}$ & $q_C$ \\
\hline 
\endfirsthead
\hline \hline  
Multipole & $r_{C,x}$ & $r_{C,y}$& $r_{C,z}$ & $q_C$ \\
\hline 
\endhead 
\hline \hline \endfoot
$\tilde{Q}_{xy}$ ($C=4$)
         & $+ \sqrt{2} D_{Q,pp}$ & $0.0$ & $0.0$ & $+0.25$ \\
         & $- \sqrt{2} D_{Q,pp}$ & $0.0$ & $0.0$ & $+0.25$ \\
         & $0.0$ & $- \sqrt{2}D_{Q,pp}$ & $0.0$ & $-0.25$ \\
         & $0.0$ & $+ \sqrt{2}D_{Q,pp}$ & $0.0$ & $-0.25$ \\
\hline 
$\tilde{Q}_{xz}$ ($C=4$)
         & $+ \sqrt{2} D_{Q,pp}$ & $0.0$ & $0.0$ & $+0.25$ \\
         & $- \sqrt{2} D_{Q,pp}$ & $0.0$ & $0.0$ & $+0.25$ \\
         & $0.0$ & $0.0$ & $-\sqrt{2}D_{Q,pp}$ & $-0.25$ \\
         & $0.0$ & $0.0$ & $+\sqrt{2}D_{Q,pp}$ & $-0.25$ \\
\hline 
$\tilde{Q}_{yz}$ ($C=4$)
         & $0.0$ & $+\sqrt{2}D_{Q,pp}$ & $0.0$ & $+0.25$ \\
         & $0.0$ & $-\sqrt{2}D_{Q,pp}$ & $0.0$ & $+0.25$ \\
         & $0.0$ & $0.0$ & $-\sqrt{2}D_{Q,pp}$ & $-0.25$ \\
         & $0.0$ & $0.0$ & $+\sqrt{2}D_{Q,pp}$ & $-0.25$ \\
\end{longtable}
\end{center}

Now we can again straightforwardly apply Eq.~(\ref{eq:epot2})
after specifying 
$\vartheta^I_{q,ss}$, $\vartheta^I_{q,pp}$, $\vartheta^I_{\mu,pp}$,
$\vartheta^I_{\mu,dd}$, $\vartheta^I_{\mu,sp}$, 
$\vartheta^I_{\mu,pd}$, $\vartheta^I_{Q,pp}$,
$\vartheta^I_{Q,sp}$, and $\vartheta^I_{Q,dd}$. 
The additive terms 
$\vartheta^I_{q,ss}$, $\vartheta^I_{q,pp}$, $\vartheta^I_{q,ss}$, 
$\vartheta^I_{\mu,pp}$, and 
$\vartheta^I_{Q,pp}$, 
are determined numerically from 
Eqs.~(\ref{eq:varthetaqss}), (\ref{eq:varthetaqpp}), (\ref{eq:varthetampp}), and 
(\ref{eq:varthetaQpp}), respectively. 
The other terms are determined from the following equations, \cite{Thiel1991,Thiel1996c}
\begin{equation}
 \vartheta^I_{q,dd} = \frac{1}{2 F_{dd}^0},
\end{equation}
\begin{equation}
 \left(\vartheta^I_{\mu,pd}\right)^{-1} - \left[\left(\vartheta^I_{\mu,pd}\right)^2 + 
(D_{\mu,pd})^2\right]^{-\frac{1}{2}} = \frac{16}{15} G_{pd}^1,
\end{equation}
\begin{equation}
 \left(\vartheta^I_{Q,sd}\right)^{-1} - 2\left[ 
\left(\vartheta^I_{Q,sd}\right)^{2} +(D_{Q,sd})^2
\right]^{-\frac{1}{2}}
+ 
\left[
\left(\vartheta^I_{Q,sd}\right)^{2} +2(D_{Q,sd})^2
\right]^{-\frac{1}{2}} 
= \frac{8}{5} G_{sd}^2,
\end{equation}
and
\begin{equation}
 \left(\vartheta^I_{Q,dd}\right)^{-1} - 2 \left[ 
\left(\vartheta^I_{Q,dd}\right)^{2} + (D_{Q,dd})^2
\right]^{-\frac{1}{2}}
+ 
\left[ 
\left(\vartheta^I_{Q,dd}\right)^{2} +2 (D_{Q,dd})^2
\right]^{-\frac{1}{2}}
 = \frac{24}{49} F_{dd}^2.
\end{equation}
The values for $\vartheta^I_{q,ss}$, $\vartheta^I_{q,pp}$, $\vartheta^I_{q,dd}$,
$\vartheta^I_{\mu,sp}$, $\vartheta^I_{\mu,pd}$, 
$\vartheta^I_{Q,pp}$, $\vartheta^I_{Q,sd}$, and $\vartheta^I_{Q,dd}$ can 
be verified against the values for \texttt{P01}, 
\texttt{P07}, \texttt{P08}, \texttt{P02}, 
\texttt{P05}, \texttt{P03}, \texttt{P04}, and \texttt{P06}, 
respectively, when specifying the keyword \texttt{Hcore} in \textsc{Mopac}. 

In Ref.~\onlinecite{Thiel1991}, $M_{20}$ is defined as 
\begin{equation}
 M_{20} = \tilde{Q}_{zx} - \frac{1}{2} \tilde{Q}_{xy}.
\end{equation}
Hence, the interaction of two multiples $M_{20}$ is calculated to be 
\begin{equation}
 [M_{20},M_{20}] = [\tilde{Q}_{zx},\tilde{Q}_{zx}] - 
\frac{1}{2} [\tilde{Q}_{zx},\tilde{Q}_{xy}] - \frac{1}{2} [\tilde{Q}_{xy},\tilde{Q}_{zx}] + \frac{1}{4} 
[\tilde{Q}_{xy},\tilde{Q}_{xy}].
\end{equation}
This formulae can be simplified to 
\begin{equation}
\label{eq:m2020_correct}
 [M_{20},M_{20}] = [\tilde{Q}_{zx},\tilde{Q}_{zx}] + \frac{3}{4} [\tilde{Q}_{xy},\tilde{Q}_{xy}].
\end{equation}
when inserting 
\begin{equation}
 [\tilde{Q}_{zx},\tilde{Q}_{xy}] = - \frac{1}{2} [\tilde{Q}_{xy},\tilde{Q}_{xy}].
\end{equation}
Note that Ref.~\onlinecite{Thiel1991} appears to contain a misprint because it 
specifies that $[\tilde{Q}_{zx},\tilde{Q}_{xy}]=0$.
Apparently, \textsc{Mopac} implements a different formula,
\begin{equation}
\label{eq:m2020_incorrect}
[M_{20},M_{20}] = [\tilde{Q}_{zx},\tilde{Q}_{zx}] - 
\frac{1}{4} [\tilde{Q}_{xy},\tilde{Q}_{xy}]
\end{equation}
which is derivable when one defines 
\begin{equation}
 M_{20} = \tilde{Q}_{zx} + \frac{1}{2} \tilde{Q}_{xy}.
\end{equation}
The values of two-center ERIs that involve at least one of the following charge distributions,
$p_z^Ip_z^I$, $p_x^Ip_x^I$, $p_y^Ip_y^I$, 
 $s^Id_{z^2}^I$, $d_{xy}^Id_{xy}^I$, 
$d_{xz}^Id_{xz}^I$, $d_{yz}^Id_{yz}^I$, $d_{z^2}^Id_{z^2}^I$, and 
$d_{x^2-y^2}^Id_{x^2-y^2}^I$, 
differ when either  Eq.~(\ref{eq:m2020_incorrect}) or 
 Eq.~(\ref{eq:m2020_correct}) is applied.
We illustrate this for the example of a Br---Cl molecule 
with an internuclear distance of 0.4 \AA\ (see Table~\ref{tab:example}).
\begin{table}[ht] 
 \caption{Value of selected two-center ERIs with PM6 parameters in eV obtained 
with \textsc{Mopac}, with Eq.~(\ref{eq:m2020_incorrect}) in \textsc{Scine}\cite{scine}, 
and with Eq.~(\ref{eq:m2020_correct}) in \textsc{Scine}.
} 
 \centering
 \begin{tabular}{cccc}
  \hline \hline 
  ERI & \textsc{Mopac} & Eq.~(\ref{eq:m2020_incorrect}) in \textsc{Scine} & 
Eq.~(\ref{eq:m2020_correct}) in \textsc{Scine} \\
  \hline
  $\left<d_{z^2}^Id_{z^2}^I|d_{xz}^Jd_{xz}^J\right>$   & 20.798 & 20.798 & 21.401 \\
  $\left<d_{z^2}^Id_{z^2}^I|d_{z^2}^Jd_{z^2}^J\right>$ & 21.475	& 21.475 & 22.681 \\ 
  $\left<p_x^Ip_x^I|d_{yz}^Jd_{yz}^J\right>$ & 10.833 & 10.833& 10.497 \\
  $\left<p_y^Ip_y^I|s^Jd_{z^2}^J\right>$ & $-0.021$ & $-0.021$ & $-0.071$ \\ 
  \hline \hline 
 \end{tabular}
 \label{tab:example}
\end{table}
Note that it is not easily possible to switch between Eq.~(\ref{eq:m2020_incorrect}) and 
Eq.~(\ref{eq:m2020_correct}).
The parameters elements appear to have been determined 
with an implementation of Eq.~(\ref{eq:m2020_incorrect}), and hence, the parameters 
cannot simply be transfered to a program implementing Eq.~(\ref{eq:m2020_correct}).
Instead, one would have to determine a new set of parameters when implementing the other 
equation (Eq.~(\ref{eq:m2020_correct})).

The implementation of the procedure to calculate those two-center ERIs 
can be compared to the implementation in \textsc{Mopac}
when invoking the keyword \texttt{Hcore} for a calculation of the electronic energy 
for a diatomic molecule which is aligned along the $z$-axis 
(for which $s$-, $p$-, and $d$-type basis functions are activated).
The values of the two-center ERIs are then listed under \texttt{TWO-ELECTRON MATRIX IN HCORE}.
The first 2025 entries are the one-center ERIs for the first atom.
The next 2025 entries are the two-center ERIs which 
arise between the first and the second atom.
The order in which the two-center ERIs are given is described in Ref.~\onlinecite{mopac2} and it is also given in the following: \\
$\left<ss|ss\right>$ $\left<ss|sp_x\right>$ $\left<ss|p_xp_x\right>$ $\left<ss|sp_y\right>$ $\left<ss|p_xp_y\right>$ $\left<ss|p_yp_y\right>$ $\left<ss|sp_z\right>$ $\left<ss|p_xp_z\right>$ $\left<ss|p_yp_z\right>$ $\left<ss|p_zp_z\right>$ 
$\left<ss|sd_{x^2-y^2}\right>$ $\left<ss|p_xd_{x^2-y^2}\right>$ $\left<ss|p_yd_{x^2-y^2}\right>$ $\left<ss|p_zd_{x^2-y^2}\right>$ $\left<ss|d_{x^2-y^2}d_{x^2-y^2}\right>$ $\left<ss|sd_{xz}\right>$ $\left<ss|p_xd_{xz}\right>$ $\left<ss|p_yd_{xz}\right>$ $\left<ss|p_zd_{xz}\right>$ $\left<ss|d_{x^2-y^2}d_{xz}\right>$ 
$\left<ss|d_{xz}d_{xz}\right>$ $\left<ss|sd_{z^2}\right>$ $\left<ss|p_xd_{z^2}\right>$ $\left<ss|p_yd_{z^2}\right>$ $\left<ss|p_zd_{z^2}\right>$ $\left<ss|d_{x^2-y^2}d_{z^2}\right>$ $\left<ss|d_{xz}d_{z^2}\right>$ $\left<ss|d_{z^2}d_{z^2}\right>$ $\left<ss|sd_{yz}\right>$ $\left<ss|p_xd_{yz}\right>$ 
$\left<ss|p_yd_{yz}\right>$ $\left<ss|p_zd_{yz}\right>$ $\left<ss|d_{x^2-y^2}d_{yz}\right>$ $\left<ss|d_{xz}d_{yz}\right>$ $\left<ss|d_{z^2}d_{yz}\right>$ $\left<ss|d_{yz}d_{yz}\right>$ $\left<ss|sd_{xy}\right>$ $\left<ss|p_xd_{xy}\right>$ $\left<ss|p_yd_{xy}\right>$ $\left<ss|p_zd_{xy}\right>$ 
$\left<ss|d_{x^2-y^2}d_{xy}\right>$ $\left<ss|d_{xz}d_{xy}\right>$ $\left<ss|d_{z^2}d_{xy}\right>$ $\left<ss|d_{yz}d_{xy}\right>$ $\left<ss|d_{xy}d_{xy}\right>$ $\left<sp_x|ss\right>$ $\left<sp_x|sp_x\right>$ $\left<sp_x|p_xp_x\right>$ $\left<sp_x|sp_y\right>$ $\left<sp_x|p_xp_y\right>$ 
$\left<sp_x|p_yp_y\right>$ $\left<sp_x|sp_z\right>$ $\left<sp_x|p_xp_z\right>$ $\left<sp_x|p_yp_z\right>$ $\left<sp_x|p_zp_z\right>$ $\left<sp_x|sd_{x^2-y^2}\right>$ $\left<sp_x|p_xd_{x^2-y^2}\right>$ $\left<sp_x|p_yd_{x^2-y^2}\right>$ $\left<sp_x|p_zd_{x^2-y^2}\right>$ $\left<sp_x|d_{x^2-y^2}d_{x^2-y^2}\right>$ 
$\left<sp_x|sd_{xz}\right>$ $\left<sp_x|p_xd_{xz}\right>$ $\left<sp_x|p_yd_{xz}\right>$ $\left<sp_x|p_zd_{xz}\right>$ $\left<sp_x|d_{x^2-y^2}d_{xz}\right>$ $\left<sp_x|d_{xz}d_{xz}\right>$ $\left<sp_x|sd_{z^2}\right>$ $\left<sp_x|p_xd_{z^2}\right>$ $\left<sp_x|p_yd_{z^2}\right>$ $\left<sp_x|p_zd_{z^2}\right>$ 
$\left<sp_x|d_{x^2-y^2}d_{z^2}\right>$ $\left<sp_x|d_{xz}d_{z^2}\right>$ $\left<sp_x|d_{z^2}d_{z^2}\right>$ $\left<sp_x|sd_{yz}\right>$ $\left<sp_x|p_xd_{yz}\right>$ $\left<sp_x|p_yd_{yz}\right>$ $\left<sp_x|p_zd_{yz}\right>$ $\left<sp_x|d_{x^2-y^2}d_{yz}\right>$ $\left<sp_x|d_{xz}d_{yz}\right>$ $\left<sp_x|d_{z^2}d_{yz}\right>$ 
$\left<sp_x|d_{yz}d_{yz}\right>$ $\left<sp_x|sd_{xy}\right>$ $\left<sp_x|p_xd_{xy}\right>$ $\left<sp_x|p_yd_{xy}\right>$ $\left<sp_x|p_zd_{xy}\right>$ $\left<sp_x|d_{x^2-y^2}d_{xy}\right>$ $\left<sp_x|d_{xz}d_{xy}\right>$ $\left<sp_x|d_{z^2}d_{xy}\right>$ $\left<sp_x|d_{yz}d_{xy}\right>$ $\left<sp_x|d_{xy}d_{xy}\right>$ 
$\left<p_xp_x|ss\right>$ $\left<p_xp_x|sp_x\right>$ $\left<p_xp_x|p_xp_x\right>$ $\left<p_xp_x|sp_y\right>$ $\left<p_xp_x|p_xp_y\right>$ $\left<p_xp_x|p_yp_y\right>$ $\left<p_xp_x|sp_z\right>$ $\left<p_xp_x|p_xp_z\right>$ $\left<p_xp_x|p_yp_z\right>$ $\left<p_xp_x|p_zp_z\right>$ 
$\left<p_xp_x|sd_{x^2-y^2}\right>$ $\left<p_xp_x|p_xd_{x^2-y^2}\right>$ $\left<p_xp_x|p_yd_{x^2-y^2}\right>$ $\left<p_xp_x|p_zd_{x^2-y^2}\right>$ $\left<p_xp_x|d_{x^2-y^2}d_{x^2-y^2}\right>$ $\left<p_xp_x|sd_{xz}\right>$ $\left<p_xp_x|p_xd_{xz}\right>$ $\left<p_xp_x|p_yd_{xz}\right>$ $\left<p_xp_x|p_zd_{xz}\right>$ $\left<p_xp_x|d_{x^2-y^2}d_{xz}\right>$ 
$\left<p_xp_x|d_{xz}d_{xz}\right>$ $\left<p_xp_x|sd_{z^2}\right>$ $\left<p_xp_x|p_xd_{z^2}\right>$ $\left<p_xp_x|p_yd_{z^2}\right>$ $\left<p_xp_x|p_zd_{z^2}\right>$ $\left<p_xp_x|d_{x^2-y^2}d_{z^2}\right>$ $\left<p_xp_x|d_{xz}d_{z^2}\right>$ $\left<p_xp_x|d_{z^2}d_{z^2}\right>$ $\left<p_xp_x|sd_{yz}\right>$ $\left<p_xp_x|p_xd_{yz}\right>$ 
$\left<p_xp_x|p_yd_{yz}\right>$ $\left<p_xp_x|p_zd_{yz}\right>$ $\left<p_xp_x|d_{x^2-y^2}d_{yz}\right>$ $\left<p_xp_x|d_{xz}d_{yz}\right>$ $\left<p_xp_x|d_{z^2}d_{yz}\right>$ $\left<p_xp_x|d_{yz}d_{yz}\right>$ $\left<p_xp_x|sd_{xy}\right>$ $\left<p_xp_x|p_xd_{xy}\right>$ $\left<p_xp_x|p_yd_{xy}\right>$ $\left<p_xp_x|p_zd_{xy}\right>$ 
$\left<p_xp_x|d_{x^2-y^2}d_{xy}\right>$ $\left<p_xp_x|d_{xz}d_{xy}\right>$ $\left<p_xp_x|d_{z^2}d_{xy}\right>$ $\left<p_xp_x|d_{yz}d_{xy}\right>$ $\left<p_xp_x|d_{xy}d_{xy}\right>$ $\left<sp_y|ss\right>$ $\left<sp_y|sp_x\right>$ $\left<sp_y|p_xp_x\right>$ $\left<sp_y|sp_y\right>$ $\left<sp_y|p_xp_y\right>$ 
$\left<sp_y|p_yp_y\right>$ $\left<sp_y|sp_z\right>$ $\left<sp_y|p_xp_z\right>$ $\left<sp_y|p_yp_z\right>$ $\left<sp_y|p_zp_z\right>$ $\left<sp_y|sd_{x^2-y^2}\right>$ $\left<sp_y|p_xd_{x^2-y^2}\right>$ $\left<sp_y|p_yd_{x^2-y^2}\right>$ $\left<sp_y|p_zd_{x^2-y^2}\right>$ $\left<sp_y|d_{x^2-y^2}d_{x^2-y^2}\right>$ 
$\left<sp_y|sd_{xz}\right>$ $\left<sp_y|p_xd_{xz}\right>$ $\left<sp_y|p_yd_{xz}\right>$ $\left<sp_y|p_zd_{xz}\right>$ $\left<sp_y|d_{x^2-y^2}d_{xz}\right>$ $\left<sp_y|d_{xz}d_{xz}\right>$ $\left<sp_y|sd_{z^2}\right>$ $\left<sp_y|p_xd_{z^2}\right>$ $\left<sp_y|p_yd_{z^2}\right>$ $\left<sp_y|p_zd_{z^2}\right>$ 
$\left<sp_y|d_{x^2-y^2}d_{z^2}\right>$ $\left<sp_y|d_{xz}d_{z^2}\right>$ $\left<sp_y|d_{z^2}d_{z^2}\right>$ $\left<sp_y|sd_{yz}\right>$ $\left<sp_y|p_xd_{yz}\right>$ $\left<sp_y|p_yd_{yz}\right>$ $\left<sp_y|p_zd_{yz}\right>$ $\left<sp_y|d_{x^2-y^2}d_{yz}\right>$ $\left<sp_y|d_{xz}d_{yz}\right>$ $\left<sp_y|d_{z^2}d_{yz}\right>$ 
$\left<sp_y|d_{yz}d_{yz}\right>$ $\left<sp_y|sd_{xy}\right>$ $\left<sp_y|p_xd_{xy}\right>$ $\left<sp_y|p_yd_{xy}\right>$ $\left<sp_y|p_zd_{xy}\right>$ $\left<sp_y|d_{x^2-y^2}d_{xy}\right>$ $\left<sp_y|d_{xz}d_{xy}\right>$ $\left<sp_y|d_{z^2}d_{xy}\right>$ $\left<sp_y|d_{yz}d_{xy}\right>$ $\left<sp_y|d_{xy}d_{xy}\right>$ 
$\left<p_xp_y|ss\right>$ $\left<p_xp_y|sp_x\right>$ $\left<p_xp_y|p_xp_x\right>$ $\left<p_xp_y|sp_y\right>$ $\left<p_xp_y|p_xp_y\right>$ $\left<p_xp_y|p_yp_y\right>$ $\left<p_xp_y|sp_z\right>$ $\left<p_xp_y|p_xp_z\right>$ $\left<p_xp_y|p_yp_z\right>$ $\left<p_xp_y|p_zp_z\right>$ 
$\left<p_xp_y|sd_{x^2-y^2}\right>$ $\left<p_xp_y|p_xd_{x^2-y^2}\right>$ $\left<p_xp_y|p_yd_{x^2-y^2}\right>$ $\left<p_xp_y|p_zd_{x^2-y^2}\right>$ $\left<p_xp_y|d_{x^2-y^2}d_{x^2-y^2}\right>$ $\left<p_xp_y|sd_{xz}\right>$ $\left<p_xp_y|p_xd_{xz}\right>$ $\left<p_xp_y|p_yd_{xz}\right>$ $\left<p_xp_y|p_zd_{xz}\right>$ $\left<p_xp_y|d_{x^2-y^2}d_{xz}\right>$ 
$\left<p_xp_y|d_{xz}d_{xz}\right>$ $\left<p_xp_y|sd_{z^2}\right>$ $\left<p_xp_y|p_xd_{z^2}\right>$ $\left<p_xp_y|p_yd_{z^2}\right>$ $\left<p_xp_y|p_zd_{z^2}\right>$ $\left<p_xp_y|d_{x^2-y^2}d_{z^2}\right>$ $\left<p_xp_y|d_{xz}d_{z^2}\right>$ $\left<p_xp_y|d_{z^2}d_{z^2}\right>$ $\left<p_xp_y|sd_{yz}\right>$ $\left<p_xp_y|p_xd_{yz}\right>$ 
$\left<p_xp_y|p_yd_{yz}\right>$ $\left<p_xp_y|p_zd_{yz}\right>$ $\left<p_xp_y|d_{x^2-y^2}d_{yz}\right>$ $\left<p_xp_y|d_{xz}d_{yz}\right>$ $\left<p_xp_y|d_{z^2}d_{yz}\right>$ $\left<p_xp_y|d_{yz}d_{yz}\right>$ $\left<p_xp_y|sd_{xy}\right>$ $\left<p_xp_y|p_xd_{xy}\right>$ $\left<p_xp_y|p_yd_{xy}\right>$ $\left<p_xp_y|p_zd_{xy}\right>$ 
$\left<p_xp_y|d_{x^2-y^2}d_{xy}\right>$ $\left<p_xp_y|d_{xz}d_{xy}\right>$ $\left<p_xp_y|d_{z^2}d_{xy}\right>$ $\left<p_xp_y|d_{yz}d_{xy}\right>$ $\left<p_xp_y|d_{xy}d_{xy}\right>$ $\left<p_yp_y|ss\right>$ $\left<p_yp_y|sp_x\right>$ $\left<p_yp_y|p_xp_x\right>$ $\left<p_yp_y|sp_y\right>$ $\left<p_yp_y|p_xp_y\right>$ 
$\left<p_yp_y|p_yp_y\right>$ $\left<p_yp_y|sp_z\right>$ $\left<p_yp_y|p_xp_z\right>$ $\left<p_yp_y|p_yp_z\right>$ $\left<p_yp_y|p_zp_z\right>$ $\left<p_yp_y|sd_{x^2-y^2}\right>$ $\left<p_yp_y|p_xd_{x^2-y^2}\right>$ $\left<p_yp_y|p_yd_{x^2-y^2}\right>$ $\left<p_yp_y|p_zd_{x^2-y^2}\right>$ $\left<p_yp_y|d_{x^2-y^2}d_{x^2-y^2}\right>$ 
$\left<p_yp_y|sd_{xz}\right>$ $\left<p_yp_y|p_xd_{xz}\right>$ $\left<p_yp_y|p_yd_{xz}\right>$ $\left<p_yp_y|p_zd_{xz}\right>$ $\left<p_yp_y|d_{x^2-y^2}d_{xz}\right>$ $\left<p_yp_y|d_{xz}d_{xz}\right>$ $\left<p_yp_y|sd_{z^2}\right>$ $\left<p_yp_y|p_xd_{z^2}\right>$ $\left<p_yp_y|p_yd_{z^2}\right>$ $\left<p_yp_y|p_zd_{z^2}\right>$ 
$\left<p_yp_y|d_{x^2-y^2}d_{z^2}\right>$ $\left<p_yp_y|d_{xz}d_{z^2}\right>$ $\left<p_yp_y|d_{z^2}d_{z^2}\right>$ $\left<p_yp_y|sd_{yz}\right>$ $\left<p_yp_y|p_xd_{yz}\right>$ $\left<p_yp_y|p_yd_{yz}\right>$ $\left<p_yp_y|p_zd_{yz}\right>$ $\left<p_yp_y|d_{x^2-y^2}d_{yz}\right>$ $\left<p_yp_y|d_{xz}d_{yz}\right>$ $\left<p_yp_y|d_{z^2}d_{yz}\right>$ 
$\left<p_yp_y|d_{yz}d_{yz}\right>$ $\left<p_yp_y|sd_{xy}\right>$ $\left<p_yp_y|p_xd_{xy}\right>$ $\left<p_yp_y|p_yd_{xy}\right>$ $\left<p_yp_y|p_zd_{xy}\right>$ $\left<p_yp_y|d_{x^2-y^2}d_{xy}\right>$ $\left<p_yp_y|d_{xz}d_{xy}\right>$ $\left<p_yp_y|d_{z^2}d_{xy}\right>$ $\left<p_yp_y|d_{yz}d_{xy}\right>$ $\left<p_yp_y|d_{xy}d_{xy}\right>$ 
$\left<sp_z|ss\right>$ $\left<sp_z|sp_x\right>$ $\left<sp_z|p_xp_x\right>$ $\left<sp_z|sp_y\right>$ $\left<sp_z|p_xp_y\right>$ $\left<sp_z|p_yp_y\right>$ $\left<sp_z|sp_z\right>$ $\left<sp_z|p_xp_z\right>$ $\left<sp_z|p_yp_z\right>$ $\left<sp_z|p_zp_z\right>$ 
$\left<sp_z|sd_{x^2-y^2}\right>$ $\left<sp_z|p_xd_{x^2-y^2}\right>$ $\left<sp_z|p_yd_{x^2-y^2}\right>$ $\left<sp_z|p_zd_{x^2-y^2}\right>$ $\left<sp_z|d_{x^2-y^2}d_{x^2-y^2}\right>$ $\left<sp_z|sd_{xz}\right>$ $\left<sp_z|p_xd_{xz}\right>$ $\left<sp_z|p_yd_{xz}\right>$ $\left<sp_z|p_zd_{xz}\right>$ $\left<sp_z|d_{x^2-y^2}d_{xz}\right>$ 
$\left<sp_z|d_{xz}d_{xz}\right>$ $\left<sp_z|sd_{z^2}\right>$ $\left<sp_z|p_xd_{z^2}\right>$ $\left<sp_z|p_yd_{z^2}\right>$ $\left<sp_z|p_zd_{z^2}\right>$ $\left<sp_z|d_{x^2-y^2}d_{z^2}\right>$ $\left<sp_z|d_{xz}d_{z^2}\right>$ $\left<sp_z|d_{z^2}d_{z^2}\right>$ $\left<sp_z|sd_{yz}\right>$ $\left<sp_z|p_xd_{yz}\right>$ 
$\left<sp_z|p_yd_{yz}\right>$ $\left<sp_z|p_zd_{yz}\right>$ $\left<sp_z|d_{x^2-y^2}d_{yz}\right>$ $\left<sp_z|d_{xz}d_{yz}\right>$ $\left<sp_z|d_{z^2}d_{yz}\right>$ $\left<sp_z|d_{yz}d_{yz}\right>$ $\left<sp_z|sd_{xy}\right>$ $\left<sp_z|p_xd_{xy}\right>$ $\left<sp_z|p_yd_{xy}\right>$ $\left<sp_z|p_zd_{xy}\right>$ 
$\left<sp_z|d_{x^2-y^2}d_{xy}\right>$ $\left<sp_z|d_{xz}d_{xy}\right>$ $\left<sp_z|d_{z^2}d_{xy}\right>$ $\left<sp_z|d_{yz}d_{xy}\right>$ $\left<sp_z|d_{xy}d_{xy}\right>$ $\left<p_xp_z|ss\right>$ $\left<p_xp_z|sp_x\right>$ $\left<p_xp_z|p_xp_x\right>$ $\left<p_xp_z|sp_y\right>$ $\left<p_xp_z|p_xp_y\right>$ 
$\left<p_xp_z|p_yp_y\right>$ $\left<p_xp_z|sp_z\right>$ $\left<p_xp_z|p_xp_z\right>$ $\left<p_xp_z|p_yp_z\right>$ $\left<p_xp_z|p_zp_z\right>$ $\left<p_xp_z|sd_{x^2-y^2}\right>$ $\left<p_xp_z|p_xd_{x^2-y^2}\right>$ $\left<p_xp_z|p_yd_{x^2-y^2}\right>$ $\left<p_xp_z|p_zd_{x^2-y^2}\right>$ $\left<p_xp_z|d_{x^2-y^2}d_{x^2-y^2}\right>$ 
$\left<p_xp_z|sd_{xz}\right>$ $\left<p_xp_z|p_xd_{xz}\right>$ $\left<p_xp_z|p_yd_{xz}\right>$ $\left<p_xp_z|p_zd_{xz}\right>$ $\left<p_xp_z|d_{x^2-y^2}d_{xz}\right>$ $\left<p_xp_z|d_{xz}d_{xz}\right>$ $\left<p_xp_z|sd_{z^2}\right>$ $\left<p_xp_z|p_xd_{z^2}\right>$ $\left<p_xp_z|p_yd_{z^2}\right>$ $\left<p_xp_z|p_zd_{z^2}\right>$ 
$\left<p_xp_z|d_{x^2-y^2}d_{z^2}\right>$ $\left<p_xp_z|d_{xz}d_{z^2}\right>$ $\left<p_xp_z|d_{z^2}d_{z^2}\right>$ $\left<p_xp_z|sd_{yz}\right>$ $\left<p_xp_z|p_xd_{yz}\right>$ $\left<p_xp_z|p_yd_{yz}\right>$ $\left<p_xp_z|p_zd_{yz}\right>$ $\left<p_xp_z|d_{x^2-y^2}d_{yz}\right>$ $\left<p_xp_z|d_{xz}d_{yz}\right>$ $\left<p_xp_z|d_{z^2}d_{yz}\right>$ 
$\left<p_xp_z|d_{yz}d_{yz}\right>$ $\left<p_xp_z|sd_{xy}\right>$ $\left<p_xp_z|p_xd_{xy}\right>$ $\left<p_xp_z|p_yd_{xy}\right>$ $\left<p_xp_z|p_zd_{xy}\right>$ $\left<p_xp_z|d_{x^2-y^2}d_{xy}\right>$ $\left<p_xp_z|d_{xz}d_{xy}\right>$ $\left<p_xp_z|d_{z^2}d_{xy}\right>$ $\left<p_xp_z|d_{yz}d_{xy}\right>$ $\left<p_xp_z|d_{xy}d_{xy}\right>$ 
$\left<p_yp_z|ss\right>$ $\left<p_yp_z|sp_x\right>$ $\left<p_yp_z|p_xp_x\right>$ $\left<p_yp_z|sp_y\right>$ $\left<p_yp_z|p_xp_y\right>$ $\left<p_yp_z|p_yp_y\right>$ $\left<p_yp_z|sp_z\right>$ $\left<p_yp_z|p_xp_z\right>$ $\left<p_yp_z|p_yp_z\right>$ $\left<p_yp_z|p_zp_z\right>$ 
$\left<p_yp_z|sd_{x^2-y^2}\right>$ $\left<p_yp_z|p_xd_{x^2-y^2}\right>$ $\left<p_yp_z|p_yd_{x^2-y^2}\right>$ $\left<p_yp_z|p_zd_{x^2-y^2}\right>$ $\left<p_yp_z|d_{x^2-y^2}d_{x^2-y^2}\right>$ $\left<p_yp_z|sd_{xz}\right>$ $\left<p_yp_z|p_xd_{xz}\right>$ $\left<p_yp_z|p_yd_{xz}\right>$ $\left<p_yp_z|p_zd_{xz}\right>$ $\left<p_yp_z|d_{x^2-y^2}d_{xz}\right>$ 
$\left<p_yp_z|d_{xz}d_{xz}\right>$ $\left<p_yp_z|sd_{z^2}\right>$ $\left<p_yp_z|p_xd_{z^2}\right>$ $\left<p_yp_z|p_yd_{z^2}\right>$ $\left<p_yp_z|p_zd_{z^2}\right>$ $\left<p_yp_z|d_{x^2-y^2}d_{z^2}\right>$ $\left<p_yp_z|d_{xz}d_{z^2}\right>$ $\left<p_yp_z|d_{z^2}d_{z^2}\right>$ $\left<p_yp_z|sd_{yz}\right>$ $\left<p_yp_z|p_xd_{yz}\right>$ 
$\left<p_yp_z|p_yd_{yz}\right>$ $\left<p_yp_z|p_zd_{yz}\right>$ $\left<p_yp_z|d_{x^2-y^2}d_{yz}\right>$ $\left<p_yp_z|d_{xz}d_{yz}\right>$ $\left<p_yp_z|d_{z^2}d_{yz}\right>$ $\left<p_yp_z|d_{yz}d_{yz}\right>$ $\left<p_yp_z|sd_{xy}\right>$ $\left<p_yp_z|p_xd_{xy}\right>$ $\left<p_yp_z|p_yd_{xy}\right>$ $\left<p_yp_z|p_zd_{xy}\right>$ 
$\left<p_yp_z|d_{x^2-y^2}d_{xy}\right>$ $\left<p_yp_z|d_{xz}d_{xy}\right>$ $\left<p_yp_z|d_{z^2}d_{xy}\right>$ $\left<p_yp_z|d_{yz}d_{xy}\right>$ $\left<p_yp_z|d_{xy}d_{xy}\right>$ $\left<p_zp_z|ss\right>$ $\left<p_zp_z|sp_x\right>$ $\left<p_zp_z|p_xp_x\right>$ $\left<p_zp_z|sp_y\right>$ $\left<p_zp_z|p_xp_y\right>$ 
$\left<p_zp_z|p_yp_y\right>$ $\left<p_zp_z|sp_z\right>$ $\left<p_zp_z|p_xp_z\right>$ $\left<p_zp_z|p_yp_z\right>$ $\left<p_zp_z|p_zp_z\right>$ $\left<p_zp_z|sd_{x^2-y^2}\right>$ $\left<p_zp_z|p_xd_{x^2-y^2}\right>$ $\left<p_zp_z|p_yd_{x^2-y^2}\right>$ $\left<p_zp_z|p_zd_{x^2-y^2}\right>$ $\left<p_zp_z|d_{x^2-y^2}d_{x^2-y^2}\right>$ 
$\left<p_zp_z|sd_{xz}\right>$ $\left<p_zp_z|p_xd_{xz}\right>$ $\left<p_zp_z|p_yd_{xz}\right>$ $\left<p_zp_z|p_zd_{xz}\right>$ $\left<p_zp_z|d_{x^2-y^2}d_{xz}\right>$ $\left<p_zp_z|d_{xz}d_{xz}\right>$ $\left<p_zp_z|sd_{z^2}\right>$ $\left<p_zp_z|p_xd_{z^2}\right>$ $\left<p_zp_z|p_yd_{z^2}\right>$ $\left<p_zp_z|p_zd_{z^2}\right>$ 
$\left<p_zp_z|d_{x^2-y^2}d_{z^2}\right>$ $\left<p_zp_z|d_{xz}d_{z^2}\right>$ $\left<p_zp_z|d_{z^2}d_{z^2}\right>$ $\left<p_zp_z|sd_{yz}\right>$ $\left<p_zp_z|p_xd_{yz}\right>$ $\left<p_zp_z|p_yd_{yz}\right>$ $\left<p_zp_z|p_zd_{yz}\right>$ $\left<p_zp_z|d_{x^2-y^2}d_{yz}\right>$ $\left<p_zp_z|d_{xz}d_{yz}\right>$ $\left<p_zp_z|d_{z^2}d_{yz}\right>$ 
$\left<p_zp_z|d_{yz}d_{yz}\right>$ $\left<p_zp_z|sd_{xy}\right>$ $\left<p_zp_z|p_xd_{xy}\right>$ $\left<p_zp_z|p_yd_{xy}\right>$ $\left<p_zp_z|p_zd_{xy}\right>$ $\left<p_zp_z|d_{x^2-y^2}d_{xy}\right>$ $\left<p_zp_z|d_{xz}d_{xy}\right>$ $\left<p_zp_z|d_{z^2}d_{xy}\right>$ $\left<p_zp_z|d_{yz}d_{xy}\right>$ $\left<p_zp_z|d_{xy}d_{xy}\right>$ 
$\left<sd_{x^2-y^2}|ss\right>$ $\left<sd_{x^2-y^2}|sp_x\right>$ $\left<sd_{x^2-y^2}|p_xp_x\right>$ $\left<sd_{x^2-y^2}|sp_y\right>$ $\left<sd_{x^2-y^2}|p_xp_y\right>$ $\left<sd_{x^2-y^2}|p_yp_y\right>$ $\left<sd_{x^2-y^2}|sp_z\right>$ $\left<sd_{x^2-y^2}|p_xp_z\right>$ $\left<sd_{x^2-y^2}|p_yp_z\right>$ $\left<sd_{x^2-y^2}|p_zp_z\right>$ 
$\left<sd_{x^2-y^2}|sd_{x^2-y^2}\right>$ $\left<sd_{x^2-y^2}|p_xd_{x^2-y^2}\right>$ $\left<sd_{x^2-y^2}|p_yd_{x^2-y^2}\right>$ $\left<sd_{x^2-y^2}|p_zd_{x^2-y^2}\right>$ $\left<sd_{x^2-y^2}|d_{x^2-y^2}d_{x^2-y^2}\right>$ $\left<sd_{x^2-y^2}|sd_{xz}\right>$ $\left<sd_{x^2-y^2}|p_xd_{xz}\right>$ $\left<sd_{x^2-y^2}|p_yd_{xz}\right>$ $\left<sd_{x^2-y^2}|p_zd_{xz}\right>$ $\left<sd_{x^2-y^2}|d_{x^2-y^2}d_{xz}\right>$ 
$\left<sd_{x^2-y^2}|d_{xz}d_{xz}\right>$ $\left<sd_{x^2-y^2}|sd_{z^2}\right>$ $\left<sd_{x^2-y^2}|p_xd_{z^2}\right>$ $\left<sd_{x^2-y^2}|p_yd_{z^2}\right>$ $\left<sd_{x^2-y^2}|p_zd_{z^2}\right>$ $\left<sd_{x^2-y^2}|d_{x^2-y^2}d_{z^2}\right>$ $\left<sd_{x^2-y^2}|d_{xz}d_{z^2}\right>$ $\left<sd_{x^2-y^2}|d_{z^2}d_{z^2}\right>$ $\left<sd_{x^2-y^2}|sd_{yz}\right>$ $\left<sd_{x^2-y^2}|p_xd_{yz}\right>$ 
$\left<sd_{x^2-y^2}|p_yd_{yz}\right>$ $\left<sd_{x^2-y^2}|p_zd_{yz}\right>$ $\left<sd_{x^2-y^2}|d_{x^2-y^2}d_{yz}\right>$ $\left<sd_{x^2-y^2}|d_{xz}d_{yz}\right>$ $\left<sd_{x^2-y^2}|d_{z^2}d_{yz}\right>$ $\left<sd_{x^2-y^2}|d_{yz}d_{yz}\right>$ $\left<sd_{x^2-y^2}|sd_{xy}\right>$ $\left<sd_{x^2-y^2}|p_xd_{xy}\right>$ $\left<sd_{x^2-y^2}|p_yd_{xy}\right>$ $\left<sd_{x^2-y^2}|p_zd_{xy}\right>$ 
$\left<sd_{x^2-y^2}|d_{x^2-y^2}d_{xy}\right>$ $\left<sd_{x^2-y^2}|d_{xz}d_{xy}\right>$ $\left<sd_{x^2-y^2}|d_{z^2}d_{xy}\right>$ $\left<sd_{x^2-y^2}|d_{yz}d_{xy}\right>$ $\left<sd_{x^2-y^2}|d_{xy}d_{xy}\right>$ $\left<p_xd_{x^2-y^2}|ss\right>$ $\left<p_xd_{x^2-y^2}|sp_x\right>$ $\left<p_xd_{x^2-y^2}|p_xp_x\right>$ $\left<p_xd_{x^2-y^2}|sp_y\right>$ $\left<p_xd_{x^2-y^2}|p_xp_y\right>$ 
$\left<p_xd_{x^2-y^2}|p_yp_y\right>$ $\left<p_xd_{x^2-y^2}|sp_z\right>$ $\left<p_xd_{x^2-y^2}|p_xp_z\right>$ $\left<p_xd_{x^2-y^2}|p_yp_z\right>$ $\left<p_xd_{x^2-y^2}|p_zp_z\right>$ $\left<p_xd_{x^2-y^2}|sd_{x^2-y^2}\right>$ $\left<p_xd_{x^2-y^2}|p_xd_{x^2-y^2}\right>$ $\left<p_xd_{x^2-y^2}|p_yd_{x^2-y^2}\right>$ $\left<p_xd_{x^2-y^2}|p_zd_{x^2-y^2}\right>$ $\left<p_xd_{x^2-y^2}|d_{x^2-y^2}d_{x^2-y^2}\right>$ 
$\left<p_xd_{x^2-y^2}|sd_{xz}\right>$ $\left<p_xd_{x^2-y^2}|p_xd_{xz}\right>$ $\left<p_xd_{x^2-y^2}|p_yd_{xz}\right>$ $\left<p_xd_{x^2-y^2}|p_zd_{xz}\right>$ $\left<p_xd_{x^2-y^2}|d_{x^2-y^2}d_{xz}\right>$ $\left<p_xd_{x^2-y^2}|d_{xz}d_{xz}\right>$ $\left<p_xd_{x^2-y^2}|sd_{z^2}\right>$ $\left<p_xd_{x^2-y^2}|p_xd_{z^2}\right>$ $\left<p_xd_{x^2-y^2}|p_yd_{z^2}\right>$ $\left<p_xd_{x^2-y^2}|p_zd_{z^2}\right>$ 
$\left<p_xd_{x^2-y^2}|d_{x^2-y^2}d_{z^2}\right>$ $\left<p_xd_{x^2-y^2}|d_{xz}d_{z^2}\right>$ $\left<p_xd_{x^2-y^2}|d_{z^2}d_{z^2}\right>$ $\left<p_xd_{x^2-y^2}|sd_{yz}\right>$ $\left<p_xd_{x^2-y^2}|p_xd_{yz}\right>$ $\left<p_xd_{x^2-y^2}|p_yd_{yz}\right>$ $\left<p_xd_{x^2-y^2}|p_zd_{yz}\right>$ $\left<p_xd_{x^2-y^2}|d_{x^2-y^2}d_{yz}\right>$ $\left<p_xd_{x^2-y^2}|d_{xz}d_{yz}\right>$ $\left<p_xd_{x^2-y^2}|d_{z^2}d_{yz}\right>$ 
$\left<p_xd_{x^2-y^2}|d_{yz}d_{yz}\right>$ $\left<p_xd_{x^2-y^2}|sd_{xy}\right>$ $\left<p_xd_{x^2-y^2}|p_xd_{xy}\right>$ $\left<p_xd_{x^2-y^2}|p_yd_{xy}\right>$ $\left<p_xd_{x^2-y^2}|p_zd_{xy}\right>$ $\left<p_xd_{x^2-y^2}|d_{x^2-y^2}d_{xy}\right>$ $\left<p_xd_{x^2-y^2}|d_{xz}d_{xy}\right>$ $\left<p_xd_{x^2-y^2}|d_{z^2}d_{xy}\right>$ $\left<p_xd_{x^2-y^2}|d_{yz}d_{xy}\right>$ $\left<p_xd_{x^2-y^2}|d_{xy}d_{xy}\right>$ 
$\left<p_yd_{x^2-y^2}|ss\right>$ $\left<p_yd_{x^2-y^2}|sp_x\right>$ $\left<p_yd_{x^2-y^2}|p_xp_x\right>$ $\left<p_yd_{x^2-y^2}|sp_y\right>$ $\left<p_yd_{x^2-y^2}|p_xp_y\right>$ $\left<p_yd_{x^2-y^2}|p_yp_y\right>$ $\left<p_yd_{x^2-y^2}|sp_z\right>$ $\left<p_yd_{x^2-y^2}|p_xp_z\right>$ $\left<p_yd_{x^2-y^2}|p_yp_z\right>$ $\left<p_yd_{x^2-y^2}|p_zp_z\right>$ 
$\left<p_yd_{x^2-y^2}|sd_{x^2-y^2}\right>$ $\left<p_yd_{x^2-y^2}|p_xd_{x^2-y^2}\right>$ $\left<p_yd_{x^2-y^2}|p_yd_{x^2-y^2}\right>$ $\left<p_yd_{x^2-y^2}|p_zd_{x^2-y^2}\right>$ $\left<p_yd_{x^2-y^2}|d_{x^2-y^2}d_{x^2-y^2}\right>$ $\left<p_yd_{x^2-y^2}|sd_{xz}\right>$ $\left<p_yd_{x^2-y^2}|p_xd_{xz}\right>$ $\left<p_yd_{x^2-y^2}|p_yd_{xz}\right>$ $\left<p_yd_{x^2-y^2}|p_zd_{xz}\right>$ $\left<p_yd_{x^2-y^2}|d_{x^2-y^2}d_{xz}\right>$ 
$\left<p_yd_{x^2-y^2}|d_{xz}d_{xz}\right>$ $\left<p_yd_{x^2-y^2}|sd_{z^2}\right>$ $\left<p_yd_{x^2-y^2}|p_xd_{z^2}\right>$ $\left<p_yd_{x^2-y^2}|p_yd_{z^2}\right>$ $\left<p_yd_{x^2-y^2}|p_zd_{z^2}\right>$ $\left<p_yd_{x^2-y^2}|d_{x^2-y^2}d_{z^2}\right>$ $\left<p_yd_{x^2-y^2}|d_{xz}d_{z^2}\right>$ $\left<p_yd_{x^2-y^2}|d_{z^2}d_{z^2}\right>$ $\left<p_yd_{x^2-y^2}|sd_{yz}\right>$ $\left<p_yd_{x^2-y^2}|p_xd_{yz}\right>$ 
$\left<p_yd_{x^2-y^2}|p_yd_{yz}\right>$ $\left<p_yd_{x^2-y^2}|p_zd_{yz}\right>$ $\left<p_yd_{x^2-y^2}|d_{x^2-y^2}d_{yz}\right>$ $\left<p_yd_{x^2-y^2}|d_{xz}d_{yz}\right>$ $\left<p_yd_{x^2-y^2}|d_{z^2}d_{yz}\right>$ $\left<p_yd_{x^2-y^2}|d_{yz}d_{yz}\right>$ $\left<p_yd_{x^2-y^2}|sd_{xy}\right>$ $\left<p_yd_{x^2-y^2}|p_xd_{xy}\right>$ $\left<p_yd_{x^2-y^2}|p_yd_{xy}\right>$ $\left<p_yd_{x^2-y^2}|p_zd_{xy}\right>$ 
$\left<p_yd_{x^2-y^2}|d_{x^2-y^2}d_{xy}\right>$ $\left<p_yd_{x^2-y^2}|d_{xz}d_{xy}\right>$ $\left<p_yd_{x^2-y^2}|d_{z^2}d_{xy}\right>$ $\left<p_yd_{x^2-y^2}|d_{yz}d_{xy}\right>$ $\left<p_yd_{x^2-y^2}|d_{xy}d_{xy}\right>$ $\left<p_zd_{x^2-y^2}|ss\right>$ $\left<p_zd_{x^2-y^2}|sp_x\right>$ $\left<p_zd_{x^2-y^2}|p_xp_x\right>$ $\left<p_zd_{x^2-y^2}|sp_y\right>$ $\left<p_zd_{x^2-y^2}|p_xp_y\right>$ 
$\left<p_zd_{x^2-y^2}|p_yp_y\right>$ $\left<p_zd_{x^2-y^2}|sp_z\right>$ $\left<p_zd_{x^2-y^2}|p_xp_z\right>$ $\left<p_zd_{x^2-y^2}|p_yp_z\right>$ $\left<p_zd_{x^2-y^2}|p_zp_z\right>$ $\left<p_zd_{x^2-y^2}|sd_{x^2-y^2}\right>$ $\left<p_zd_{x^2-y^2}|p_xd_{x^2-y^2}\right>$ $\left<p_zd_{x^2-y^2}|p_yd_{x^2-y^2}\right>$ $\left<p_zd_{x^2-y^2}|p_zd_{x^2-y^2}\right>$ $\left<p_zd_{x^2-y^2}|d_{x^2-y^2}d_{x^2-y^2}\right>$ 
$\left<p_zd_{x^2-y^2}|sd_{xz}\right>$ $\left<p_zd_{x^2-y^2}|p_xd_{xz}\right>$ $\left<p_zd_{x^2-y^2}|p_yd_{xz}\right>$ $\left<p_zd_{x^2-y^2}|p_zd_{xz}\right>$ $\left<p_zd_{x^2-y^2}|d_{x^2-y^2}d_{xz}\right>$ $\left<p_zd_{x^2-y^2}|d_{xz}d_{xz}\right>$ $\left<p_zd_{x^2-y^2}|sd_{z^2}\right>$ $\left<p_zd_{x^2-y^2}|p_xd_{z^2}\right>$ $\left<p_zd_{x^2-y^2}|p_yd_{z^2}\right>$ $\left<p_zd_{x^2-y^2}|p_zd_{z^2}\right>$ 
$\left<p_zd_{x^2-y^2}|d_{x^2-y^2}d_{z^2}\right>$ $\left<p_zd_{x^2-y^2}|d_{xz}d_{z^2}\right>$ $\left<p_zd_{x^2-y^2}|d_{z^2}d_{z^2}\right>$ $\left<p_zd_{x^2-y^2}|sd_{yz}\right>$ $\left<p_zd_{x^2-y^2}|p_xd_{yz}\right>$ $\left<p_zd_{x^2-y^2}|p_yd_{yz}\right>$ $\left<p_zd_{x^2-y^2}|p_zd_{yz}\right>$ $\left<p_zd_{x^2-y^2}|d_{x^2-y^2}d_{yz}\right>$ $\left<p_zd_{x^2-y^2}|d_{xz}d_{yz}\right>$ $\left<p_zd_{x^2-y^2}|d_{z^2}d_{yz}\right>$ 
$\left<p_zd_{x^2-y^2}|d_{yz}d_{yz}\right>$ $\left<p_zd_{x^2-y^2}|sd_{xy}\right>$ $\left<p_zd_{x^2-y^2}|p_xd_{xy}\right>$ $\left<p_zd_{x^2-y^2}|p_yd_{xy}\right>$ $\left<p_zd_{x^2-y^2}|p_zd_{xy}\right>$ $\left<p_zd_{x^2-y^2}|d_{x^2-y^2}d_{xy}\right>$ $\left<p_zd_{x^2-y^2}|d_{xz}d_{xy}\right>$ $\left<p_zd_{x^2-y^2}|d_{z^2}d_{xy}\right>$ $\left<p_zd_{x^2-y^2}|d_{yz}d_{xy}\right>$ $\left<p_zd_{x^2-y^2}|d_{xy}d_{xy}\right>$ 
$\left<d_{x^2-y^2}d_{x^2-y^2}|ss\right>$ $\left<d_{x^2-y^2}d_{x^2-y^2}|sp_x\right>$ $\left<d_{x^2-y^2}d_{x^2-y^2}|p_xp_x\right>$ $\left<d_{x^2-y^2}d_{x^2-y^2}|sp_y\right>$ $\left<d_{x^2-y^2}d_{x^2-y^2}|p_xp_y\right>$ $\left<d_{x^2-y^2}d_{x^2-y^2}|p_yp_y\right>$ $\left<d_{x^2-y^2}d_{x^2-y^2}|sp_z\right>$ $\left<d_{x^2-y^2}d_{x^2-y^2}|p_xp_z\right>$ $\left<d_{x^2-y^2}d_{x^2-y^2}|p_yp_z\right>$ $\left<d_{x^2-y^2}d_{x^2-y^2}|p_zp_z\right>$ 
$\left<d_{x^2-y^2}d_{x^2-y^2}|sd_{x^2-y^2}\right>$ $\left<d_{x^2-y^2}d_{x^2-y^2}|p_xd_{x^2-y^2}\right>$ $\left<d_{x^2-y^2}d_{x^2-y^2}|p_yd_{x^2-y^2}\right>$ $\left<d_{x^2-y^2}d_{x^2-y^2}|p_zd_{x^2-y^2}\right>$ $\left<d_{x^2-y^2}d_{x^2-y^2}|d_{x^2-y^2}d_{x^2-y^2}\right>$ $\left<d_{x^2-y^2}d_{x^2-y^2}|sd_{xz}\right>$ $\left<d_{x^2-y^2}d_{x^2-y^2}|p_xd_{xz}\right>$ $\left<d_{x^2-y^2}d_{x^2-y^2}|p_yd_{xz}\right>$ $\left<d_{x^2-y^2}d_{x^2-y^2}|p_zd_{xz}\right>$ $\left<d_{x^2-y^2}d_{x^2-y^2}|d_{x^2-y^2}d_{xz}\right>$ 
$\left<d_{x^2-y^2}d_{x^2-y^2}|d_{xz}d_{xz}\right>$ $\left<d_{x^2-y^2}d_{x^2-y^2}|sd_{z^2}\right>$ $\left<d_{x^2-y^2}d_{x^2-y^2}|p_xd_{z^2}\right>$ $\left<d_{x^2-y^2}d_{x^2-y^2}|p_yd_{z^2}\right>$ $\left<d_{x^2-y^2}d_{x^2-y^2}|p_zd_{z^2}\right>$ $\left<d_{x^2-y^2}d_{x^2-y^2}|d_{x^2-y^2}d_{z^2}\right>$ $\left<d_{x^2-y^2}d_{x^2-y^2}|d_{xz}d_{z^2}\right>$ $\left<d_{x^2-y^2}d_{x^2-y^2}|d_{z^2}d_{z^2}\right>$ $\left<d_{x^2-y^2}d_{x^2-y^2}|sd_{yz}\right>$ $\left<d_{x^2-y^2}d_{x^2-y^2}|p_xd_{yz}\right>$ 
$\left<d_{x^2-y^2}d_{x^2-y^2}|p_yd_{yz}\right>$ $\left<d_{x^2-y^2}d_{x^2-y^2}|p_zd_{yz}\right>$ $\left<d_{x^2-y^2}d_{x^2-y^2}|d_{x^2-y^2}d_{yz}\right>$ $\left<d_{x^2-y^2}d_{x^2-y^2}|d_{xz}d_{yz}\right>$ $\left<d_{x^2-y^2}d_{x^2-y^2}|d_{z^2}d_{yz}\right>$ $\left<d_{x^2-y^2}d_{x^2-y^2}|d_{yz}d_{yz}\right>$ $\left<d_{x^2-y^2}d_{x^2-y^2}|sd_{xy}\right>$ $\left<d_{x^2-y^2}d_{x^2-y^2}|p_xd_{xy}\right>$ $\left<d_{x^2-y^2}d_{x^2-y^2}|p_yd_{xy}\right>$ $\left<d_{x^2-y^2}d_{x^2-y^2}|p_zd_{xy}\right>$ 
$\left<d_{x^2-y^2}d_{x^2-y^2}|d_{x^2-y^2}d_{xy}\right>$ $\left<d_{x^2-y^2}d_{x^2-y^2}|d_{xz}d_{xy}\right>$ $\left<d_{x^2-y^2}d_{x^2-y^2}|d_{z^2}d_{xy}\right>$ $\left<d_{x^2-y^2}d_{x^2-y^2}|d_{yz}d_{xy}\right>$ $\left<d_{x^2-y^2}d_{x^2-y^2}|d_{xy}d_{xy}\right>$ $\left<sd_{xz}|ss\right>$ $\left<sd_{xz}|sp_x\right>$ $\left<sd_{xz}|p_xp_x\right>$ $\left<sd_{xz}|sp_y\right>$ $\left<sd_{xz}|p_xp_y\right>$ 
$\left<sd_{xz}|p_yp_y\right>$ $\left<sd_{xz}|sp_z\right>$ $\left<sd_{xz}|p_xp_z\right>$ $\left<sd_{xz}|p_yp_z\right>$ $\left<sd_{xz}|p_zp_z\right>$ $\left<sd_{xz}|sd_{x^2-y^2}\right>$ $\left<sd_{xz}|p_xd_{x^2-y^2}\right>$ $\left<sd_{xz}|p_yd_{x^2-y^2}\right>$ $\left<sd_{xz}|p_zd_{x^2-y^2}\right>$ $\left<sd_{xz}|d_{x^2-y^2}d_{x^2-y^2}\right>$ 
$\left<sd_{xz}|sd_{xz}\right>$ $\left<sd_{xz}|p_xd_{xz}\right>$ $\left<sd_{xz}|p_yd_{xz}\right>$ $\left<sd_{xz}|p_zd_{xz}\right>$ $\left<sd_{xz}|d_{x^2-y^2}d_{xz}\right>$ $\left<sd_{xz}|d_{xz}d_{xz}\right>$ $\left<sd_{xz}|sd_{z^2}\right>$ $\left<sd_{xz}|p_xd_{z^2}\right>$ $\left<sd_{xz}|p_yd_{z^2}\right>$ $\left<sd_{xz}|p_zd_{z^2}\right>$ 
$\left<sd_{xz}|d_{x^2-y^2}d_{z^2}\right>$ $\left<sd_{xz}|d_{xz}d_{z^2}\right>$ $\left<sd_{xz}|d_{z^2}d_{z^2}\right>$ $\left<sd_{xz}|sd_{yz}\right>$ $\left<sd_{xz}|p_xd_{yz}\right>$ $\left<sd_{xz}|p_yd_{yz}\right>$ $\left<sd_{xz}|p_zd_{yz}\right>$ $\left<sd_{xz}|d_{x^2-y^2}d_{yz}\right>$ $\left<sd_{xz}|d_{xz}d_{yz}\right>$ $\left<sd_{xz}|d_{z^2}d_{yz}\right>$ 
$\left<sd_{xz}|d_{yz}d_{yz}\right>$ $\left<sd_{xz}|sd_{xy}\right>$ $\left<sd_{xz}|p_xd_{xy}\right>$ $\left<sd_{xz}|p_yd_{xy}\right>$ $\left<sd_{xz}|p_zd_{xy}\right>$ $\left<sd_{xz}|d_{x^2-y^2}d_{xy}\right>$ $\left<sd_{xz}|d_{xz}d_{xy}\right>$ $\left<sd_{xz}|d_{z^2}d_{xy}\right>$ $\left<sd_{xz}|d_{yz}d_{xy}\right>$ $\left<sd_{xz}|d_{xy}d_{xy}\right>$ 
$\left<p_xd_{xz}|ss\right>$ $\left<p_xd_{xz}|sp_x\right>$ $\left<p_xd_{xz}|p_xp_x\right>$ $\left<p_xd_{xz}|sp_y\right>$ $\left<p_xd_{xz}|p_xp_y\right>$ $\left<p_xd_{xz}|p_yp_y\right>$ $\left<p_xd_{xz}|sp_z\right>$ $\left<p_xd_{xz}|p_xp_z\right>$ $\left<p_xd_{xz}|p_yp_z\right>$ $\left<p_xd_{xz}|p_zp_z\right>$ 
$\left<p_xd_{xz}|sd_{x^2-y^2}\right>$ $\left<p_xd_{xz}|p_xd_{x^2-y^2}\right>$ $\left<p_xd_{xz}|p_yd_{x^2-y^2}\right>$ $\left<p_xd_{xz}|p_zd_{x^2-y^2}\right>$ $\left<p_xd_{xz}|d_{x^2-y^2}d_{x^2-y^2}\right>$ $\left<p_xd_{xz}|sd_{xz}\right>$ $\left<p_xd_{xz}|p_xd_{xz}\right>$ $\left<p_xd_{xz}|p_yd_{xz}\right>$ $\left<p_xd_{xz}|p_zd_{xz}\right>$ $\left<p_xd_{xz}|d_{x^2-y^2}d_{xz}\right>$ 
$\left<p_xd_{xz}|d_{xz}d_{xz}\right>$ $\left<p_xd_{xz}|sd_{z^2}\right>$ $\left<p_xd_{xz}|p_xd_{z^2}\right>$ $\left<p_xd_{xz}|p_yd_{z^2}\right>$ $\left<p_xd_{xz}|p_zd_{z^2}\right>$ $\left<p_xd_{xz}|d_{x^2-y^2}d_{z^2}\right>$ $\left<p_xd_{xz}|d_{xz}d_{z^2}\right>$ $\left<p_xd_{xz}|d_{z^2}d_{z^2}\right>$ $\left<p_xd_{xz}|sd_{yz}\right>$ $\left<p_xd_{xz}|p_xd_{yz}\right>$ 
$\left<p_xd_{xz}|p_yd_{yz}\right>$ $\left<p_xd_{xz}|p_zd_{yz}\right>$ $\left<p_xd_{xz}|d_{x^2-y^2}d_{yz}\right>$ $\left<p_xd_{xz}|d_{xz}d_{yz}\right>$ $\left<p_xd_{xz}|d_{z^2}d_{yz}\right>$ $\left<p_xd_{xz}|d_{yz}d_{yz}\right>$ $\left<p_xd_{xz}|sd_{xy}\right>$ $\left<p_xd_{xz}|p_xd_{xy}\right>$ $\left<p_xd_{xz}|p_yd_{xy}\right>$ $\left<p_xd_{xz}|p_zd_{xy}\right>$ 
$\left<p_xd_{xz}|d_{x^2-y^2}d_{xy}\right>$ $\left<p_xd_{xz}|d_{xz}d_{xy}\right>$ $\left<p_xd_{xz}|d_{z^2}d_{xy}\right>$ $\left<p_xd_{xz}|d_{yz}d_{xy}\right>$ $\left<p_xd_{xz}|d_{xy}d_{xy}\right>$ $\left<p_yd_{xz}|ss\right>$ $\left<p_yd_{xz}|sp_x\right>$ $\left<p_yd_{xz}|p_xp_x\right>$ $\left<p_yd_{xz}|sp_y\right>$ $\left<p_yd_{xz}|p_xp_y\right>$ 
$\left<p_yd_{xz}|p_yp_y\right>$ $\left<p_yd_{xz}|sp_z\right>$ $\left<p_yd_{xz}|p_xp_z\right>$ $\left<p_yd_{xz}|p_yp_z\right>$ $\left<p_yd_{xz}|p_zp_z\right>$ $\left<p_yd_{xz}|sd_{x^2-y^2}\right>$ $\left<p_yd_{xz}|p_xd_{x^2-y^2}\right>$ $\left<p_yd_{xz}|p_yd_{x^2-y^2}\right>$ $\left<p_yd_{xz}|p_zd_{x^2-y^2}\right>$ $\left<p_yd_{xz}|d_{x^2-y^2}d_{x^2-y^2}\right>$ 
$\left<p_yd_{xz}|sd_{xz}\right>$ $\left<p_yd_{xz}|p_xd_{xz}\right>$ $\left<p_yd_{xz}|p_yd_{xz}\right>$ $\left<p_yd_{xz}|p_zd_{xz}\right>$ $\left<p_yd_{xz}|d_{x^2-y^2}d_{xz}\right>$ $\left<p_yd_{xz}|d_{xz}d_{xz}\right>$ $\left<p_yd_{xz}|sd_{z^2}\right>$ $\left<p_yd_{xz}|p_xd_{z^2}\right>$ $\left<p_yd_{xz}|p_yd_{z^2}\right>$ $\left<p_yd_{xz}|p_zd_{z^2}\right>$ 
$\left<p_yd_{xz}|d_{x^2-y^2}d_{z^2}\right>$ $\left<p_yd_{xz}|d_{xz}d_{z^2}\right>$ $\left<p_yd_{xz}|d_{z^2}d_{z^2}\right>$ $\left<p_yd_{xz}|sd_{yz}\right>$ $\left<p_yd_{xz}|p_xd_{yz}\right>$ $\left<p_yd_{xz}|p_yd_{yz}\right>$ $\left<p_yd_{xz}|p_zd_{yz}\right>$ $\left<p_yd_{xz}|d_{x^2-y^2}d_{yz}\right>$ $\left<p_yd_{xz}|d_{xz}d_{yz}\right>$ $\left<p_yd_{xz}|d_{z^2}d_{yz}\right>$ 
$\left<p_yd_{xz}|d_{yz}d_{yz}\right>$ $\left<p_yd_{xz}|sd_{xy}\right>$ $\left<p_yd_{xz}|p_xd_{xy}\right>$ $\left<p_yd_{xz}|p_yd_{xy}\right>$ $\left<p_yd_{xz}|p_zd_{xy}\right>$ $\left<p_yd_{xz}|d_{x^2-y^2}d_{xy}\right>$ $\left<p_yd_{xz}|d_{xz}d_{xy}\right>$ $\left<p_yd_{xz}|d_{z^2}d_{xy}\right>$ $\left<p_yd_{xz}|d_{yz}d_{xy}\right>$ $\left<p_yd_{xz}|d_{xy}d_{xy}\right>$ 
$\left<p_zd_{xz}|ss\right>$ $\left<p_zd_{xz}|sp_x\right>$ $\left<p_zd_{xz}|p_xp_x\right>$ $\left<p_zd_{xz}|sp_y\right>$ $\left<p_zd_{xz}|p_xp_y\right>$ $\left<p_zd_{xz}|p_yp_y\right>$ $\left<p_zd_{xz}|sp_z\right>$ $\left<p_zd_{xz}|p_xp_z\right>$ $\left<p_zd_{xz}|p_yp_z\right>$ $\left<p_zd_{xz}|p_zp_z\right>$ 
$\left<p_zd_{xz}|sd_{x^2-y^2}\right>$ $\left<p_zd_{xz}|p_xd_{x^2-y^2}\right>$ $\left<p_zd_{xz}|p_yd_{x^2-y^2}\right>$ $\left<p_zd_{xz}|p_zd_{x^2-y^2}\right>$ $\left<p_zd_{xz}|d_{x^2-y^2}d_{x^2-y^2}\right>$ $\left<p_zd_{xz}|sd_{xz}\right>$ $\left<p_zd_{xz}|p_xd_{xz}\right>$ $\left<p_zd_{xz}|p_yd_{xz}\right>$ $\left<p_zd_{xz}|p_zd_{xz}\right>$ $\left<p_zd_{xz}|d_{x^2-y^2}d_{xz}\right>$ 
$\left<p_zd_{xz}|d_{xz}d_{xz}\right>$ $\left<p_zd_{xz}|sd_{z^2}\right>$ $\left<p_zd_{xz}|p_xd_{z^2}\right>$ $\left<p_zd_{xz}|p_yd_{z^2}\right>$ $\left<p_zd_{xz}|p_zd_{z^2}\right>$ $\left<p_zd_{xz}|d_{x^2-y^2}d_{z^2}\right>$ $\left<p_zd_{xz}|d_{xz}d_{z^2}\right>$ $\left<p_zd_{xz}|d_{z^2}d_{z^2}\right>$ $\left<p_zd_{xz}|sd_{yz}\right>$ $\left<p_zd_{xz}|p_xd_{yz}\right>$ 
$\left<p_zd_{xz}|p_yd_{yz}\right>$ $\left<p_zd_{xz}|p_zd_{yz}\right>$ $\left<p_zd_{xz}|d_{x^2-y^2}d_{yz}\right>$ $\left<p_zd_{xz}|d_{xz}d_{yz}\right>$ $\left<p_zd_{xz}|d_{z^2}d_{yz}\right>$ $\left<p_zd_{xz}|d_{yz}d_{yz}\right>$ $\left<p_zd_{xz}|sd_{xy}\right>$ $\left<p_zd_{xz}|p_xd_{xy}\right>$ $\left<p_zd_{xz}|p_yd_{xy}\right>$ $\left<p_zd_{xz}|p_zd_{xy}\right>$ 
$\left<p_zd_{xz}|d_{x^2-y^2}d_{xy}\right>$ $\left<p_zd_{xz}|d_{xz}d_{xy}\right>$ $\left<p_zd_{xz}|d_{z^2}d_{xy}\right>$ $\left<p_zd_{xz}|d_{yz}d_{xy}\right>$ $\left<p_zd_{xz}|d_{xy}d_{xy}\right>$ $\left<d_{x^2-y^2}d_{xz}|ss\right>$ $\left<d_{x^2-y^2}d_{xz}|sp_x\right>$ $\left<d_{x^2-y^2}d_{xz}|p_xp_x\right>$ $\left<d_{x^2-y^2}d_{xz}|sp_y\right>$ $\left<d_{x^2-y^2}d_{xz}|p_xp_y\right>$ 
$\left<d_{x^2-y^2}d_{xz}|p_yp_y\right>$ $\left<d_{x^2-y^2}d_{xz}|sp_z\right>$ $\left<d_{x^2-y^2}d_{xz}|p_xp_z\right>$ $\left<d_{x^2-y^2}d_{xz}|p_yp_z\right>$ $\left<d_{x^2-y^2}d_{xz}|p_zp_z\right>$ $\left<d_{x^2-y^2}d_{xz}|sd_{x^2-y^2}\right>$ $\left<d_{x^2-y^2}d_{xz}|p_xd_{x^2-y^2}\right>$ $\left<d_{x^2-y^2}d_{xz}|p_yd_{x^2-y^2}\right>$ $\left<d_{x^2-y^2}d_{xz}|p_zd_{x^2-y^2}\right>$ $\left<d_{x^2-y^2}d_{xz}|d_{x^2-y^2}d_{x^2-y^2}\right>$ 
$\left<d_{x^2-y^2}d_{xz}|sd_{xz}\right>$ $\left<d_{x^2-y^2}d_{xz}|p_xd_{xz}\right>$ $\left<d_{x^2-y^2}d_{xz}|p_yd_{xz}\right>$ $\left<d_{x^2-y^2}d_{xz}|p_zd_{xz}\right>$ $\left<d_{x^2-y^2}d_{xz}|d_{x^2-y^2}d_{xz}\right>$ $\left<d_{x^2-y^2}d_{xz}|d_{xz}d_{xz}\right>$ $\left<d_{x^2-y^2}d_{xz}|sd_{z^2}\right>$ $\left<d_{x^2-y^2}d_{xz}|p_xd_{z^2}\right>$ $\left<d_{x^2-y^2}d_{xz}|p_yd_{z^2}\right>$ $\left<d_{x^2-y^2}d_{xz}|p_zd_{z^2}\right>$ 
$\left<d_{x^2-y^2}d_{xz}|d_{x^2-y^2}d_{z^2}\right>$ $\left<d_{x^2-y^2}d_{xz}|d_{xz}d_{z^2}\right>$ $\left<d_{x^2-y^2}d_{xz}|d_{z^2}d_{z^2}\right>$ $\left<d_{x^2-y^2}d_{xz}|sd_{yz}\right>$ $\left<d_{x^2-y^2}d_{xz}|p_xd_{yz}\right>$ $\left<d_{x^2-y^2}d_{xz}|p_yd_{yz}\right>$ $\left<d_{x^2-y^2}d_{xz}|p_zd_{yz}\right>$ $\left<d_{x^2-y^2}d_{xz}|d_{x^2-y^2}d_{yz}\right>$ $\left<d_{x^2-y^2}d_{xz}|d_{xz}d_{yz}\right>$ $\left<d_{x^2-y^2}d_{xz}|d_{z^2}d_{yz}\right>$ 
$\left<d_{x^2-y^2}d_{xz}|d_{yz}d_{yz}\right>$ $\left<d_{x^2-y^2}d_{xz}|sd_{xy}\right>$ $\left<d_{x^2-y^2}d_{xz}|p_xd_{xy}\right>$ $\left<d_{x^2-y^2}d_{xz}|p_yd_{xy}\right>$ $\left<d_{x^2-y^2}d_{xz}|p_zd_{xy}\right>$ $\left<d_{x^2-y^2}d_{xz}|d_{x^2-y^2}d_{xy}\right>$ $\left<d_{x^2-y^2}d_{xz}|d_{xz}d_{xy}\right>$ $\left<d_{x^2-y^2}d_{xz}|d_{z^2}d_{xy}\right>$ $\left<d_{x^2-y^2}d_{xz}|d_{yz}d_{xy}\right>$ $\left<d_{x^2-y^2}d_{xz}|d_{xy}d_{xy}\right>$ 
$\left<d_{xz}d_{xz}|ss\right>$ $\left<d_{xz}d_{xz}|sp_x\right>$ $\left<d_{xz}d_{xz}|p_xp_x\right>$ $\left<d_{xz}d_{xz}|sp_y\right>$ $\left<d_{xz}d_{xz}|p_xp_y\right>$ $\left<d_{xz}d_{xz}|p_yp_y\right>$ $\left<d_{xz}d_{xz}|sp_z\right>$ $\left<d_{xz}d_{xz}|p_xp_z\right>$ $\left<d_{xz}d_{xz}|p_yp_z\right>$ $\left<d_{xz}d_{xz}|p_zp_z\right>$ 
$\left<d_{xz}d_{xz}|sd_{x^2-y^2}\right>$ $\left<d_{xz}d_{xz}|p_xd_{x^2-y^2}\right>$ $\left<d_{xz}d_{xz}|p_yd_{x^2-y^2}\right>$ $\left<d_{xz}d_{xz}|p_zd_{x^2-y^2}\right>$ $\left<d_{xz}d_{xz}|d_{x^2-y^2}d_{x^2-y^2}\right>$ $\left<d_{xz}d_{xz}|sd_{xz}\right>$ $\left<d_{xz}d_{xz}|p_xd_{xz}\right>$ $\left<d_{xz}d_{xz}|p_yd_{xz}\right>$ $\left<d_{xz}d_{xz}|p_zd_{xz}\right>$ $\left<d_{xz}d_{xz}|d_{x^2-y^2}d_{xz}\right>$ 
$\left<d_{xz}d_{xz}|d_{xz}d_{xz}\right>$ $\left<d_{xz}d_{xz}|sd_{z^2}\right>$ $\left<d_{xz}d_{xz}|p_xd_{z^2}\right>$ $\left<d_{xz}d_{xz}|p_yd_{z^2}\right>$ $\left<d_{xz}d_{xz}|p_zd_{z^2}\right>$ $\left<d_{xz}d_{xz}|d_{x^2-y^2}d_{z^2}\right>$ $\left<d_{xz}d_{xz}|d_{xz}d_{z^2}\right>$ $\left<d_{xz}d_{xz}|d_{z^2}d_{z^2}\right>$ $\left<d_{xz}d_{xz}|sd_{yz}\right>$ $\left<d_{xz}d_{xz}|p_xd_{yz}\right>$ 
$\left<d_{xz}d_{xz}|p_yd_{yz}\right>$ $\left<d_{xz}d_{xz}|p_zd_{yz}\right>$ $\left<d_{xz}d_{xz}|d_{x^2-y^2}d_{yz}\right>$ $\left<d_{xz}d_{xz}|d_{xz}d_{yz}\right>$ $\left<d_{xz}d_{xz}|d_{z^2}d_{yz}\right>$ $\left<d_{xz}d_{xz}|d_{yz}d_{yz}\right>$ $\left<d_{xz}d_{xz}|sd_{xy}\right>$ $\left<d_{xz}d_{xz}|p_xd_{xy}\right>$ $\left<d_{xz}d_{xz}|p_yd_{xy}\right>$ $\left<d_{xz}d_{xz}|p_zd_{xy}\right>$ 
$\left<d_{xz}d_{xz}|d_{x^2-y^2}d_{xy}\right>$ $\left<d_{xz}d_{xz}|d_{xz}d_{xy}\right>$ $\left<d_{xz}d_{xz}|d_{z^2}d_{xy}\right>$ $\left<d_{xz}d_{xz}|d_{yz}d_{xy}\right>$ $\left<d_{xz}d_{xz}|d_{xy}d_{xy}\right>$ $\left<sd_{z^2}|ss\right>$ $\left<sd_{z^2}|sp_x\right>$ $\left<sd_{z^2}|p_xp_x\right>$ $\left<sd_{z^2}|sp_y\right>$ $\left<sd_{z^2}|p_xp_y\right>$ 
$\left<sd_{z^2}|p_yp_y\right>$ $\left<sd_{z^2}|sp_z\right>$ $\left<sd_{z^2}|p_xp_z\right>$ $\left<sd_{z^2}|p_yp_z\right>$ $\left<sd_{z^2}|p_zp_z\right>$ $\left<sd_{z^2}|sd_{x^2-y^2}\right>$ $\left<sd_{z^2}|p_xd_{x^2-y^2}\right>$ $\left<sd_{z^2}|p_yd_{x^2-y^2}\right>$ $\left<sd_{z^2}|p_zd_{x^2-y^2}\right>$ $\left<sd_{z^2}|d_{x^2-y^2}d_{x^2-y^2}\right>$ 
$\left<sd_{z^2}|sd_{xz}\right>$ $\left<sd_{z^2}|p_xd_{xz}\right>$ $\left<sd_{z^2}|p_yd_{xz}\right>$ $\left<sd_{z^2}|p_zd_{xz}\right>$ $\left<sd_{z^2}|d_{x^2-y^2}d_{xz}\right>$ $\left<sd_{z^2}|d_{xz}d_{xz}\right>$ $\left<sd_{z^2}|sd_{z^2}\right>$ $\left<sd_{z^2}|p_xd_{z^2}\right>$ $\left<sd_{z^2}|p_yd_{z^2}\right>$ $\left<sd_{z^2}|p_zd_{z^2}\right>$ 
$\left<sd_{z^2}|d_{x^2-y^2}d_{z^2}\right>$ $\left<sd_{z^2}|d_{xz}d_{z^2}\right>$ $\left<sd_{z^2}|d_{z^2}d_{z^2}\right>$ $\left<sd_{z^2}|sd_{yz}\right>$ $\left<sd_{z^2}|p_xd_{yz}\right>$ $\left<sd_{z^2}|p_yd_{yz}\right>$ $\left<sd_{z^2}|p_zd_{yz}\right>$ $\left<sd_{z^2}|d_{x^2-y^2}d_{yz}\right>$ $\left<sd_{z^2}|d_{xz}d_{yz}\right>$ $\left<sd_{z^2}|d_{z^2}d_{yz}\right>$ 
$\left<sd_{z^2}|d_{yz}d_{yz}\right>$ $\left<sd_{z^2}|sd_{xy}\right>$ $\left<sd_{z^2}|p_xd_{xy}\right>$ $\left<sd_{z^2}|p_yd_{xy}\right>$ $\left<sd_{z^2}|p_zd_{xy}\right>$ $\left<sd_{z^2}|d_{x^2-y^2}d_{xy}\right>$ $\left<sd_{z^2}|d_{xz}d_{xy}\right>$ $\left<sd_{z^2}|d_{z^2}d_{xy}\right>$ $\left<sd_{z^2}|d_{yz}d_{xy}\right>$ $\left<sd_{z^2}|d_{xy}d_{xy}\right>$ 
$\left<p_xd_{z^2}|ss\right>$ $\left<p_xd_{z^2}|sp_x\right>$ $\left<p_xd_{z^2}|p_xp_x\right>$ $\left<p_xd_{z^2}|sp_y\right>$ $\left<p_xd_{z^2}|p_xp_y\right>$ $\left<p_xd_{z^2}|p_yp_y\right>$ $\left<p_xd_{z^2}|sp_z\right>$ $\left<p_xd_{z^2}|p_xp_z\right>$ $\left<p_xd_{z^2}|p_yp_z\right>$ $\left<p_xd_{z^2}|p_zp_z\right>$ 
$\left<p_xd_{z^2}|sd_{x^2-y^2}\right>$ $\left<p_xd_{z^2}|p_xd_{x^2-y^2}\right>$ $\left<p_xd_{z^2}|p_yd_{x^2-y^2}\right>$ $\left<p_xd_{z^2}|p_zd_{x^2-y^2}\right>$ $\left<p_xd_{z^2}|d_{x^2-y^2}d_{x^2-y^2}\right>$ $\left<p_xd_{z^2}|sd_{xz}\right>$ $\left<p_xd_{z^2}|p_xd_{xz}\right>$ $\left<p_xd_{z^2}|p_yd_{xz}\right>$ $\left<p_xd_{z^2}|p_zd_{xz}\right>$ $\left<p_xd_{z^2}|d_{x^2-y^2}d_{xz}\right>$ 
$\left<p_xd_{z^2}|d_{xz}d_{xz}\right>$ $\left<p_xd_{z^2}|sd_{z^2}\right>$ $\left<p_xd_{z^2}|p_xd_{z^2}\right>$ $\left<p_xd_{z^2}|p_yd_{z^2}\right>$ $\left<p_xd_{z^2}|p_zd_{z^2}\right>$ $\left<p_xd_{z^2}|d_{x^2-y^2}d_{z^2}\right>$ $\left<p_xd_{z^2}|d_{xz}d_{z^2}\right>$ $\left<p_xd_{z^2}|d_{z^2}d_{z^2}\right>$ $\left<p_xd_{z^2}|sd_{yz}\right>$ $\left<p_xd_{z^2}|p_xd_{yz}\right>$ 
$\left<p_xd_{z^2}|p_yd_{yz}\right>$ $\left<p_xd_{z^2}|p_zd_{yz}\right>$ $\left<p_xd_{z^2}|d_{x^2-y^2}d_{yz}\right>$ $\left<p_xd_{z^2}|d_{xz}d_{yz}\right>$ $\left<p_xd_{z^2}|d_{z^2}d_{yz}\right>$ $\left<p_xd_{z^2}|d_{yz}d_{yz}\right>$ $\left<p_xd_{z^2}|sd_{xy}\right>$ $\left<p_xd_{z^2}|p_xd_{xy}\right>$ $\left<p_xd_{z^2}|p_yd_{xy}\right>$ $\left<p_xd_{z^2}|p_zd_{xy}\right>$ 
$\left<p_xd_{z^2}|d_{x^2-y^2}d_{xy}\right>$ $\left<p_xd_{z^2}|d_{xz}d_{xy}\right>$ $\left<p_xd_{z^2}|d_{z^2}d_{xy}\right>$ $\left<p_xd_{z^2}|d_{yz}d_{xy}\right>$ $\left<p_xd_{z^2}|d_{xy}d_{xy}\right>$ $\left<p_yd_{z^2}|ss\right>$ $\left<p_yd_{z^2}|sp_x\right>$ $\left<p_yd_{z^2}|p_xp_x\right>$ $\left<p_yd_{z^2}|sp_y\right>$ $\left<p_yd_{z^2}|p_xp_y\right>$ 
$\left<p_yd_{z^2}|p_yp_y\right>$ $\left<p_yd_{z^2}|sp_z\right>$ $\left<p_yd_{z^2}|p_xp_z\right>$ $\left<p_yd_{z^2}|p_yp_z\right>$ $\left<p_yd_{z^2}|p_zp_z\right>$ $\left<p_yd_{z^2}|sd_{x^2-y^2}\right>$ $\left<p_yd_{z^2}|p_xd_{x^2-y^2}\right>$ $\left<p_yd_{z^2}|p_yd_{x^2-y^2}\right>$ $\left<p_yd_{z^2}|p_zd_{x^2-y^2}\right>$ $\left<p_yd_{z^2}|d_{x^2-y^2}d_{x^2-y^2}\right>$ 
$\left<p_yd_{z^2}|sd_{xz}\right>$ $\left<p_yd_{z^2}|p_xd_{xz}\right>$ $\left<p_yd_{z^2}|p_yd_{xz}\right>$ $\left<p_yd_{z^2}|p_zd_{xz}\right>$ $\left<p_yd_{z^2}|d_{x^2-y^2}d_{xz}\right>$ $\left<p_yd_{z^2}|d_{xz}d_{xz}\right>$ $\left<p_yd_{z^2}|sd_{z^2}\right>$ $\left<p_yd_{z^2}|p_xd_{z^2}\right>$ $\left<p_yd_{z^2}|p_yd_{z^2}\right>$ $\left<p_yd_{z^2}|p_zd_{z^2}\right>$ 
$\left<p_yd_{z^2}|d_{x^2-y^2}d_{z^2}\right>$ $\left<p_yd_{z^2}|d_{xz}d_{z^2}\right>$ $\left<p_yd_{z^2}|d_{z^2}d_{z^2}\right>$ $\left<p_yd_{z^2}|sd_{yz}\right>$ $\left<p_yd_{z^2}|p_xd_{yz}\right>$ $\left<p_yd_{z^2}|p_yd_{yz}\right>$ $\left<p_yd_{z^2}|p_zd_{yz}\right>$ $\left<p_yd_{z^2}|d_{x^2-y^2}d_{yz}\right>$ $\left<p_yd_{z^2}|d_{xz}d_{yz}\right>$ $\left<p_yd_{z^2}|d_{z^2}d_{yz}\right>$ 
$\left<p_yd_{z^2}|d_{yz}d_{yz}\right>$ $\left<p_yd_{z^2}|sd_{xy}\right>$ $\left<p_yd_{z^2}|p_xd_{xy}\right>$ $\left<p_yd_{z^2}|p_yd_{xy}\right>$ $\left<p_yd_{z^2}|p_zd_{xy}\right>$ $\left<p_yd_{z^2}|d_{x^2-y^2}d_{xy}\right>$ $\left<p_yd_{z^2}|d_{xz}d_{xy}\right>$ $\left<p_yd_{z^2}|d_{z^2}d_{xy}\right>$ $\left<p_yd_{z^2}|d_{yz}d_{xy}\right>$ $\left<p_yd_{z^2}|d_{xy}d_{xy}\right>$ 
$\left<p_zd_{z^2}|ss\right>$ $\left<p_zd_{z^2}|sp_x\right>$ $\left<p_zd_{z^2}|p_xp_x\right>$ $\left<p_zd_{z^2}|sp_y\right>$ $\left<p_zd_{z^2}|p_xp_y\right>$ $\left<p_zd_{z^2}|p_yp_y\right>$ $\left<p_zd_{z^2}|sp_z\right>$ $\left<p_zd_{z^2}|p_xp_z\right>$ $\left<p_zd_{z^2}|p_yp_z\right>$ $\left<p_zd_{z^2}|p_zp_z\right>$ 
$\left<p_zd_{z^2}|sd_{x^2-y^2}\right>$ $\left<p_zd_{z^2}|p_xd_{x^2-y^2}\right>$ $\left<p_zd_{z^2}|p_yd_{x^2-y^2}\right>$ $\left<p_zd_{z^2}|p_zd_{x^2-y^2}\right>$ $\left<p_zd_{z^2}|d_{x^2-y^2}d_{x^2-y^2}\right>$ $\left<p_zd_{z^2}|sd_{xz}\right>$ $\left<p_zd_{z^2}|p_xd_{xz}\right>$ $\left<p_zd_{z^2}|p_yd_{xz}\right>$ $\left<p_zd_{z^2}|p_zd_{xz}\right>$ $\left<p_zd_{z^2}|d_{x^2-y^2}d_{xz}\right>$ 
$\left<p_zd_{z^2}|d_{xz}d_{xz}\right>$ $\left<p_zd_{z^2}|sd_{z^2}\right>$ $\left<p_zd_{z^2}|p_xd_{z^2}\right>$ $\left<p_zd_{z^2}|p_yd_{z^2}\right>$ $\left<p_zd_{z^2}|p_zd_{z^2}\right>$ $\left<p_zd_{z^2}|d_{x^2-y^2}d_{z^2}\right>$ $\left<p_zd_{z^2}|d_{xz}d_{z^2}\right>$ $\left<p_zd_{z^2}|d_{z^2}d_{z^2}\right>$ $\left<p_zd_{z^2}|sd_{yz}\right>$ $\left<p_zd_{z^2}|p_xd_{yz}\right>$ 
$\left<p_zd_{z^2}|p_yd_{yz}\right>$ $\left<p_zd_{z^2}|p_zd_{yz}\right>$ $\left<p_zd_{z^2}|d_{x^2-y^2}d_{yz}\right>$ $\left<p_zd_{z^2}|d_{xz}d_{yz}\right>$ $\left<p_zd_{z^2}|d_{z^2}d_{yz}\right>$ $\left<p_zd_{z^2}|d_{yz}d_{yz}\right>$ $\left<p_zd_{z^2}|sd_{xy}\right>$ $\left<p_zd_{z^2}|p_xd_{xy}\right>$ $\left<p_zd_{z^2}|p_yd_{xy}\right>$ $\left<p_zd_{z^2}|p_zd_{xy}\right>$ 
$\left<p_zd_{z^2}|d_{x^2-y^2}d_{xy}\right>$ $\left<p_zd_{z^2}|d_{xz}d_{xy}\right>$ $\left<p_zd_{z^2}|d_{z^2}d_{xy}\right>$ $\left<p_zd_{z^2}|d_{yz}d_{xy}\right>$ $\left<p_zd_{z^2}|d_{xy}d_{xy}\right>$ $\left<d_{x^2-y^2}d_{z^2}|ss\right>$ $\left<d_{x^2-y^2}d_{z^2}|sp_x\right>$ $\left<d_{x^2-y^2}d_{z^2}|p_xp_x\right>$ $\left<d_{x^2-y^2}d_{z^2}|sp_y\right>$ $\left<d_{x^2-y^2}d_{z^2}|p_xp_y\right>$ 
$\left<d_{x^2-y^2}d_{z^2}|p_yp_y\right>$ $\left<d_{x^2-y^2}d_{z^2}|sp_z\right>$ $\left<d_{x^2-y^2}d_{z^2}|p_xp_z\right>$ $\left<d_{x^2-y^2}d_{z^2}|p_yp_z\right>$ $\left<d_{x^2-y^2}d_{z^2}|p_zp_z\right>$ $\left<d_{x^2-y^2}d_{z^2}|sd_{x^2-y^2}\right>$ $\left<d_{x^2-y^2}d_{z^2}|p_xd_{x^2-y^2}\right>$ $\left<d_{x^2-y^2}d_{z^2}|p_yd_{x^2-y^2}\right>$ $\left<d_{x^2-y^2}d_{z^2}|p_zd_{x^2-y^2}\right>$ $\left<d_{x^2-y^2}d_{z^2}|d_{x^2-y^2}d_{x^2-y^2}\right>$ 
$\left<d_{x^2-y^2}d_{z^2}|sd_{xz}\right>$ $\left<d_{x^2-y^2}d_{z^2}|p_xd_{xz}\right>$ $\left<d_{x^2-y^2}d_{z^2}|p_yd_{xz}\right>$ $\left<d_{x^2-y^2}d_{z^2}|p_zd_{xz}\right>$ $\left<d_{x^2-y^2}d_{z^2}|d_{x^2-y^2}d_{xz}\right>$ $\left<d_{x^2-y^2}d_{z^2}|d_{xz}d_{xz}\right>$ $\left<d_{x^2-y^2}d_{z^2}|sd_{z^2}\right>$ $\left<d_{x^2-y^2}d_{z^2}|p_xd_{z^2}\right>$ $\left<d_{x^2-y^2}d_{z^2}|p_yd_{z^2}\right>$ $\left<d_{x^2-y^2}d_{z^2}|p_zd_{z^2}\right>$ 
$\left<d_{x^2-y^2}d_{z^2}|d_{x^2-y^2}d_{z^2}\right>$ $\left<d_{x^2-y^2}d_{z^2}|d_{xz}d_{z^2}\right>$ $\left<d_{x^2-y^2}d_{z^2}|d_{z^2}d_{z^2}\right>$ $\left<d_{x^2-y^2}d_{z^2}|sd_{yz}\right>$ $\left<d_{x^2-y^2}d_{z^2}|p_xd_{yz}\right>$ $\left<d_{x^2-y^2}d_{z^2}|p_yd_{yz}\right>$ $\left<d_{x^2-y^2}d_{z^2}|p_zd_{yz}\right>$ $\left<d_{x^2-y^2}d_{z^2}|d_{x^2-y^2}d_{yz}\right>$ $\left<d_{x^2-y^2}d_{z^2}|d_{xz}d_{yz}\right>$ $\left<d_{x^2-y^2}d_{z^2}|d_{z^2}d_{yz}\right>$ 
$\left<d_{x^2-y^2}d_{z^2}|d_{yz}d_{yz}\right>$ $\left<d_{x^2-y^2}d_{z^2}|sd_{xy}\right>$ $\left<d_{x^2-y^2}d_{z^2}|p_xd_{xy}\right>$ $\left<d_{x^2-y^2}d_{z^2}|p_yd_{xy}\right>$ $\left<d_{x^2-y^2}d_{z^2}|p_zd_{xy}\right>$ $\left<d_{x^2-y^2}d_{z^2}|d_{x^2-y^2}d_{xy}\right>$ $\left<d_{x^2-y^2}d_{z^2}|d_{xz}d_{xy}\right>$ $\left<d_{x^2-y^2}d_{z^2}|d_{z^2}d_{xy}\right>$ $\left<d_{x^2-y^2}d_{z^2}|d_{yz}d_{xy}\right>$ $\left<d_{x^2-y^2}d_{z^2}|d_{xy}d_{xy}\right>$ 
$\left<d_{xz}d_{z^2}|ss\right>$ $\left<d_{xz}d_{z^2}|sp_x\right>$ $\left<d_{xz}d_{z^2}|p_xp_x\right>$ $\left<d_{xz}d_{z^2}|sp_y\right>$ $\left<d_{xz}d_{z^2}|p_xp_y\right>$ $\left<d_{xz}d_{z^2}|p_yp_y\right>$ $\left<d_{xz}d_{z^2}|sp_z\right>$ $\left<d_{xz}d_{z^2}|p_xp_z\right>$ $\left<d_{xz}d_{z^2}|p_yp_z\right>$ $\left<d_{xz}d_{z^2}|p_zp_z\right>$ 
$\left<d_{xz}d_{z^2}|sd_{x^2-y^2}\right>$ $\left<d_{xz}d_{z^2}|p_xd_{x^2-y^2}\right>$ $\left<d_{xz}d_{z^2}|p_yd_{x^2-y^2}\right>$ $\left<d_{xz}d_{z^2}|p_zd_{x^2-y^2}\right>$ $\left<d_{xz}d_{z^2}|d_{x^2-y^2}d_{x^2-y^2}\right>$ $\left<d_{xz}d_{z^2}|sd_{xz}\right>$ $\left<d_{xz}d_{z^2}|p_xd_{xz}\right>$ $\left<d_{xz}d_{z^2}|p_yd_{xz}\right>$ $\left<d_{xz}d_{z^2}|p_zd_{xz}\right>$ $\left<d_{xz}d_{z^2}|d_{x^2-y^2}d_{xz}\right>$ 
$\left<d_{xz}d_{z^2}|d_{xz}d_{xz}\right>$ $\left<d_{xz}d_{z^2}|sd_{z^2}\right>$ $\left<d_{xz}d_{z^2}|p_xd_{z^2}\right>$ $\left<d_{xz}d_{z^2}|p_yd_{z^2}\right>$ $\left<d_{xz}d_{z^2}|p_zd_{z^2}\right>$ $\left<d_{xz}d_{z^2}|d_{x^2-y^2}d_{z^2}\right>$ $\left<d_{xz}d_{z^2}|d_{xz}d_{z^2}\right>$ $\left<d_{xz}d_{z^2}|d_{z^2}d_{z^2}\right>$ $\left<d_{xz}d_{z^2}|sd_{yz}\right>$ $\left<d_{xz}d_{z^2}|p_xd_{yz}\right>$ 
$\left<d_{xz}d_{z^2}|p_yd_{yz}\right>$ $\left<d_{xz}d_{z^2}|p_zd_{yz}\right>$ $\left<d_{xz}d_{z^2}|d_{x^2-y^2}d_{yz}\right>$ $\left<d_{xz}d_{z^2}|d_{xz}d_{yz}\right>$ $\left<d_{xz}d_{z^2}|d_{z^2}d_{yz}\right>$ $\left<d_{xz}d_{z^2}|d_{yz}d_{yz}\right>$ $\left<d_{xz}d_{z^2}|sd_{xy}\right>$ $\left<d_{xz}d_{z^2}|p_xd_{xy}\right>$ $\left<d_{xz}d_{z^2}|p_yd_{xy}\right>$ $\left<d_{xz}d_{z^2}|p_zd_{xy}\right>$ 
$\left<d_{xz}d_{z^2}|d_{x^2-y^2}d_{xy}\right>$ $\left<d_{xz}d_{z^2}|d_{xz}d_{xy}\right>$ $\left<d_{xz}d_{z^2}|d_{z^2}d_{xy}\right>$ $\left<d_{xz}d_{z^2}|d_{yz}d_{xy}\right>$ $\left<d_{xz}d_{z^2}|d_{xy}d_{xy}\right>$ $\left<d_{z^2}d_{z^2}|ss\right>$ $\left<d_{z^2}d_{z^2}|sp_x\right>$ $\left<d_{z^2}d_{z^2}|p_xp_x\right>$ $\left<d_{z^2}d_{z^2}|sp_y\right>$ $\left<d_{z^2}d_{z^2}|p_xp_y\right>$ 
$\left<d_{z^2}d_{z^2}|p_yp_y\right>$ $\left<d_{z^2}d_{z^2}|sp_z\right>$ $\left<d_{z^2}d_{z^2}|p_xp_z\right>$ $\left<d_{z^2}d_{z^2}|p_yp_z\right>$ $\left<d_{z^2}d_{z^2}|p_zp_z\right>$ $\left<d_{z^2}d_{z^2}|sd_{x^2-y^2}\right>$ $\left<d_{z^2}d_{z^2}|p_xd_{x^2-y^2}\right>$ $\left<d_{z^2}d_{z^2}|p_yd_{x^2-y^2}\right>$ $\left<d_{z^2}d_{z^2}|p_zd_{x^2-y^2}\right>$ $\left<d_{z^2}d_{z^2}|d_{x^2-y^2}d_{x^2-y^2}\right>$ 
$\left<d_{z^2}d_{z^2}|sd_{xz}\right>$ $\left<d_{z^2}d_{z^2}|p_xd_{xz}\right>$ $\left<d_{z^2}d_{z^2}|p_yd_{xz}\right>$ $\left<d_{z^2}d_{z^2}|p_zd_{xz}\right>$ $\left<d_{z^2}d_{z^2}|d_{x^2-y^2}d_{xz}\right>$ $\left<d_{z^2}d_{z^2}|d_{xz}d_{xz}\right>$ $\left<d_{z^2}d_{z^2}|sd_{z^2}\right>$ $\left<d_{z^2}d_{z^2}|p_xd_{z^2}\right>$ $\left<d_{z^2}d_{z^2}|p_yd_{z^2}\right>$ $\left<d_{z^2}d_{z^2}|p_zd_{z^2}\right>$ 
$\left<d_{z^2}d_{z^2}|d_{x^2-y^2}d_{z^2}\right>$ $\left<d_{z^2}d_{z^2}|d_{xz}d_{z^2}\right>$ $\left<d_{z^2}d_{z^2}|d_{z^2}d_{z^2}\right>$ $\left<d_{z^2}d_{z^2}|sd_{yz}\right>$ $\left<d_{z^2}d_{z^2}|p_xd_{yz}\right>$ $\left<d_{z^2}d_{z^2}|p_yd_{yz}\right>$ $\left<d_{z^2}d_{z^2}|p_zd_{yz}\right>$ $\left<d_{z^2}d_{z^2}|d_{x^2-y^2}d_{yz}\right>$ $\left<d_{z^2}d_{z^2}|d_{xz}d_{yz}\right>$ $\left<d_{z^2}d_{z^2}|d_{z^2}d_{yz}\right>$ 
$\left<d_{z^2}d_{z^2}|d_{yz}d_{yz}\right>$ $\left<d_{z^2}d_{z^2}|sd_{xy}\right>$ $\left<d_{z^2}d_{z^2}|p_xd_{xy}\right>$ $\left<d_{z^2}d_{z^2}|p_yd_{xy}\right>$ $\left<d_{z^2}d_{z^2}|p_zd_{xy}\right>$ $\left<d_{z^2}d_{z^2}|d_{x^2-y^2}d_{xy}\right>$ $\left<d_{z^2}d_{z^2}|d_{xz}d_{xy}\right>$ $\left<d_{z^2}d_{z^2}|d_{z^2}d_{xy}\right>$ $\left<d_{z^2}d_{z^2}|d_{yz}d_{xy}\right>$ $\left<d_{z^2}d_{z^2}|d_{xy}d_{xy}\right>$ 
$\left<sd_{yz}|ss\right>$ $\left<sd_{yz}|sp_x\right>$ $\left<sd_{yz}|p_xp_x\right>$ $\left<sd_{yz}|sp_y\right>$ $\left<sd_{yz}|p_xp_y\right>$ $\left<sd_{yz}|p_yp_y\right>$ $\left<sd_{yz}|sp_z\right>$ $\left<sd_{yz}|p_xp_z\right>$ $\left<sd_{yz}|p_yp_z\right>$ $\left<sd_{yz}|p_zp_z\right>$ 
$\left<sd_{yz}|sd_{x^2-y^2}\right>$ $\left<sd_{yz}|p_xd_{x^2-y^2}\right>$ $\left<sd_{yz}|p_yd_{x^2-y^2}\right>$ $\left<sd_{yz}|p_zd_{x^2-y^2}\right>$ $\left<sd_{yz}|d_{x^2-y^2}d_{x^2-y^2}\right>$ $\left<sd_{yz}|sd_{xz}\right>$ $\left<sd_{yz}|p_xd_{xz}\right>$ $\left<sd_{yz}|p_yd_{xz}\right>$ $\left<sd_{yz}|p_zd_{xz}\right>$ $\left<sd_{yz}|d_{x^2-y^2}d_{xz}\right>$ 
$\left<sd_{yz}|d_{xz}d_{xz}\right>$ $\left<sd_{yz}|sd_{z^2}\right>$ $\left<sd_{yz}|p_xd_{z^2}\right>$ $\left<sd_{yz}|p_yd_{z^2}\right>$ $\left<sd_{yz}|p_zd_{z^2}\right>$ $\left<sd_{yz}|d_{x^2-y^2}d_{z^2}\right>$ $\left<sd_{yz}|d_{xz}d_{z^2}\right>$ $\left<sd_{yz}|d_{z^2}d_{z^2}\right>$ $\left<sd_{yz}|sd_{yz}\right>$ $\left<sd_{yz}|p_xd_{yz}\right>$ 
$\left<sd_{yz}|p_yd_{yz}\right>$ $\left<sd_{yz}|p_zd_{yz}\right>$ $\left<sd_{yz}|d_{x^2-y^2}d_{yz}\right>$ $\left<sd_{yz}|d_{xz}d_{yz}\right>$ $\left<sd_{yz}|d_{z^2}d_{yz}\right>$ $\left<sd_{yz}|d_{yz}d_{yz}\right>$ $\left<sd_{yz}|sd_{xy}\right>$ $\left<sd_{yz}|p_xd_{xy}\right>$ $\left<sd_{yz}|p_yd_{xy}\right>$ $\left<sd_{yz}|p_zd_{xy}\right>$ 
$\left<sd_{yz}|d_{x^2-y^2}d_{xy}\right>$ $\left<sd_{yz}|d_{xz}d_{xy}\right>$ $\left<sd_{yz}|d_{z^2}d_{xy}\right>$ $\left<sd_{yz}|d_{yz}d_{xy}\right>$ $\left<sd_{yz}|d_{xy}d_{xy}\right>$ $\left<p_xd_{yz}|ss\right>$ $\left<p_xd_{yz}|sp_x\right>$ $\left<p_xd_{yz}|p_xp_x\right>$ $\left<p_xd_{yz}|sp_y\right>$ $\left<p_xd_{yz}|p_xp_y\right>$ 
$\left<p_xd_{yz}|p_yp_y\right>$ $\left<p_xd_{yz}|sp_z\right>$ $\left<p_xd_{yz}|p_xp_z\right>$ $\left<p_xd_{yz}|p_yp_z\right>$ $\left<p_xd_{yz}|p_zp_z\right>$ $\left<p_xd_{yz}|sd_{x^2-y^2}\right>$ $\left<p_xd_{yz}|p_xd_{x^2-y^2}\right>$ $\left<p_xd_{yz}|p_yd_{x^2-y^2}\right>$ $\left<p_xd_{yz}|p_zd_{x^2-y^2}\right>$ $\left<p_xd_{yz}|d_{x^2-y^2}d_{x^2-y^2}\right>$ 
$\left<p_xd_{yz}|sd_{xz}\right>$ $\left<p_xd_{yz}|p_xd_{xz}\right>$ $\left<p_xd_{yz}|p_yd_{xz}\right>$ $\left<p_xd_{yz}|p_zd_{xz}\right>$ $\left<p_xd_{yz}|d_{x^2-y^2}d_{xz}\right>$ $\left<p_xd_{yz}|d_{xz}d_{xz}\right>$ $\left<p_xd_{yz}|sd_{z^2}\right>$ $\left<p_xd_{yz}|p_xd_{z^2}\right>$ $\left<p_xd_{yz}|p_yd_{z^2}\right>$ $\left<p_xd_{yz}|p_zd_{z^2}\right>$ 
$\left<p_xd_{yz}|d_{x^2-y^2}d_{z^2}\right>$ $\left<p_xd_{yz}|d_{xz}d_{z^2}\right>$ $\left<p_xd_{yz}|d_{z^2}d_{z^2}\right>$ $\left<p_xd_{yz}|sd_{yz}\right>$ $\left<p_xd_{yz}|p_xd_{yz}\right>$ $\left<p_xd_{yz}|p_yd_{yz}\right>$ $\left<p_xd_{yz}|p_zd_{yz}\right>$ $\left<p_xd_{yz}|d_{x^2-y^2}d_{yz}\right>$ $\left<p_xd_{yz}|d_{xz}d_{yz}\right>$ $\left<p_xd_{yz}|d_{z^2}d_{yz}\right>$ 
$\left<p_xd_{yz}|d_{yz}d_{yz}\right>$ $\left<p_xd_{yz}|sd_{xy}\right>$ $\left<p_xd_{yz}|p_xd_{xy}\right>$ $\left<p_xd_{yz}|p_yd_{xy}\right>$ $\left<p_xd_{yz}|p_zd_{xy}\right>$ $\left<p_xd_{yz}|d_{x^2-y^2}d_{xy}\right>$ $\left<p_xd_{yz}|d_{xz}d_{xy}\right>$ $\left<p_xd_{yz}|d_{z^2}d_{xy}\right>$ $\left<p_xd_{yz}|d_{yz}d_{xy}\right>$ $\left<p_xd_{yz}|d_{xy}d_{xy}\right>$ 
$\left<p_yd_{yz}|ss\right>$ $\left<p_yd_{yz}|sp_x\right>$ $\left<p_yd_{yz}|p_xp_x\right>$ $\left<p_yd_{yz}|sp_y\right>$ $\left<p_yd_{yz}|p_xp_y\right>$ $\left<p_yd_{yz}|p_yp_y\right>$ $\left<p_yd_{yz}|sp_z\right>$ $\left<p_yd_{yz}|p_xp_z\right>$ $\left<p_yd_{yz}|p_yp_z\right>$ $\left<p_yd_{yz}|p_zp_z\right>$ 
$\left<p_yd_{yz}|sd_{x^2-y^2}\right>$ $\left<p_yd_{yz}|p_xd_{x^2-y^2}\right>$ $\left<p_yd_{yz}|p_yd_{x^2-y^2}\right>$ $\left<p_yd_{yz}|p_zd_{x^2-y^2}\right>$ $\left<p_yd_{yz}|d_{x^2-y^2}d_{x^2-y^2}\right>$ $\left<p_yd_{yz}|sd_{xz}\right>$ $\left<p_yd_{yz}|p_xd_{xz}\right>$ $\left<p_yd_{yz}|p_yd_{xz}\right>$ $\left<p_yd_{yz}|p_zd_{xz}\right>$ $\left<p_yd_{yz}|d_{x^2-y^2}d_{xz}\right>$ 
$\left<p_yd_{yz}|d_{xz}d_{xz}\right>$ $\left<p_yd_{yz}|sd_{z^2}\right>$ $\left<p_yd_{yz}|p_xd_{z^2}\right>$ $\left<p_yd_{yz}|p_yd_{z^2}\right>$ $\left<p_yd_{yz}|p_zd_{z^2}\right>$ $\left<p_yd_{yz}|d_{x^2-y^2}d_{z^2}\right>$ $\left<p_yd_{yz}|d_{xz}d_{z^2}\right>$ $\left<p_yd_{yz}|d_{z^2}d_{z^2}\right>$ $\left<p_yd_{yz}|sd_{yz}\right>$ $\left<p_yd_{yz}|p_xd_{yz}\right>$ 
$\left<p_yd_{yz}|p_yd_{yz}\right>$ $\left<p_yd_{yz}|p_zd_{yz}\right>$ $\left<p_yd_{yz}|d_{x^2-y^2}d_{yz}\right>$ $\left<p_yd_{yz}|d_{xz}d_{yz}\right>$ $\left<p_yd_{yz}|d_{z^2}d_{yz}\right>$ $\left<p_yd_{yz}|d_{yz}d_{yz}\right>$ $\left<p_yd_{yz}|sd_{xy}\right>$ $\left<p_yd_{yz}|p_xd_{xy}\right>$ $\left<p_yd_{yz}|p_yd_{xy}\right>$ $\left<p_yd_{yz}|p_zd_{xy}\right>$ 
$\left<p_yd_{yz}|d_{x^2-y^2}d_{xy}\right>$ $\left<p_yd_{yz}|d_{xz}d_{xy}\right>$ $\left<p_yd_{yz}|d_{z^2}d_{xy}\right>$ $\left<p_yd_{yz}|d_{yz}d_{xy}\right>$ $\left<p_yd_{yz}|d_{xy}d_{xy}\right>$ $\left<p_zd_{yz}|ss\right>$ $\left<p_zd_{yz}|sp_x\right>$ $\left<p_zd_{yz}|p_xp_x\right>$ $\left<p_zd_{yz}|sp_y\right>$ $\left<p_zd_{yz}|p_xp_y\right>$ 
$\left<p_zd_{yz}|p_yp_y\right>$ $\left<p_zd_{yz}|sp_z\right>$ $\left<p_zd_{yz}|p_xp_z\right>$ $\left<p_zd_{yz}|p_yp_z\right>$ $\left<p_zd_{yz}|p_zp_z\right>$ $\left<p_zd_{yz}|sd_{x^2-y^2}\right>$ $\left<p_zd_{yz}|p_xd_{x^2-y^2}\right>$ $\left<p_zd_{yz}|p_yd_{x^2-y^2}\right>$ $\left<p_zd_{yz}|p_zd_{x^2-y^2}\right>$ $\left<p_zd_{yz}|d_{x^2-y^2}d_{x^2-y^2}\right>$ 
$\left<p_zd_{yz}|sd_{xz}\right>$ $\left<p_zd_{yz}|p_xd_{xz}\right>$ $\left<p_zd_{yz}|p_yd_{xz}\right>$ $\left<p_zd_{yz}|p_zd_{xz}\right>$ $\left<p_zd_{yz}|d_{x^2-y^2}d_{xz}\right>$ $\left<p_zd_{yz}|d_{xz}d_{xz}\right>$ $\left<p_zd_{yz}|sd_{z^2}\right>$ $\left<p_zd_{yz}|p_xd_{z^2}\right>$ $\left<p_zd_{yz}|p_yd_{z^2}\right>$ $\left<p_zd_{yz}|p_zd_{z^2}\right>$ 
$\left<p_zd_{yz}|d_{x^2-y^2}d_{z^2}\right>$ $\left<p_zd_{yz}|d_{xz}d_{z^2}\right>$ $\left<p_zd_{yz}|d_{z^2}d_{z^2}\right>$ $\left<p_zd_{yz}|sd_{yz}\right>$ $\left<p_zd_{yz}|p_xd_{yz}\right>$ $\left<p_zd_{yz}|p_yd_{yz}\right>$ $\left<p_zd_{yz}|p_zd_{yz}\right>$ $\left<p_zd_{yz}|d_{x^2-y^2}d_{yz}\right>$ $\left<p_zd_{yz}|d_{xz}d_{yz}\right>$ $\left<p_zd_{yz}|d_{z^2}d_{yz}\right>$ 
$\left<p_zd_{yz}|d_{yz}d_{yz}\right>$ $\left<p_zd_{yz}|sd_{xy}\right>$ $\left<p_zd_{yz}|p_xd_{xy}\right>$ $\left<p_zd_{yz}|p_yd_{xy}\right>$ $\left<p_zd_{yz}|p_zd_{xy}\right>$ $\left<p_zd_{yz}|d_{x^2-y^2}d_{xy}\right>$ $\left<p_zd_{yz}|d_{xz}d_{xy}\right>$ $\left<p_zd_{yz}|d_{z^2}d_{xy}\right>$ $\left<p_zd_{yz}|d_{yz}d_{xy}\right>$ $\left<p_zd_{yz}|d_{xy}d_{xy}\right>$ 
$\left<d_{x^2-y^2}d_{yz}|ss\right>$ $\left<d_{x^2-y^2}d_{yz}|sp_x\right>$ $\left<d_{x^2-y^2}d_{yz}|p_xp_x\right>$ $\left<d_{x^2-y^2}d_{yz}|sp_y\right>$ $\left<d_{x^2-y^2}d_{yz}|p_xp_y\right>$ $\left<d_{x^2-y^2}d_{yz}|p_yp_y\right>$ $\left<d_{x^2-y^2}d_{yz}|sp_z\right>$ $\left<d_{x^2-y^2}d_{yz}|p_xp_z\right>$ $\left<d_{x^2-y^2}d_{yz}|p_yp_z\right>$ $\left<d_{x^2-y^2}d_{yz}|p_zp_z\right>$ 
$\left<d_{x^2-y^2}d_{yz}|sd_{x^2-y^2}\right>$ $\left<d_{x^2-y^2}d_{yz}|p_xd_{x^2-y^2}\right>$ $\left<d_{x^2-y^2}d_{yz}|p_yd_{x^2-y^2}\right>$ $\left<d_{x^2-y^2}d_{yz}|p_zd_{x^2-y^2}\right>$ $\left<d_{x^2-y^2}d_{yz}|d_{x^2-y^2}d_{x^2-y^2}\right>$ $\left<d_{x^2-y^2}d_{yz}|sd_{xz}\right>$ $\left<d_{x^2-y^2}d_{yz}|p_xd_{xz}\right>$ $\left<d_{x^2-y^2}d_{yz}|p_yd_{xz}\right>$ $\left<d_{x^2-y^2}d_{yz}|p_zd_{xz}\right>$ $\left<d_{x^2-y^2}d_{yz}|d_{x^2-y^2}d_{xz}\right>$ 
$\left<d_{x^2-y^2}d_{yz}|d_{xz}d_{xz}\right>$ $\left<d_{x^2-y^2}d_{yz}|sd_{z^2}\right>$ $\left<d_{x^2-y^2}d_{yz}|p_xd_{z^2}\right>$ $\left<d_{x^2-y^2}d_{yz}|p_yd_{z^2}\right>$ $\left<d_{x^2-y^2}d_{yz}|p_zd_{z^2}\right>$ $\left<d_{x^2-y^2}d_{yz}|d_{x^2-y^2}d_{z^2}\right>$ $\left<d_{x^2-y^2}d_{yz}|d_{xz}d_{z^2}\right>$ $\left<d_{x^2-y^2}d_{yz}|d_{z^2}d_{z^2}\right>$ $\left<d_{x^2-y^2}d_{yz}|sd_{yz}\right>$ $\left<d_{x^2-y^2}d_{yz}|p_xd_{yz}\right>$ 
$\left<d_{x^2-y^2}d_{yz}|p_yd_{yz}\right>$ $\left<d_{x^2-y^2}d_{yz}|p_zd_{yz}\right>$ $\left<d_{x^2-y^2}d_{yz}|d_{x^2-y^2}d_{yz}\right>$ $\left<d_{x^2-y^2}d_{yz}|d_{xz}d_{yz}\right>$ $\left<d_{x^2-y^2}d_{yz}|d_{z^2}d_{yz}\right>$ $\left<d_{x^2-y^2}d_{yz}|d_{yz}d_{yz}\right>$ $\left<d_{x^2-y^2}d_{yz}|sd_{xy}\right>$ $\left<d_{x^2-y^2}d_{yz}|p_xd_{xy}\right>$ $\left<d_{x^2-y^2}d_{yz}|p_yd_{xy}\right>$ $\left<d_{x^2-y^2}d_{yz}|p_zd_{xy}\right>$ 
$\left<d_{x^2-y^2}d_{yz}|d_{x^2-y^2}d_{xy}\right>$ $\left<d_{x^2-y^2}d_{yz}|d_{xz}d_{xy}\right>$ $\left<d_{x^2-y^2}d_{yz}|d_{z^2}d_{xy}\right>$ $\left<d_{x^2-y^2}d_{yz}|d_{yz}d_{xy}\right>$ $\left<d_{x^2-y^2}d_{yz}|d_{xy}d_{xy}\right>$ $\left<d_{xz}d_{yz}|ss\right>$ $\left<d_{xz}d_{yz}|sp_x\right>$ $\left<d_{xz}d_{yz}|p_xp_x\right>$ $\left<d_{xz}d_{yz}|sp_y\right>$ $\left<d_{xz}d_{yz}|p_xp_y\right>$ 
$\left<d_{xz}d_{yz}|p_yp_y\right>$ $\left<d_{xz}d_{yz}|sp_z\right>$ $\left<d_{xz}d_{yz}|p_xp_z\right>$ $\left<d_{xz}d_{yz}|p_yp_z\right>$ $\left<d_{xz}d_{yz}|p_zp_z\right>$ $\left<d_{xz}d_{yz}|sd_{x^2-y^2}\right>$ $\left<d_{xz}d_{yz}|p_xd_{x^2-y^2}\right>$ $\left<d_{xz}d_{yz}|p_yd_{x^2-y^2}\right>$ $\left<d_{xz}d_{yz}|p_zd_{x^2-y^2}\right>$ $\left<d_{xz}d_{yz}|d_{x^2-y^2}d_{x^2-y^2}\right>$ 
$\left<d_{xz}d_{yz}|sd_{xz}\right>$ $\left<d_{xz}d_{yz}|p_xd_{xz}\right>$ $\left<d_{xz}d_{yz}|p_yd_{xz}\right>$ $\left<d_{xz}d_{yz}|p_zd_{xz}\right>$ $\left<d_{xz}d_{yz}|d_{x^2-y^2}d_{xz}\right>$ $\left<d_{xz}d_{yz}|d_{xz}d_{xz}\right>$ $\left<d_{xz}d_{yz}|sd_{z^2}\right>$ $\left<d_{xz}d_{yz}|p_xd_{z^2}\right>$ $\left<d_{xz}d_{yz}|p_yd_{z^2}\right>$ $\left<d_{xz}d_{yz}|p_zd_{z^2}\right>$ 
$\left<d_{xz}d_{yz}|d_{x^2-y^2}d_{z^2}\right>$ $\left<d_{xz}d_{yz}|d_{xz}d_{z^2}\right>$ $\left<d_{xz}d_{yz}|d_{z^2}d_{z^2}\right>$ $\left<d_{xz}d_{yz}|sd_{yz}\right>$ $\left<d_{xz}d_{yz}|p_xd_{yz}\right>$ $\left<d_{xz}d_{yz}|p_yd_{yz}\right>$ $\left<d_{xz}d_{yz}|p_zd_{yz}\right>$ $\left<d_{xz}d_{yz}|d_{x^2-y^2}d_{yz}\right>$ $\left<d_{xz}d_{yz}|d_{xz}d_{yz}\right>$ $\left<d_{xz}d_{yz}|d_{z^2}d_{yz}\right>$ 
$\left<d_{xz}d_{yz}|d_{yz}d_{yz}\right>$ $\left<d_{xz}d_{yz}|sd_{xy}\right>$ $\left<d_{xz}d_{yz}|p_xd_{xy}\right>$ $\left<d_{xz}d_{yz}|p_yd_{xy}\right>$ $\left<d_{xz}d_{yz}|p_zd_{xy}\right>$ $\left<d_{xz}d_{yz}|d_{x^2-y^2}d_{xy}\right>$ $\left<d_{xz}d_{yz}|d_{xz}d_{xy}\right>$ $\left<d_{xz}d_{yz}|d_{z^2}d_{xy}\right>$ $\left<d_{xz}d_{yz}|d_{yz}d_{xy}\right>$ $\left<d_{xz}d_{yz}|d_{xy}d_{xy}\right>$ 
$\left<d_{z^2}d_{yz}|ss\right>$ $\left<d_{z^2}d_{yz}|sp_x\right>$ $\left<d_{z^2}d_{yz}|p_xp_x\right>$ $\left<d_{z^2}d_{yz}|sp_y\right>$ $\left<d_{z^2}d_{yz}|p_xp_y\right>$ $\left<d_{z^2}d_{yz}|p_yp_y\right>$ $\left<d_{z^2}d_{yz}|sp_z\right>$ $\left<d_{z^2}d_{yz}|p_xp_z\right>$ $\left<d_{z^2}d_{yz}|p_yp_z\right>$ $\left<d_{z^2}d_{yz}|p_zp_z\right>$ 
$\left<d_{z^2}d_{yz}|sd_{x^2-y^2}\right>$ $\left<d_{z^2}d_{yz}|p_xd_{x^2-y^2}\right>$ $\left<d_{z^2}d_{yz}|p_yd_{x^2-y^2}\right>$ $\left<d_{z^2}d_{yz}|p_zd_{x^2-y^2}\right>$ $\left<d_{z^2}d_{yz}|d_{x^2-y^2}d_{x^2-y^2}\right>$ $\left<d_{z^2}d_{yz}|sd_{xz}\right>$ $\left<d_{z^2}d_{yz}|p_xd_{xz}\right>$ $\left<d_{z^2}d_{yz}|p_yd_{xz}\right>$ $\left<d_{z^2}d_{yz}|p_zd_{xz}\right>$ $\left<d_{z^2}d_{yz}|d_{x^2-y^2}d_{xz}\right>$ 
$\left<d_{z^2}d_{yz}|d_{xz}d_{xz}\right>$ $\left<d_{z^2}d_{yz}|sd_{z^2}\right>$ $\left<d_{z^2}d_{yz}|p_xd_{z^2}\right>$ $\left<d_{z^2}d_{yz}|p_yd_{z^2}\right>$ $\left<d_{z^2}d_{yz}|p_zd_{z^2}\right>$ $\left<d_{z^2}d_{yz}|d_{x^2-y^2}d_{z^2}\right>$ $\left<d_{z^2}d_{yz}|d_{xz}d_{z^2}\right>$ $\left<d_{z^2}d_{yz}|d_{z^2}d_{z^2}\right>$ $\left<d_{z^2}d_{yz}|sd_{yz}\right>$ $\left<d_{z^2}d_{yz}|p_xd_{yz}\right>$ 
$\left<d_{z^2}d_{yz}|p_yd_{yz}\right>$ $\left<d_{z^2}d_{yz}|p_zd_{yz}\right>$ $\left<d_{z^2}d_{yz}|d_{x^2-y^2}d_{yz}\right>$ $\left<d_{z^2}d_{yz}|d_{xz}d_{yz}\right>$ $\left<d_{z^2}d_{yz}|d_{z^2}d_{yz}\right>$ $\left<d_{z^2}d_{yz}|d_{yz}d_{yz}\right>$ $\left<d_{z^2}d_{yz}|sd_{xy}\right>$ $\left<d_{z^2}d_{yz}|p_xd_{xy}\right>$ $\left<d_{z^2}d_{yz}|p_yd_{xy}\right>$ $\left<d_{z^2}d_{yz}|p_zd_{xy}\right>$ 
$\left<d_{z^2}d_{yz}|d_{x^2-y^2}d_{xy}\right>$ $\left<d_{z^2}d_{yz}|d_{xz}d_{xy}\right>$ $\left<d_{z^2}d_{yz}|d_{z^2}d_{xy}\right>$ $\left<d_{z^2}d_{yz}|d_{yz}d_{xy}\right>$ $\left<d_{z^2}d_{yz}|d_{xy}d_{xy}\right>$ $\left<d_{yz}d_{yz}|ss\right>$ $\left<d_{yz}d_{yz}|sp_x\right>$ $\left<d_{yz}d_{yz}|p_xp_x\right>$ $\left<d_{yz}d_{yz}|sp_y\right>$ $\left<d_{yz}d_{yz}|p_xp_y\right>$ 
$\left<d_{yz}d_{yz}|p_yp_y\right>$ $\left<d_{yz}d_{yz}|sp_z\right>$ $\left<d_{yz}d_{yz}|p_xp_z\right>$ $\left<d_{yz}d_{yz}|p_yp_z\right>$ $\left<d_{yz}d_{yz}|p_zp_z\right>$ $\left<d_{yz}d_{yz}|sd_{x^2-y^2}\right>$ $\left<d_{yz}d_{yz}|p_xd_{x^2-y^2}\right>$ $\left<d_{yz}d_{yz}|p_yd_{x^2-y^2}\right>$ $\left<d_{yz}d_{yz}|p_zd_{x^2-y^2}\right>$ $\left<d_{yz}d_{yz}|d_{x^2-y^2}d_{x^2-y^2}\right>$ 
$\left<d_{yz}d_{yz}|sd_{xz}\right>$ $\left<d_{yz}d_{yz}|p_xd_{xz}\right>$ $\left<d_{yz}d_{yz}|p_yd_{xz}\right>$ $\left<d_{yz}d_{yz}|p_zd_{xz}\right>$ $\left<d_{yz}d_{yz}|d_{x^2-y^2}d_{xz}\right>$ $\left<d_{yz}d_{yz}|d_{xz}d_{xz}\right>$ $\left<d_{yz}d_{yz}|sd_{z^2}\right>$ $\left<d_{yz}d_{yz}|p_xd_{z^2}\right>$ $\left<d_{yz}d_{yz}|p_yd_{z^2}\right>$ $\left<d_{yz}d_{yz}|p_zd_{z^2}\right>$ 
$\left<d_{yz}d_{yz}|d_{x^2-y^2}d_{z^2}\right>$ $\left<d_{yz}d_{yz}|d_{xz}d_{z^2}\right>$ $\left<d_{yz}d_{yz}|d_{z^2}d_{z^2}\right>$ $\left<d_{yz}d_{yz}|sd_{yz}\right>$ $\left<d_{yz}d_{yz}|p_xd_{yz}\right>$ $\left<d_{yz}d_{yz}|p_yd_{yz}\right>$ $\left<d_{yz}d_{yz}|p_zd_{yz}\right>$ $\left<d_{yz}d_{yz}|d_{x^2-y^2}d_{yz}\right>$ $\left<d_{yz}d_{yz}|d_{xz}d_{yz}\right>$ $\left<d_{yz}d_{yz}|d_{z^2}d_{yz}\right>$ 
$\left<d_{yz}d_{yz}|d_{yz}d_{yz}\right>$ $\left<d_{yz}d_{yz}|sd_{xy}\right>$ $\left<d_{yz}d_{yz}|p_xd_{xy}\right>$ $\left<d_{yz}d_{yz}|p_yd_{xy}\right>$ $\left<d_{yz}d_{yz}|p_zd_{xy}\right>$ $\left<d_{yz}d_{yz}|d_{x^2-y^2}d_{xy}\right>$ $\left<d_{yz}d_{yz}|d_{xz}d_{xy}\right>$ $\left<d_{yz}d_{yz}|d_{z^2}d_{xy}\right>$ $\left<d_{yz}d_{yz}|d_{yz}d_{xy}\right>$ $\left<d_{yz}d_{yz}|d_{xy}d_{xy}\right>$ 
$\left<sd_{xy}|ss\right>$ $\left<sd_{xy}|sp_x\right>$ $\left<sd_{xy}|p_xp_x\right>$ $\left<sd_{xy}|sp_y\right>$ $\left<sd_{xy}|p_xp_y\right>$ $\left<sd_{xy}|p_yp_y\right>$ $\left<sd_{xy}|sp_z\right>$ $\left<sd_{xy}|p_xp_z\right>$ $\left<sd_{xy}|p_yp_z\right>$ $\left<sd_{xy}|p_zp_z\right>$ 
$\left<sd_{xy}|sd_{x^2-y^2}\right>$ $\left<sd_{xy}|p_xd_{x^2-y^2}\right>$ $\left<sd_{xy}|p_yd_{x^2-y^2}\right>$ $\left<sd_{xy}|p_zd_{x^2-y^2}\right>$ $\left<sd_{xy}|d_{x^2-y^2}d_{x^2-y^2}\right>$ $\left<sd_{xy}|sd_{xz}\right>$ $\left<sd_{xy}|p_xd_{xz}\right>$ $\left<sd_{xy}|p_yd_{xz}\right>$ $\left<sd_{xy}|p_zd_{xz}\right>$ $\left<sd_{xy}|d_{x^2-y^2}d_{xz}\right>$ 
$\left<sd_{xy}|d_{xz}d_{xz}\right>$ $\left<sd_{xy}|sd_{z^2}\right>$ $\left<sd_{xy}|p_xd_{z^2}\right>$ $\left<sd_{xy}|p_yd_{z^2}\right>$ $\left<sd_{xy}|p_zd_{z^2}\right>$ $\left<sd_{xy}|d_{x^2-y^2}d_{z^2}\right>$ $\left<sd_{xy}|d_{xz}d_{z^2}\right>$ $\left<sd_{xy}|d_{z^2}d_{z^2}\right>$ $\left<sd_{xy}|sd_{yz}\right>$ $\left<sd_{xy}|p_xd_{yz}\right>$ 
$\left<sd_{xy}|p_yd_{yz}\right>$ $\left<sd_{xy}|p_zd_{yz}\right>$ $\left<sd_{xy}|d_{x^2-y^2}d_{yz}\right>$ $\left<sd_{xy}|d_{xz}d_{yz}\right>$ $\left<sd_{xy}|d_{z^2}d_{yz}\right>$ $\left<sd_{xy}|d_{yz}d_{yz}\right>$ $\left<sd_{xy}|sd_{xy}\right>$ $\left<sd_{xy}|p_xd_{xy}\right>$ $\left<sd_{xy}|p_yd_{xy}\right>$ $\left<sd_{xy}|p_zd_{xy}\right>$ 
$\left<sd_{xy}|d_{x^2-y^2}d_{xy}\right>$ $\left<sd_{xy}|d_{xz}d_{xy}\right>$ $\left<sd_{xy}|d_{z^2}d_{xy}\right>$ $\left<sd_{xy}|d_{yz}d_{xy}\right>$ $\left<sd_{xy}|d_{xy}d_{xy}\right>$ $\left<p_xd_{xy}|ss\right>$ $\left<p_xd_{xy}|sp_x\right>$ $\left<p_xd_{xy}|p_xp_x\right>$ $\left<p_xd_{xy}|sp_y\right>$ $\left<p_xd_{xy}|p_xp_y\right>$ 
$\left<p_xd_{xy}|p_yp_y\right>$ $\left<p_xd_{xy}|sp_z\right>$ $\left<p_xd_{xy}|p_xp_z\right>$ $\left<p_xd_{xy}|p_yp_z\right>$ $\left<p_xd_{xy}|p_zp_z\right>$ $\left<p_xd_{xy}|sd_{x^2-y^2}\right>$ $\left<p_xd_{xy}|p_xd_{x^2-y^2}\right>$ $\left<p_xd_{xy}|p_yd_{x^2-y^2}\right>$ $\left<p_xd_{xy}|p_zd_{x^2-y^2}\right>$ $\left<p_xd_{xy}|d_{x^2-y^2}d_{x^2-y^2}\right>$ 
$\left<p_xd_{xy}|sd_{xz}\right>$ $\left<p_xd_{xy}|p_xd_{xz}\right>$ $\left<p_xd_{xy}|p_yd_{xz}\right>$ $\left<p_xd_{xy}|p_zd_{xz}\right>$ $\left<p_xd_{xy}|d_{x^2-y^2}d_{xz}\right>$ $\left<p_xd_{xy}|d_{xz}d_{xz}\right>$ $\left<p_xd_{xy}|sd_{z^2}\right>$ $\left<p_xd_{xy}|p_xd_{z^2}\right>$ $\left<p_xd_{xy}|p_yd_{z^2}\right>$ $\left<p_xd_{xy}|p_zd_{z^2}\right>$ 
$\left<p_xd_{xy}|d_{x^2-y^2}d_{z^2}\right>$ $\left<p_xd_{xy}|d_{xz}d_{z^2}\right>$ $\left<p_xd_{xy}|d_{z^2}d_{z^2}\right>$ $\left<p_xd_{xy}|sd_{yz}\right>$ $\left<p_xd_{xy}|p_xd_{yz}\right>$ $\left<p_xd_{xy}|p_yd_{yz}\right>$ $\left<p_xd_{xy}|p_zd_{yz}\right>$ $\left<p_xd_{xy}|d_{x^2-y^2}d_{yz}\right>$ $\left<p_xd_{xy}|d_{xz}d_{yz}\right>$ $\left<p_xd_{xy}|d_{z^2}d_{yz}\right>$ 
$\left<p_xd_{xy}|d_{yz}d_{yz}\right>$ $\left<p_xd_{xy}|sd_{xy}\right>$ $\left<p_xd_{xy}|p_xd_{xy}\right>$ $\left<p_xd_{xy}|p_yd_{xy}\right>$ $\left<p_xd_{xy}|p_zd_{xy}\right>$ $\left<p_xd_{xy}|d_{x^2-y^2}d_{xy}\right>$ $\left<p_xd_{xy}|d_{xz}d_{xy}\right>$ $\left<p_xd_{xy}|d_{z^2}d_{xy}\right>$ $\left<p_xd_{xy}|d_{yz}d_{xy}\right>$ $\left<p_xd_{xy}|d_{xy}d_{xy}\right>$ 
$\left<p_yd_{xy}|ss\right>$ $\left<p_yd_{xy}|sp_x\right>$ $\left<p_yd_{xy}|p_xp_x\right>$ $\left<p_yd_{xy}|sp_y\right>$ $\left<p_yd_{xy}|p_xp_y\right>$ $\left<p_yd_{xy}|p_yp_y\right>$ $\left<p_yd_{xy}|sp_z\right>$ $\left<p_yd_{xy}|p_xp_z\right>$ $\left<p_yd_{xy}|p_yp_z\right>$ $\left<p_yd_{xy}|p_zp_z\right>$ 
$\left<p_yd_{xy}|sd_{x^2-y^2}\right>$ $\left<p_yd_{xy}|p_xd_{x^2-y^2}\right>$ $\left<p_yd_{xy}|p_yd_{x^2-y^2}\right>$ $\left<p_yd_{xy}|p_zd_{x^2-y^2}\right>$ $\left<p_yd_{xy}|d_{x^2-y^2}d_{x^2-y^2}\right>$ $\left<p_yd_{xy}|sd_{xz}\right>$ $\left<p_yd_{xy}|p_xd_{xz}\right>$ $\left<p_yd_{xy}|p_yd_{xz}\right>$ $\left<p_yd_{xy}|p_zd_{xz}\right>$ $\left<p_yd_{xy}|d_{x^2-y^2}d_{xz}\right>$ 
$\left<p_yd_{xy}|d_{xz}d_{xz}\right>$ $\left<p_yd_{xy}|sd_{z^2}\right>$ $\left<p_yd_{xy}|p_xd_{z^2}\right>$ $\left<p_yd_{xy}|p_yd_{z^2}\right>$ $\left<p_yd_{xy}|p_zd_{z^2}\right>$ $\left<p_yd_{xy}|d_{x^2-y^2}d_{z^2}\right>$ $\left<p_yd_{xy}|d_{xz}d_{z^2}\right>$ $\left<p_yd_{xy}|d_{z^2}d_{z^2}\right>$ $\left<p_yd_{xy}|sd_{yz}\right>$ $\left<p_yd_{xy}|p_xd_{yz}\right>$ 
$\left<p_yd_{xy}|p_yd_{yz}\right>$ $\left<p_yd_{xy}|p_zd_{yz}\right>$ $\left<p_yd_{xy}|d_{x^2-y^2}d_{yz}\right>$ $\left<p_yd_{xy}|d_{xz}d_{yz}\right>$ $\left<p_yd_{xy}|d_{z^2}d_{yz}\right>$ $\left<p_yd_{xy}|d_{yz}d_{yz}\right>$ $\left<p_yd_{xy}|sd_{xy}\right>$ $\left<p_yd_{xy}|p_xd_{xy}\right>$ $\left<p_yd_{xy}|p_yd_{xy}\right>$ $\left<p_yd_{xy}|p_zd_{xy}\right>$ 
$\left<p_yd_{xy}|d_{x^2-y^2}d_{xy}\right>$ $\left<p_yd_{xy}|d_{xz}d_{xy}\right>$ $\left<p_yd_{xy}|d_{z^2}d_{xy}\right>$ $\left<p_yd_{xy}|d_{yz}d_{xy}\right>$ $\left<p_yd_{xy}|d_{xy}d_{xy}\right>$ $\left<p_zd_{xy}|ss\right>$ $\left<p_zd_{xy}|sp_x\right>$ $\left<p_zd_{xy}|p_xp_x\right>$ $\left<p_zd_{xy}|sp_y\right>$ $\left<p_zd_{xy}|p_xp_y\right>$ 
$\left<p_zd_{xy}|p_yp_y\right>$ $\left<p_zd_{xy}|sp_z\right>$ $\left<p_zd_{xy}|p_xp_z\right>$ $\left<p_zd_{xy}|p_yp_z\right>$ $\left<p_zd_{xy}|p_zp_z\right>$ $\left<p_zd_{xy}|sd_{x^2-y^2}\right>$ $\left<p_zd_{xy}|p_xd_{x^2-y^2}\right>$ $\left<p_zd_{xy}|p_yd_{x^2-y^2}\right>$ $\left<p_zd_{xy}|p_zd_{x^2-y^2}\right>$ $\left<p_zd_{xy}|d_{x^2-y^2}d_{x^2-y^2}\right>$ 
$\left<p_zd_{xy}|sd_{xz}\right>$ $\left<p_zd_{xy}|p_xd_{xz}\right>$ $\left<p_zd_{xy}|p_yd_{xz}\right>$ $\left<p_zd_{xy}|p_zd_{xz}\right>$ $\left<p_zd_{xy}|d_{x^2-y^2}d_{xz}\right>$ $\left<p_zd_{xy}|d_{xz}d_{xz}\right>$ $\left<p_zd_{xy}|sd_{z^2}\right>$ $\left<p_zd_{xy}|p_xd_{z^2}\right>$ $\left<p_zd_{xy}|p_yd_{z^2}\right>$ $\left<p_zd_{xy}|p_zd_{z^2}\right>$ 
$\left<p_zd_{xy}|d_{x^2-y^2}d_{z^2}\right>$ $\left<p_zd_{xy}|d_{xz}d_{z^2}\right>$ $\left<p_zd_{xy}|d_{z^2}d_{z^2}\right>$ $\left<p_zd_{xy}|sd_{yz}\right>$ $\left<p_zd_{xy}|p_xd_{yz}\right>$ $\left<p_zd_{xy}|p_yd_{yz}\right>$ $\left<p_zd_{xy}|p_zd_{yz}\right>$ $\left<p_zd_{xy}|d_{x^2-y^2}d_{yz}\right>$ $\left<p_zd_{xy}|d_{xz}d_{yz}\right>$ $\left<p_zd_{xy}|d_{z^2}d_{yz}\right>$ 
$\left<p_zd_{xy}|d_{yz}d_{yz}\right>$ $\left<p_zd_{xy}|sd_{xy}\right>$ $\left<p_zd_{xy}|p_xd_{xy}\right>$ $\left<p_zd_{xy}|p_yd_{xy}\right>$ $\left<p_zd_{xy}|p_zd_{xy}\right>$ $\left<p_zd_{xy}|d_{x^2-y^2}d_{xy}\right>$ $\left<p_zd_{xy}|d_{xz}d_{xy}\right>$ $\left<p_zd_{xy}|d_{z^2}d_{xy}\right>$ $\left<p_zd_{xy}|d_{yz}d_{xy}\right>$ $\left<p_zd_{xy}|d_{xy}d_{xy}\right>$ 
$\left<d_{x^2-y^2}d_{xy}|ss\right>$ $\left<d_{x^2-y^2}d_{xy}|sp_x\right>$ $\left<d_{x^2-y^2}d_{xy}|p_xp_x\right>$ $\left<d_{x^2-y^2}d_{xy}|sp_y\right>$ $\left<d_{x^2-y^2}d_{xy}|p_xp_y\right>$ $\left<d_{x^2-y^2}d_{xy}|p_yp_y\right>$ $\left<d_{x^2-y^2}d_{xy}|sp_z\right>$ $\left<d_{x^2-y^2}d_{xy}|p_xp_z\right>$ $\left<d_{x^2-y^2}d_{xy}|p_yp_z\right>$ $\left<d_{x^2-y^2}d_{xy}|p_zp_z\right>$ 
$\left<d_{x^2-y^2}d_{xy}|sd_{x^2-y^2}\right>$ $\left<d_{x^2-y^2}d_{xy}|p_xd_{x^2-y^2}\right>$ $\left<d_{x^2-y^2}d_{xy}|p_yd_{x^2-y^2}\right>$ $\left<d_{x^2-y^2}d_{xy}|p_zd_{x^2-y^2}\right>$ $\left<d_{x^2-y^2}d_{xy}|d_{x^2-y^2}d_{x^2-y^2}\right>$ $\left<d_{x^2-y^2}d_{xy}|sd_{xz}\right>$ $\left<d_{x^2-y^2}d_{xy}|p_xd_{xz}\right>$ $\left<d_{x^2-y^2}d_{xy}|p_yd_{xz}\right>$ $\left<d_{x^2-y^2}d_{xy}|p_zd_{xz}\right>$ $\left<d_{x^2-y^2}d_{xy}|d_{x^2-y^2}d_{xz}\right>$ 
$\left<d_{x^2-y^2}d_{xy}|d_{xz}d_{xz}\right>$ $\left<d_{x^2-y^2}d_{xy}|sd_{z^2}\right>$ $\left<d_{x^2-y^2}d_{xy}|p_xd_{z^2}\right>$ $\left<d_{x^2-y^2}d_{xy}|p_yd_{z^2}\right>$ $\left<d_{x^2-y^2}d_{xy}|p_zd_{z^2}\right>$ $\left<d_{x^2-y^2}d_{xy}|d_{x^2-y^2}d_{z^2}\right>$ $\left<d_{x^2-y^2}d_{xy}|d_{xz}d_{z^2}\right>$ $\left<d_{x^2-y^2}d_{xy}|d_{z^2}d_{z^2}\right>$ $\left<d_{x^2-y^2}d_{xy}|sd_{yz}\right>$ $\left<d_{x^2-y^2}d_{xy}|p_xd_{yz}\right>$ 
$\left<d_{x^2-y^2}d_{xy}|p_yd_{yz}\right>$ $\left<d_{x^2-y^2}d_{xy}|p_zd_{yz}\right>$ $\left<d_{x^2-y^2}d_{xy}|d_{x^2-y^2}d_{yz}\right>$ $\left<d_{x^2-y^2}d_{xy}|d_{xz}d_{yz}\right>$ $\left<d_{x^2-y^2}d_{xy}|d_{z^2}d_{yz}\right>$ $\left<d_{x^2-y^2}d_{xy}|d_{yz}d_{yz}\right>$ $\left<d_{x^2-y^2}d_{xy}|sd_{xy}\right>$ $\left<d_{x^2-y^2}d_{xy}|p_xd_{xy}\right>$ $\left<d_{x^2-y^2}d_{xy}|p_yd_{xy}\right>$ $\left<d_{x^2-y^2}d_{xy}|p_zd_{xy}\right>$ 
$\left<d_{x^2-y^2}d_{xy}|d_{x^2-y^2}d_{xy}\right>$ $\left<d_{x^2-y^2}d_{xy}|d_{xz}d_{xy}\right>$ $\left<d_{x^2-y^2}d_{xy}|d_{z^2}d_{xy}\right>$ $\left<d_{x^2-y^2}d_{xy}|d_{yz}d_{xy}\right>$ $\left<d_{x^2-y^2}d_{xy}|d_{xy}d_{xy}\right>$ $\left<d_{xz}d_{xy}|ss\right>$ $\left<d_{xz}d_{xy}|sp_x\right>$ $\left<d_{xz}d_{xy}|p_xp_x\right>$ $\left<d_{xz}d_{xy}|sp_y\right>$ $\left<d_{xz}d_{xy}|p_xp_y\right>$ 
$\left<d_{xz}d_{xy}|p_yp_y\right>$ $\left<d_{xz}d_{xy}|sp_z\right>$ $\left<d_{xz}d_{xy}|p_xp_z\right>$ $\left<d_{xz}d_{xy}|p_yp_z\right>$ $\left<d_{xz}d_{xy}|p_zp_z\right>$ $\left<d_{xz}d_{xy}|sd_{x^2-y^2}\right>$ $\left<d_{xz}d_{xy}|p_xd_{x^2-y^2}\right>$ $\left<d_{xz}d_{xy}|p_yd_{x^2-y^2}\right>$ $\left<d_{xz}d_{xy}|p_zd_{x^2-y^2}\right>$ $\left<d_{xz}d_{xy}|d_{x^2-y^2}d_{x^2-y^2}\right>$ 
$\left<d_{xz}d_{xy}|sd_{xz}\right>$ $\left<d_{xz}d_{xy}|p_xd_{xz}\right>$ $\left<d_{xz}d_{xy}|p_yd_{xz}\right>$ $\left<d_{xz}d_{xy}|p_zd_{xz}\right>$ $\left<d_{xz}d_{xy}|d_{x^2-y^2}d_{xz}\right>$ $\left<d_{xz}d_{xy}|d_{xz}d_{xz}\right>$ $\left<d_{xz}d_{xy}|sd_{z^2}\right>$ $\left<d_{xz}d_{xy}|p_xd_{z^2}\right>$ $\left<d_{xz}d_{xy}|p_yd_{z^2}\right>$ $\left<d_{xz}d_{xy}|p_zd_{z^2}\right>$ 
$\left<d_{xz}d_{xy}|d_{x^2-y^2}d_{z^2}\right>$ $\left<d_{xz}d_{xy}|d_{xz}d_{z^2}\right>$ $\left<d_{xz}d_{xy}|d_{z^2}d_{z^2}\right>$ $\left<d_{xz}d_{xy}|sd_{yz}\right>$ $\left<d_{xz}d_{xy}|p_xd_{yz}\right>$ $\left<d_{xz}d_{xy}|p_yd_{yz}\right>$ $\left<d_{xz}d_{xy}|p_zd_{yz}\right>$ $\left<d_{xz}d_{xy}|d_{x^2-y^2}d_{yz}\right>$ $\left<d_{xz}d_{xy}|d_{xz}d_{yz}\right>$ $\left<d_{xz}d_{xy}|d_{z^2}d_{yz}\right>$ 
$\left<d_{xz}d_{xy}|d_{yz}d_{yz}\right>$ $\left<d_{xz}d_{xy}|sd_{xy}\right>$ $\left<d_{xz}d_{xy}|p_xd_{xy}\right>$ $\left<d_{xz}d_{xy}|p_yd_{xy}\right>$ $\left<d_{xz}d_{xy}|p_zd_{xy}\right>$ $\left<d_{xz}d_{xy}|d_{x^2-y^2}d_{xy}\right>$ $\left<d_{xz}d_{xy}|d_{xz}d_{xy}\right>$ $\left<d_{xz}d_{xy}|d_{z^2}d_{xy}\right>$ $\left<d_{xz}d_{xy}|d_{yz}d_{xy}\right>$ $\left<d_{xz}d_{xy}|d_{xy}d_{xy}\right>$ 
$\left<d_{z^2}d_{xy}|ss\right>$ $\left<d_{z^2}d_{xy}|sp_x\right>$ $\left<d_{z^2}d_{xy}|p_xp_x\right>$ $\left<d_{z^2}d_{xy}|sp_y\right>$ $\left<d_{z^2}d_{xy}|p_xp_y\right>$ $\left<d_{z^2}d_{xy}|p_yp_y\right>$ $\left<d_{z^2}d_{xy}|sp_z\right>$ $\left<d_{z^2}d_{xy}|p_xp_z\right>$ $\left<d_{z^2}d_{xy}|p_yp_z\right>$ $\left<d_{z^2}d_{xy}|p_zp_z\right>$ 
$\left<d_{z^2}d_{xy}|sd_{x^2-y^2}\right>$ $\left<d_{z^2}d_{xy}|p_xd_{x^2-y^2}\right>$ $\left<d_{z^2}d_{xy}|p_yd_{x^2-y^2}\right>$ $\left<d_{z^2}d_{xy}|p_zd_{x^2-y^2}\right>$ $\left<d_{z^2}d_{xy}|d_{x^2-y^2}d_{x^2-y^2}\right>$ $\left<d_{z^2}d_{xy}|sd_{xz}\right>$ $\left<d_{z^2}d_{xy}|p_xd_{xz}\right>$ $\left<d_{z^2}d_{xy}|p_yd_{xz}\right>$ $\left<d_{z^2}d_{xy}|p_zd_{xz}\right>$ $\left<d_{z^2}d_{xy}|d_{x^2-y^2}d_{xz}\right>$ 
$\left<d_{z^2}d_{xy}|d_{xz}d_{xz}\right>$ $\left<d_{z^2}d_{xy}|sd_{z^2}\right>$ $\left<d_{z^2}d_{xy}|p_xd_{z^2}\right>$ $\left<d_{z^2}d_{xy}|p_yd_{z^2}\right>$ $\left<d_{z^2}d_{xy}|p_zd_{z^2}\right>$ $\left<d_{z^2}d_{xy}|d_{x^2-y^2}d_{z^2}\right>$ $\left<d_{z^2}d_{xy}|d_{xz}d_{z^2}\right>$ $\left<d_{z^2}d_{xy}|d_{z^2}d_{z^2}\right>$ $\left<d_{z^2}d_{xy}|sd_{yz}\right>$ $\left<d_{z^2}d_{xy}|p_xd_{yz}\right>$ 
$\left<d_{z^2}d_{xy}|p_yd_{yz}\right>$ $\left<d_{z^2}d_{xy}|p_zd_{yz}\right>$ $\left<d_{z^2}d_{xy}|d_{x^2-y^2}d_{yz}\right>$ $\left<d_{z^2}d_{xy}|d_{xz}d_{yz}\right>$ $\left<d_{z^2}d_{xy}|d_{z^2}d_{yz}\right>$ $\left<d_{z^2}d_{xy}|d_{yz}d_{yz}\right>$ $\left<d_{z^2}d_{xy}|sd_{xy}\right>$ $\left<d_{z^2}d_{xy}|p_xd_{xy}\right>$ $\left<d_{z^2}d_{xy}|p_yd_{xy}\right>$ $\left<d_{z^2}d_{xy}|p_zd_{xy}\right>$ 
$\left<d_{z^2}d_{xy}|d_{x^2-y^2}d_{xy}\right>$ $\left<d_{z^2}d_{xy}|d_{xz}d_{xy}\right>$ $\left<d_{z^2}d_{xy}|d_{z^2}d_{xy}\right>$ $\left<d_{z^2}d_{xy}|d_{yz}d_{xy}\right>$ $\left<d_{z^2}d_{xy}|d_{xy}d_{xy}\right>$ $\left<d_{yz}d_{xy}|ss\right>$ $\left<d_{yz}d_{xy}|sp_x\right>$ $\left<d_{yz}d_{xy}|p_xp_x\right>$ $\left<d_{yz}d_{xy}|sp_y\right>$ $\left<d_{yz}d_{xy}|p_xp_y\right>$ 
$\left<d_{yz}d_{xy}|p_yp_y\right>$ $\left<d_{yz}d_{xy}|sp_z\right>$ $\left<d_{yz}d_{xy}|p_xp_z\right>$ $\left<d_{yz}d_{xy}|p_yp_z\right>$ $\left<d_{yz}d_{xy}|p_zp_z\right>$ $\left<d_{yz}d_{xy}|sd_{x^2-y^2}\right>$ $\left<d_{yz}d_{xy}|p_xd_{x^2-y^2}\right>$ $\left<d_{yz}d_{xy}|p_yd_{x^2-y^2}\right>$ $\left<d_{yz}d_{xy}|p_zd_{x^2-y^2}\right>$ $\left<d_{yz}d_{xy}|d_{x^2-y^2}d_{x^2-y^2}\right>$ 
$\left<d_{yz}d_{xy}|sd_{xz}\right>$ $\left<d_{yz}d_{xy}|p_xd_{xz}\right>$ $\left<d_{yz}d_{xy}|p_yd_{xz}\right>$ $\left<d_{yz}d_{xy}|p_zd_{xz}\right>$ $\left<d_{yz}d_{xy}|d_{x^2-y^2}d_{xz}\right>$ $\left<d_{yz}d_{xy}|d_{xz}d_{xz}\right>$ $\left<d_{yz}d_{xy}|sd_{z^2}\right>$ $\left<d_{yz}d_{xy}|p_xd_{z^2}\right>$ $\left<d_{yz}d_{xy}|p_yd_{z^2}\right>$ $\left<d_{yz}d_{xy}|p_zd_{z^2}\right>$ 
$\left<d_{yz}d_{xy}|d_{x^2-y^2}d_{z^2}\right>$ $\left<d_{yz}d_{xy}|d_{xz}d_{z^2}\right>$ $\left<d_{yz}d_{xy}|d_{z^2}d_{z^2}\right>$ $\left<d_{yz}d_{xy}|sd_{yz}\right>$ $\left<d_{yz}d_{xy}|p_xd_{yz}\right>$ $\left<d_{yz}d_{xy}|p_yd_{yz}\right>$ $\left<d_{yz}d_{xy}|p_zd_{yz}\right>$ $\left<d_{yz}d_{xy}|d_{x^2-y^2}d_{yz}\right>$ $\left<d_{yz}d_{xy}|d_{xz}d_{yz}\right>$ $\left<d_{yz}d_{xy}|d_{z^2}d_{yz}\right>$ 
$\left<d_{yz}d_{xy}|d_{yz}d_{yz}\right>$ $\left<d_{yz}d_{xy}|sd_{xy}\right>$ $\left<d_{yz}d_{xy}|p_xd_{xy}\right>$ $\left<d_{yz}d_{xy}|p_yd_{xy}\right>$ $\left<d_{yz}d_{xy}|p_zd_{xy}\right>$ $\left<d_{yz}d_{xy}|d_{x^2-y^2}d_{xy}\right>$ $\left<d_{yz}d_{xy}|d_{xz}d_{xy}\right>$ $\left<d_{yz}d_{xy}|d_{z^2}d_{xy}\right>$ $\left<d_{yz}d_{xy}|d_{yz}d_{xy}\right>$ $\left<d_{yz}d_{xy}|d_{xy}d_{xy}\right>$ 
$\left<d_{xy}d_{xy}|ss\right>$ $\left<d_{xy}d_{xy}|sp_x\right>$ $\left<d_{xy}d_{xy}|p_xp_x\right>$ $\left<d_{xy}d_{xy}|sp_y\right>$ $\left<d_{xy}d_{xy}|p_xp_y\right>$ $\left<d_{xy}d_{xy}|p_yp_y\right>$ $\left<d_{xy}d_{xy}|sp_z\right>$ $\left<d_{xy}d_{xy}|p_xp_z\right>$ $\left<d_{xy}d_{xy}|p_yp_z\right>$ $\left<d_{xy}d_{xy}|p_zp_z\right>$ 
$\left<d_{xy}d_{xy}|sd_{x^2-y^2}\right>$ $\left<d_{xy}d_{xy}|p_xd_{x^2-y^2}\right>$ $\left<d_{xy}d_{xy}|p_yd_{x^2-y^2}\right>$ $\left<d_{xy}d_{xy}|p_zd_{x^2-y^2}\right>$ $\left<d_{xy}d_{xy}|d_{x^2-y^2}d_{x^2-y^2}\right>$ $\left<d_{xy}d_{xy}|sd_{xz}\right>$ $\left<d_{xy}d_{xy}|p_xd_{xz}\right>$ $\left<d_{xy}d_{xy}|p_yd_{xz}\right>$ $\left<d_{xy}d_{xy}|p_zd_{xz}\right>$ $\left<d_{xy}d_{xy}|d_{x^2-y^2}d_{xz}\right>$ 
$\left<d_{xy}d_{xy}|d_{xz}d_{xz}\right>$ $\left<d_{xy}d_{xy}|sd_{z^2}\right>$ $\left<d_{xy}d_{xy}|p_xd_{z^2}\right>$ $\left<d_{xy}d_{xy}|p_yd_{z^2}\right>$ $\left<d_{xy}d_{xy}|p_zd_{z^2}\right>$ $\left<d_{xy}d_{xy}|d_{x^2-y^2}d_{z^2}\right>$ $\left<d_{xy}d_{xy}|d_{xz}d_{z^2}\right>$ $\left<d_{xy}d_{xy}|d_{z^2}d_{z^2}\right>$ $\left<d_{xy}d_{xy}|sd_{yz}\right>$ $\left<d_{xy}d_{xy}|p_xd_{yz}\right>$ 
$\left<d_{xy}d_{xy}|p_yd_{yz}\right>$ $\left<d_{xy}d_{xy}|p_zd_{yz}\right>$ $\left<d_{xy}d_{xy}|d_{x^2-y^2}d_{yz}\right>$ $\left<d_{xy}d_{xy}|d_{xz}d_{yz}\right>$ $\left<d_{xy}d_{xy}|d_{z^2}d_{yz}\right>$ $\left<d_{xy}d_{xy}|d_{yz}d_{yz}\right>$ $\left<d_{xy}d_{xy}|sd_{xy}\right>$ $\left<d_{xy}d_{xy}|p_xd_{xy}\right>$ $\left<d_{xy}d_{xy}|p_yd_{xy}\right>$ $\left<d_{xy}d_{xy}|p_zd_{xy}\right>$ 
$\left<d_{xy}d_{xy}|d_{x^2-y^2}d_{xy}\right>$ $\left<d_{xy}d_{xy}|d_{xz}d_{xy}\right>$ $\left<d_{xy}d_{xy}|d_{z^2}d_{xy}\right>$ $\left<d_{xy}d_{xy}|d_{yz}d_{xy}\right>$ $\left<d_{xy}d_{xy}|d_{xy}d_{xy}\right>$

\subsection{Transformation from the Local to the Global Coordinate System}
\label{sec:localcoordinate}

Within the MNDO-type models, we calculate 
the multipole-multipole interactions in a local coordinate system 
as visualized in Figure~5 in the main text.
A local coordinate system is defined for each pair of atoms $I$ and $J\ne I$ which are positioned 
at $\tilde{\fett{R}}_I = (\tilde{R}_{I,x},\tilde{R}_{I,y},\tilde{R}_{I,z})$ 
and $\tilde{\fett{R}}_J= (\tilde{R}_{J,x},\tilde{R}_{J,y},\tilde{R}_{J,z})$, respectively.
The standard basis $\fett{x}^{\text{loc}},\fett{y}^{\text{loc}},\fett{z}^{\text{loc}}$ of the 
local coordinate system 
can then be determined, e.g., as described in Ref.~\onlinecite{Glaeske1987} or in 
Ref.~\onlinecite{Steinmann2013}.
The first unit vector $\fett{z}^{\text{loc}}$ is defined as the normalized vector connecting $\tilde{\fett{R}}_I$ and 
$\tilde{\fett{R}}_J$, 
\begin{equation}
 \fett{z}^{\text{loc}} = \frac{\tilde{\fett{R}}_I-\tilde{\fett{R}}_J}{|\tilde{\fett{R}}_I-\tilde{\fett{R}}_J|} = 
({z}^{\text{loc}}_x, 
{z}^{\text{loc}}_y, {z}^{\text{loc}}_z).
\end{equation}
The vector $\fett{z}^{\text{loc}}$ can be applied to construct a perpendicular vector
$\fett{y}^{\text{loc}}$, 
\begin{equation}
 \fett{y}^{\text{loc}} = \frac{1}{\sqrt{{z^{\text{loc}}_x}^2+z{^{\text{loc}}_y}^2}}(-z^{\text{loc}}_y, 
z^{\text{loc}}_x, 0).
\end{equation}
The cross product of $\fett{z}^{\text{loc}}$ and $\fett{y}^{\text{loc}}$ yields $\fett{x}^{\text{loc}}$, 
\begin{equation}
 \fett{x}^{\text{loc}} = \fett{z}^{\text{loc}} \times \fett{y}^{\text{loc}}.
\end{equation}
We can then construct the rotation matrices which transform the results from the local 
to the global coordinate system. \cite{Glaeske1987}

The implementation of the transformation procedure 
can be compared to the implementation in \textsc{Mopac}
when invoking the keyword \texttt{Hcore} the calculation of the electronic energy 
of a diatomic molecule which is \textit{not} aligned along the $z$-axis.
The one hundred first to two hundredth entry 
 listed under \texttt{TWO-ELECTRON MATRIX IN HCORE}
are the 
transformed two-center ERIs. 

\subsection{Modification to the PM6 Core-Core Repulsion Energy}
\label{subsec:pm6mod}

For certain element pairs, a scaling factor different from 
$f_{IJ}^{\text{PM6}}$ (Eq.~(62) in the main text) is applied in the PM6 model. \cite{Stewart2007} 
The scaling function $f_{IJ}^{',\text{PM6}}$ to calculate the core-core repulsion energy between 
 two carbon atoms is given by 
\begin{equation}
\label{eq:pm6scal2}
\begin{split}
f_{IJ}^{',\text{PM6}} =&  1 + 2 { x^{Z_I,Z_J}} 
\exp{\left(- {  \alpha^{Z_I,Z_J}} \left(\tilde{R}_{IJ} + 
0.0003 \text{\AA}^{-5} \tilde{R}_{IJ}^6\right)\right)}  \\ & 
+9.28 \exp{\left(-5.98 \text{\AA}^{-1} \tilde{R}_{IJ} \right)}  + K^{Z_I} 
               \exp{\left(-  L^{Z_I} 
\left( \tilde{R}_{IJ} - 
 M^{Z_I} \right)^2 \right)}   \\ & +  
K^{Z_J} 
               \exp{\left(-  L^{Z_J} 
\left( \tilde{R}_{IJ} - 
 M^{Z_J} \right)^2 \right)}. \\
\end{split}
\end{equation}
For N--H and O--H interactions, the scaling function $f_{IJ}^{'',\text{PM6}}$ reads, 
\begin{equation}
\label{eq:pm6eccoh}
\begin{split}
f_{IJ}^{'',\text{PM6}} =& 1+2{ x^{Z_I,Z_J}}
\exp{\left(-{\alpha^{Z_I,Z_J}} \text{\AA}^{-1} \tilde{R}_{IJ}^2\right)} + K^{Z_I} 
               \exp{\left(-  L^{Z_I} 
\left( \tilde{R}_{IJ} - 
 M^{Z_I} \right)^2 \right)}   \\ & +  
K^{Z_J} 
               \exp{\left(-  L^{Z_J} 
\left( \tilde{R}_{IJ} - 
 M^{Z_J} \right)^2 \right)}. \\
\end{split}
\end{equation}
Note that \textsc{Mopac} applies this equation not only for 
N--H and O--H interactions, but also for C--H interactions, which, however, 
is not the intended use according 
to the original publication in Ref.~\onlinecite{Stewart2007}.
The scaling function $f_{IJ}^{''',\text{PM6}}$ for Si--O interactions is given by
\begin{equation}
\label{eq:pm6eccsio}
\begin{split}
f_{IJ}^{''',\text{PM6}} =& 1 + 2 { x^{Z_I,Z_J}} 
\exp{\left(- {  \alpha^{Z_I,Z_J}} \left(\tilde{R}_{IJ} + 
0.0003 \text{\AA}^{-5} \tilde{R}_{IJ}^6\right)\right)}  \\ & 
- 0.0007 \exp{\left(- \text{\AA}^{-2} \left(\tilde{R}_{IJ}-2.9 \text{\AA}\right)^2\right)}\\ &  + K^{Z_I} 
               \exp{\left(-  L^{Z_I} 
\left( \tilde{R}_{IJ} - 
 M^{Z_I} \right)^2 \right)}     +  
K^{Z_J} 
               \exp{\left(-  L^{Z_J} 
\left( \tilde{R}_{IJ} - 
 M^{Z_J} \right)^2 \right)}. \\
\end{split}
\end{equation}
As there exists no theoretical foundation for the introduction of the modified expressions 
to calculate the core-core repulsion energy,  we do not know 
why the application of these modified scaling factors yield a better 
agreement with experimental data.

\subsection{Constraints on Parameters} 
\label{par:optimizationstragety}

During the parameter optimization, certain constraints have to be imposed on the parameter values
to keep the parametric expressions sensible from a physical point of view.
Several parameters, for instance, determine the sign of an exponential functions which depends 
on the internuclear distance $\tilde{R}_{IJ}$.
The sign of the argument of the exponential function must be negative so that 
it does not become infinite for large  $\tilde{R}_{IJ}$ 
which means that $\zeta_{l(\mu)}^{Z_I}>0$, ${\zeta'}_{l(\mu)}^{Z_I}>0$, $\alpha^{Z_I}>0$, 
$\alpha^{Z_I,Z_J}>0$, $a_{l(\mu)}^{Z_I}+a_{l(\lambda)}^{Z_J}>0$, and
$L_a^{Z_I}>0$. 
Furthermore, the scaling factors for the core-core repulsion must not become 
negative (which would correspond to an attractive interaction between two cores),
i.e., $x^{Z_I,Z_J}$ and $K_a^{Z_I}$ have to be constrained such that 
the scaling factors are positive.
These constraints are fulfilled by all NDDO-SEMO models.
To our understanding, it is not simply possible to constrain the values of 
the other parameters in a meaningful way.

\begin{figure}[ht]
 \centering 
 \includegraphics[width=.75\textwidth]{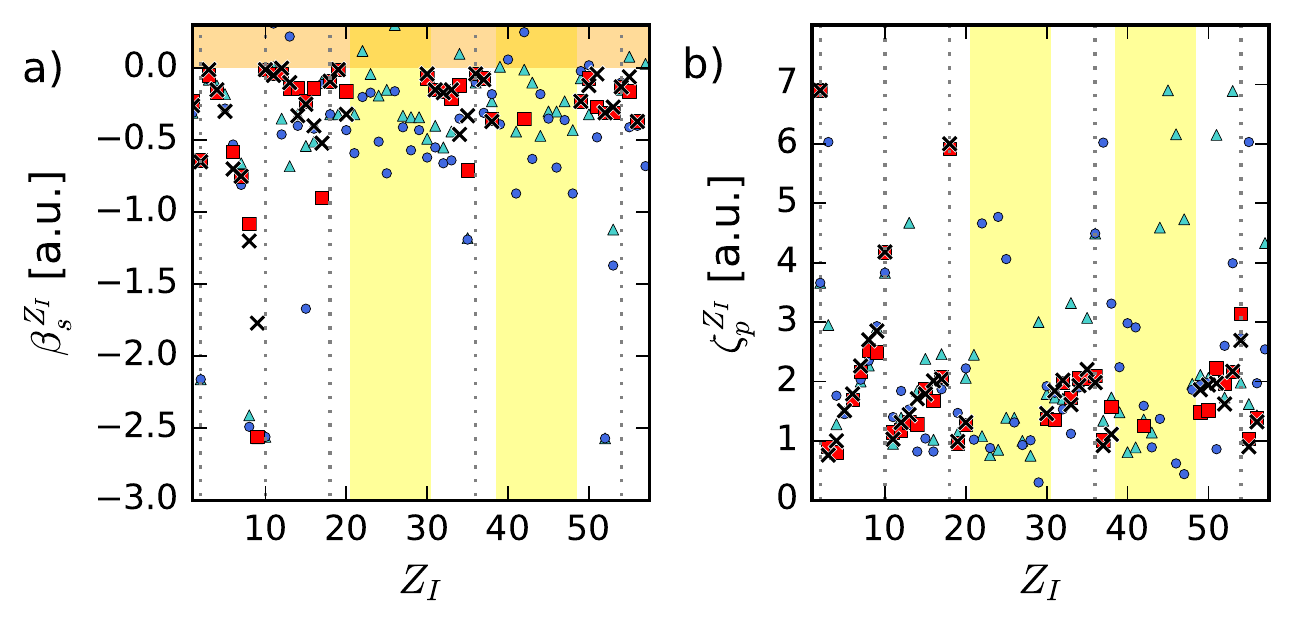}
 \caption{Variation of a) $\beta_{s}^{Z_I}$ 
and b) $\zeta_p^{Z_I}$ 
in a.u.\ with 
the atomic number $Z_I$ in a range of $Z_I$ = 1--57:
MNDO values \cite{Dewar1977,Stewart2004} (black crosses)
PM6 values \cite{Stewart2007} (light blue triangles),
PM7 values \cite{Stewart2012} (dark blue circles), and 
AM1 values \cite{Dewar1985,Stewart2004} (red squares).
We highlight the transition metal blocks ($21\le Z_I\le 30$ and 
$39\le Z_I\le 48$) by a yellow background
and indicate $Z_I$ of rare gases by gray dashed lines.
The orange area marks the parameter values for which 
$\beta_{s}^{Z_I}>0$.
}
\label{fig:parameter_reg}
\end{figure}

Interestingly, the parameters appear to vary regularly with the 
atomic number for lighter elements. 
Dewar and Thiel noted that the MNDO parameters change in a remarkably regular manner with the 
atomic number of hydrogen, carbon, nitrogen, and oxygen \cite{Dewar1977} 
(see also Figure~\ref{fig:parameter_reg}).
This regular behavior is so pronounced that it was, for example, possible 
to estimate the parameters for fluorine to good accuracy based on the ones for hydrogen, carbon, nitrogen, 
and oxygen.\cite{Dewar1978}
Dewar and co-workers remarked that this `suggest[s] that the MNDO 
method as a whole is suitably self-consistent'. \cite{Dewar1978}
For heavier elements, the regularity is lost  (e.g., for the second 
transition-metal block). Additionally, one can also see that the parameters 
also vary significantly between the different NDDO-SEMO models.

\newpage

\subsection{Parametrization of the MNDO Model}
\label{sec:mndoparam}

\begin{figure}[ht]
 \centering 
 \includegraphics[width=.45\textwidth]{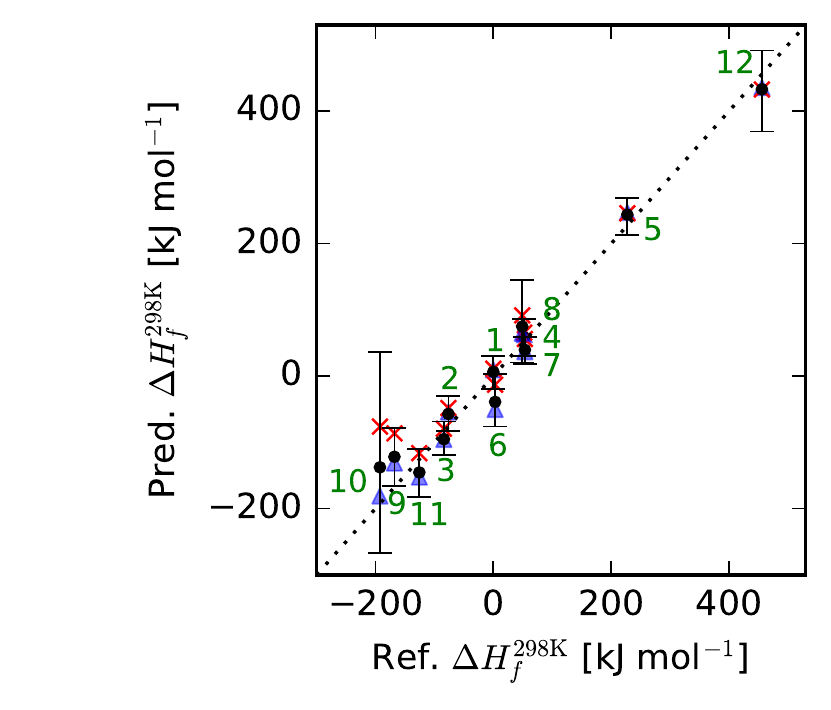}
 \caption{ 
Comparison of the 
reference $\Delta H_{f}^{\text{298K}}$ with the 
predicted values for $\Delta H_{f}^{\text{298K}}$ in kJ~mol$^{-1}$.
$\Delta H_{f}^{\text{298K}}$ was predicted with the  
MNDO values for $\fett{p}^m$ (red crosses), the $\fett{p}^m_{\mathcal{D}^m}$ values 
(blue triangles), and the $\bar{\fett{p}}^m_{\mathcal{D}^m_b}$ values (black circles). 
We provide the 95\% confidence interval for the bootstrapped $\Delta H_{f}^{\text{298K}}$
(gray bars). The uncertainties for the experimental data are too small to be visible 
in this figure.
It is denoted for each datapoint in green font to which entry in $\mathcal{D}^m$
it belongs (cf. Figure~16 in the main text).
}
\label{fig:bootstrapsampling_hof}
\end{figure}

For this work, we re-optimized the $\beta_{l(\mu)}^{Z_I}$ parameters in the MNDO model for carbon and hydrogen.
The applied reference data set $\mathcal{D}^m$ contains 12 heats of formation 
at 298 K ($\Delta H_{f}^{\text{298K}}$) of hydrocarbon compounds
which are also present in the original reference data set of MNDO \cite{Dewar1977}
(see Table~\ref{tab:dm_hof}).
\begin{center}
\begin{longtable}{crc}
  \caption{Reference data set $\mathcal{D}^m$ consisting of 
$\Delta H_{f}^{\text{298K}}$ (and the standard deviation of 
$\Delta H_{f}^{\text{298K}}$ )
 for twelve compounds in kJ~mol$^{-1}$. 
} \label{tab:dm_hof} \\
\hline
\hline 
Compound & $\Delta H_{f}^{\text{298K}}$ & Ref. \\
\hline 
\endfirsthead 
\hline
\hline 
Compound & $\Delta H_{f}^{\text{298K}}$ & Ref. \\
\hline 
\endhead 
\hline \hline \endfoot
dihydrogen	& $0.0\pm0.0$ 		& --- \\ 
methane		& $-76.3\pm0.3$ 	& \onlinecite{Manion2002} \\
ethane		& $-84.0\pm0.4$		& \onlinecite{Manion2002} \\
ethene		& $52.4\pm0.5$		& \onlinecite{Manion2002} \\
ethyne		& $227.4\pm0.8$		& \onlinecite{Manion2002} \\
cyclopropane	& $53.5\pm0.6$		& \onlinecite{Knowlton1949} \\
cyclobutane	& $3.0\pm0.1$		& \onlinecite{Kaarsemaker2010} \\
benzene		& $49.0\pm0.9$		& \onlinecite{Roux2008} \\
neopentane 	& $-167.9\pm0.6$	& \onlinecite{Good1970} \\
n-butane	& $-125.6\pm0.7$	& \onlinecite{Pittam1972} \\
adamantane 	& $-192.5\pm0.4$	& \onlinecite{Clark1979} \\
1,3-butadyne	& $455.8\pm2.0$		& \onlinecite{Rogers2003} \\
\end{longtable}
\end{center}
We list the values for the parameters 
$\beta_{s}^{1}$, $\beta_{s}^{6}$, and $\beta_{p}^{6}$ which are applied 
within MNDO \cite{Dewar1977} and which we obtained in our parametrizations 
in Table~\ref{tab:dm_par}. We supply the parameter files which can be read in through 
the keyword \texttt{External} in \textsc{Mopac} to reproduce our results with standard software.
\begin{center}
\begin{longtable}{cccc}
  \caption{Values for the parameters 
$\beta_{s}^{1}$, $\beta_{s}^{6}$, and $\beta_{p}^{6}$ which are applied 
within MNDO \cite{Dewar1977} and which we obtained in our parametrizations 
in a.u.
} \label{tab:dm_par} \\
\hline
\hline 
Parametrization & $\beta_{s}^{1}$ & $\beta_{s}^{6}$ & $\beta_{p}^{6}$ \\
\hline 
\endfirsthead 
\hline
\hline 
Parametrization & $\beta_{s}^{1}$ & $\beta_{s}^{6}$ & $\beta_{p}^{6}$ \\
\hline 
\endhead 
\hline \hline \endfoot
MNDO                                 & $-0.26$ & $-0.70$ & $-0.29$ \\
$\fett{p}^m_{\mathcal{D}^m}$         & $-0.26$ & $-0.65$ & $-0.32$ \\ 
$\bar{\fett{p}}^m_{\mathcal{D}^m_b}$ & $-0.26$ & $-0.68$ & $-0.30$ \\
\end{longtable}
\end{center}
The results for $\Delta H_{f}^{\text{298K}}$ obtained 
with the MNDO values, the $\fett{p}^m_{\mathcal{D}^m}$ values, and 
the $\bar{\fett{p}}^m_{\mathcal{D}^m_b}$ values for 
$\beta_{s}^{1}$, 
$\beta_{s}^{6}$, 
and $\beta_{p}^{6}$ are given in Table~\ref{tab:dm_hof_pred} and illustrated 
in Figure~\ref{fig:bootstrapsampling_hof}.
\begin{center}
\begin{longtable}{crrrrr}
  \caption{$\Delta H_{f}^{\text{298K}}$ obtained 
with the original MNDO values, the $\fett{p}^m_{\mathcal{D}^m}$ values, and 
the bootstrapped minimum, mean, and maximum $\bar{\fett{p}}^m_{\mathcal{D}^m_b}$ values for 
$\beta_{s}^{1}$, 
$\beta_{s}^{6}$, 
and $\beta_{p}^{6}$ 
in kJ~mol$^{-1}$. 
} \label{tab:dm_hof_pred} \\
\hline
\hline 
Compound & MNDO & $\fett{p}^m_{\mathcal{D}^m}$ & \multicolumn{3}{c}{$\bar{\fett{p}}^m_{\mathcal{D}^m_b}$} \\
& & & Min. & Mean & Max. \\
\hline 
\endfirsthead 
\hline
\hline 
Compound & MNDO & $\fett{p}^m_{\mathcal{D}^m}$ & \multicolumn{3}{c}{$\bar{\fett{p}}^m_{\mathcal{D}^m_b}$} \\
& & & Min. & Mean & Max. \\
\hline 
\endhead 
\hline \hline \endfoot
dihydrogen	& $11.2$	& $9.5$ 	& $-19.9$  & $6.4$    & $30.1$ \\
methane		& $-48.3$	& $-54.1$	& $-83.2$  & $-57.2$  & $-30.0$ \\
ethane		& $-79.5$	& $-95.6$	& $-119.6$ & $-95.2$  & $-68.5$ \\
ethene		& $65.6$	& $62.5$	& $29.8$   & $58.4$   & $86.7$ \\
ethyne		& $245.7$	& $247.1$	& $213.3$  & $243.7$  & $268.3$ \\
cyclopropane	& $55.9$	& $36.9$	& $18.6$   & $39.7$   & $58.6$ \\
cyclobutane	& $-13.0$	& $-50.3$	& $-76.0$  & $-38.9$  & $3.3$ \\
benzene		& $91.7$	& $64.7$	& $20.6$   & $74.9$   & $145.0$ \\
neopentane 	& $-86.5$	& $-131.3$	& $-166.2$ & $-121.8$ & $-78.3$ \\
n-butane	& $-116.3$	& $-152.5$	& $-182.6$ & $-145.5$ & $-109.8$ \\
adamantane 	& $-76.2$	& $-181.0$ 	& $-266.6$ & $-137.7$ & $36.5$ \\
1,3-butadyne	& $432.7$	& $435.6$	& $369.1$  & $432.9$  & $491.5$ \\
\end{longtable}
\end{center}

\newpage

\bibliographystyle{achemso2}
\bibliography{Literature}

\providecommand{\refin}[1]{\\ \textbf{Referenced in:} #1}

\end{document}